\documentclass{article}
\usepackage[utf8]{inputenc}
\usepackage{enumerate,amsmath,amsthm, amsfonts, float, amssymb,physics, graphicx, authblk, marvosym, hyperref, listings, setspace, placeins, caption, dsfont, setspace, authblk}
\usepackage[usenames, dvipsnames]{color}

\usepackage[left=2cm, right=2cm, top=2cm, headheight=18pt, headsep=4ex, includehead]{geometry}

\usepackage{fancyhdr}
\date{}

\author[1]{Taylor Okonek}
\author[1,2]{Jon Wakefield}
\affil[1]{Department of Biostatistics, University of Washington, 3980 15th Ave NE, Box 351617, Seattle, WA 98195, USA. Email: tokonek@uw.edu}
\affil[2]{Department of Statistics, University of Washington, Padelford Hall, NE Stevens Way, Seattle, WA 98195, USA. Email: jonno@uw.edu}

\title{A Computationally Efficient Approach to Fully Bayesian Benchmarking}

\usepackage[english]{babel}
\usepackage{csquotes}
\usepackage[authordate, backend=biber, bookpages=false, isbn=false, numbermonth=false, mincrossrefs=10, longcrossref=true, maxcitenames=2, cmsdate=both, sorting = nyt, urldate=long,  giveninits=true]{biblatex-chicago}
\DeclareFieldFormat{url}{Available at\addcolon\space\url{#1}}
\DeclareFieldFormat{doi}{DOI\addcolon\space https://doi.org/\url{#1}}

\newbibmacro*{begentry:fake}{%
  \savefield{addendum}{\savedaddendumfield}%
  \clearfield{addendum}}

\renewbibmacro*{bibindex}{%
  \ifbibindex
    {\indexnames{labelname}%
     \indexfield{indextitle}}
    {}%
  \usebibmacro{begentry:fake}}

\AtUsedriver{\letbibmacro{bibindex}{begentry:fake}}

\renewbibmacro*{finentry}{%
  \ifboolexpr{togl {blx@bibliography} or togl {cms@addendum}}
    {\restorefield{addendum}{\savedaddendumfield}%
     \printfield{addendum}}
    {}%
  \finentry}

\DefineBibliographyStrings{english}{%
  urlseen = {accessed},
}
\DeclareFieldFormat{urldate}{\printfield{url} \mkbibparens{\bibstring{urlseen}\space#1}}
\DeclareFieldFormat{url}{\space}

\usepackage{url}
\bibliography{manuscript.bib}

\begin{document}

\maketitle

\begin{abstract}
    In small area estimation, it is sometimes necessary to use model-based methods to produce estimates in areas with little or no data. In official statistics, we often require that some aggregate of small area estimates agree with a national estimate for internal consistency purposes. Enforcing this agreement is referred to as benchmarking, and while methods currently exist to perform benchmarking, few are ideal for applications with non-normal outcomes and benchmarks with uncertainty. Fully Bayesian benchmarking is a theoretically appealing approach insofar as we can obtain posterior distributions conditional on a benchmarking constraint. However, existing implementations may be computationally prohibitive. In this paper, we critically review benchmarking methods in the context of small area estimation in low- and middle-income countries with binary outcomes and uncertain benchmarks, and propose a novel approach in which an unbenchmarked method that produces area-level samples can be combined with a rejection sampler or Metropolis-Hastings algorithm to produce benchmarked posterior distributions in a computationally efficient way. To illustrate the flexibility and efficiency of our approach, we provide comparisons to an existing benchmarking approach in a simulation, and applications to HIV prevalence and under-5 mortality estimation. Code implementing our methodology is available in the R package \texttt{stbench}.\\
    
\end{abstract}


\newpage

\section{Introduction}

In a public health context, small area estimates, where small areas are defined as domains with little to no data, are often produced by official statistics agencies for the purpose of informing targeted public health interventions. In low- and middle-income countries (LMICs), the most reliable source of public health information are household surveys such as Demographic and Health Surveys (DHS) and Multiple Indicator Cluster Surveys (MICS), and as such, direct (weighted) survey estimates are considered the gold standard when they are precise enough to be practically useful. In small area estimation, where we have by definition little to no data in certain domains, it may not be possible to obtain direct, survey estimates with reasonable precision. In these settings, model-based methods are often used when estimates at small levels of aggregation are required \autocite{rao2015small}. Model-based methods that incorporate spatial random effects, in the absence of reliable covariate information, allow for estimates in all small areas to be produced with little to no data, introducing some bias into the estimates in exchange for tighter interval estimates \autocite{knorr2000bayesian, riebler2016intuitive, wakefield2020small}. Area-level models, such as the popular Fay-Herriot model \autocite{fay1979estimates}, consider data at the level of the small area, whereas unit-level models consider data at the level of the individual or cluster \autocite{battese1988error}. When area-level sample sizes are particularly small, unit-level models may be more desirable. For a review of both area-level and unit-level models, see Chapters 4-7 of \textcite{rao2015small}.

It is often required that small area estimates agree with estimates at a higher level of aggregation. For example, subnational estimates may be required to aggregate to a national level estimate. These higher level estimates are referred to as benchmarks, and are frequently considered more reliable than small area estimates since more data are available to inform them, or they are direct (weighted) estimates and therefore less dependent on model assumptions. Note that benchmarking as we consider it in this paper is distinct from calibration weighting \autocite{sarndal2003model, sizhou2020}. The latter involves incorporating known, population-level \textit{covariate} information into a model to adjust for potential bias in a subnational model, while the former involves incorporating known, population-level \textit{outcome} information into a model to enforce internal consistency in an official statistics setting.

Benchmarks may be from the same data source that was used to produce small area estimates or from an outside source, referred to as internal benchmarking and external benchmarking, respectively \autocite{bell2013benchmarking}. In an internal benchmarking setting, the use of the same data twice---once to calculate estimates at a higher level of aggregation, and once to calculate estimates at a subnational level---leads to an understating of statistical uncertainty. This is an important limitation of internal benchmarking, and all benchmarking approaches that rely on internal benchmarking must be used with caution. We note that the particular applications we consider in this paper are in an external benchmarking setting, and that this setting is relatively common in global health applications \autocite{osgood2018mapping, eaton2021naomi} as well as more recent applications in agriculture \autocite{chen2022hierarchical}. Internal benchmarking settings and their limitations are outside the scope of this paper. We direct the reader to \textcite{pfeffermann2006small} for one approach to assess uncertainty of internally benchmarked estimators.

Many existing benchmarking approaches treat benchmarking as a constraint problem, where estimates at smaller areas are constrained to agree with estimates at a higher level of aggregation. Approaches vary by the ways in which constraints have been incorporated into a modeling framework, and as such, the interpretation of resulting benchmarked estimates varies by approach as well. There are many ways to categorize benchmarking approaches, but one we may consider is a one-step versus two-step procedure. In a two-step procedure, estimates and uncertainty are first obtained from a model that is agnostic to the benchmarking constraint and are then adjusted to satisfy the benchmarking constraint. \textcite{datta2011bayesian, ghosh2013two, steorts2020smoothing, wang2008small, patra2018constrained, patra2019constrained, ghosh2015benchmarked, williams2013incorporating} and \textcite{berg2018benchmarked} all consider a two-step procedure where benchmarked estimates are calculated by minimizing posterior expected loss (for a given loss function) subject to a benchmarking constraint. Unbenchmarked estimates are first obtained, and then projected into a constrained space. 

Other two-step procedures include difference benchmarking and ratio benchmarking (henceforth referred to as raking), described in \textcite{erciulescu2019model}. Again, a model that is agnostic to the benchmarking constraint is first fit, and then estimates are adjusted by a constant so that the benchmarking constraint is satisfied \autocite{erciulescu2019model, erciulescu2018benchmarking, erciulescu2020statistical}. \textcite{you_rao_dick} consider a hierarchical Bayesian model for unbenchmarked estimates, but obtain benchmarked estimates using the raking method, and quantify uncertainty via posterior MSE as opposed to estimating full posterior distributions for benchmarked estimates.

Other benchmarking approaches incorporate the benchmarking constraint into the data likelihood---referred to as an augmented model in \textcite{bell2013benchmarking}, \textcite{berg2018benchmarked}, and \textcite{stefan2021small}---and thus produce automatically benchmarked estimates in a single modeling step. \textcite{you2002pseudo} also follow this approach, and refer to this as a ``self-benchmarking" property. 

Others \autocite{nandram2011bayesian, janicki2017benchmarking, erciulescu2019model, nandram2019bayesian, zhang2020fully} propose benchmarking approaches that produce full, benchmarked posterior distributions for small area estimates, making uncertainty quantification straightforward in an external benchmarking setting. Following \textcite{zhang2020fully}, we refer to such approaches as fully Bayesian benchmarking approaches. Note that we consider this type of approach to be distinct from fitting a Bayesian model and benchmarking only point estimates or variances. \textcite{nandram2011bayesian} develop a fully Bayesian benchmarking approach specifically for betabinomial models. \textcite{janicki2017benchmarking} perform fully Bayesian benchmarking by minimizing the KL divergence between a benchmarked and unbenchmarked posterior using moment constraints. \textcite{nandram2019bayesian} perform fully Bayesian benchmarking using a transformation of the benchmarking constraint that corresponds to “deleting” a single small area, but note that benchmarked estimates vary based on which area is deleted. \textcite{erciulescu2019model} consider four different benchmarking approaches, one of which is the Bayesian method described in \textcite{nandram2011bayesian}, and three of which involve fitting a hierarchical Bayesian model and benchmarking point estimates using ratio or difference benchmarking. \textcite{zhang2020fully} develop a fully Bayesian benchmarking approach that produces full posterior distributions conditional on either a soft or hard benchmarking constraint.

Benchmarking approaches have also been developed in a time series context, with certain similarities to small area estimation in requiring finer estimates in space/time to agree with an aggregate estimate \autocite{dagum2006benchmarking}. Different benchmarking approaches are more or less appealing than others---depending on context---with regards to obtaining measures of uncertainty, computational tractability, and the way in which the benchmarking constraint is enforced.

Benchmarking methods may also differ depending on whether benchmarking must be \textit{exact} or \textit{inexact}. In exact benchmarking, as the name suggests, the benchmarking constraint must hold exactly, whereas in inexact benchmarking the constraint must hold within some margin of error. The latter can be viewed as a soft constraint as opposed to a hard constraint. Exact benchmarking may be appropriate if the benchmarks are unbiased and have little to no uncertainty, which can occur when they come from a census \autocite{hillmer1987benchmarking, trabelsi1990bench}.

The estimation of under-5 mortality rates (U5MR) in LMICs at a subnational level motivates our desire for benchmarking. The UN Inter-agency Group for Child Mortality Estimation (IGME) produces annual, national level estimates of U5MR for all countries using a Bayesian B-spline bias-reduction (B3) method \autocite{alkema2014global}. Various data are used to produce B3 estimates, including vital registration, census, and household surveys, and many of these sources cannot be used for producing subnational estimates because geographic information is lacking, or the data type is not amenable to incorporate into a small area model. Subnational estimates of U5MR are of interest in addition to national level estimates, in accordance with the Sustainable Development Goals (https://sustainabledevelopment.un.org/post2015/transformingourworld). As the methods at a subnational level are model-based and incorporate spatio-temporal smoothing terms in order to produce precise estimates in each small area \textit{and} incorporate only data sources with geographic information, benchmarking is required to align subnational estimates with national estimates from the B3 model for internal consistency within the production of UN IGME estimates. Subnational estimates are currently produced for a handful of countries using a Beta-binomial model described in \textcite{wu2021}, but benchmarking approaches in this context have not yet been rigorously explored. As national estimates are model-based (with uncertainty) in this context, we aim to use a benchmarking approach that incorporates national level uncertainty (i.e.,~is \textit{inexact}). Additionally, the approach we use should be fast and computationally flexible, theoretically justified, and should produce benchmarked estimates that lie between 0 and 1 as our outcome is a proportion. Computational speed and flexibility are particularly important in an LMIC setting, as practitioners and national statistics offices in LMICs may not have the same computational resources as those in high-income countries. 

In this article, we present two novel implementations of the fully Bayesian benchmarking approach described in \textcite{zhang2020fully} that are more flexible and computationally tractable in many settings. Our approaches combine an unbenchmarked model with either a rejection sampler or Metropolis-Hastings algorithm to produce fully Bayesian benchmarked posteriors, which we describe in subsection \ref{sec:proposed_method}. We compare our method to that described in \textcite{zhang2020fully} in the setting of modeling HIV prevalence in South Africa, as well as modeling U5MR in Namibia. These applications were chosen to demonstrate the flexibility of the proposed fully Bayesian approaches in estimating outcomes that lie between 0 and 1 with unique benchmarking constraints, the efficiency of the method in cases where the benchmarks are very consistent or inconsistent with the small area estimates, and the flexibility of the approaches to handle both area-level and unit-level models. To emphasize the novelty of our approaches and computational advantages of the proposed method over the Markov chain Monte Carlo (MCMC) samplers used in \textcite{zhang2020fully} in terms of flexibility, we use integrated nested Laplace approximation (INLA) and Template Model Builder (TMB) as alternative ways to conduct Bayesian inference using Laplace approximations, that are fast and do not require users to code model-specific fitting routines \autocite{rue2009approximate, kristensen2016tmb}. We additionally compare run times for our proposed approaches to that of \textcite{zhang2020fully} in a simulation, and show that our proposed approaches provide not only increased flexibility in terms of modeling for Bayesian inference, but computational speed gains as well. All code for fitting the models described in this paper is available via the R package \texttt{stbench}, found at github.com/taylorokonek/stbench.

\section{Small area models in low- and middle-income countries}

Model-based small area estimation has a long history in LMICs, where we typically rely on nationally representative surveys to estimate public health outcomes as opposed to using censuses or vital registration systems. While small area estimates that incorporate the survey design directly are preferred, they are often impractical at a small area level, with either too little precision to be practically useful or no data available in some small areas \autocite{lehtonen2009design, wakefield2020small}. This lack of data primarily comes from a disconnect between the administrative level at which these surveys are designed to produce reliable estimates (often administrative level 1, or state) and the administrative level at which public health interventions are made (often administrative level 2, or county). Model-based methods with spatial smoothing terms allow us to obtain precise estimates in \textit{all} small areas, as required for public health estimates \autocite{datta2009model, wakefield2020small}. 

Many public health outcomes, including HIV prevalence and U5MR that we consider for our applications, are estimated from Bernoulli or binomial counts. As such, binomial models are a typical choice in an LMIC setting \autocite{eaton2021naomi, wu2021}.

Similarly to small area estimates, national estimates of public health outcomes typically come from nationally representative household surveys in LMICs as well. Practically, this means that national benchmarks are estimated with uncertainty, as both model-based approaches and approaches that directly incorporate the survey design are used to produce national estimates. With regard to benchmarking, this means that in a LMIC context, we are primarily concerned with \textit{inexact} benchmarking.

While our motivation for benchmarking primarily comes from subnational estimation of U5MR, the application to HIV prevalence is similarly important in a LMIC context, in that small area estimates of HIV prevalence are produced in an official statistics setting by UNAIDS and require benchmarking for internal consistency \autocite{eaton2021naomi}. Similarly to U5MR, subnational estimates of HIV prevalence inform public health interventions and allow countries to monitor their progress towards the Sustainable Development Goals. The theoretical properties, computational speed, and flexibility of our proposed approaches are relevant to HIV prevalence estimation in LMICs, as (similarly to U5MR) national estimates are produced with uncertainty, and benchmarked estimates should lie between 0 and 1.

\subsection{Small area models}

Below we describe the unbenchmarked models that we consider for the HIV application, but they are generally appropriate for a binary outcome. The unbenchmarked models for the U5MR application are described in Section 2.2 of the Supplemental data. 

If there are sufficient data in each target area, then weighted (direct) estimates are reliable. Such estimates include the Horvitz-Thompson (HT) \autocite{horvitz1952generalization} and H\'ajek estimators \autocite{hajek1971discussion}. We denote the weighted estimate $\hat{\theta}_i^{w}$ and its design-based variance $\hat{V}_i^{*,DES}$ for each area $i$. 

In situations where the direct estimates can be calculated but have unacceptably high design variance, one may smooth using a popular area-based model, which is often referred to as a Fay-Herriot model \autocite{fay1979estimates}. When the data are sparser still, the area-based models cannot be used, because $\hat{\theta}_i^{HT}$ and/or $\hat{V}_i^{*,DES}$ cannot be reliably calculated. In these cases, the raw data (counts, in the binary context) are modelled (in our HIV and U5MR examples, at the cluster level) to give a unit-level model.

\subsubsection{Area-level model: Spatial Fay-Herriot}
\label{sec:area-levelmodel}

Under the Fay-Herriot model, a working likelihood may be based on the distribution,
\begin{align*}
	y_{1i}^*= \text{logit}(\hat{\theta}^{HT}_i) \sim \text{N} (\eta_i, \hat{V}^{DES}_{i})
\end{align*}
where $\hat{\theta}_i^{HT}$ is the HT estimator for area $i$, $V_i^{*, DES}$ is the variance of $\hat{\theta}_i^{HT}$, and $\hat{V}^{DES}_i = V_i^{*, DES} / (\hat{\theta}^{2HT}_i(1 - \hat{\theta}^{2HT}_i))^2$ is obtained via the delta method. Treating $y_{1i}^*$ and $\hat{V}^{DES}_{i}$ as fixed, as in the Fay-Herriot model, we then consider the area-level model
\begin{align*}
	y_{1i}^* \mid \eta_{i} & \sim \text{N} (\eta_i, \hat{V}^{DES}_{i}) \\
	\eta_i & = \beta_0 + b_i
\end{align*}
where $\beta_0$ is an intercept term, and $b_i$ follows a BYM2 spatial model \autocite{riebler2016intuitive}, denoted $b_i \sim \text{BYM2}(\tau_{b}, \phi)$, where $\tau_b$ is the total precision and $\phi$ is the proportion of the variance that is spatial. The BYM2 model is a reparameterization of the BYM spatial random effect developed by \textcite{besag1991bayesian}, which consists of an unstructured (iid) random effect and a structured, ICAR spatial random effect \autocite{besag1974spatial}. A sum-to-zero constraint is placed on the structured component of the BYM2 random effect to ensure identifiability \autocite{besag1991bayesian}. The above model is often used in the estimation of U5MR in LMICs as it directly accounts for survey design and incorporates spatial smoothing random effects \autocite{li2019changes, wakefield2020small}, and is readily applicable to applications of HIV prevalence as well \autocite{wakefield2020small}. 

For priors, we set $\beta_0 \sim \text{N}(0, 0.001^{-1})$, $\phi \sim \text{Beta}(0.5, 0.5)$. We use penalized complexity (PC) priors for the precision $\tau_{b}$ with values $U = 1$ and $\alpha = 0.01$, corresponding to a prior belief that the probability $\sigma_b = 1/\sqrt{\tau_b}$ is greater than $1$ is $1$\% \autocite{simpson2017penalising}. Note that although the variance in the Fay-Herriot model is typically fixed, there is a growing literature where the variance is instead estimated \autocite{you2006small, liu2007hierarchical, sugasawa2017bayesian, gao2022spatial}.

\subsubsection{Unit-level model: Binomial}
\label{sec:unit-levelmodel}

We may assume binomial observations of the outcome for each cluster $c$ within area $i$. Let $y_{1i[c]}$ be the number of cases observed in $n_{i[c]}$ total people in cluster $c$ within area $i$. We consider the unit-level model
\begin{align*}
	y_{1i[c]}  \mid n_{i[c]}, \theta_{i[c]} & \sim \text{Binomial}(n_{i[c]}, \theta_{i[c]}), \\
	\eta_{i[c]} = \text{logit}(\theta_{i[c]}) & = \beta_0 + b_i + e_{i[c]},
\end{align*} 
where $\theta_{i[c]}$ is prevalence in cluster $c$ of area $i$, $\beta_0$ is an intercept term, and $b_i$ again follows a BYM2 model \autocite{riebler2016intuitive}, denoted $b_i \sim \text{BYM2}(\tau_{b}, \phi)$, and $e_i[c] \overset{iid}{\sim} \text{N}(0, \sigma^2_e)$ is an iid cluster-level random effect. Constraints and priors are the same as for the area-level model, with the addition of a PC prior on $\sigma_e$ with values $U = 1$ and $\alpha = 0.01$.

Area-level predictions $\theta_i$ are obtained by marginalizing over the cluster random effect to get
\begin{align*}
	\theta_i & = \text{expit} \left( \frac{\beta_0 + b_i}{\sqrt{1 + h^2 \sigma^2_e}} \right),
\end{align*}
where $h = \frac{16 \sqrt{3}}{15 \pi}$. The cluster-level random effect is excluded from predictions, and a correction is done using $h$ and $\sigma^2_e$, as described in \textcite{dong2021modeling} and detailed in Section 9.13.1 and Exercise 9.1 of \textcite{wakefield2013bayesian}. The correction accounts for within-cluster variation that induces cluster-level overdispersion. If we instead believed the cluster-level random effect reflected true \textit{between}-cluster differences, we could obtain predictions such as those obtained in \textcite{dong2021modeling}.

\section{Methods}
\label{sec:Methods}

Let $y_2$ be a national level estimate of the outcome of interest, possibly with uncertainty given by a confidence interval or standard error, $\boldsymbol{\hat{\theta}} = (\hat{\theta}_1, \dots, \hat{\theta}_n)^\top$ be small area estimates for areas $i = 1, \dots, n$, and $w_i$ be population size weights that do not depend on $\boldsymbol{\theta}$, standardized so that $\sum_{i = 1}^n w_i = 1$. Often, these weights are set to $w_i = N_i / \sum_{j = 1}^n N_j$, where $N_j$ is the population size for the $n$ small areas. Note that the population sizes used should correspond to the population under study. For example, when estimating HIV prevalence using survey data, $N_j$ should be population counts for individuals within the age range in the sampling frame. For DHS surveys, this consists of individuals aged 15-49. To our knowledge, incorporating uncertainty from the population counts into benchmarking approaches has not yet been considered in the benchmarking literature. 

Throughout the paper, we consider benchmarking constraints of the most basic form, $\sum_{i = 1}^n w_i \hat{\theta}_i = y_2$. While more complex benchmarking constraints, or even multiple benchmarking constraints as noted in \textcite{zhang2020fully}, may be useful in certain settings, this simple equality constraint is reasonable for the HIV prevalence and U5MR applications we consider, and is the constraint currently used in practice by both UNAIDS and the UN IGME \autocite{eaton2021naomi, wu2021}. For examples of benchmarking applications with inequality constraints, see \textcite{steorts2020smoothing} or \textcite{nandram2022bayesian}.


In the following subsections we describe a subset of existing approaches to benchmarking and propose two novel approaches to fully Bayesian benchmarking. The methods we describe in detail were chosen either because they are commonly used benchmarking methods in an official statistics setting, or because they are particularly relevant to our motivating application. One method is not \textit{inherently} preferable, though we argue that some are more appropriate than others in the context of estimating outcomes between 0 and 1 in LMICs. 

\subsection{Benchmarked Bayes Estimate Approach}
\label{subsec:bayesestapproach}

The first approach we describe was developed in a Bayesian, decision theoretic framework \autocite{datta2011bayesian, steorts2020smoothing}. This involves minimizing expected posterior MSE loss subject to the benchmarking constraint, which results in a projection of the unbenchmarked estimates into a benchmarked (constrained) space. Since it's development, others have extended this decision theoretic approach with different loss functions \autocite{ghosh2015benchmarked, williams2013incorporating, berg2018benchmarked}. Methods have also recently been developed to obtain benchmarked uncertainty for these estimates under this decision theoretic framework \autocite{patra2018constrained, patra2019constrained}. As it involves minimizing expected posterior loss subject to a benchmarking consraint, we call this approach the Benchmarked Bayes Estimate approach. Though we will argue that this approach is not appropriate for our motivating application, we detail it here as it is theoretically appealing, fast, and commonly used in many benchmarking applications. 

For $n$ small areas, let $\hat{\boldsymbol{\theta}} = (\hat{\theta}_1, \dots, \hat{\theta}_n)^\top$ be the direct estimators of the small area means $\boldsymbol{\theta} = (\theta_1, \dots, \theta_n)^\top$. We are interested in computing the benchmarked Bayes estimator $\hat{\boldsymbol{\theta}}^{BM} = (\hat{\theta}_1^{BM}, \dots, \hat{\theta}_n^{BM})^\top$ of $\boldsymbol{\theta}$ such that the constraint $\sum_{i = 1}^n w_i \hat{\theta}_i^{BM} = y_2$ is satisfied. 

As \textcite{datta2011bayesian} are interested in an estimate of the benchmarked posterior mean, they consider minimizing the posterior expectation of the weighted squared error loss $\sum_{i = 1}^n \phi_i E[(\theta_i - e_i)^2 \mid \textbf{y}_1]$ under the constraint $\bar{e}_w := \sum_{i= 1}^n w_i e_i = y_2$, where $\phi_i$ are weights not necessarily equal to $w_i$, and $\textbf{y}_1$ is a vector of area-level observations. They note that the weights $\phi_i$ could be different for different policy makers, and a simple default is setting $\phi_i = 1$ for all $i = 1, \dots, n$. The resulting benchmarked Bayes estimate (solution to the constrained minimization of the posterior expectation of the weighted squared error loss) is 

\begin{equation}
	\label{eq:benchedmean}
	\hat{\boldsymbol{\theta}}^{BM} = \hat{\boldsymbol{\theta}}^B + s^{-1} (y_2 - \bar{\hat{\theta}}^B_w) \textbf{r},
\end{equation}
where $\hat{\boldsymbol{\theta}}^B = (\hat{\theta}_1^B, \dots, \hat{\theta}_n^B)^\top$  is a vector of unbenchmarked posterior means $E[\theta_i \mid \textbf{y}_1]$ under a given prior, $\bar{\hat{\theta}}^B_w = \sum_{i = 1}^n w_i \hat{\theta}_i^B$, $\textbf{r} = (r_1, \dots, r_n)^\top$, $r_i = w_i / \phi_i$, and $s = \sum_{i = 1}^n w_i^2 / \phi_i$. Note that if $\phi_i = 1$ for all $i = 1, \dots, n$, the benchmarked Bayes estimate becomes $\hat{\boldsymbol{\theta}}^{BM} = \hat{\boldsymbol{\theta}}^B + (\textbf{w}^\top \textbf{w})^{-1} (y_2 - \bar{\hat{\theta}}^B_w) \textbf{w}$. Thus from Equation (\ref{eq:benchedmean}) we can see that the benchmarked Bayes estimate is a function of the unbenchmarked Bayes estimate under mean squared error (MSE) loss (posterior means) and user-specified weights. This is computationally appealing, as obtaining benchmarked estimates with this method is done via a quick post-processing step. 

To obtain uncertainty around the benchmarked estimates, posterior samples can be projected into the space defined by the benchmarking constraint \autocite{patra2018constrained, patra2019constrained}. Geometrically, we can interpret this benchmarked Bayes estimate as the point estimate within the space defined by the benchmarking constraint that is as close to the unbenchmarked Bayes estimate as possible, where closeness is measured in terms of expected weighted squared error \autocite{steorts2020smoothing}. 


The benchmarked estimate from Equation (\ref{eq:benchedmean}) is an exactly benchmarked estimate, i.e., the benchmarking constraint holds exactly. Alternatively, in inexact benchmarking, the benchmarking constraint need not hold exactly. If inexact benchmarking is desired in the benchmarked Bayes estimate framework, a penalty term $\lambda > 0$ (pre-specified by the user) is introduced, and a slightly different loss function is considered,
\begin{align*}
	L(\boldsymbol{\theta}, \textbf{e}) = \lambda(y_2 - \bar{e}_w)^2 + \sum_{i = 1}^n \phi_i (\theta_i - e_i)^2.
\end{align*}
The Bayes estimate associated with this loss is 
\begin{align*}
	\hat{\boldsymbol{\theta}}^B_\lambda = \hat{\boldsymbol{\theta}}^B + (s + \lambda^{-1})^{-1} (y_2 - \bar{\hat{\theta}}^B_w) \textbf{r},
\end{align*}
where we note that as $\lambda \to \infty$, $\hat{\boldsymbol{\theta}}^B_\lambda$ approaches the exactly benchmarked Bayes estimate in Equation (\ref{eq:benchedmean}). Considering a loss function with the addition of a penalty term for the benchmarking constraint allows the user to incorporate a predetermined level of agreement between the benchmarks and aggregated, unbenchmarked model estimates that must be met, whether that be exact benchmarking (corresponding to $\lambda \to \infty$) or inexact, where the resulting estimate $\hat{\boldsymbol{\theta}}^B_\lambda$ is a compromise between the exactly benchmarked and unbenchmarked Bayes estimate.

In the context of U5MR and HIV prevalence, the outcome of interest lies between 0 and 1, and may be close to 0. Any estimate that falls outside of $[0,1]$ would be invalid. In applications where the outcome of interest lies in a restricted space, the benchmarked Bayes estimate approach described above can return invalid benchmarked estimates. In particular, note that the benchmarked Bayes estimates may possibly lie below zero when $\bar{\hat{\theta}}_w^B > y_2$. \textcite{ghosh2015benchmarked} note this issue and consider a variant of the Kullback-Leibler loss function rather than weighted MSE loss, which addresses the issue of estimates falling below $0$ in the case of positive small area estimates. \textcite{williams2013incorporating} and \textcite{berg2018benchmarked} also note this issue, and the latter propose using a specific form for the weights $\phi_i$ in Equation (1) to deal with unbenchmarked estimates that are close to the boundary. Their approach does not guarantee, however, that benchmarked estimates will lie within the required restricted space. While they note that in many situations benchmarked estimates that lie outside the restricted space will be rare---and in such cases their approach may be sufficient---when estimating rare disease prevalence or mortality this boundary issue is a concern.

The benchmarked Bayes estimate approach is fast and theoretically justified in settings where estimates fall on the real line, are positive as in \textcite{ghosh2015benchmarked}, or lie well within the boundary of $[0,1]$. However, with the loss functions considered thus far in the literature, the approach will often fall short for targets that are on a restricted range.

\subsection{Raking approach}
\label{subsec:rakingapproach}

The second benchmarking approach we consider is simple and popular: raking, also referred to as the ratio-adjustment method \autocite{datta2011bayesian, zhang2020fully, ghosh2015benchmarked}. A version of the raking approach is used by the Institute for Health Metrics and Evaluation (IHME) in a variety of applications (e.g., \textcite{osgood2018mapping, local2021mapping}), in the Naomi model for estimating HIV prevalence and incidence \autocite{eaton2021naomi}, and in the subnational U5MR estimates currently produced by UN IGME \autocite{wu2021}. This approach is commonly applied post-hoc. The key feature of the raking approach is a ratio comparing an unbenchmarked national estimate to the national level benchmark 
\begin{align}
	\label{eq:benchratio}
	R = \frac{\hat{\theta}^N}{y_2},
\end{align}
where $\hat{\theta}^N$ denotes some unbenchmarked national level estimate, and $y_2$ again denotes the national level benchmark. For the unbenchmarked national estimate, one could plug in the weighted sum of unbenchmarked small area estimates as $\hat{\theta}^N = \sum_{i = 1}^n w_i \hat{\theta}_i^M$, where $w_i$ again denote population count weights and must satisfy $\sum_{i= 1}^n w_i = 1$, and $\hat{\theta}_i^M$ denote the posterior estimates (means or medians) in each area from our unbenchmarked model. 

In a post-hoc raking approach and with a sampling-based method, the posterior draws of $\hat{\theta}_i$ from an unbenchmarked model are multiplied by $1/R$ so that the constraint $\sum_{i = 1}^n w_i \hat{\theta}_i^M = y_2$ holds. Of note, this benchmarking approach will treat the unbenchmarked estimates in every area in the same fashion, regardless of the uncertainty in the unbenchmarked estimates. As such, the ranking of regions based on posterior medians/means will be preserved between unbenchmarked and benchmarked estimates. This behavior follows because of the ad hoc nature of the raking adjustment, as noted by \textcite{datta2011bayesian}. It may be preferable to instead treat unbenchmarked estimates with more uncertainty differently than those with less.  

The raking approach to benchmarking can also be approximated via the inclusion of a log offset term for the ratio comparing an unbenchmarked national estimate to the national level benchmark in logistic models when the outcome of interest is rare, as in under-5 mortality estimation. In the supplement of \textcite{wakefield2019estimating} they show that, for rare outcomes, including a log offset for $R$ in a logistic regression model corresponds approximately to the same multiplicative bias adjustment that would be done in the post-hoc raking approach. This approach is currently used in the UN IGME's subnational U5MR estimates. We note that the form of raking involving the inclusion of a log offset for $R$ does in fact produce fully Bayesian estimates, in the sense that a full posterior distribution for the benchmarked estimates is produced. However, the approach differs from the fully Bayesian approach described in Section \ref{subsec:fullybayesapproach} in that a likelihood is not specified for the benchmarks themselves.

\subsection{Fully Bayesian benchmarking approach}
\label{subsec:fullybayesapproach}

We define the fully Bayesian approach to benchmarking, in the same way as \textcite{zhang2020fully}, to be an approach that provides full posterior distributions for benchmarked estimates. Of note, \textcite{nandram2011bayesian} also consider a fully Bayesian benchmarking approach for data with binomial outcomes specific to beta-binomial models, but their approach is not directly generalizable to other models, and relies on Gibbs sampling for obtaining posterior estimates which may be computationally prohibitive. \textcite{janicki2017benchmarking} perform fully Bayesian benchmarking by minimizing the KL divergence between a benchmarked and unbenchmarked posterior using moment constraints, though they note that their approach may be computationally intractable for non-normal outcomes. \textcite{nandram2019bayesian} perform fully Bayesian benchmarking using a transformation of the benchmarking constraint that corresponds to ``deleting" a single small area, but note that benchmarked estimates vary based on which area is deleted. For a more complete review of the existing fully Bayesian benchmarking approaches listed above, see Section 3.2 of \textcite{zhang2020fully}. In the following, we describe only the fully Bayesian approach of \textcite{zhang2020fully} as it is general to a wide class of models, allows for nonlinear constraints, and as noted by \textcite{zhang2020fully}, has fewer implementation limitations than previous fully Bayesian approaches.

Consider an area-level, Bayesian hierarchical model for small area estimation. For $n$ small areas, let the area-level parameters we wish to estimate be denoted by $\boldsymbol{\theta} = (\theta_1, \dots, \theta_n)^\top$, with a hierarchical structure specified through a model $\pi(\boldsymbol{\theta} \mid \boldsymbol{\phi})$ with a vector of hyperparameters $\boldsymbol{\phi}$. The area-level observations are denoted by $\textbf{y}_1 = (y_{11}, \dots, y_{1n})^\top$. For example, the values $y_{1i}$ may be binomial counts of HIV status, with the parameters $\theta_i$ corresponding to HIV prevalence in area $i$. We can write the posterior distribution as
\begin{align*}
	\pi(\boldsymbol{\theta}, \boldsymbol{\phi} \mid \textbf{y}_1) \propto \pi(\textbf{y}_1 \mid \boldsymbol{\theta}, \boldsymbol{\phi}) \pi(\boldsymbol{\theta} \mid \boldsymbol{\phi}) \pi(\boldsymbol{\phi}).
\end{align*} 
Though \textcite{zhang2020fully} consider more complex forms of benchmarking constraints, for simplicity we again consider the constraint $\sum_{i = 1}^n w_{i} \theta_i = y_2$, where $w_i$ and $y_2$ are defined previously. The method we describe is generally applicable to a variety of benchmarking constraints, however. 

To incorporate the benchmarking constraint into their hierarchical model, \textcite{zhang2020fully} define an additional likelihood term for the benchmarks, $\pi(y_2 \mid \boldsymbol{\theta})$. This results in the benchmarked posterior distribution 
\begin{align}
	\label{eq:fullybayespost}
	\pi(\boldsymbol{\theta}, \boldsymbol{\phi} \mid \textbf{y}_1, y_2) \propto \pi(\textbf{y}_1 \mid \boldsymbol{\theta}, \boldsymbol{\phi}) \pi(y_2 \mid \boldsymbol{\theta}) \pi(\boldsymbol{\theta} \mid \boldsymbol{\phi}) \pi(\boldsymbol{\phi}),
\end{align}
with the assumption that $\textbf{y}_1$ and $y_2$ are conditionally independent given $\boldsymbol{\theta}$. Notably, this assumption does not hold in internal benchmarking settings. The likelihood term for the benchmarks, $\pi(y_2 \mid \boldsymbol{\theta})$, pulls the likelihood for the area-level observations, $\pi(\textbf{y}_1 \mid \boldsymbol{\theta}, \boldsymbol{\phi})$, towards the benchmarks. Uncertainty quantification is thus straightforward, as we obtain an entire benchmarked posterior distribution as opposed to simply a benchmarked point estimate.  

Under exact benchmarking, we set $\pi(y_2 \mid \boldsymbol{\theta}) = \text{I}\left[ \sum_{i = 1}^n w_{i} \theta_i = y_2 \right]$. Under inexact benchmarking, $\pi(y_2 \mid \boldsymbol{\theta})$ is a non-degenerate distribution specified by the user. One can incorporate a discrepancy parameter $\lambda$ into this distribution to allow for varying levels of agreement between the benchmarks and the aggregate parameters if desired. For example, in the application we consider setting $\pi(y_2 \mid \boldsymbol{\theta})$ equal to a normal distribution with mean $\sum_{i = 1}^n w_{i} \theta_i$ and variance equal to the variance of the national benchmark. This would be equivalent to setting $1/\lambda$ equal to the variance of the national estimate in the Bayes estimate approach described in Section \ref{subsec:bayesestapproach}. 

A benefit of this benchmarking approach is that it allows for nonlinear benchmarking constraints. This is particularly relevant for logistic models, where the benchmarking constraint may take the form $\sum_{i = 1}^n w_i \theta_i = \sum_{i = 1}^n w_i \text{expit}(\eta_i)$ for a linear predictor $\eta_i$. Unlike the estimates from a benchmarked Bayes estimate approach, the fully Bayesian benchmarking approach ensures that benchmarked estimates remain within the boundaries of the parameter space.

The fully Bayesian benchmarking approach can be used with both area-level and unit-level models. Though \textcite{zhang2020fully} describe an extension of their fully Bayesian approach to unit-level models, an implementation of this does not currently exist. \textcite{zhang2020fully} provide code in their R package \texttt{demest} for Poisson, binomial, multinomial, and normal models, with optional point mass, Poisson, binomial, and normal distributions for the likelihood for the benchmark, found at \href{https://github.com/StatisticsNZ/demest}{github.com/StatisticsNZ/demest}. They provide an outline for a MCMC scheme for a general model. However, implementation of this MCMC scheme will be model-specific, which can limit the uptake of the method. Statistical programs such as INLA and TMB provide alternative ways to conduct Bayesian inference which have great computational advantages over MCMC samplers. These computational advantages are described in detail for INLA in \textcite{rue2009approximate} and for TMB in \textcite{kristensen2016tmb}, and involve the use of Laplace approximations to obtain posterior distributions at a fraction of the time it would take using MCMC algorithms. The methods are particularly suited to space and space-time modeling with Markov random field models---situations in which MCMC can be especially computationally demanding because of the dependence in the posterior \autocite{margossian2020hamiltonian}.

The benchmarking approach we propose in the following section is a more general implementation of the inexact, fully Bayesian approach described by \textcite{zhang2020fully}, and allows us to obtain fully Bayesian benchmarked estimates from any method that produces area-level samples from an unbenchmarked model. The approach we propose is readily applicable to both area- and unit-level models, so long as area-level samples can be produced, and can be used in conjunction with statistical programs such as INLA and TMB.

\subsection{Proposed approach} 
\label{sec:proposed_method}

We propose an external benchmarking approach that combines an unbenchmarked model with either a rejection sampler or Metropolis-Hastings algorithm to produce fully Benchmarked posterior distributions conditional on a benchmarking constraint under \textit{inexact} benchmarking. The key to our proposed approach is that from Equation (\ref{eq:fullybayespost}) we can write
\begin{align*}
	\pi(\boldsymbol{\theta}, \boldsymbol{\phi} \mid \textbf{y}_1, y_2) & \propto \pi(\textbf{y}_1 \mid \boldsymbol{\theta}, \boldsymbol{\phi}) \pi(y_2 \mid \boldsymbol{\theta}) \pi(\boldsymbol{\theta} \mid \boldsymbol{\phi}) \pi(\boldsymbol{\phi}), \\
	& \propto \pi(\boldsymbol{\theta} , \boldsymbol{\phi} \mid \textbf{y}_1)  \pi(y_2 \mid \boldsymbol{\theta}).
\end{align*}
Intuitively, we can think of $\pi(y_2 \mid \boldsymbol{\theta})$ as a likelihood for the benchmarks and $\pi(\boldsymbol{\theta}, \boldsymbol{\phi} \mid \textbf{y}_1)$ as a ``prior" that corresponds to the posterior based on area-level observations $\textbf{y}_1$. For concreteness we consider a normal distribution for $\pi(y_2 \mid \boldsymbol{\theta})$, i.e.,
$$
y_2 \mid \boldsymbol{\theta} \sim N \left( \sum_{i = 1}^n w_i \theta_i, \sigma^2_{y_2} \right),
$$ 
where $w_i$ are population weights that sum to one across all regions, and $\sigma^2_{y_2}$ is treated as the known, national level variance for $y_2$. One could consider other distributions for the benchmark likelihood, however we focus on the normal case in the following derivations.

Importantly, $\pi(\boldsymbol{\theta} , \boldsymbol{\phi} \mid \textbf{y}_1)$ is the posterior distribution from an unbenchmarked model, i.e., a model that is agnostic to the benchmarking constraint. In practice, this is the small area, subnational model fit without consideration of the national benchmarks. The most appropriate unbenchmarked subnational model will depend on the context of the statistical problem, though we note that if unbenchmarked subnational estimates are particularly far from reliable national benchmarks, this may indicate that the unbenchmarked model is inappropriate for the data. The approach we describe below will produce samples from the benchmarked posterior conditional on the benchmarking constraint for a given unbenchmarked model.  

We note that though we target the same constrained posterior distribution as \textcite{zhang2020fully}, our approach is distinct in the implementation in that we do not obtain samples from this posterior distribution in a single step, such as with an MCMC algorithm as implemented in \textcite{zhang2020fully}. Rather, our approach allows for first obtaining samples from an unbenchmarked model (not necessarily using MCMC, hence greater flexibility and potential speed gains) and benchmarking in a second step involving either a rejection sampler or a Metropolis-Hastings algorithm.

\subsubsection{Rejection sampler}
\label{subsubsec: rejectsamp}

In a rejection sampling framework, we can obtain samples from $\pi(\boldsymbol{\theta}, \boldsymbol{\phi} \mid \textbf{y}_1, y_2)$ by filtering samples from $\pi(\boldsymbol{\theta} , \boldsymbol{\phi} \mid \textbf{y}_1)$ through the information provided by $\pi(y_2 \mid \boldsymbol{\theta})$. This filtering is done via the following rejection sampler \autocite{smith1992bayesian}:
\begin{enumerate}
	\item Generate $U \sim \text{Uniform}(0,1)$ and $(\boldsymbol{\theta}, \boldsymbol{\phi}) \sim \pi(\boldsymbol{\theta}, \boldsymbol{\phi} \mid \textbf{y}_1)$ independently.
	\item Accept $(\boldsymbol{\theta}, \boldsymbol{\phi})$ if
	$$
	U < \frac{\pi(y_2 \mid \boldsymbol{\theta})}{\sup_{\boldsymbol{\theta}} \pi(y_2 \mid \boldsymbol{\theta})} = \exp(-\frac{1}{2\sigma^2_{y_2}} \left(\sum_{i = 1}^m w_i \theta_i - y_2 \right)^2),
	$$
	Otherwise, return to Step 1.
\end{enumerate}
This rejection sampling approach targets the same constrained posterior distribution as the fully Bayesian benchmarking approach of \textcite{zhang2020fully} but in a computationally straightforward manner. As the rejection sampling approach only requires posterior draws from an unbenchmarked posterior distribution, this allows practitioners to use a wider array of computation tools to conduct fully Bayesian benchmarking, even when benchmarking constraints are nonlinear. A relevant example of such a computational tool is INLA, which does not allow for non-linear predictors, and therefore the likelihood corresponding to the benchmark cannot be directly incorporated into the model \autocite{rue2009approximate}. Crucially, posterior samples can be generated from an INLA analysis.

\subsubsection{Metropolis-Hastings}
\label{subsubsec:mh}

The rejection sampling approach to fully Bayesian benchmarking may be inefficient in cases where benchmarks are far from the population aggregated estimates from an unbenchmarked model. One potential way to address this inefficiency is to instead use an independence Metropolis-Hastings (MH) algorithm \autocite{tierney1994markov}. Intuitively, we can fit an \textit{adjusted} unbenchmarked distribution that has been shifted towards the national benchmarks and adjust for this shift in the acceptance rate. This serves to increase the proportion of accepted samples relative to using the unbenchmarked distribution as a proposal distribution by moving the proposal distribution closer to where the constrained posterior should be. 

Here we present one example of an adjusted unbenchmarked distribution that can be used as a proposal distribution, and show how the shift can be corrected for in the acceptance rate. Note that we assume in this example that the unadjusted, unbenchmarked model we would want to fit has a fixed intercept $\beta$ with a flat prior, $\pi(\beta) \propto 1$. This assumption is needed for terms to cancel in the acceptance rate, and is not an unreasonable assumption given that the models fit in our motivating application typically satisfy this assumption \autocite{wu2021}. We note that this assumption can be relaxed if other proposal distributions are considered, which may improve the acceptance rate, but that the acceptance probability may not cancel as conveniently in these cases. 

As in subsection \ref{subsec:fullybayesapproach}, for $n$ small areas, let the area-level parameters we wish to estimate be denoted by $\boldsymbol{\theta} = (\theta_1, \dots, \theta_n)^\top$, with a hierarchical structure specified through a model $\pi(\boldsymbol{\theta} \mid \boldsymbol{\phi})$ with a vector of hyperparameters $\boldsymbol{\phi}$. The area-level observations are denoted by $\textbf{y}_1 = (y_{11}, \dots, y_{1n})^\top$, and a likelihood term for the benchmark $y_2$ is given by $\pi(y_2 \mid \boldsymbol{\theta})$. Suppose that the area-level parameters are linked to a mean model $g(\boldsymbol{\theta}) = \beta + \boldsymbol{\eta}$ via a link function $g$, where $\beta$ is a fixed intercept term with $\pi(\beta) \propto 1$, and $\boldsymbol{\eta}$ denoting a summation of any remaining fixed or random effect terms in the unbenchmarked model (e.g.,~spatio-temporal smoothing terms, covariates, etc.). Let $\pi^+(\beta) \sim N(g(y_2), \sigma^2_+)$ denote an alternative prior for the intercept, centered at $g(y_2)$ with fixed variance $\sigma^2_+$. Then we have
\begin{align*}
	\pi(\boldsymbol{\theta}, \boldsymbol{\phi} \mid \textbf{y}_1, y_2) & \propto \pi(\textbf{y}_1 \mid \boldsymbol{\eta}, \beta, \boldsymbol{\phi}) \pi(y_2 \mid \boldsymbol{\eta}, \beta) \pi(\boldsymbol{\eta} \mid \boldsymbol{\phi}) \pi(\boldsymbol{\phi}) \pi(\beta), \\
	& = \pi(\textbf{y}_1 \mid \boldsymbol{\eta}, \beta, \boldsymbol{\phi}) \pi(y_2 \mid \boldsymbol{\eta}, \beta) \pi(\boldsymbol{\eta} \mid \boldsymbol{\phi}) \pi(\boldsymbol{\phi}) \pi(\beta) \left( \frac{\pi^+(\beta)}{\pi^+(\beta)} \right), \\
	& \propto \pi^+(\boldsymbol{\theta}, \boldsymbol{\phi} \mid \textbf{y}_1) \left( \frac{\pi(y_2 \mid \boldsymbol{\eta}, \beta)}{\pi^+(\beta)} \right),
\end{align*}
where $\pi^+(\boldsymbol{\theta}, \boldsymbol{\phi} \mid \textbf{y}_1)$ is an adjusted unbenchmarked distribution with $\pi^+(\beta)$ specified as the prior for the intercept.

In this framework, we can obtain samples from the same benchmarked posterior distribution $\pi(\boldsymbol{\theta}, \boldsymbol{\phi} \mid \textbf{y}_1, y_2)$ using independent samples from the adjusted unbenchmarked posterior distribution $\pi^+(\boldsymbol{\theta}, \boldsymbol{\phi} \mid \textbf{y}_1)$. The algorithm is executed as follows:

\begin{enumerate}
	\item Initalize $(\beta^0, \boldsymbol{\eta}^0)$.
	\item Sample $(\beta', \boldsymbol{\eta}') \sim \pi^+(\boldsymbol{\theta}, \boldsymbol{\phi} \mid \textbf{y}_1)$.
	\item Compute the acceptance probability 
	\begin{align*}
		A & = \text{min} \left( 1, \frac{\pi(\boldsymbol{\theta}', \boldsymbol{\phi}' \mid \textbf{y}_1, y_2) \pi^+(\boldsymbol{\theta}^0, \boldsymbol{\phi}^0 \mid \textbf{y}_1)}{\pi(\boldsymbol{\theta}^0, \boldsymbol{\phi}^0 \mid \textbf{y}_1, y_2) \pi^+(\boldsymbol{\theta}', \boldsymbol{\phi}' \mid \textbf{y}_1)}\right) \\
		& = \text{min} \left( 1, \frac{\pi(y_2 \mid \boldsymbol{\eta}', \beta') \pi^+(\beta^0)}{\pi(y_2 \mid \boldsymbol{\eta}^0 , \beta^0) \pi^+(\beta')} \right) 
	\end{align*}
	\item With probability $A$, accept the proposed value $(\beta', \boldsymbol{\eta}')$ and set $\beta^0, \boldsymbol{\eta}^0) = \beta', \boldsymbol{\eta}')$. Otherwise, set $(\beta^0, \boldsymbol{\eta}^0) = (\beta^0, \boldsymbol{\eta}^0)$.
\end{enumerate}

\noindent Just as with the rejection sampling approach, the MH approach targets the same constrained posterior distribution as the fully Bayesian benchmarking approach of \textcite{zhang2020fully} but in a computationally convenient manner. 

\subsubsection{Convergence diagnostics}
\label{subsubsec: convergence}

While the MH algorithm may have a higher acceptance rate than the rejection sampling approach (and therefore greater computational speed), additional care must be taken to ensure that the algorithm has mixed and converged properly. Though not a convergence diagnostic, one basic check is to aim for an acceptance rate near 23.4\%, as suggested in \textcite{gelman1997weak} as the optimal acceptance rate under normal posteriors for random walk Metropolis algorithms. It may be desirable to vary the prior $\pi^+(\beta)$ in the MH algorithm to obtain an acceptance rate close to this optimal value.

A common convergence diagnostic for MCMC sampling is the potential scale reduction factor $\hat{R}$, introduced by \textcite{gelman1992inference}. Intuitively, $\hat{R}$ (or split-$\hat{R}$ as it is extended in \textcite{gelman2013bayesian}) compares the variance of each individual Markov chain to the variance of all of the chains combined. If the sampler has mixed appropriately, $\hat{R}$ should be close to 1. More recently, new convergence diagnostics have been developed for MCMC methods that go beyond $\hat{R}$, as there are certain situations where $\hat{R}$ can fail to correctly diagnose poor mixing. In brief, \textcite{vehtari2021rank} introduce rank-normalized split-$\hat{R}$ and bulk effective sample size (bulk-ESS) as measures that have good asymptotic efficiency through avoiding a normality assumption in non-rank-normalized measures. For more detail see \textcite{vehtari2021rank} for an overview of existing convergence diagnostics and details on those they propose.

Following the guidelines in \textcite{vehtari2021rank}, we recommend running at least four chains if using the MH approach to benchmarking that we propose. Aiming for a rank-normalized split-$\hat{R}$ less than 1.01 and a bulk-ESS that exceeds 400 (with 4, 1000-iteration chains after warmup) provide default guidelines.

\subsubsection{Limitations}
\label{subsubsec: limitations}

Despite the potential improvements in speed from the MH approach over the rejection sampler approach, both methods may still be inefficient in cases where benchmarks are far from the population aggregated estimates from an unbenchmarked model. It may be necessary in such cases to use an approach such as \textcite{zhang2020fully}'s in order to produce benchmarked estimates. However, we caution that if the proportion of accepted samples is very small this may be indicative of inconsistencies between the two data sources, and model elaboration may be required in this case. 

Additionally, the rejection sampling and MH approaches cannot benchmark estimates with zero variance, as the likelihood $\pi(y_2 \mid \boldsymbol{\theta})$ would be a point mass. We note this benchmarking scenario could be approximated by setting $\sigma^2_{y_2}$ to be very small, but the rejection sampling approach would likely be very inefficient due to the likelihood for the benchmark being very concentrated. The scenario where benchmarks have positive uncertainty is the most practically relevant in our context, as situations where the national benchmark does not have some degree of uncertainty in an LMIC context is rare. For fully Bayesian benchmarking to benchmarks with zero variance, we refer the reader to the MCMC methods described in \textcite{zhang2020fully}.

A final consideration with the MH approach is that using this algorithm with the suggested proposal distribution will not allow for comparison between the resulting benchmarked estimates and the unbenchmarked estimates from the unadjusted, unbenchmarked model, without separately fitting an unadjusted, unbenchmarked model. In some official statistics settings it may be desirable or even required to compare benchmarked estimates to unbenchmarked estimates, and if the time it takes to separately fit an unadjusted, unbenchmarked model is time-consuming, the MH approach may be impractical.

\section{Simulation}
\label{sec: simulation}

To demonstrate the improved computational speed of our approach compared to \textcite{zhang2020fully}, we compare run-times for our approach used with INLA to that of \textcite{zhang2020fully} implemented in STAN; a probabilistic programming language that uses a variant of Hamiltonian Monte Carlo to do full Bayesian inference \autocite{carpenter2017stan}. We chose not to compare the computational speed of our approach to that of \textcite{datta2011bayesian} or the raking approach, as both of those benchmarking approaches target different benchmarked estimates than our proposed method. Note however that both the raking and benchmarked Bayes estimate approach will be the fastest approach to benchmarking in general, as they involve a very quick step to adjusting unbenchmarked draws/estimates, and do not rely on acceptance rates (as our methods do to achieve a reasonable number of effective samples) or MCMC methods.

We simulate unit-level (cluster) binomial observations, using the nine provinces of South Africa as our spatial structure. For each simulation setting, binomial probabilities were given by $p_i = 0.28, 0.29, \dots, 0.35, 0.36$, with $100$ binomial trials in each cluster. Equal probability weights were given to each province. We varied: the number of samples taken in each area, $n = \{5, 10, 100, 1000\}$, the national-level benchmark, $y_2 = \{0.29, 0.3\}$, and the variance of the national-level benchmark, $\sigma^2_{y_2} = \{0.01, 0.0001\}$. For each simulation setting, we generated 10 unique datasets using the given parameters.

The unbenchmarked model we fit to the generated data is the unit-level model used in our application to South Africa (and described in Section \ref{sec:unit-levelmodel}), where we have binomial observations for clusters $c$ within area $i$. Let $y_{1i[c]}$ be the number of cases observed in $n_{i[c]}$ total observations in cluster $c$ within area $i$. We consider the unbenchmarked, unit-level model
\begin{align*}
	y_{1i[c]}  \mid n_{i[c]}, \theta_{i[c]} & \sim \text{Binomial}(n_{i[c]}, \theta_{i[c]}), \\
	\eta_{i[c]} = \text{logit}(\theta_{i[c]}) & = \beta_0 + b_i + e_{i[c]}
\end{align*} 
where $\theta_{i[c]}$ is case prevalence in area $i$ and cluster $c$, $\beta_0$ is an intercept term, $b_i$ follows a BYM2 model \autocite{riebler2016intuitive}, denoted $b_i \sim \text{BYM2}(\tau_{b}, \phi)$, and $e_i[c] \overset{iid}{\sim} \text{N}(0, \sigma^2_e)$ is an independent and identically distributed (iid) cluster-level random effect. Note that although the model we fit was not used to generate the data, as we are interested only in comparing run times from each method, and ensuring that the benchmarked distributions are the same from each method, this does not affect the validity of our simulation results. More details on hyperprior specification and simulating the data can be found in Section 3 of the Supplemental data.

The modified unbenchmarked model, used for the MH algorithm is the same as above with the exception of the prior for the intercept being $\pi^+(\beta_0) \overset{d}{=} \text{N}(\text{logit}(y_2), \sqrt{0.1})$, where $y_2$ is the national-level benchmark. The benchmarked model we fit using the \textcite{zhang2020fully} approach includes the additional likelihood
$$
y_2 \mid \boldsymbol{\theta}, \sigma^2_{y_2} \sim \text{N} \left( \sum_{i = 1}^n w_i \theta_i, \sigma^2_{y_2} \right),
$$
and the proposed rejection sampler and MH algorithm approaches are carried out as described in subsections \ref{subsubsec: rejectsamp} and \ref{subsubsec:mh}. Code for reproducing the simulation can be found at github.com/taylorokonek/benchmarking-paper-sim.

To fairly compare run-times, we compare the time it takes to:
\begin{enumerate}
	\item Fit the fully Benchmarked model per \textcite{zhang2020fully} in STAN, and obtain a bulk-ESS of $1000$.
	\item Fit the unbenchmarked model in INLA, draw posterior samples, and obtain $1000$ accepted samples using the rejection sampler.
	\item Fit the modified unbenchmarked model in INLA, draw posterior samples, and obtain a bulk-ESS of $1000$ using the MH algorithm.
\end{enumerate}
For both STAN and the MH algorithm, we run four chains with $1000$ burn-in samples each, and the appropriate number of samples after to obtain the desired bulk-ESS. 

\begin{figure}[H]
	\centering
	\includegraphics[scale = 0.4]{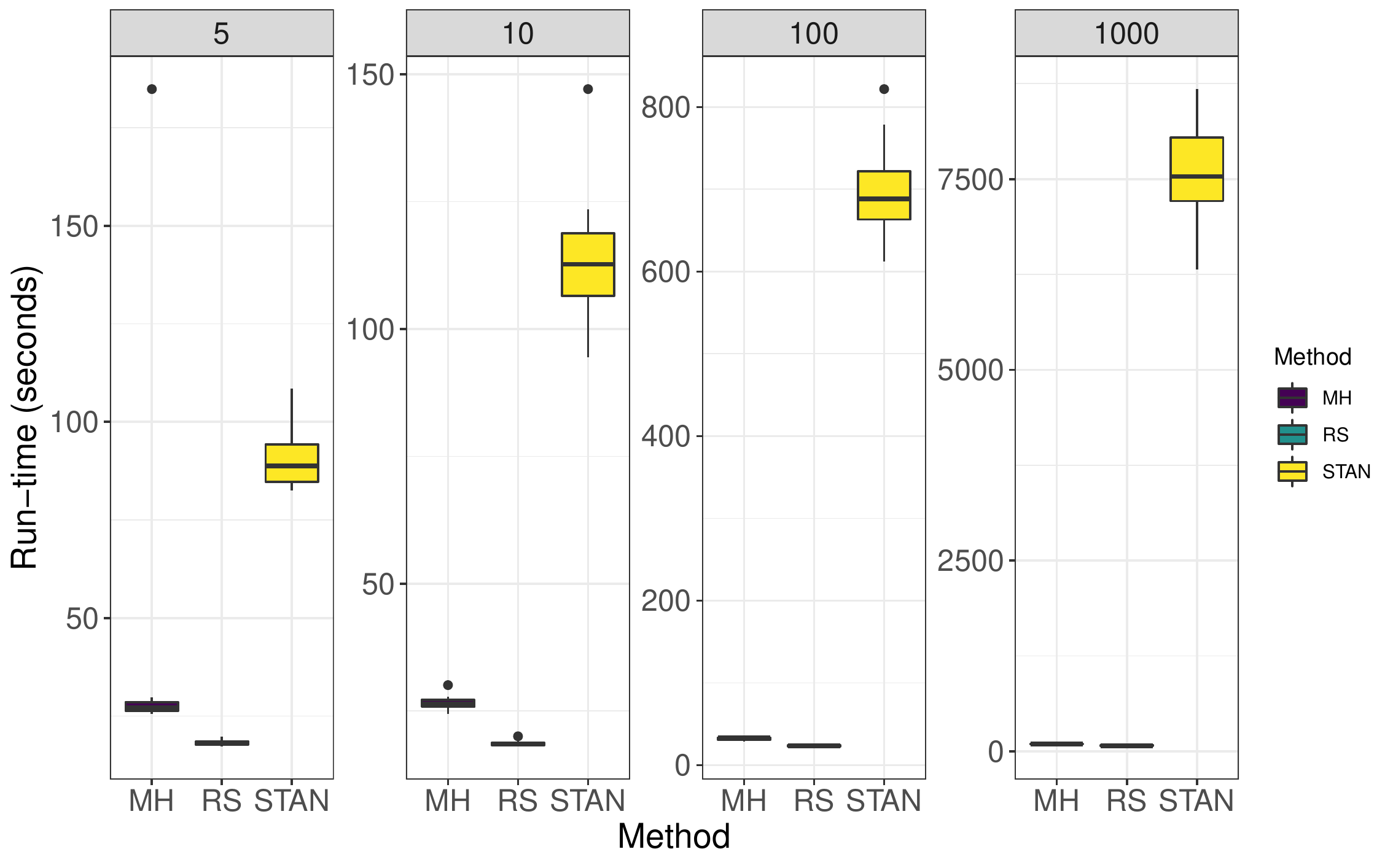}
	\caption{Setting: $y_2 = 0.29$, $\sigma^2_{y_2} = 0.01$}  Comparative, total run-time (in seconds) needed to obtain $1000$ samples from the rejection sampling (RS) approach, or $1000$ bulk-ESS from the Metropolis-Hastings (MH) approach or the approach of \textcite{zhang2020fully} (STAN). Each boxplot contains $10$ observations, with data generated under the given simulation setting for $10$ different seeds. 
	\label{fig:runtimes1}
\end{figure}

\begin{figure}[H]
	\centering
	\includegraphics[scale = 0.4]{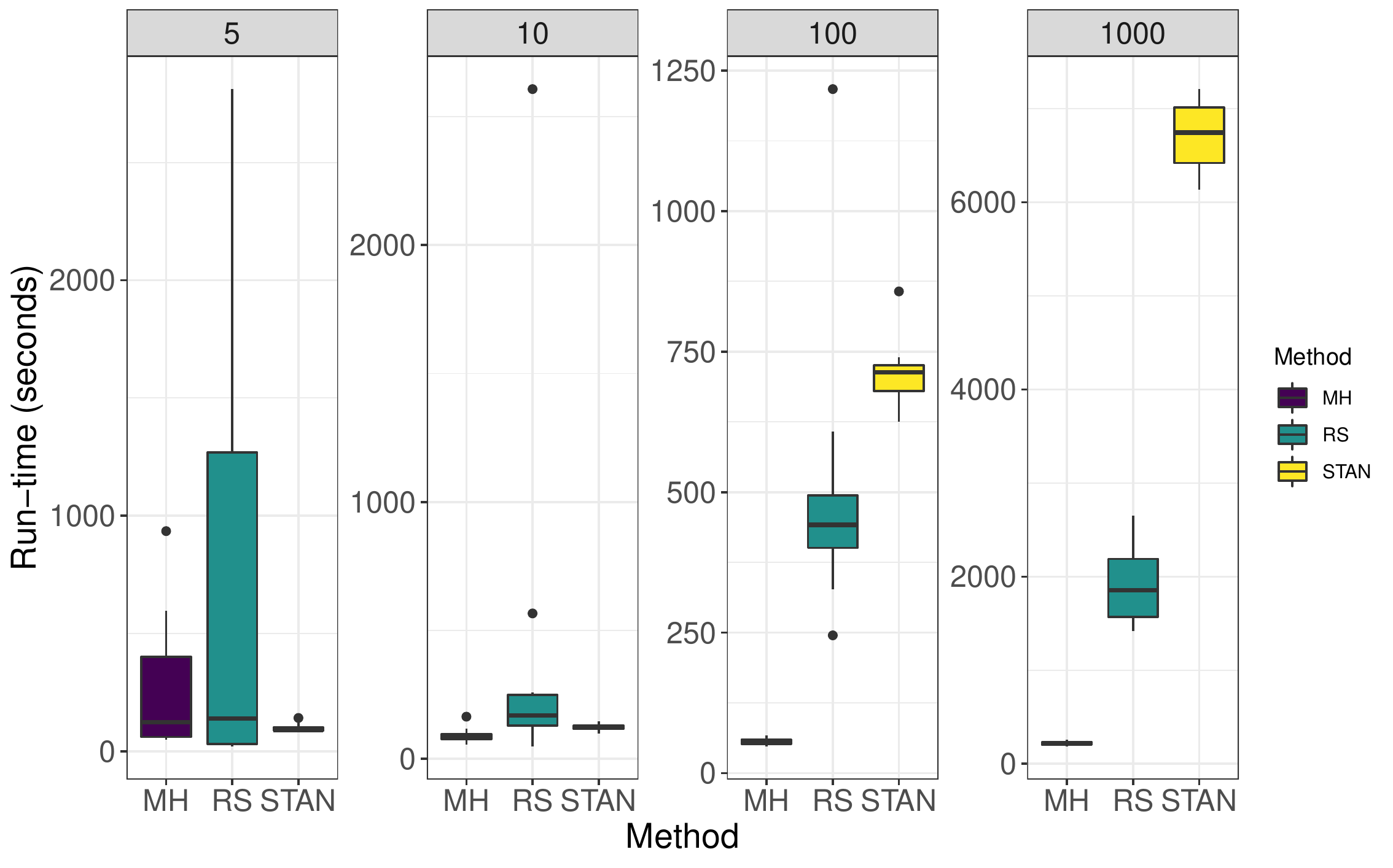}
	\caption{ Setting: $y_2 = 0.29$, $\sigma^2_{y_2} = 0.0001$}  Comparative, total run-time (in seconds) needed to obtain $1000$ samples from the rejection sampling (RS) approach, or $1000$ bulk-ESS from the Metropolis-Hastings (MH) approach or the approach of \textcite{zhang2020fully} (STAN). Each boxplot contains $10$ observations, with data generated under the given simulation setting for $10$ different seeds.
	\label{fig:runtimes2}
\end{figure}

In Figure \ref{fig:runtimes1} we compare methods under the setting $\{y_2 = 0.29, \sigma^2_y = 0.01\}$, and in Figure \ref{fig:runtimes2} we compare methods under the setting $\{y_2 = 0.29, \sigma^2_y = 0.0001\}$. As expected, our proposed approaches outperform that of \textcite{zhang2020fully} in most simulation settings, as noted in Figures \ref{fig:runtimes1} and \ref{fig:runtimes2}. In particular, the amount of time needed to obtain $1000$ samples from the benchmarked posterior distribution is much lower for both of our proposed approaches than that of \textcite{zhang2020fully} when the number of samples in each area is large. When the national variance is smaller, as in Figure \ref{fig:runtimes2}, the \textcite{zhang2020fully} tends to outperform both the rejection sampler and MH algorithm at low sample sizes ($5$ and $10$ per area), but not at larger sample sizes. Finally, note that, in settings with smaller national variance, the Metropolis-Hastings algorithm outperforms the rejection sampler approach in terms of computational speed, while the reverse is true when national variance is large. Under the setting $\{y_2 = 0.29, \sigma^2_y = 0.01, n = 5\}$, our proposed methods are 3-5 times faster than that of \textcite{zhang2020fully} on average, and under the setting $\{y_2 = 0.29, \sigma^2_y = 0.01, n = 1000\}$, our proposed methods are 75-100 times faster that that of \textcite{zhang2020fully} on average. The speed of the MH algorithm could potentially be optimized by modifying prior for the intercept, $\pi^+(\beta_0)$, and we suggest that when possible, multiple priors should be tested if the acceptance rate in the MH algorithm is lower than desired. The included figures and above summary are representative of the simulation results, and additional results from the simulation can be found in Section 3 of the Supplemental data.

\section{Application}
\label{sec:application}

Below we describe the data, models, and software used in the HIV prevalence application. The application to U5MR is in Section 2 of the Supplemental Data. 

We fit both area-level and unit-level models to demonstrate the flexibility of our proposed method over existing computational software. As noted in subsection \ref{subsec:fullybayesapproach}, there is currently no available implementation of the fully Bayesian benchmarking approach described in \textcite{zhang2020fully} to unit-level models. As unit-level models are commonly used in LMICs when estimating public health outcomes in an official statistics setting \autocite{wakefield2020small, wu2021}, it is important for practitioners to have a method available to conduct fully Bayesian benchmarking with unit-level models without needing to write their own MCMC algorithm. We demonstrate that our proposed method is readily applicable to both area- and unit-level models. 

For our unbenchmarked area-level model, we consider the spatial Fay-Herriot model defined in \textcite{mercer2015space}. The model is detailed in Section \ref{sec:area-levelmodel}. The Fay-Herriot model is one of the most well-known small area models, that directly incorporates aspects of the survey design using a transformed survey-weighted estimate and an iid random effect to increase precision of the resulting estimates \autocite{fay1979estimates}. The spatial extension incorporates both an iid random effect \textit{and} an additional strucutred spatial random effect to further capture spatial dependency in the observations \autocite{mercer2015space}. 

For our unbenchmarked unit-level model, we consider a hierarchical Bayesian model with a spatial random effect term for the application to HIV prevalence, and the model currently used to produce subnational estimates of U5MR for the UN IGME \autocite{wu2021}. The model is detailed in Section \ref{sec:unit-levelmodel}. Of note, these unit-level models do not directly account for the survey design like the area-level model. It has been suggested that, when possible, the inclusion of covariates related to the survey design could be included in such a model-based framework to account for the survey design \autocite{wakefield2013bayesian, wu2021}. One example of this can be found in \textcite{paige2022design}, though as noted in \textcite{wakefield2020small}, in many surveys conducted in LMICs the covariates corresponding to the survey design may be unavailable. 

\subsection{Data}

Spatial boundary files for South Africa are obtained from GADM, the Database of Global Administrative Areas \autocite{GADM}. 

We estimate HIV prevalence in South Africa from the 2016 South Africa DHS survey. The survey followed a multi-stage, stratified design, and was designed to provide estimates at the Administrative 1 (admin1) level, which consists of nine provinces, which we considered to be our small areas for this application. These nine provinces were stratified by urban/farm/traditional area status, and therefore resulted in 26 strata, as the Western Cape province does not have a traditional area geotype. The sampling frame was established from the 2011 census, and 750 enumeration areas (primary sampling units, PSUs) were selected across strata. The second stage of sampling sampled dwelling units, or households, from the enumeration areas, and every individual within the household (if available) was included in the survey. Only men and women aged 15-49 were included in the HIV dataset. Households within a given enumeration area are given a single geographic location, and we denote these as clusters from here on. GPS coordinates are displaced by up to 2km for urban clusters and 5km for rural clusters, but are never displaced outside of their area of stratification. 

\begin{figure}[H]
	\centering
	\includegraphics[scale = 0.4]{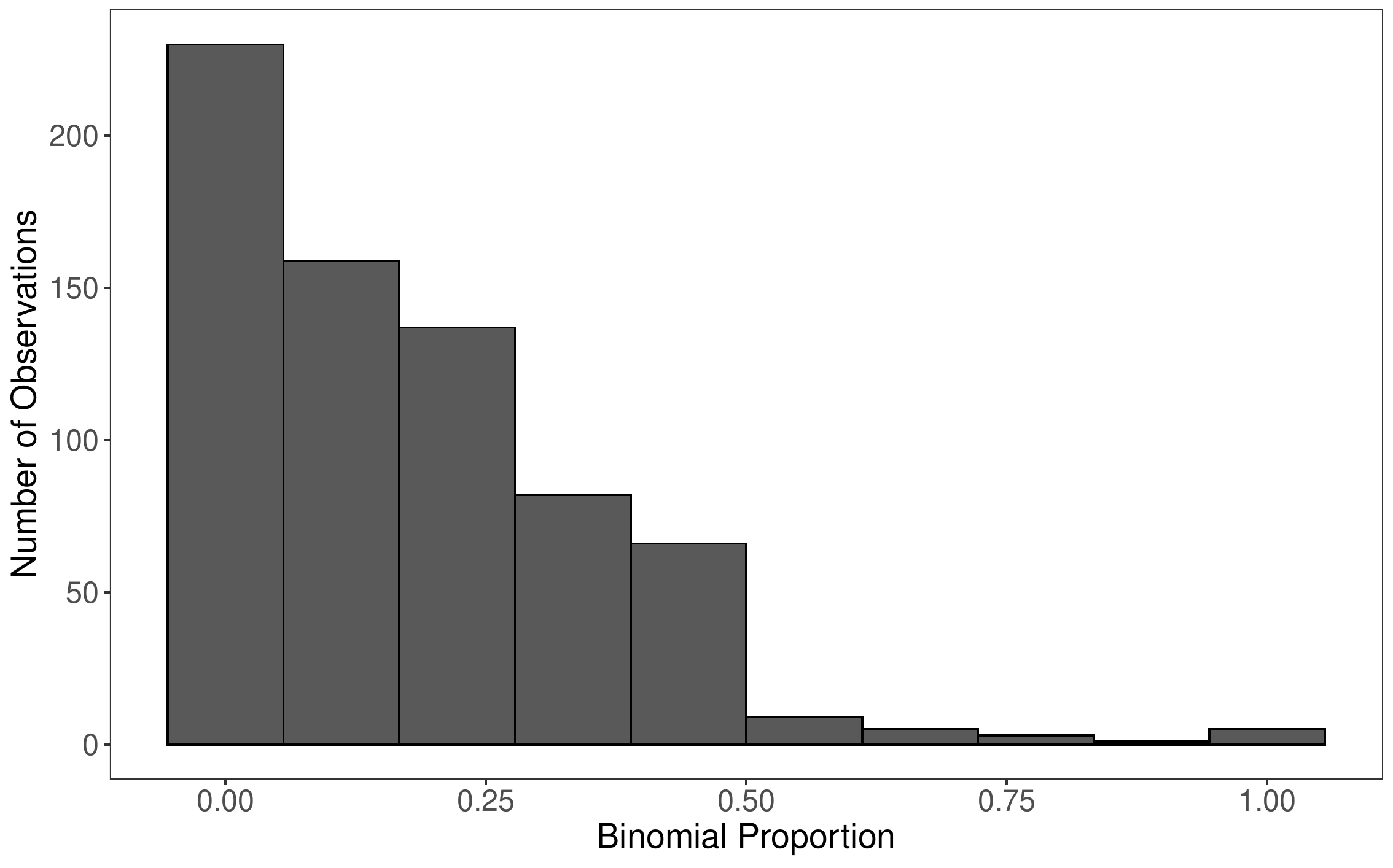}
	\caption{Observed binomial proportions in each PSU.}
	\label{fig:descrip}
\end{figure}

Of note, all nine small areas (provinces) used in model fit for this application contained at least one binomial observation. The number of PSUs in each small area ranged from 56 to 88. The binomial counts within each primary sampling unit ranged from 0 to 11, with totals in each primary sampling unit ranging from 1 to 30. The distribution of observed binomial proportions in each PSU can be seen in Figure \ref{fig:descrip}.

We obtain a national level estimate for 2016 in South Africa from the national Thembisa model, an HIV epidemic projection model that produces the official estimates published by UNAIDS for South Africa \autocite{johnson2017estimating, mahy2019hiv, stover2019updates}. The national level estimate of HIV prevalence for 2016 in South Africa is $17.1$\%, with a 95\% confidence interval of ($15.6$\%, $18.3$\%). For the benchmark likelihood for the fully Bayesian benchmarking approaches we need a standard error for this benchmarked estimate, which we take to be 0.61 based on the assumption that the national level estimate is asymptotically normally distributed. The Thembisa model incorporates data from five Human Sciences Research Council (HSRC) surveys conducted from 2002 to 2017, the 2016 DHS survey, Antenatal Sentinel HIV and Syphilis Surveys, and antiretroviral therapy coverage data. Although the DHS survey is included in the model for the benchmarks, we treat this as an external benchmarking scenario as multiple other data sources are included in the national model as well.

\begin{figure}[H]
	\centering
	\includegraphics[scale = 0.5]{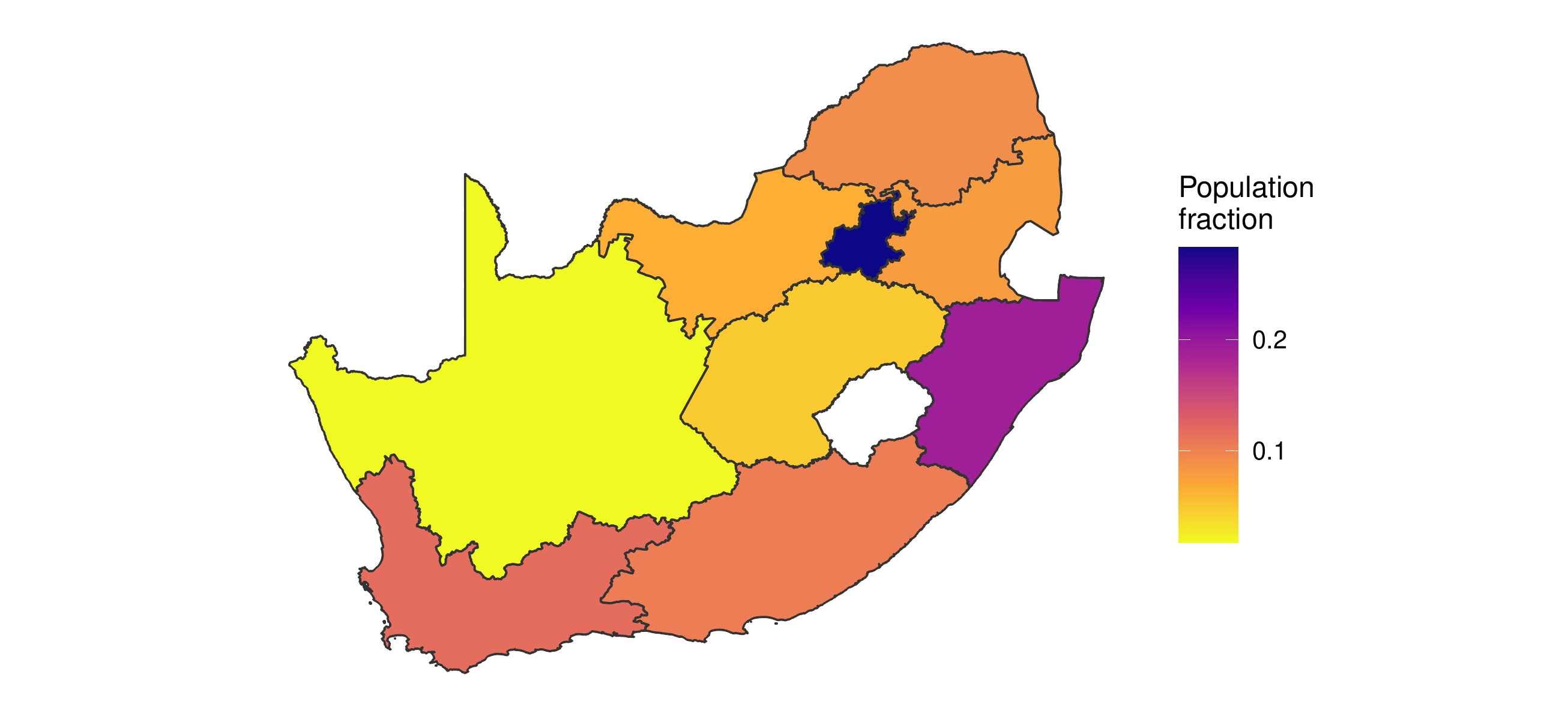}
	\caption{Proportion of the 15-49 population living in each province. These proportions are the weights used to aggregate province level estimates to the national level.}
	\label{fig:SApop}
\end{figure}

The population count data we use for our HIV application comes from the provincial Thembisa model \autocite{johnson2017progress}. We use this specific population data as they are currently used in UNAIDS models of subnational HIV prevalence in South Africa \autocite{eaton2021naomi}. In Figure \ref{fig:SApop}, we plot the spatial distribution of individuals aged 15-49 (the population in our HIV example) in 2016 in South Africa, i.e., out of all people aged 15-49 in South Africa, the proportion of that sub-population who live in each region is displayed.  

\subsection{Benchmarked models}

We compute benchmarked estimates using the Fully Bayesian approach of \textcite{zhang2020fully}, the proposed approaches, and the raking approach. This allows us to show: (1) the benchmarked estimates from our proposed approaches are identical to those using the \textcite{zhang2020fully} method, and (2) a comparison of the benchmarked estimates from the proposed approach to what is currently used in practice (raking) by the UN when estimating both HIV Prevalence and U5MR subnationally \autocite{eaton2021naomi, wu2021}. We do not compute benchmarked estimates using the benchmarked Bayes estimate approach, as we have noted in Section \ref{subsec:bayesestapproach} that is inappropriate for these applications. We obtain credible intervals for our benchmarked and unbenchmarked estimates using draws from the posterior distributions.

\subsubsection{Approaches}



\noindent \textbf{Raking:}

\noindent We obtain posterior medians $\hat{\theta}_i^M$ from an unbenchmarked model for each area $i$, and compute $R$ from Equation (\ref{eq:benchratio}), using weights $w_i$ and our benchmark $y_2$, setting $\hat{\theta}^N := \sum_{i = 1}^n w_i \hat{\theta}_i^M$. We then adjust unbenchmarked posterior draws $\hat{\theta}_i^{(k)}$, $k = 1, \dots, K$, to obtain benchmarked posterior draws $\hat{\theta}_i^{R(k)} = \hat{\theta}_i^{(k)} / R$. \\

\noindent \textbf{Fully Bayesian: \textcite{zhang2020fully}}

\noindent Using weights $w_i$ and our benchmark $y_2$ with associated standard error $\sigma^2_{y_2}$, we fit the unbenchmarked model with the additional likelihood 
\begin{align*}
	y_2 \mid \boldsymbol{\theta} & \sim \text{Normal} \left( \sum_{i = 1}^m w_i \theta_i, \sigma^2_{y_2} \right), \\
\end{align*}
and denote benchmarked posterior draws by $\hat{\theta}_{i}^{FB1(k)}$. \\

\noindent \textbf{Fully Bayesian: Rejection sampler:}

\noindent We obtain $k = 1, \dots, K$ posterior draws $\hat{\theta}_{i}^{(k)}$ from an unbenchmarked model for each area $i$, and apply the algorithm described in Section \ref{subsubsec: rejectsamp} using weights $w_i$, and the benchmark $y_2$ with associated standard error $\sigma^2_{y_2}$, to obtain benchmarked posterior draws $\hat{\theta}_i^{FB2(k)}$. \\

\noindent \textbf{Fully Bayesian: Metropolis-Hastings algorithm:}

\noindent We obtain $k = 1, \dots, K$ posterior draws $\hat{\theta}_{i}^{(k)}$ from an \textit{adjusted} unbenchmarked model for each area $i$, and apply the algorithm described in Section \ref{subsubsec:mh} using weights $w_i$, and the benchmark $y_2$ with associated standard error $\sigma^2_{y_2}$, to obtain benchmarked posterior draws $\hat{\theta}_i^{FB3(k)}$. The adjusted unbenchmarked model uses the prior $\pi^+(\beta_0) \sim \text{Normal}(\text{logit}(y_2), 0.001^{-1}).$

\subsection{Model Validation}

As we view benchmarking as a necessary adjustment needed to an existing unbenchmarked model in an official statistics setting, we choose to validate the unbenchmarked models we fit \textit{prior} to benchmarking. Note that if benchmarking is instead viewed as a way to reduce potential bias in subnational estimates, model validation may need to be done to the resulting benchmarked models as opposed to the unbenchmarked models. 

\textcite{wakefield2020small}, among others \autocite{wu2021, osgood2018mapping}, suggest that one way to approach model validation in small area models in a survey setting is to check whether the direct estimates in each area lie within an uncertainty interval surrounding the posterior means of the predictive distribution in that area. In a survey setting, the direct estimates are considered the gold standard; hence, model-based estimates should not stray too far from the direct estimates. This also gives one an idea of average coverage across areas: 80\% of the direct estimates (left out data) should lie within 80\% credible intervals based on the predictive distribution. 

We take this approach to model validation for both our unbenchmarked area-level and unit-level models. Details and figures containing the results of model validation for the HIV application can be found in Section 1.4 of the Supplemental data, and Section 2.6 of the Supplemental data for the U5MR application.

\subsection{Computation}

As the novelty of our approach is computational, we chose the statistical programmes used in the application to demonstrate the flexibility of our approach compared to the fully Bayesian approach of \textcite{zhang2020fully}. 

For the approach described in \textcite{zhang2020fully}, we implement models in Template Model Builder (TMB) \autocite{kristensen2016tmb}. TMB is a flexible modeling tool that takes advantage of Laplace approximations for computational efficiency, and allows users to specify a wide variety of models in \texttt{C++}. For a review of TMB, see Section 4 of \textcite{osgood2021statistical}. Importantly, TMB allows users to specify nonlinear predictors, unlike INLA, which is needed for implementing the fully Bayesian benchmarking approach of \textcite{zhang2020fully} in settings with binary outcomes. Of note, TMB still requires model-specific implementations, and therefore does not have as nice of an interface as other programs (such as INLA, which we will describe below).

To show that the results from our proposed method are identical to those from the \textcite{zhang2020fully} method, we also fit the unbenchmarked HIV models in our application using TMB. This ensures that the posterior distributions are directly comparable. Not that our proposed method could have been implemented using INLA to fit the unbenchmarked model as well, but as INLA uses a different Laplace approximation than TMB, the results would not have been exactly, directly comparable.

For our U5MR application, we fit all unbenchmarked models in INLA to show that our proposed method is compatible with the software currently used to produce official estimates of U5MR for the UN IGME. INLA is an appealing, alternative Laplace approximation programme for obtaining posterior distributions for latent Gaussian models \autocite{rue2009approximate}. Unlike TMB, INLA does not allow users to specify nonlinear predictors. Since our proposed approach does not require the use of such nonlinear predictors for the benchmarking constraint by taking advantage of a two-step procedure, we implement the models for our U5MR application using INLA to demonstrate the benefit of the increased flexibility of our approach to accommodate fast Laplace approximation techniques such as INLA.

\subsection{Results}

\begin{figure}[H]
	\centering
	\includegraphics[scale = 0.6]{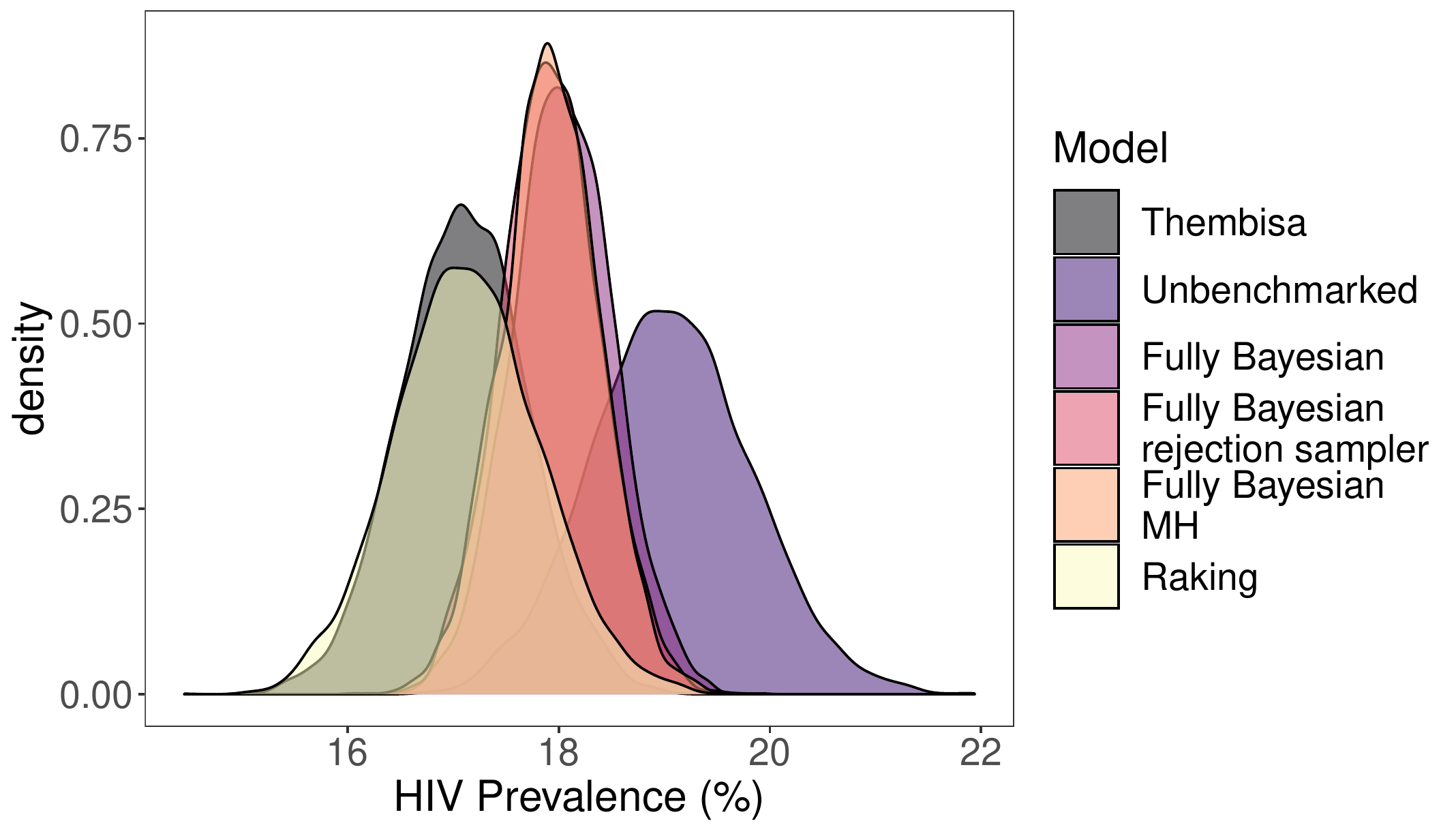}
	\caption{Aggregated national level HIV prevalence estimates from Thembisa, unbenchmarked, and benchmarked results from the unit-level model. The Fully Bayesian model refers to the approach described in \textcite{zhang2020fully}, and the Fully Bayesian rejection sampler and Fully Bayesian MH (Metropolis-Hastings) refer to the proposed approach. All densities are based on 5000 samples.}
	\label{fig:hivnatlresults}
\end{figure}

In Figure \ref{fig:hivnatlresults}, we display national level estimates from Thembisa, the unbenchmarked, and benchmarked results from the unit-level model. Results from the area-level model are very similar to those from the unit-level model, and can be found in Section 1.2 of the Supplemental Data, with direct comparisons between the unit-level and area-level models found in Section 1.3 of the Supplemental Data. We see that the Fully Bayesian benchmarking approaches produce essentially identical results, and that they are a compromise between the likelihood given by the national level benchmark and the unbenchmarked posterior estimate. The raking approach enforces exact benchmarking, which is evidenced by the overlap between the density of the Thembisa national estimate and that of the benchmarked method's national aggregated estimate. The fully Bayesian rejection sampler approach accepts 7.6\% of unbenchmarked samples using the unit-level model, and 8.6\% of unbenchmarked samples using the area-level model. The fully Bayesian Metropolis-Hastings approach accepts 18.8\% of unbenchmarked samples using the unit-level model, and 18.9\% of unbenchmarked samples using the area-level model. 

\begin{figure}[H]
	\centering
	\includegraphics[scale = 0.65]{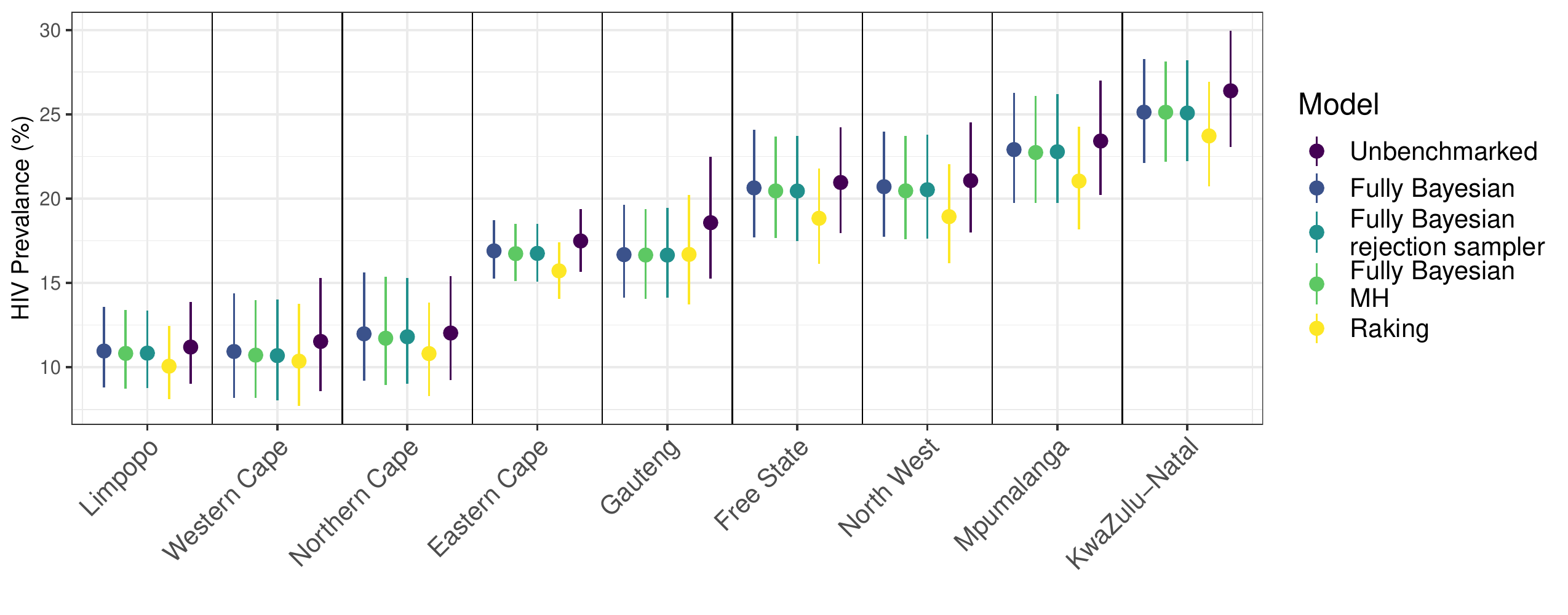}
	\caption{Comparison of HIV Prevalence estimates from benchmarked and unbenchmarked unit-level models at a subnational level. Error bars correspond to 95\% credible intervals, and provinces are arranged from left to right in order of unbenchmarked median.}
	\label{fig:ranking}
\end{figure}

Figure \ref{fig:ranking} displays the subnational breakdown of benchmarked and unbenchmarked estimates from the unit-level model. Of note, the benchmarked posterior medians and credible intervals for the Fully Bayesian approaches look roughly identical (as they should), and the posterior distributions for each province from the raking approach are all slightly lower than those from the Fully Bayesian approaches. This difference is due to the raking approach enforcing the benchmarking constraint exactly, thus pulling the unbenchmarked estimates down farther than the Fully Bayesian approaches, as the national benchmark is lower than the unbenchmarked national estimate. Though more difficult to see, the Fully Bayesian approach treats regions with more uncertainty differently than those with less uncertainty. In particular, note that in the Eastern Cape province the fully Bayesian benchmarked posterior medians are not as different from the unbenchmarked median as in the Gauteng province, where uncertainty is much higher in the unbenchmarked estimate. This is opposed to the Raking approach, which shifts all estimates by the same multiplicative constant, regardless of uncertainty.

The model validation results for the HIV application for both unit- and area-level models suggest that both models are reasonably well-suited to the data. We compare posterior medians of the predictive distribution in each area (having left that area out of model fitting) to the direct estimate in each area, respectively. The 80\% credible intervals capture $8/9$ direct estimates in each model, and the 50\% credible intervals capture $3/9$ and $4/9$ in the unit- and area-level models, respectively. This suggests that the area-level model may be slightly more suited to the data, which is unsurprising given that the area-level model accounts for the survey design directly. 

Additional plots and tables, including those for the area-level model, can be found in Section 1 of the Supplemental Data. 

\subsection{U5MR}

The results for the U5MR application are in Section 2 of the Supplemental Data. 

\section{Discussion}


In this paper we have summarized existing benchmarking approaches and their benefits and drawbacks with regard to data with rare binary outcomes in LMICs. We consider benchmarking methods that make use of a benchmarking constraint via one-step or two-step approaches, and pay particular attention to the resulting interpretation of the benchmarked estimates, the ease of uncertainty quantification, the acknowledgement of uncertainty in national estimates, the use of non-linear constraints, and computational tractability. We believe that the proposed rejection sampling and MH approaches to fully Bayesian benchmarking provide a desirable balance all of these concerns, and provide alternative computational approaches to the one-step, fully Bayesian benchmarking method developed by \textcite{zhang2020fully}, which in many cases may be inflexible with regards to modeling choice and computational tractability, while targeting the same benchmarked, posterior distribution. 

We show via an application of various benchmarking approaches to an HIV prevalence example that the proposed two-step approaches to fully Bayesian benchmarking produce the same benchmarked estimates as the one-step \textcite{zhang2020fully} approach, and that the resulting estimates are a compromise between the national level estimate and unbenchmarked estimates. The rejection sampling and MH approaches allow us to take advantage of potentially faster computational programs than traditional MCMC samplers, such as INLA or TMB, as evidenced via our application to U5MR. Inference for spatial models (continuous models especially) using MCMC methods is computationally challenging, and our approach allows users to conduct fully Bayesian benchmarking while relying on Laplace approximation methods for inference that are more suited for such models \autocite{osgood2021statistical}. Additionally, the rejection sampling and MH approaches are easily applied to both unit-level and area-level models, as evidenced in both our HIV and U5MR applications. Though the purpose of this paper was not to provide an in-depth review of small area models in LMICs, one reviewer noted that our unit-level models do not directly account for the survey design. This point should not go unnoticed, and as with any application involving survey data, careful consideration should to be given to the survey design even if a model-based approach is required for adequate precision. 

There are several limitations with the proposed approaches to fully Bayesian benchmarking. The first and potentially most pressing is the inability to conduct exact benchmarking, which may be required in some settings and is especially relevant if national benchmarks come from a census, though censuses have uncertainty in practice. Scenarios where true exact benchmarking is required are rare in an LMIC context. While we note that exact benchmarking can be approximated via the proposed approaches, it will likely be computationally inefficient. The one-step approach to fully Bayesian benchmarking may be a more useful approach if exact benchmarking is required, or a posterior projection approach using a loss function that respects the bounds on the parameters that are used for the weights. 

Additionally, our method does not account for uncertainty in the population count data. The population data used in our applications did not have reported uncertainty. If uncertainty for population count data were available, this would ideally be incorporated into the likelihood for the benchmarking constraint. Worldpop has recently started quantifying uncertainty in population data for select countries in their bottom-up Bayesian models, which may be of interest to include in future applications \autocite{leasure2020national}. 

{\raggedright \printbibliography}

\end{document}


\maketitle
	
	\section{HIV Application}
	
	\subsection{Unit-level figures and tables}
	
\begin{figure}[!h]
		\centering
		\includegraphics[scale = 0.75]{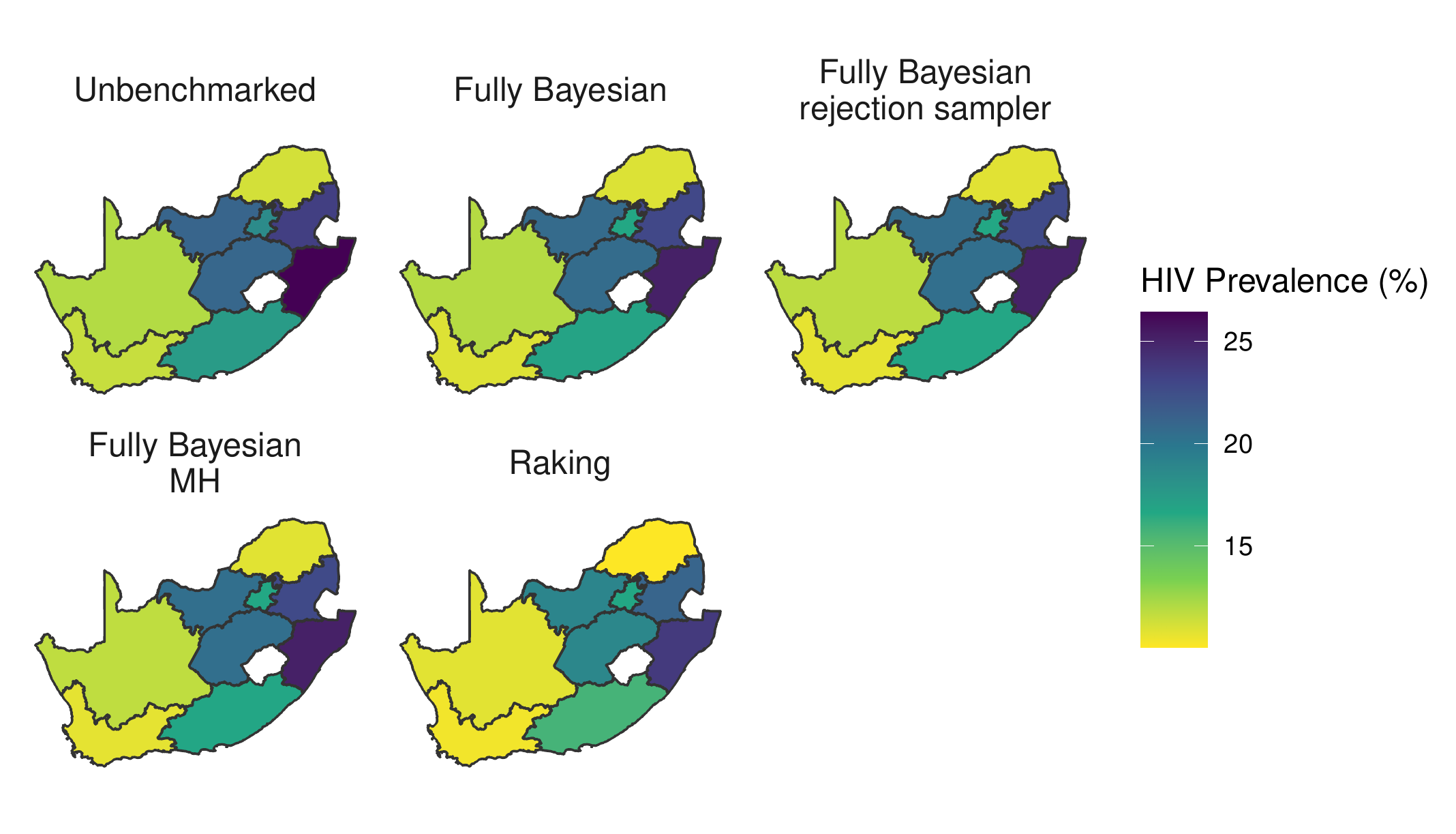}
		\caption{Comparison of HIV prevalence estimates from benchmarked and unbenchmarked unit-level models.}
		\label{fig:mapcomparemed_unitlevel_SA}
	\end{figure}

	The subnational breakdown of unbenchmarked and benchmarked, unit-level models can be seen in Figure \ref{fig:mapcomparemed_unitlevel_SA} and Table \ref{tab:hivadmin1results}.
	
	\begin{table}[!ht]
		\centering
		\caption{Admin1 HIV prevalence estimates from unbenchmarked and benchmarked unit-level models. Point estimates provided are medians, with 95\% credible intervals. Provinces are arranged in the order of lowest unbenchmarked median to highest.}
		\begin{tabular}{l|l|l|l|l|l}
			\textbf{Province} & \textbf{Unbenched (\%)} & \textbf{FB (\%)}  & \textbf{\begin{tabular}[c]{@{}l@{}}FB: Rejection \\ sampler (\%)\end{tabular}} & \textbf{FB: MH (\%)} & \textbf{Raking (\%)} \\ \hline
			Limpopo           & 11.2 (9.0, 13.9)        & 11.0 (8.8, 13.6)  & 10.8 (8.8, 13.4)                                                               & 10.8 (8.7, 13.4)     & 10.1 (8.1, 12.5)     \\ \hline
			Western Cape      & 11.5 (8.6, 15.3)        & 10.9 (8.2, 14.4)  & 10.7 (8.0, 14.0)                                                               & 10.7 (8.2, 14.0)     & 10.4 (7.7, 13.8)     \\ \hline
			Northern Cape     & 12.0 (9.2, 15.4)        & 12.0 (9.2, 15.6)  & 11.8 (9.0, 15.3)                                                               & 11.7 (8.9, 15.4)     & 10.8 (8.3, 13.8)     \\ \hline
			Eastern Cape      & 17.5 (15.6, 19.4)       & 16.9 (15.3, 18.7) & 16.8 (15.1, 18.5)                                                              & 16.7 (15.1, 18.5)    & 15.7 (14.1, 17.4)    \\ \hline
			Gauteng           & 18.6 (15.3, 22.5)       & 16.7 (14.1, 19.6) & 16.7 (14.1, 19.5)                                                              & 16.7 (14.1, 19.4)    & 16.7 (13.7, 20.2)    \\ \hline
			Free State        & 21.0 (17.9, 24.2)       & 20.6 (17.7, 24.1) & 20.5 (17.5, 23.7)                                                              & 20.5 (17.7, 23.7)    & 18.8 (16.1, 21.8)    \\ \hline
			North West        & 21.1 (18.0, 24.5)       & 20.7 (17.8, 24.0) & 20.5 (17.6, 23.8)                                                              & 20.5 (17.6, 23.7)    & 18.9 (16.2, 22.0)    \\ \hline
			Mpumalanga        & 23.4 (20.2, 27.0)       & 22.9 (19.8, 26.3) & 22.8 (19.7, 26.2)                                                              & 22.7 (19.8, 26.1)    & 21.0 (18.2, 24.3)    \\ \hline
			KwaZulu-Natal     & 26.4 (23.1, 30.0)       & 25.1 (22.1, 28.3) & 25.1 (22.2, 28.2)                                                              & 25.1 (22.2, 28.1)    & 23.7 (20.7, 26.9)   
		\end{tabular}
		\label{tab:hivadmin1results}
	\end{table}
	
	A table represetation of the aggregated national estimates presented in Figure 2 in the main body of the paper can be found in Figure \ref{tab:hivnatlresults}.
	
	\begin{table}[!ht]
		\centering
		\caption{Aggregated national level HIV prevalence estimates from Thembisa, unbenchmarked, and benchmarked results from the unit-level model. 95\% credible intervals are given next to posterior medians.}
		\begin{tabular}{l|l|l}
			\textbf{Model}           & \multicolumn{1}{l|}{\textbf{Median (\%)}} & \multicolumn{1}{l}{\textbf{SD (\%)}} \\ \hline
			Thembisa                 & 17.1 (15.6, 18.3)                         & 0.61                                 \\ \hline
			Unbenched            & 19.1 (17.6, 20.6)                         & 0.76                                 \\ \hline
			FB & 18.0 (17.1, 19.0)                         & 0.48                                 \\ \hline
			FB: Rejection sampler & 17.9 (17.0, 18.9)                         & 0.46                                 \\ \hline
			FB: MH & 17.9 (17.0, 18.8) & 0.47 \\ \hline
			Raking         & 17.1 (15.8, 18.5)                         & 0.68                                 \\ 
		\end{tabular}
		\label{tab:hivnatlresults}
	\end{table}

	\subsection{Area-level figures and tables}
	
\begin{figure}[!h]
		\centering
		\includegraphics[scale = 0.5]{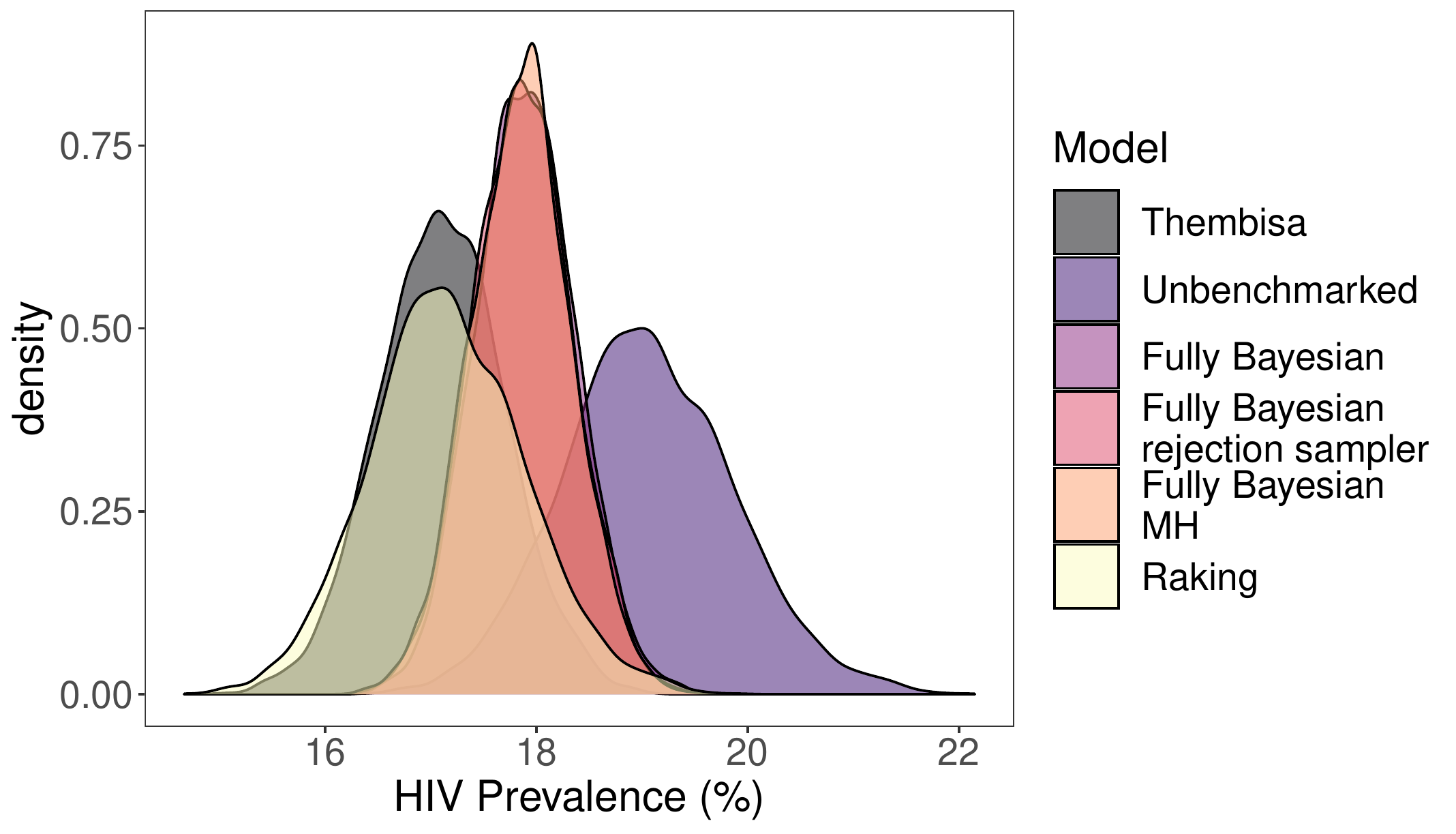}
		\caption{Aggregated national level HIV prevalence estimates from Thembisa, unbenchmarked, and benchmarked results from the area-level model. All densities are based on 5000 samples}
		\label{fig:densitynatlcompareSA_arealevel}
	\end{figure}

	In Figure \ref{fig:densitynatlcompareSA_arealevel}, we display national level estimates from Thembisa, the unbenchmarked, and benchmarked results from the area-level model. Results from this model are very similar to results from the unit-level model, found in the main body of the paper. 
	
	\begin{table}[!ht]
		\centering
		\caption{Aggregated national level HIV prevalence estimates from Thembisa, unbenchmarked, and benchmarked results from the area-level model. 95\% credible intervals are given next to posterior medians.}
		\begin{tabular}{l|l|l}
			\textbf{Model}        & \textbf{Median (\%)} & \textbf{SD (\%)} \\ \hline
			Thembisa              & 17.1 (15.6, 18.3)    & 0.61             \\ \hline
			Unbenched             & 19.0 (17.5, 20.7)      & 0.81             \\ \hline
			FB                    & 17.9 (17.0, 18.8)      & 0.47             \\ \hline
			FB: Rejection sampler & 17.9 (17.0, 18.8)      & 0.47             \\ \hline
			FB: MH & 17.9 (16.9, 18.8)      & 0.47             \\ \hline
			Raking                & 17.1 (15.8, 18.7)    & 0.73                  
		\end{tabular}
		\label{tab:hivnatlresults_arealevel}
	\end{table}
	
	\begin{figure}[!h]
		\centering
		\includegraphics[scale = 0.75]{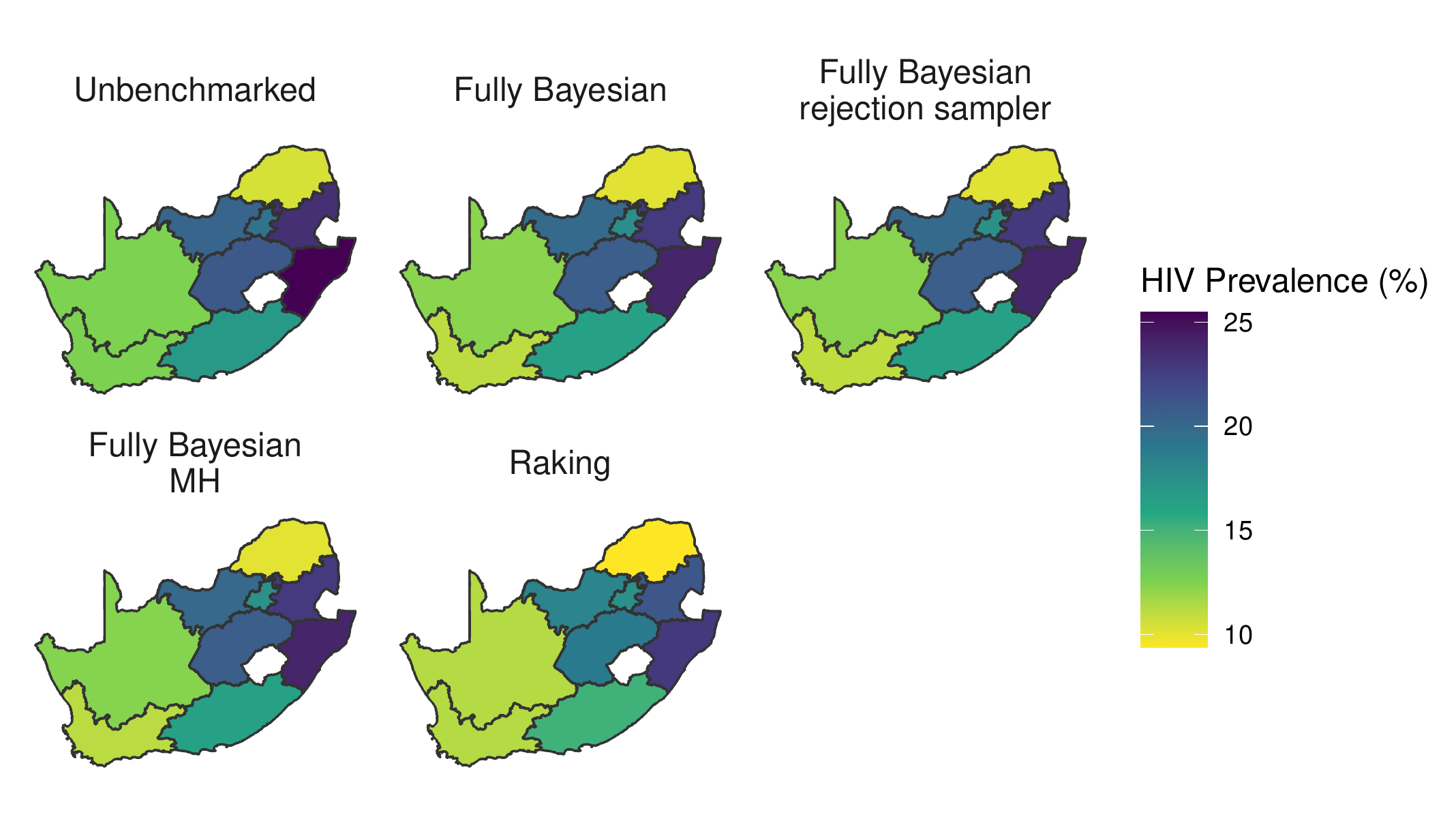}
		\caption{Comparison of HIV prevalence estimates from benchmarked and unbenchmarked area-level models.}
		\label{fig:mapcomparemed_arealevel_SA}
	\end{figure}
	
	\begin{table}[H]
		\centering
		\caption{Admin1 HIV prevalence estimates from unbenchmarked and benchmarked area-level models. Point estimates provided are medians, with 95\% credible intervals. Provinces are arranged in the order of lowest unbenchmarked median to highest.}
		\begin{tabular}{l|l|l|l|l|l}
			\textbf{Province} & \textbf{Unbenched (\%)} & \textbf{FB (\%)}  & \textbf{\begin{tabular}[c]{@{}l@{}}FB: Rejection\\ sampler (\%)\end{tabular}} & \textbf{FB: MH (\%)} & \textbf{Raking (\%)} \\ \hline
			Limpopo           & 10.4 (8.3, 13.1)        & 10.1 (8.0, 12.8)  & 10.2 (8.0, 12.8)                                                              & 10.1 (8.0, 12.8)     & 9.4 (7.5, 11.8)      \\ \hline
			Northern Cape     & 12.5 (8.9, 17.4)        & 12.2 (8.6, 17.0)  & 12.3 (8.7, 17.1)                                                              & 12.4 (8.8, 17.3)     & 11.3 (8.0, 15.7)     \\ \hline
			Western Cape      & 12.6 (8.2, 18.8)        & 11.1 (7.5, 16.4)  & 11.0 (7.4, 16.2)                                                              & 11.1 (7.5, 16.0)     & 11.3 (7.4, 17.0)     \\ \hline
			Eastern Cape      & 16.8 (14.6, 19.4)       & 16.3 (14.3, 18.5) & 16.3 (14.1, 18.7)                                                             & 16.2 (14.2, 18.6)    & 15.1 (13.1, 17.4)    \\ \hline
			Gauteng           & 19.3 (15.9, 23.6)       & 17.5 (14.7, 20.6) & 17.4 (14.7, 20.5)                                                             & 17.3 (14.5, 20.3)    & 17.4 (14.3, 21.2)    \\ \hline
			North West        & 20.2 (17.5, 23.3)       & 19.9 (17.2, 22.8) & 19.9 (17.2, 22.9)                                                             & 20.0 (17.3, 22.9)    & 18.2 (15.7, 21.0)    \\ \hline
			Free State        & 21.0 (17.9, 24.4)       & 20.6 (17.7, 24.1) & 20.7 (17.6, 24.0)                                                             & 20.7 (17.6, 24.0)    & 18.9 (16.1, 21.9)    \\ \hline
			Mpumalanga        & 23.4 (20.2, 27)         & 22.9 (19.7, 26.4) & 22.9 (19.6, 26.4)                                                             & 22.8 (19.6, 26.3)    & 21.1 (18.2, 24.3)    \\ \hline
			KwaZulu-Natal     & 25.5 (21.7, 29.4)       & 24.0 (20.7, 27.6) & 24.0 (20.7, 27.4)                                                             & 24.0 (20.6, 27.5)    & 22.9 (19.5, 26.5)   
		\end{tabular}
		\label{tab:hivadmin1results_arealevel}
	\end{table}
	
	\newpage
	
	\subsection{Area-level and unit-level comparison plots}

\begin{figure}[H]
		\centering
		\includegraphics[scale = 0.7]{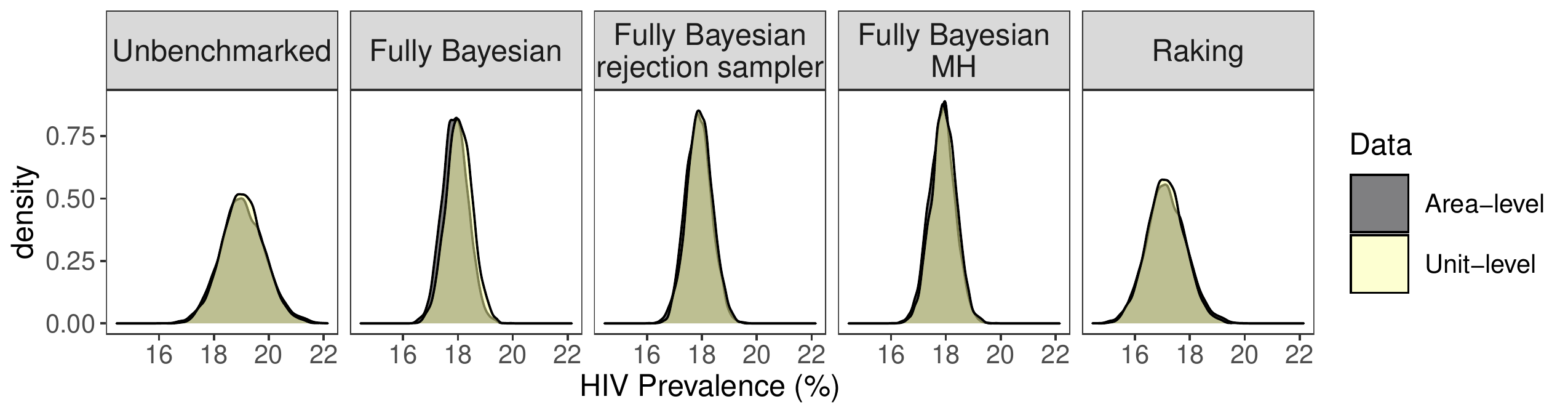}
		\caption{Aggregated national level HIV prevalence estimates from unbenchmarked and benchmarked models. All densities are based on 5000 samples.}
		\label{fig:densitynatlHIVarea}
	\end{figure}

	In Figure \ref{fig:densitynatlHIVarea} we display the densities for the national aggregated HIV prevalence estimates from unbenchmarked and benchmarked models, comparing across unit-level and area-level models. As noted in the main text, the differences between unit-level and area-level estimates are very small, although there are relatively large differences between the unbenchmarked and benchmarked models themselves.
	
\begin{figure}[H]
		\centering
		\includegraphics[scale = 0.7]{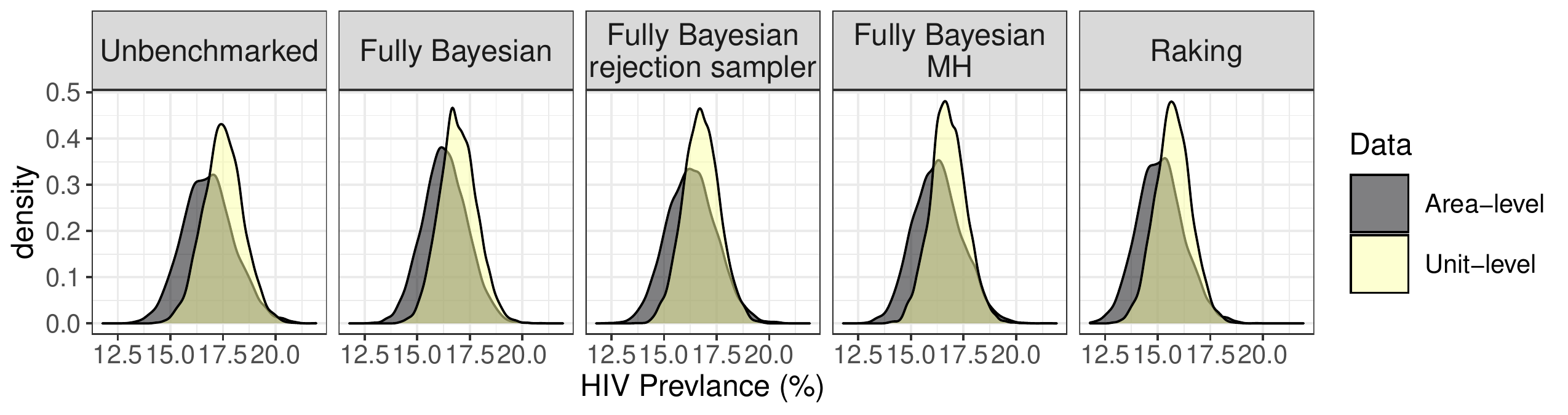}
		\caption{Subnational HIV prevalence estimates for the Eastern Cape province from unbenchmarked and benchmarked models. All densities are based on 5000 samples.}
		\label{fig:densitysubnatlHIVeasterncape}
	\end{figure}

\begin{figure}[H]
		\centering
		\includegraphics[scale = 0.7]{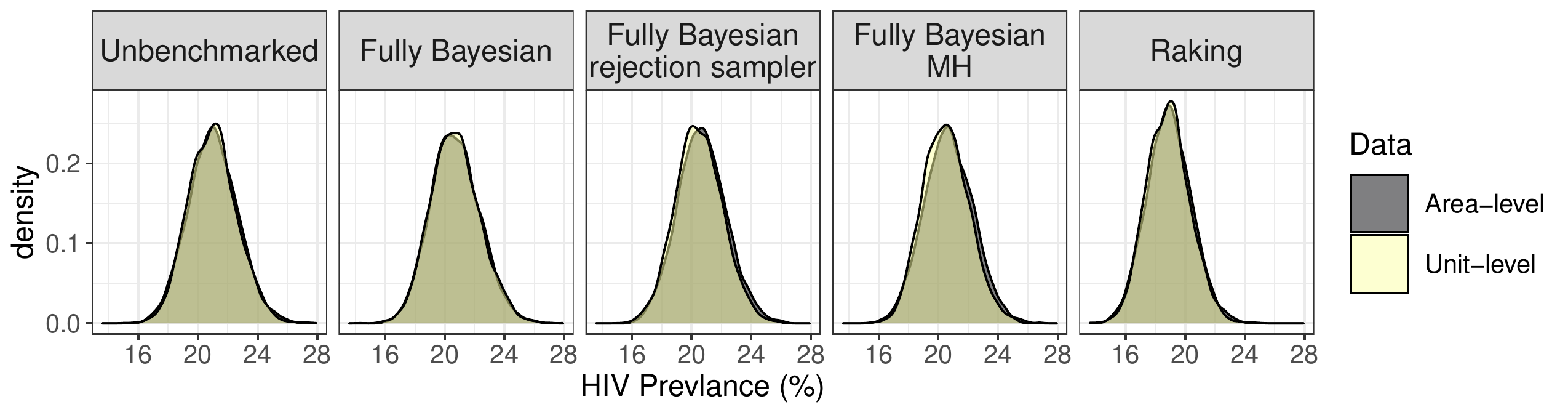}
		\caption{Subnational HIV prevalence estimates for the Free State province from unbenchmarked and benchmarked models. All densities are based on 5000 samples.}
		\label{fig:densitysubnatlHIVfreestate}
	\end{figure}

\begin{figure}[H]
		\centering
		\includegraphics[scale = 0.7]{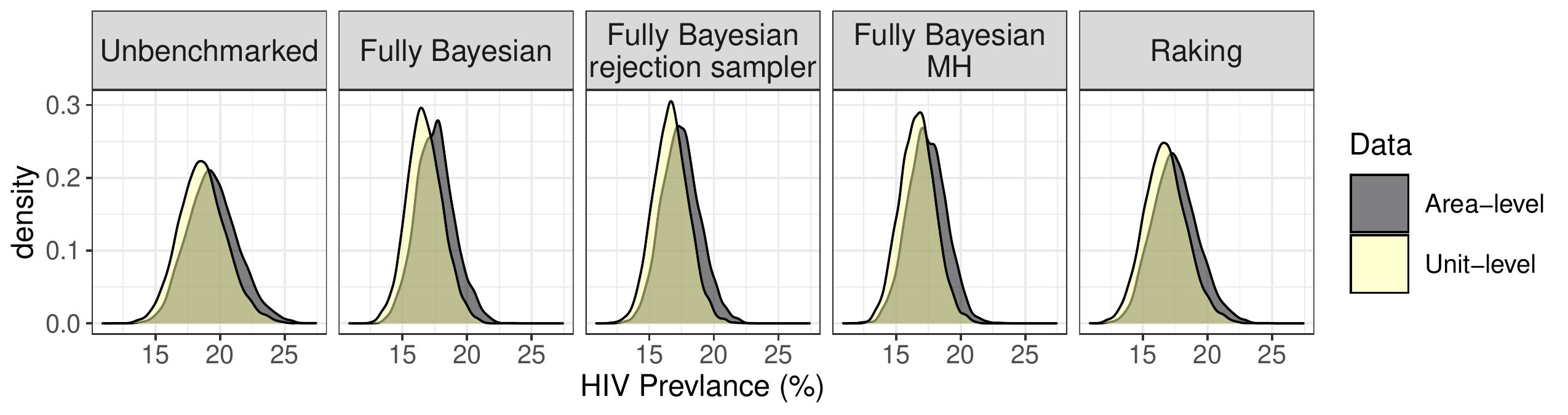}
		\caption{Subnational HIV prevalence estimates for the Gauteng province from unbenchmarked and benchmarked models. All densities are based on 5000 samples.}
		\label{fig:densitysubnatlHIVgauteng}
	\end{figure}

\begin{figure}[H]
		\centering
		\includegraphics[scale = 0.7]{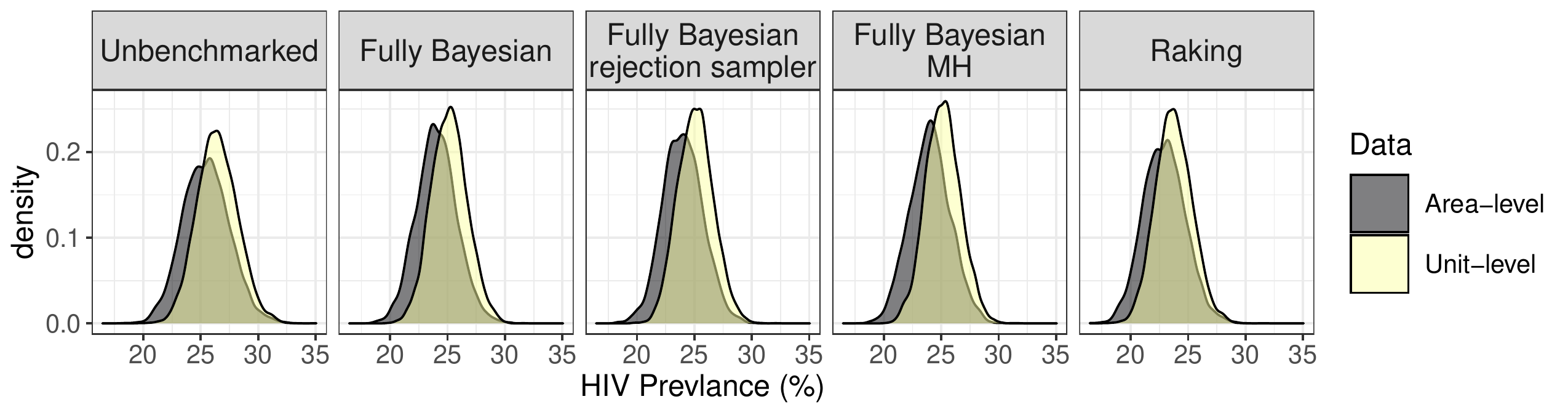}
		\caption{Subnational HIV prevalence estimates for the KwaZulu-Natal province from unbenchmarked and benchmarked models. All densities are based on 5000 samples.}
		\label{fig:densitysubnatlHIVKwaZuluNatal}
	\end{figure}

\begin{figure}[H]
		\centering
		\includegraphics[scale = 0.7]{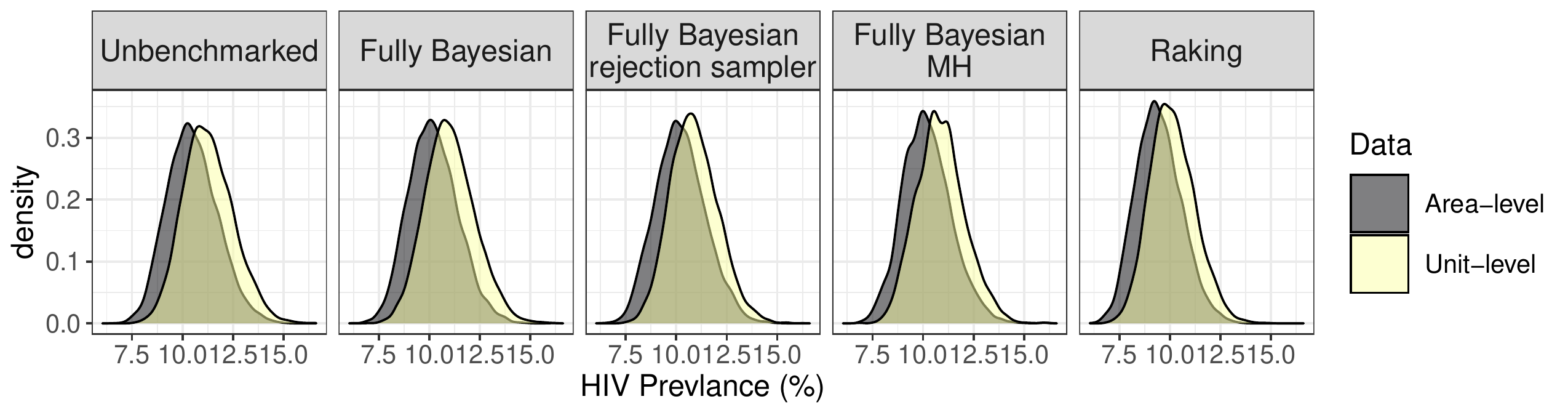}
		\caption{Subnational HIV prevalence estimates for the Limpopo province from unbenchmarked and benchmarked models. All densities are based on 5000 samples.}
		\label{fig:densitysubnatlHIVLimpopo}
	\end{figure}

\begin{figure}[H]
		\centering
		\includegraphics[scale = 0.7]{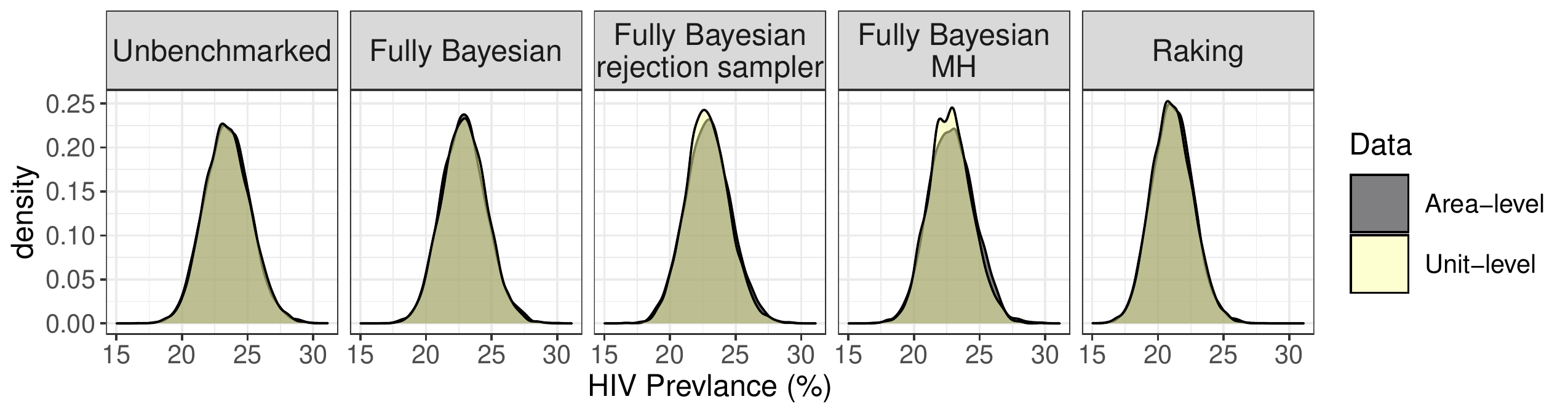}
		\caption{Subnational HIV prevalence estimates for the Mpumalanga province from unbenchmarked and benchmarked models. All densities are based on 5000 samples.}
		\label{fig:densitysubnatlHIVMpumalanga}
	\end{figure}

\begin{figure}[H]
		\centering
		\includegraphics[scale = 0.7]{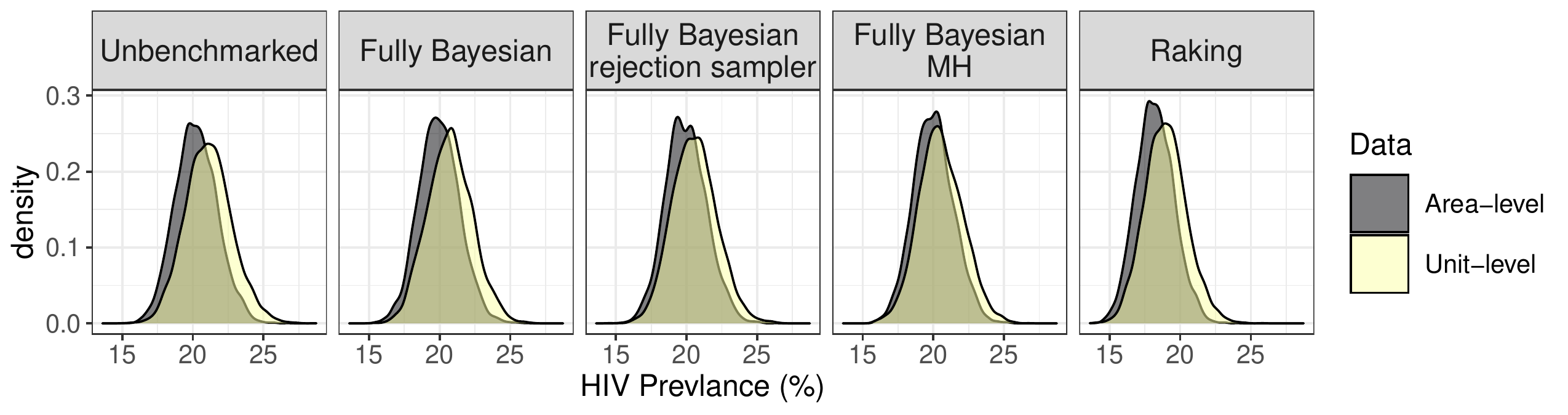}
		\caption{Subnational HIV prevalence estimates for the North West province from unbenchmarked and benchmarked models. All densities are based on 5000 samples.}
		\label{fig:densitysubnatlHIVnorthwest}
	\end{figure}

\begin{figure}[H]
		\centering
		\includegraphics[scale = 0.7]{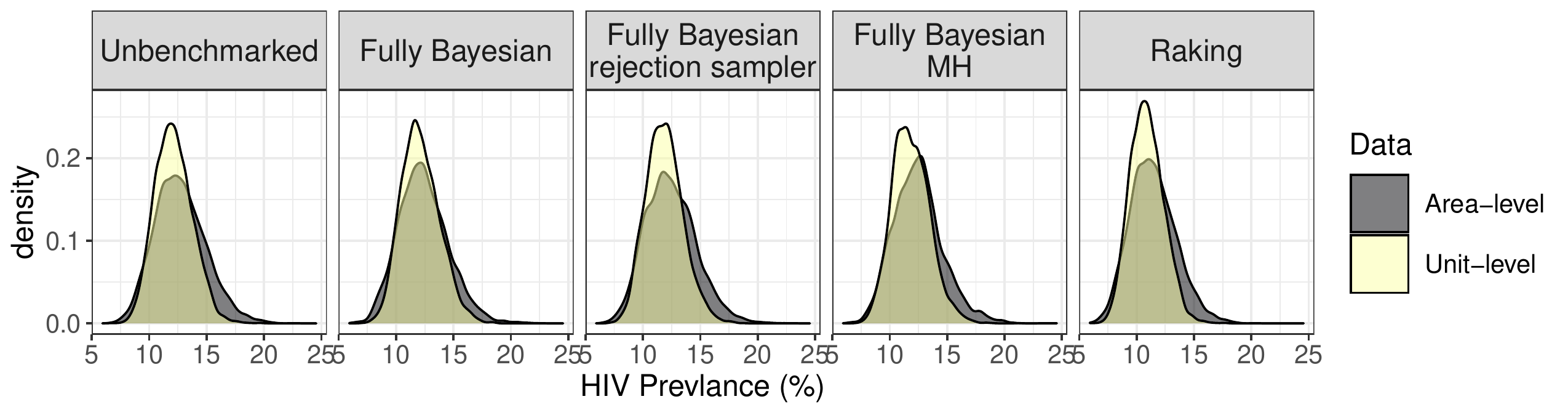}
		\caption{Subnational HIV prevalence estimates for the Northern Cape province from unbenchmarked and benchmarked models. All densities are based on 5000 samples.}
		\label{fig:densitysubnatlHIVNorthernCape}
	\end{figure}

\begin{figure}[H]
		\centering
		\includegraphics[scale = 0.7]{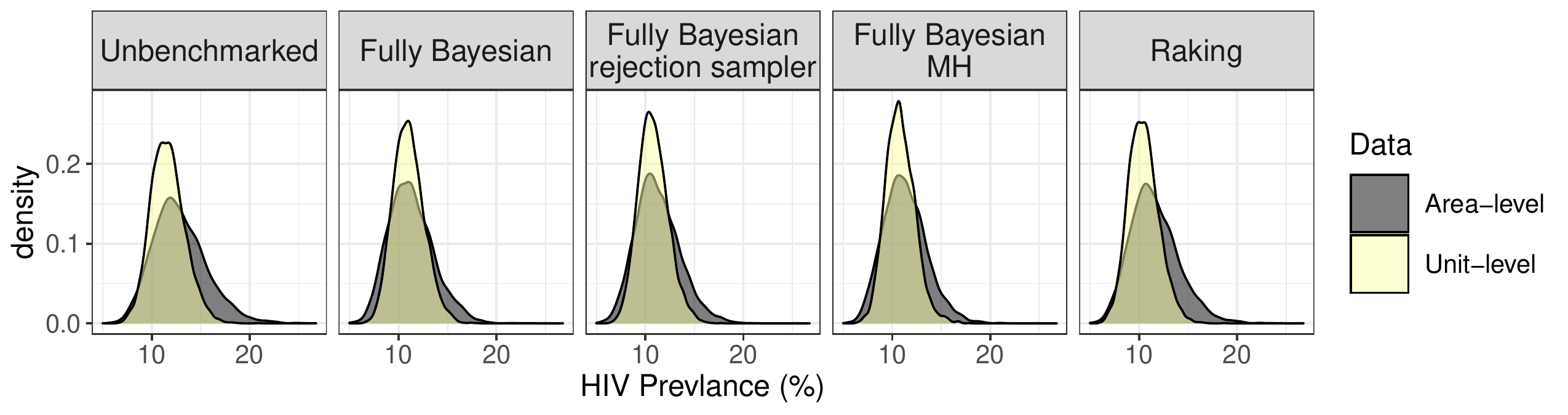}
		\caption{Subnational HIV prevalence estimates for the Western Cape province from unbenchmarked and benchmarked models. All densities are based on 5000 samples.}
		\label{fig:densitysubnatlHIVWesternCape}
	\end{figure}

	In Figures \ref{fig:densitysubnatlHIVeasterncape} through \ref{fig:densitysubnatlHIVWesternCape}, we show the differences between the unit-level and area-level, benchmarked and unbenchmarked sub-national estimates for each of the nine provinces. For most provinces, the differences between area-level and unit-level models are small, though we note that estimates from the unit-level models may have higher precision.
	
\begin{figure}[H]
		\centering
		\includegraphics[scale = 0.5]{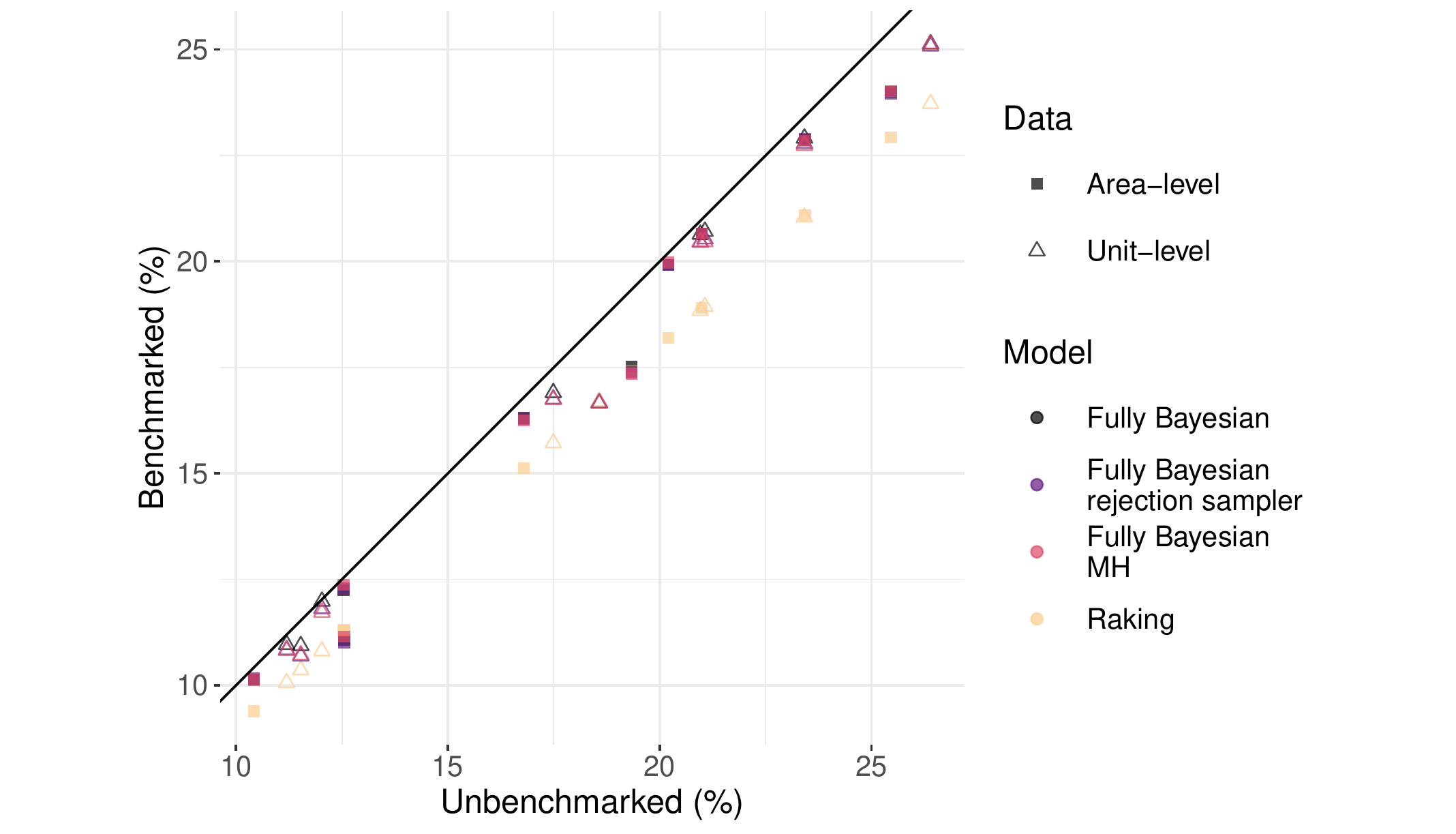}
		\caption{Comparison of unbenchmarked and benchmarked posterior medians by model and data type at a subnational level.}
		\label{fig:scattercomparemedHIV}
	\end{figure}

\begin{figure}[H]
		\centering
		\includegraphics[scale = 0.5]{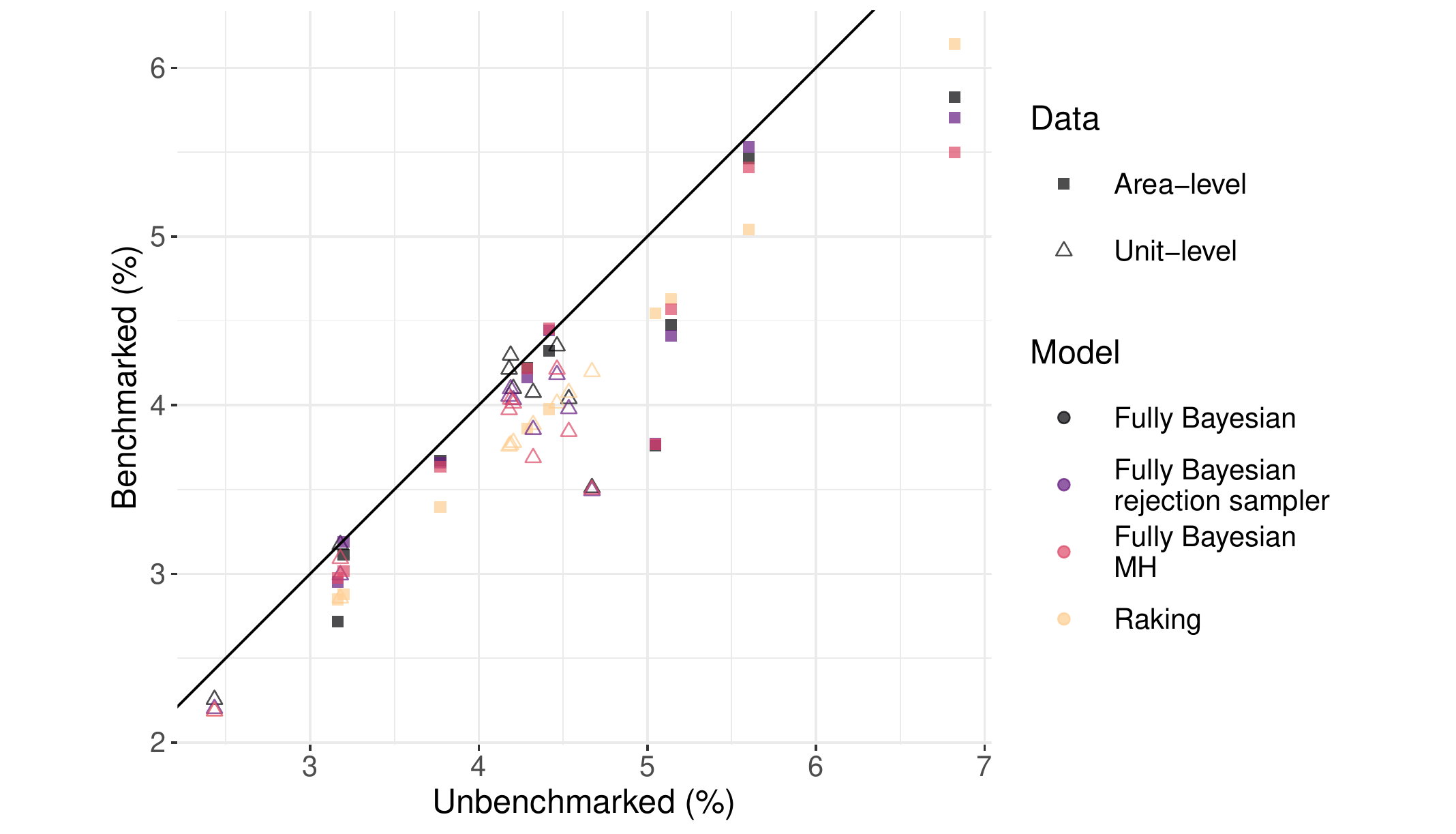}
		\caption{Comparison of unbenchmarked and benchmarked posterior 95\% CI widths by model and data type at a subnational level.}
		\label{fig:scattercompareciHIV}
	\end{figure}

\begin{figure}[H]
		\centering
		\includegraphics[scale = 0.5]{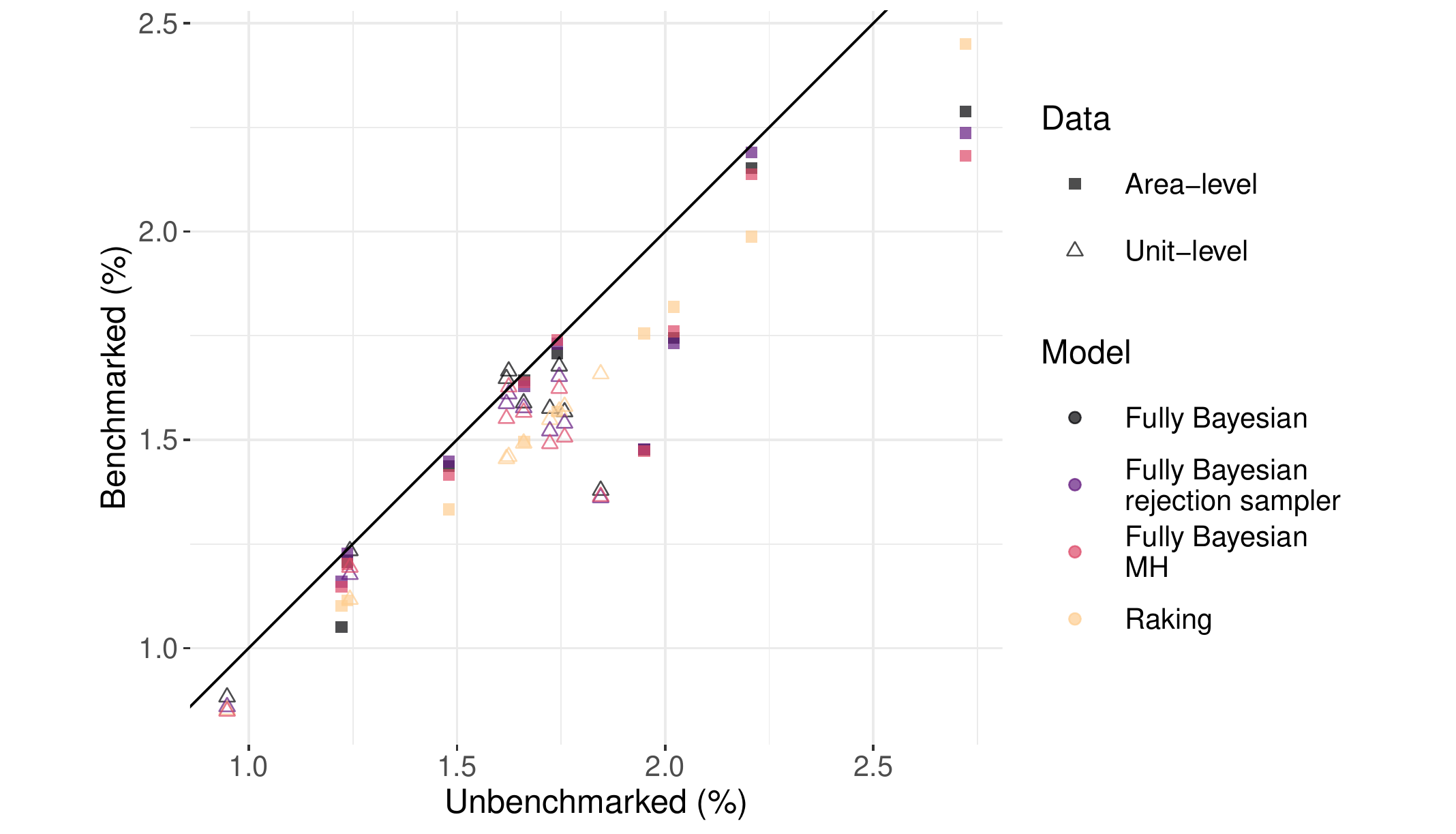}
		\caption{Comparison of unbenchmarked and benchmarked posterior standard deviations by model and data type at a subnational level.}
		\label{fig:scattercomparesdHIV}
	\end{figure}

	In Figures \ref{fig:scattercomparemedHIV} through \ref{fig:scattercomparesdHIV}, we compare benchmarked and unbenchmarked estimates and their uncertainty for area-level and unit-level models.
	
\begin{figure}[H]
		\centering
		\includegraphics[scale = 0.75]{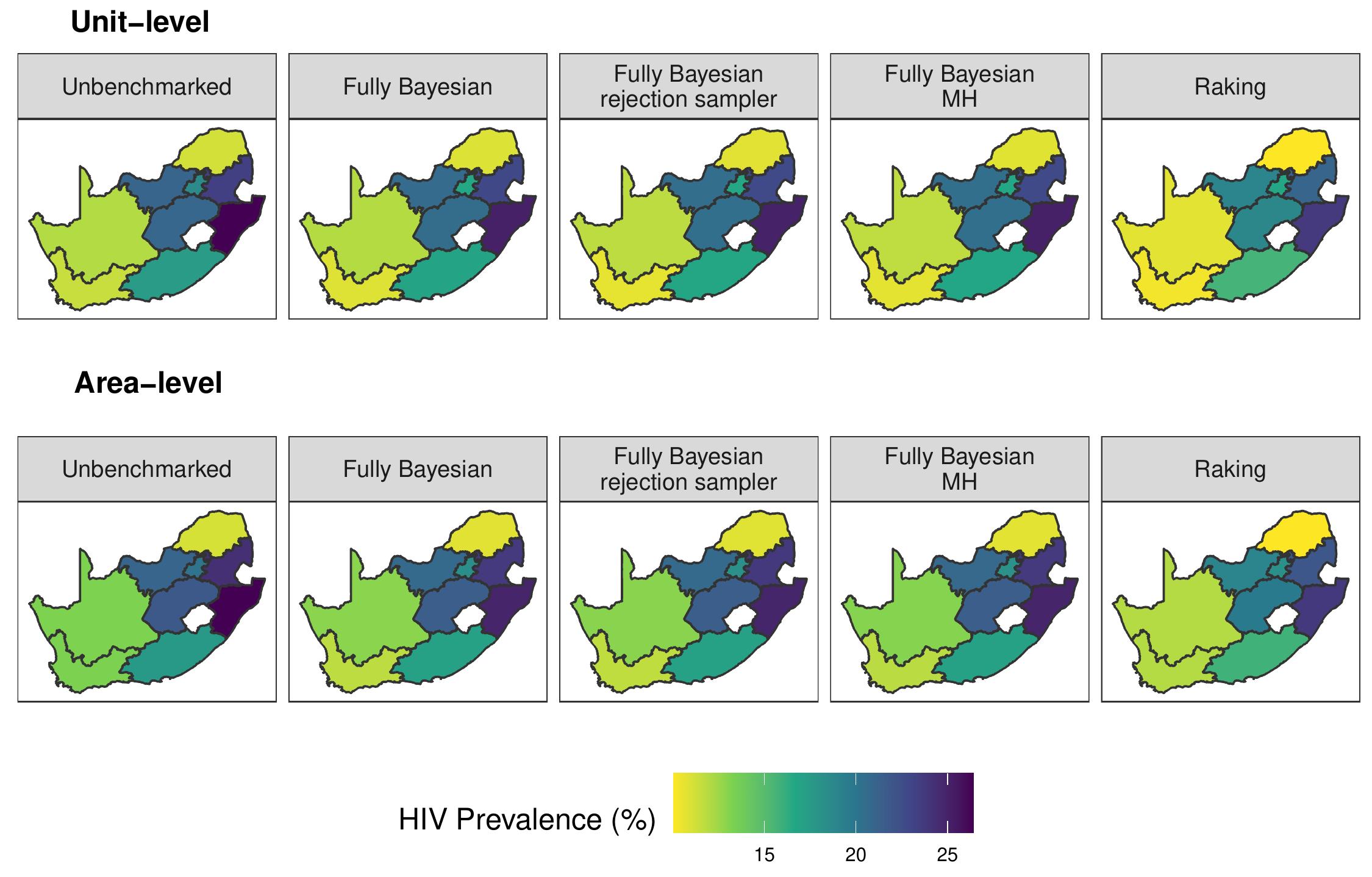}
		\caption{Unbenchmarked and benchmarked posterior medians by model and data type at a subnational level.}
		\label{fig:mapcomparemedHIVboth}
	\end{figure}

\begin{figure}[H]
		\centering
		\includegraphics[scale = 0.75]{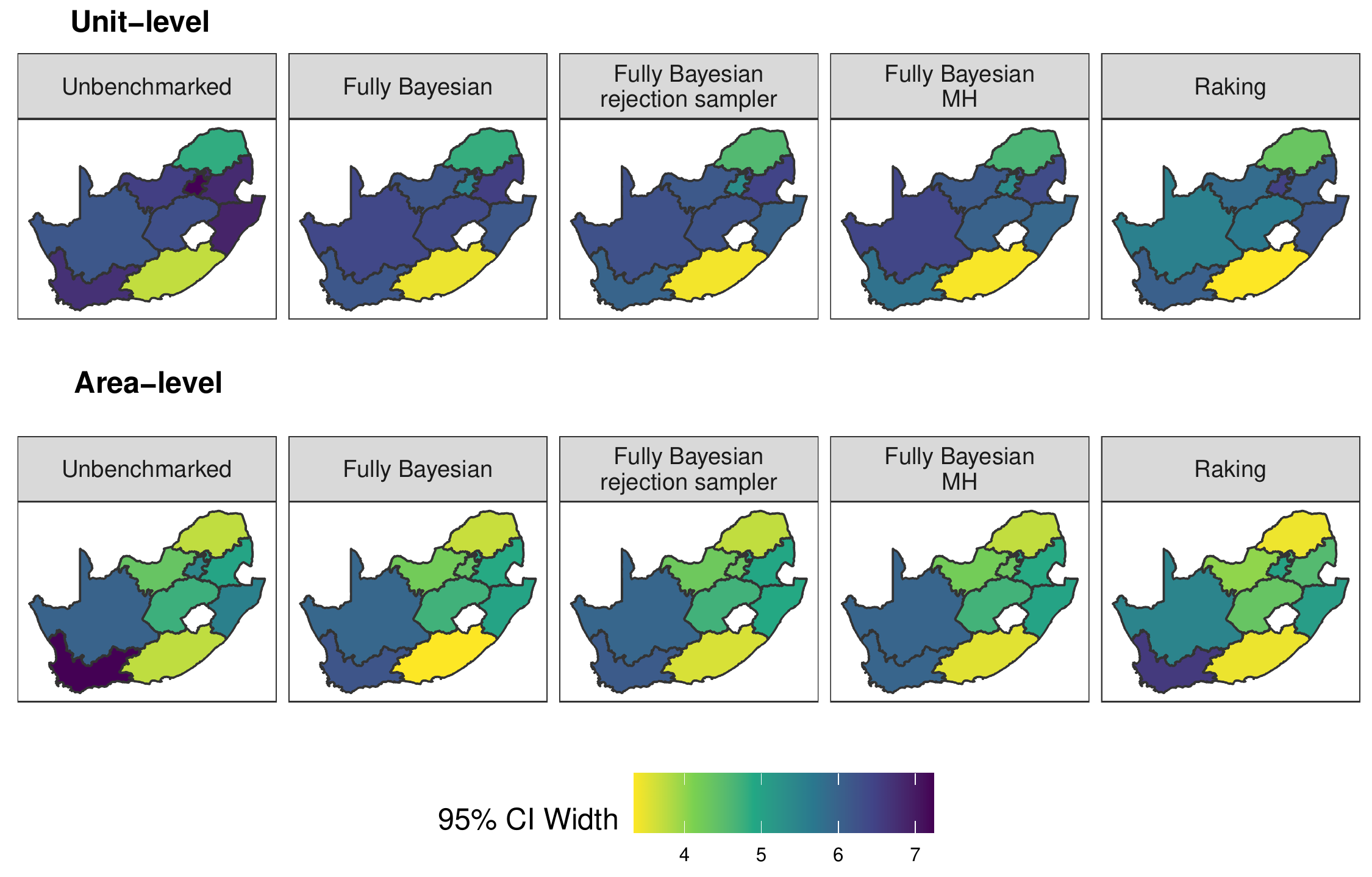}
		\caption{Unbenchmarked and benchmarked posterior 95\% CI widths by model and data type at a subnational level.}
		\label{fig:mapcompareciHIVboth}
	\end{figure}

\begin{figure}[H]
		\centering
		\includegraphics[scale = 0.75]{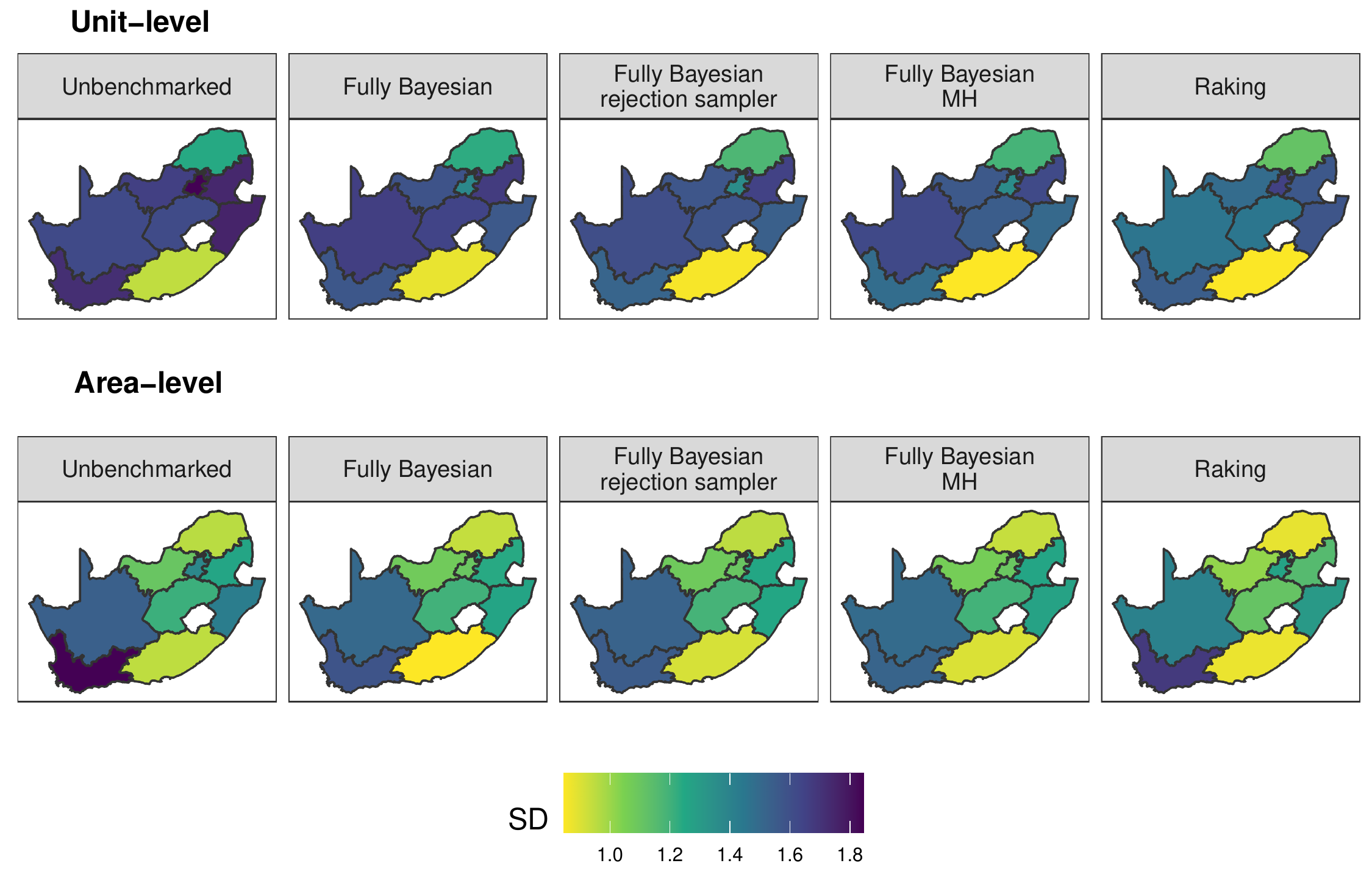}
		\caption{Unbenchmarked and benchmarked posterior standard deviations by model and data type at a subnational level.}
		\label{fig:mapcomparesdHIVboth}
	\end{figure}

	In Figures \ref{fig:mapcomparemedHIVboth} through \ref{fig:mapcomparesdHIVboth}, we display benchmarked and unbenchmarked estimates and their uncertainty for area-level and unit-level models.
	
	\subsection{Model Validation}
	
	Below we display the model validation results for the HIV application for both unit- and area-level models. We compare posterior medians of the predictive distribution in each area (having left that area out of model fitting) to the direct estimate in each area, respectively. The direct estimates in each area are simply the Horvitz-Thompons estimators in each are $i$, with design variance on the logit scale calculated as $\hat{V}_i^{DES} = \hat{\sigma}_i^{2HT} / (\hat{\theta}_i^{2HT} (1 - \hat{\theta}_i^{2HT} ))^2$. We obtain the posterior predictive distribution in each area $i$ by fitting the unbenchmarked model having left out the data for area $i$, and then adding the design variance to those samples on the logit scale. The resulting posterior predictive samples, on the logit scale, are thus a sum of a posterior draw from $\text{logit}(\hat{\theta}_i)$ and $N(0,\hat{V}_i^{DES})$. Credible intervals are calculated from the draws of the posterior predictive distributions.
	
	We note that both the 50\% and 80\% intervals suggest that both unit- and area-level models are reasonably well-suited to our data, with the 80\% credible intervals capturing $8/9$ direct estimates in each model, and the 50\% credible intervals capturing $3/9$ and $4/9$ in the unit- and area-level models, respectively. This suggests that the area-level model may be more suited to the data, which is unsurprising given that the area-level model accounts for the survey design directly. For the unit-level model (as noted in the discussion in the main paper) that there are likely variables in the survey design that are not being accounted for in the unit-level model that may lead to undercoverage. 
	
	\begin{figure}[H]
		\centering
		\includegraphics[scale = 0.6]{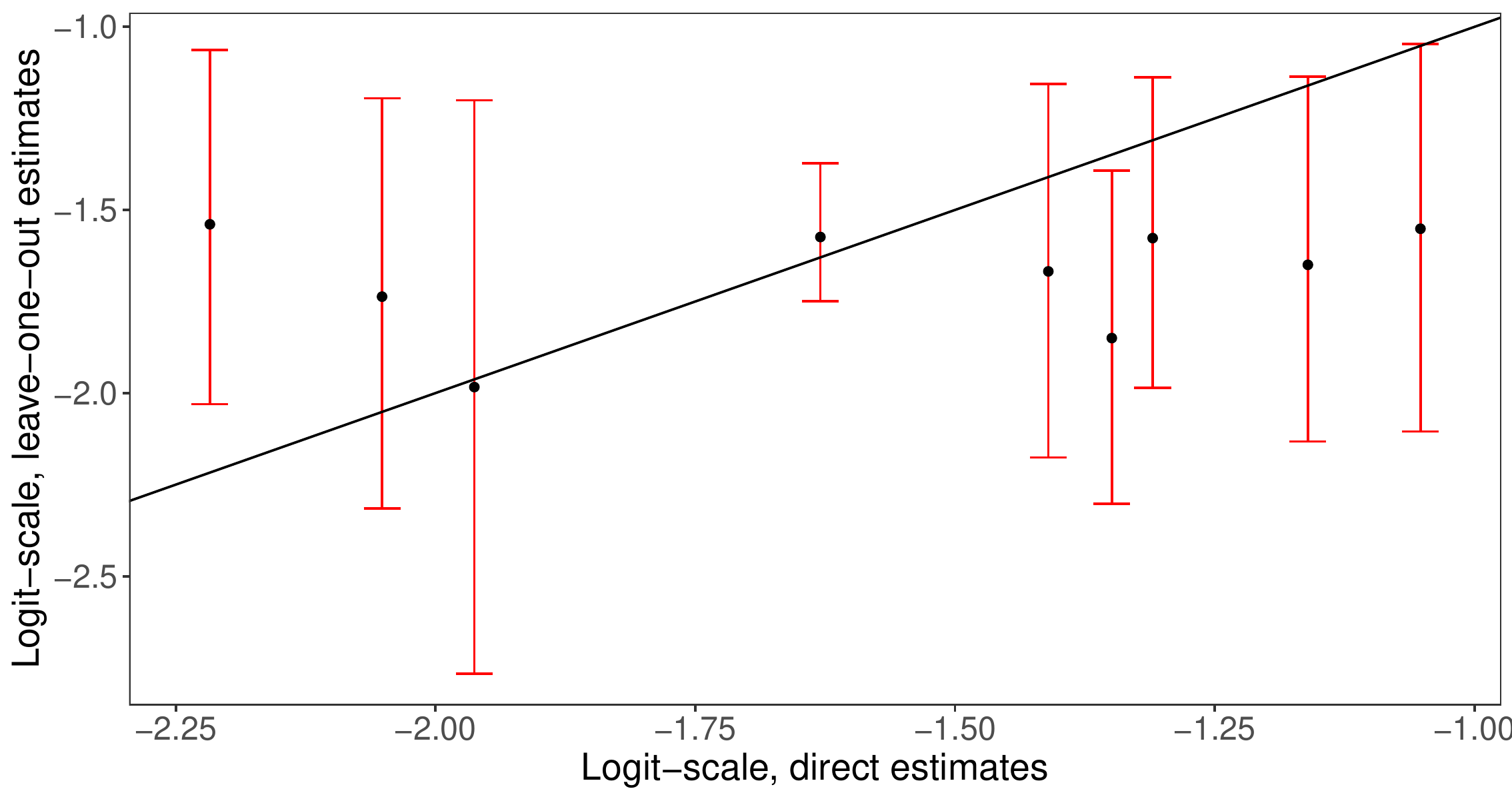}
		\caption{Scatterplot comparing leave-one-out posterior predictive estimates and direct estimates on the logit scale for the unit-level HIV model, with 80\% confidence intervals.}
		\label{fig:loounit80}
	\end{figure}
	
	\begin{figure}[H]
		\centering
		\includegraphics[scale = 0.6]{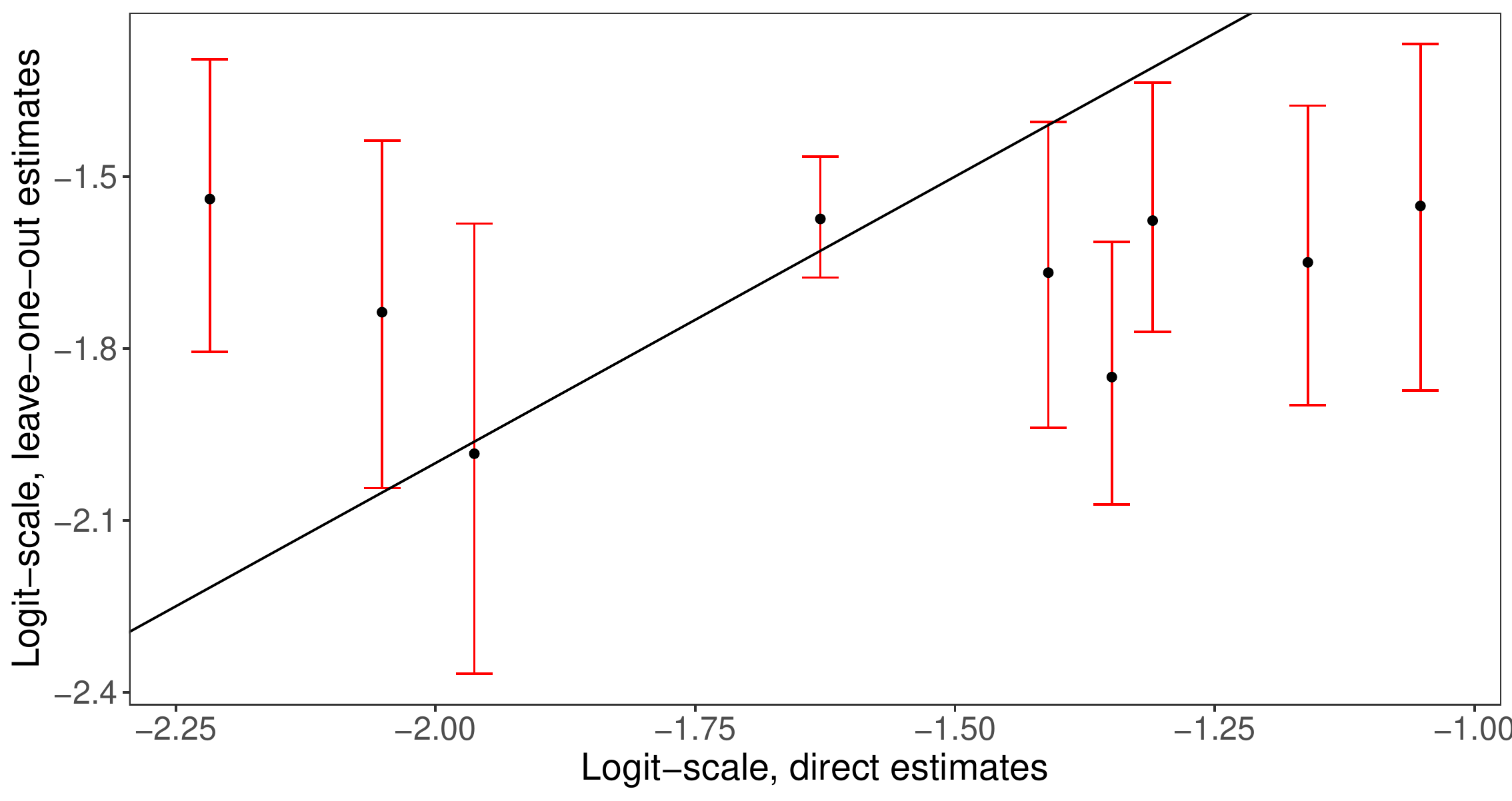}
		\caption{Scatterplot comparing leave-one-out posterior predictive estimates and direct estimates on the logit scale for the unit-level HIV model, with 50\% confidence intervals.}
		\label{fig:loounit50}
	\end{figure}
	
	\begin{figure}[H]
		\centering
		\includegraphics[scale = 0.6]{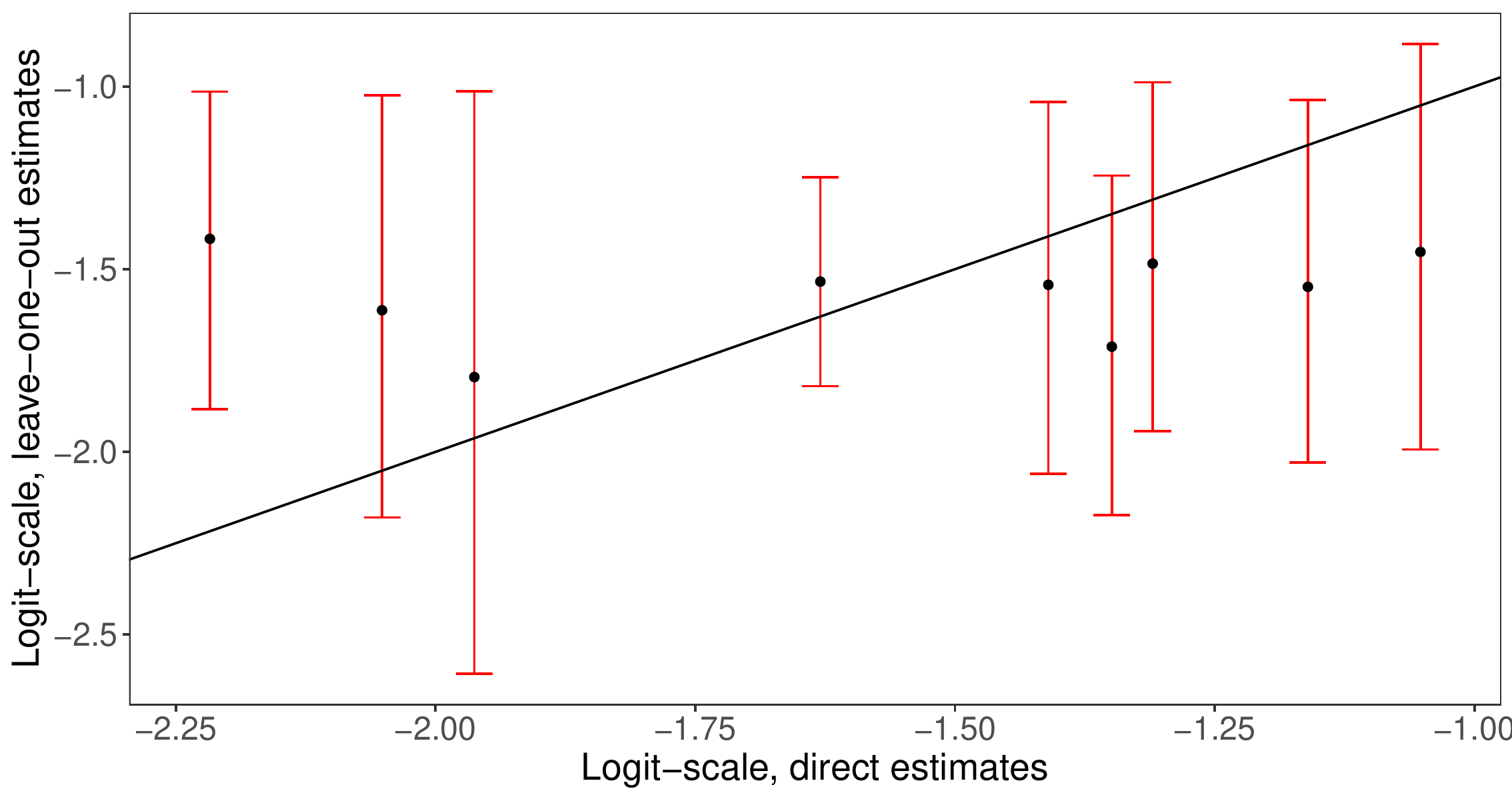}
		\caption{Scatterplot comparing leave-one-out posterior predictive estimates and direct estimates on the logit scale for the area-level HIV model, with 80\% confidence intervals.}
		\label{fig:looarea80}
	\end{figure}
	
	\begin{figure}[H]
		\centering
		\includegraphics[scale = 0.6]{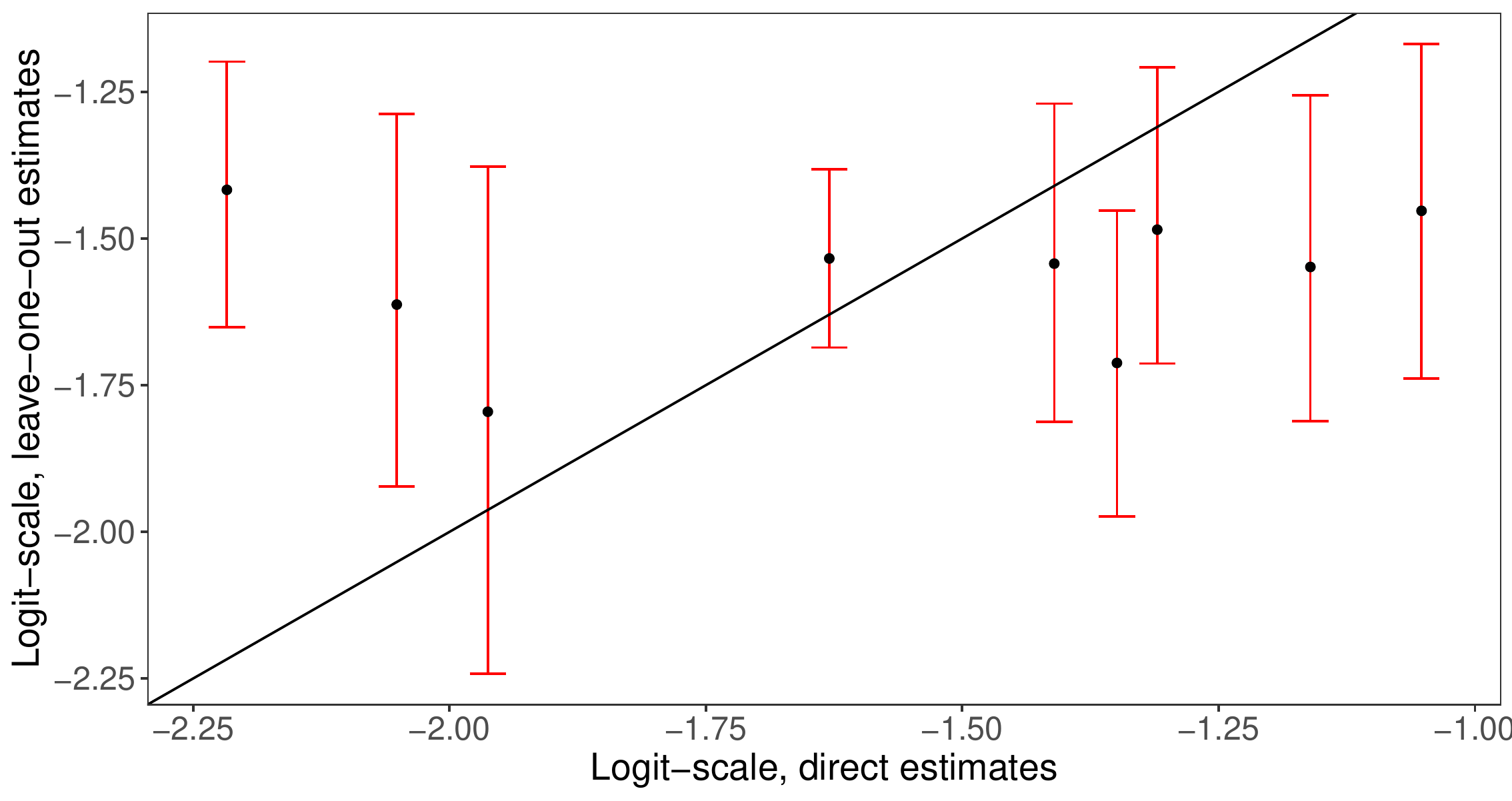}
		\caption{Scatterplot comparing leave-one-out posterior predictive estimates and direct estimates on the logit scale for the area-level HIV model, with 50\% confidence intervals.}
		\label{fig:looarea50}
	\end{figure}
	
	\section{U5MR Application}
	
	Below we describe the data, models, and software used in the U5MR application. 
	
	\subsection{Data}
	
	Spatial boundary files for Namibia are obtained from GADM, the Database of Global Administrative Areas \autocite{GADM}. We use the 2013 administrative boundary (the most recent DHS survey year) for all years in our U5MR model for Namibia, and note that this implies 13 Administrative 1 (admin1) regions. Of note, the Kavango region split in 2013 to form Kavango East and Kavango West. We obtain estimates for the Kavango region as a whole, as this is consistent with the 2013 GADM file, the most recent DHS survey we have available for Namibia, and current subnational UN IGME estimates for Namibia. 
	
	For our U5MR application, we use data from the 2000, 2006-2007, and 2013 DHS surveys for Namibia. These three surveys were chosen for their GPS data availability, and to align with the analysis carried out to produce official subnational U5MR estimates by UN IGME \autocite{IGME2020}, and therefore allow for comparison between the benchmarking methods used in that paper and those proposed here. The surveys followed a multi-stage, stratified design and were designed to provide estimates at the admin1 level, which consists of 13 regions. Again for consistency with official UN IGME estimates, we make predictions for years 2014-2019. 
	
	We obtain national level benchmarks for 2000-2019 in Namibia from the UN IGME national estimates from the B3 model \autocite{alkema2014global}. The national level U5MR estimates and their 90\% confidence intervals can be found in Table \ref{tab:namibiaworldpop}. The likelihood for our benchmarks for the fully Bayesian benchmarking approach requires standard errors for the benchmarks, which we approximate via the assumption that the benchmark is asymptotically normally distributed. These standard errors are presented in Table \ref{tab:namibiaworldpop}. The data sources used to produce the B3 estimates include surveys/censuses with full birth histories, summary birth histories, and household deaths. A full table of the data sources used is presented in Table \ref{tab:b3sources}. Of note, the three full birth history DHS surveys we use for our model (2000, 2006-2007, 2013) are included in the B3 model. As many additional data sources went into the UN IGME estimates, we consider this setting to be external benchmarking and therefore treat our estimates as independent of the national level benchmarks despite this inclusion. Implications of the potential internal benchmarking scenario are outside the scope of this paper.
	
	\begin{table}[!ht]
		\centering
		\caption{Data sources included in the B3 model for Namibia. MM = maternal mortality.}
		\label{tab:b3sources}
		\begin{tabular}{l|r|r}
			\multicolumn{1}{l|}{\textbf{Survey/Census}} & \multicolumn{1}{l|}{\textbf{Year}} & \multicolumn{1}{l}{\textbf{Collection Method}} \\ \hline
			Census                                      & 1991                               & Summary Birth History                          \\ \hline
			DHS (MM adjusted)                           & 1992                               & Full Birth History                             \\ \hline
			DHS                                         & 1992                               & Summary Birth History                          \\ \hline
			DHS                                         & 1992                               & Full Birth History                             \\ \hline
			DHS (MM adjusted)                           & 2000                               & Full Birth History                             \\ \hline
			DHS                                         & 2000                               & Summary Birth History                          \\ \hline
			DHS                                         & 2000                               & Full Birth History                             \\ \hline
			Census                                      & 2001                               & Household Deaths                               \\ \hline
			DHS (MM adjusted)                           & 2006-2007                          & Full Birth History                             \\ \hline
			DHS                                         & 2006-2007                          & Summary Birth History                          \\ \hline
			DHS                                         & 2006-2007                          & Full Birth History                             \\ \hline
			DHS                                         & 2006-2007                          & Household Deaths                               \\ \hline
			Census                                      & 2011                               & Summary Birth History                          \\ \hline
			Census                                      & 2011                               & Household Deaths                               \\ \hline
			DHS (MM adjusted)                           & 2013                               & Full Birth History                             \\ \hline
			DHS                                         & 2013                               & Full Birth History                             \\ \hline
			Inter-censal Demographic Survey             & 2016                               & Summary Birth History                          \\ \hline
			Inter-censal Demographic Survey             & 2016                               & Household Deaths                              
		\end{tabular}
	\end{table}

	\begin{table}[!ht]
		\centering
		\caption{National benchmarks for U5MR from the UN IGME B3 model for Namibia. Standard errors are computed using the upper bound of the 90\% confidence interval via the assumption that the benchmark is normally distributed. U5MR is reported as deaths per 1000 live births.}
		\label{tab:namibiaworldpop}
		\begin{tabular}{r|r|l|r}
			\textbf{Year} & \textbf{U5MR} & \textbf{CI (90\%)} & \textbf{SE (\%)*} \\ \hline
			2000          & 75.4          & (67.9, 84.4)       & 5.5               \\ \hline
			2001          & 74.9          & (67.5, 83.8)       & 5.4               \\ \hline
			2002          & 74.3          & (67.0, 83.2)       & 5.4               \\ \hline
			2003          & 74.0          & (66.5, 82.8)       & 5.3               \\ \hline
			2004          & 73.7          & (66.1, 82.3)       & 5.2               \\ \hline
			2005          & 70.2          & (62.8, 78.6)       & 5.1               \\ \hline
			2006          & 65.6          & (58.4, 73.8)       & 5.0               \\ \hline
			2007          & 60.7          & (53.3, 69.4)       & 5.3               \\ \hline
			2008          & 55.8          & (47.9, 65.0)       & 5.6               \\ \hline
			2009          & 51.8          & (43.5, 62.3)       & 6.4               \\ \hline
			2010          & 49.7          & (40.4, 62.0)       & 7.5               \\ \hline
			2011          & 50.4          & (39.4, 66.0)       & 9.5               \\ \hline
			2012          & 52.2          & (39.3, 71.4)       & 11.7               \\ \hline
			2013          & 50.1          & (36.3, 71.8)       & 13.2              \\ \hline
			2014          & 48.1          & (33.7, 72.3)       & 14.7              \\ \hline
			2015          & 47.8          & (32.3, 74.7)       & 16.3              \\ \hline
			2016          & 46.3          & (30.0, 75.6)       & 17.8              \\ \hline
			2017          & 44.5          & (28.0, 75.1)       & 18.6              \\ \hline
			2018          & 43.3          & (26.2, 76.0)       & 19.9              \\ \hline
			2019          & 42.4          & (24.8, 77.0)       & 21.1             
		\end{tabular}
	\end{table}
	
\begin{figure}[H]
		\centering
		\includegraphics[scale = 0.84]{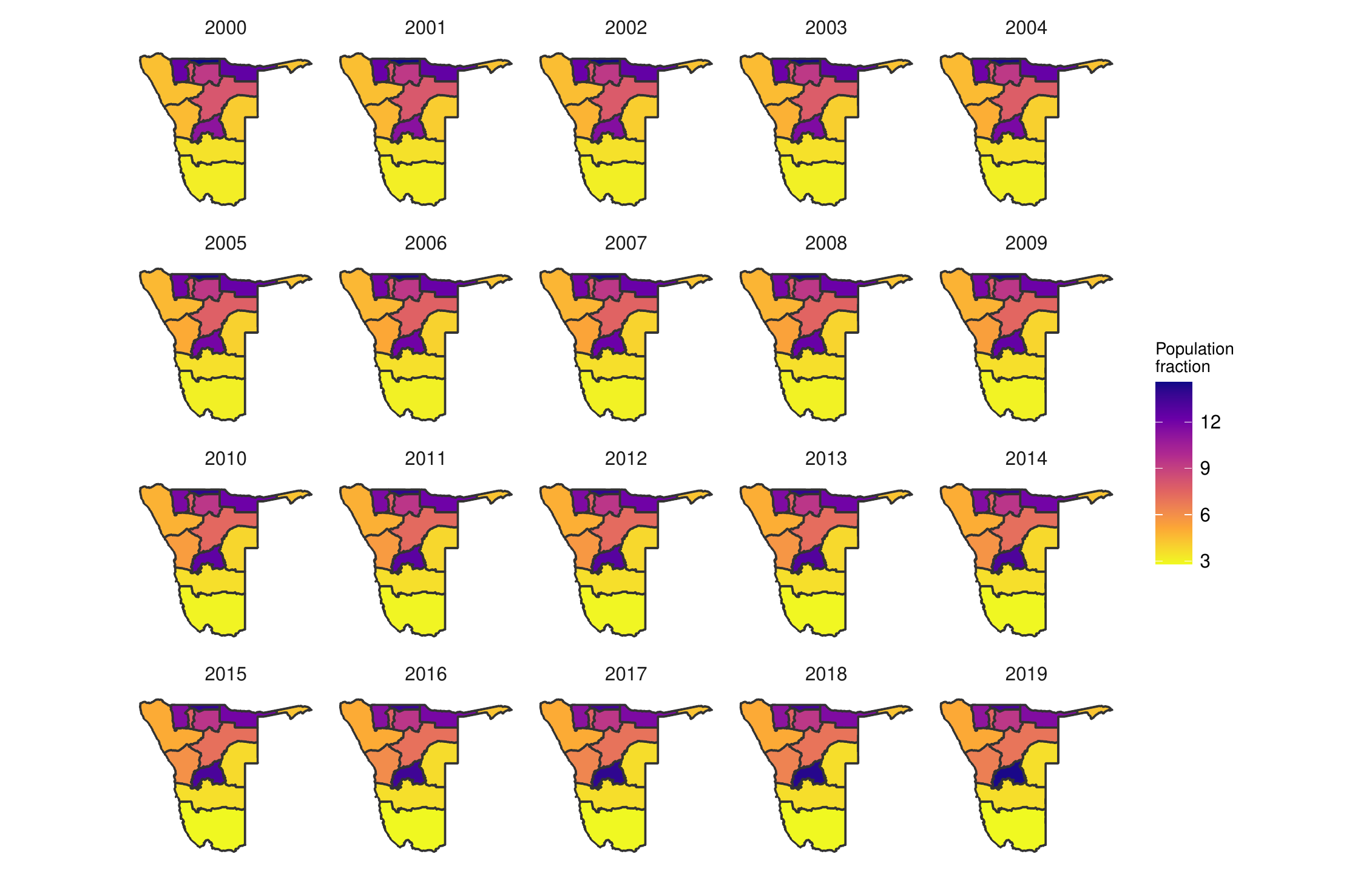}
		\caption{Percentage of the under-5 population living in each region by year.}
		\label{fig:NAMpop}
	\end{figure}

\begin{figure}[H]
		\centering
		\includegraphics[scale = 0.7]{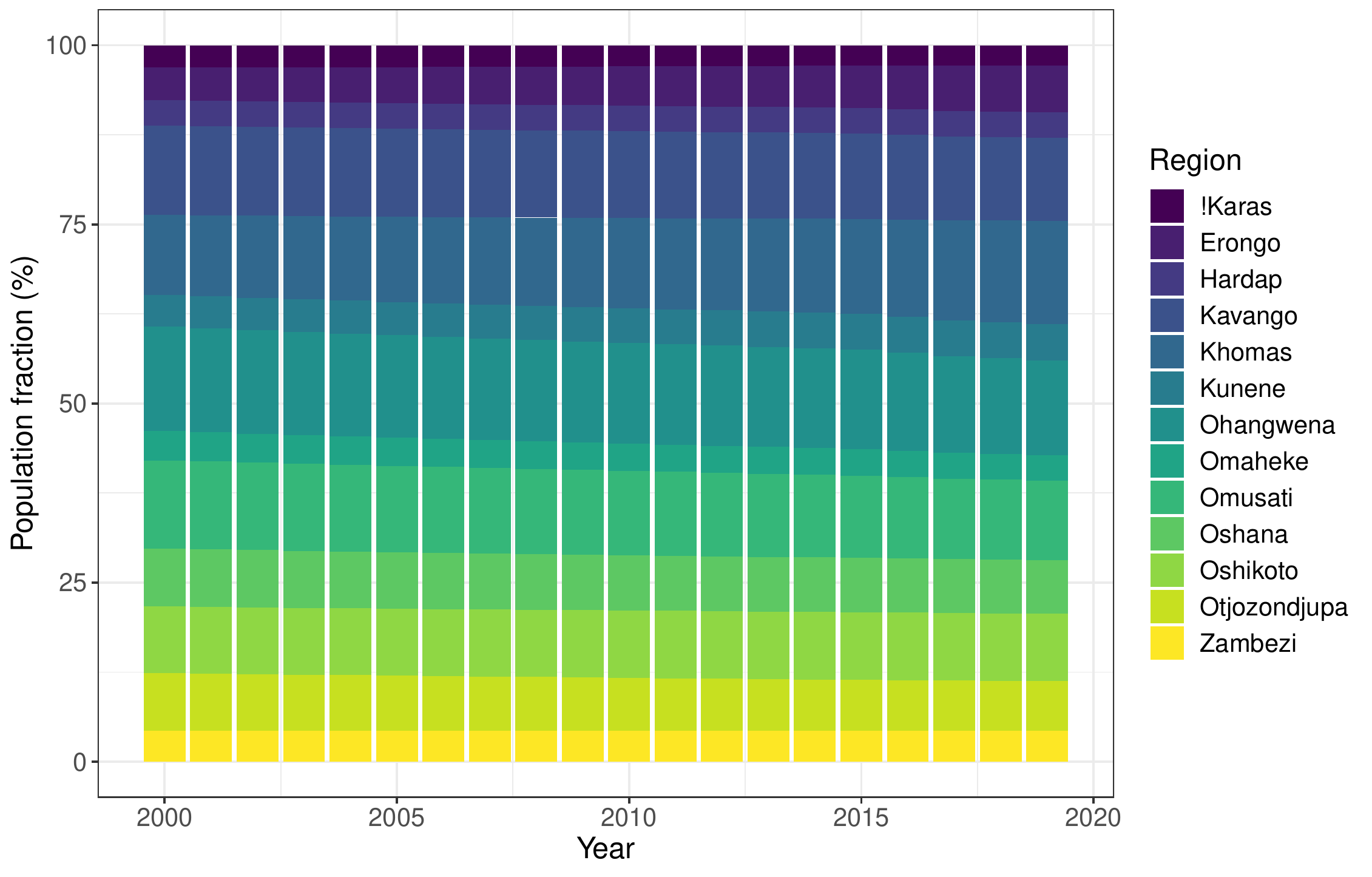}
		\caption{Percentage of the under-5 population living in each region by year.}
		\label{fig:NAMpop_bar}
	\end{figure}

	We use population count data from WorldPop for our U5MR application \autocite{tatem2017worldpop, stevens2015disaggregating}. Population counts are available at a 100 meter resolution, and using spatial boundary files from GADM, we aggregate population counts to the admin1 level. The population count data from WorldPop prior to 2020 is not provided with any measure of uncertainty, and so we treat the estimated population proportions as fixed. The spatial distribution of the under five population from 2000-2019 in Namibia can be seen in Figure \ref{fig:NAMpop}. In Figure \ref{fig:NAMpop_bar}, the same data is displayed as a stacked bar chart over time.
	
	\subsection{Unbenchmarked models}
	
	\subsubsection{Area-level: Spatial Fay-Herriot}
	
	We obtain design-based, direct estimates ${}_{60} \hat{q}_{0it}^{HT}$ of U5MR and their variances for each area $i$ and time $t$, and smooth the direct estimates via a spatio-temporal extension to the classic Fay-Herriot model described in \cite{li2019changes} \autocite{fay1979estimates, mercer2015space}. Direct estimates are combined across surveys using a standard meta-analysis approach and an HIV bias correction is done, both of which are described in \cite{li2019changes} in detail.
	
	\subsubsection{Unit-level: Binomial}
	
	From each DHS survey, we have records of birth and death dates for each child born to a mother sampled in the survey. These records are expanded into monthly binary outcomes that indicate whether death occurs for a given child in a given month, allowing us to perform a discrete time survival analysis with discrete hazards by age groups. In this way, children who survive longer contribute more observations to our data. The binary outcomes are collapsed to binomial observations in clusters $c$ within area $i$ in year $t$ from survey $s$ for age groups defined as 0-1, 1-11, 12-23, 24-35, 36-47, and 48-59 months. Let $m = 0, 1, \dots, 59$ denote age in months, and let
	
	\begin{alignat*}{3}
		\begin{aligned} 
			a[m] = & \begin{cases}
				1, 0 \leq m < 1 \\
				2, 1 \leq m < 12 \\
				3, 12 \leq m < 24 \\
				4, 24 \leq m < 36 \\
				5, 36 \leq m < 48 \\
				6, 48 \leq m < 60
			\end{cases}
		\end{aligned} & \hskip 6em &
		\begin{aligned}
			a^*[m] = & \begin{cases}
				1, 0 \leq m < 1 \\
				2, 1 \leq m < 12 \\
				3, 12 \leq m < 60 
			\end{cases}
		\end{aligned}
	\end{alignat*}
	We have number of deaths observed $y_{1m, i[c]ts}$ in $n_{m,i[c]ts}$ months of life in cluster $c$ within area $i$, year $t$ in agemonth $m$ from survey $s$. We consider the unbenchmarked model
	\begin{align}
		\label{eq:u5mrunbenched}
		\begin{split}
			y_{1m,i[c]ts} \mid n_{m,i[c]ts}, {}_1q_{m,i[c]ts} & \sim \text{Binomial}(n_{m,i[c]ts},{}_1q_{m,i[c]ts}) \\
			\eta_{m,i[c]ts} = \text{logit}({_1q_{m,i[c]ts}}) & = \mu_{a[m]} + \alpha_t + \gamma_{a^*[m]t} + b_i \\
			& + \beta_i \times t + \delta_{it} + \nu_s + e_{i[c]} + \log(\text{HIV}_{ts}), 
		\end{split}
	\end{align} 
	and aggregate to a space/time/age level via
	$$
	{_1q_{m,it}} = \text{expit} \left( \frac{\mu_{a[m]} + \alpha_t + \gamma_{a^*[m]t} + b_i + \beta_i \times t + \delta_{it}}{\sqrt{1 + h^2 \sigma^2_e}} \right),
	$$
	where ${}_1q_{m,it}$ is the monthly hazard of death for age $m$. The linear predictor contains fixed intercepts for age bands $\mu_{a[m]}$, an iid temporal random effect $\alpha_t \sim N(0, \tau_\alpha)$, and age group-specific temporal random effects following a random walk 2 (RW2) $\gamma_{a^*[m]t}$ \autocite{lindgren2008second}. The term $b_i$ is a spatial random effect following a BYM2 prior \autocite{riebler2016intuitive, besag1991bayesian}. A sum-to-zero constraint is placed on the structured component of the BYM2 random effect to ensure identifiability. Two separate terms allow for space-time interactions: $\beta_i$ are random, area-specific slopes, and $\delta_{it}$ is a Type IV Knorr-Held interaction term \autocite{knorr2000bayesian}. Finally we have a survey fixed effect $\nu_s$ which we constrain to sum to zero, and a log offset for HIV, varying by survey and time. These log offsets account for the differing survival probabilities of children born to HIV-positive mothers \autocite{walker2012child}. The inclusion of a log offset term in our linear predictor corresponds approximately to a multiplicative bias correction for mothers who could not be sampled due to death from HIV \autocite{wakefield2019estimating}. 
	
	We obtain predictions of U5MR at each time point ${}_{60}q_{0t}$ by calculating posterior draws $k = 1, \dots, K$ of all terms from Equation \ref{eq:u5mrunbenched} and the formula
	\begin{align*}
		{}_{60}q_{0it}^{(k)} & = 1 - \prod_{m = 1}^{59} \left( 1 - \text{expit} \left( \frac{\mu_{a[m]}^{(k)} + \alpha_t^{(k)} + \gamma_{a^*[m]t}^{(k)} + b_i^{(k)} + \beta_i^{(k)} \times t + \delta_{it}^{(k)}}{\sqrt{1 + h^2 \sigma^{2(k)}_e}} \right) \right).
	\end{align*}
	Note that we do not include the survey fixed effect, cluster random effect, or log offset in our predictions. 
	
	For the variance parameters in all random effects, we use PC priors with parameters $U = 1$, $\alpha = 0.01$. For the age-specific intercepts, we use the default INLA prior, which is normal, centered at $0$ with precision $0.001$. For the $\phi$ parameter in the BYM2 spatial random effect, we use a PC prior with parameters $U = 0.5$, $\alpha = 2/3$. These prior and hyperprior choices are used in all benchmarked models below, when applicable. 
	
	\subsection{Benchmarked models}
	
	\subsubsection{Approaches}

	\noindent \textbf{Raking:}
	
	\noindent For the unit-level model, we obtain posterior medians ${}_{60}\hat{q}_{0it}^M$ from the unbenchmarked model for each area $i$ and time point $t$, and compute
	\begin{align*}
		R^{R1}_{t} & = \frac{\sum_{i = 1}^m w_{it} \times  {}_{60}\hat{q}_{0it}^M}{y_{2t}}
	\end{align*}
	using weights $w_{it}$ and national benchmarks are each time point $y_{2t}$. We then refit the unbenchmarked model including $R^{R1}_t$ as a log offset term in the linear predictor to obtain benchmarked posterior draws ${}_{60}\hat{q}^{R1(k)}_{0it}$. 
	
	For the area-level model, we adjust the unbenchmarked direct estimates with the ratio $R^{R1}_t$ in the same manner we adjust for HIV, as described in \cite{li2019changes}, and then proceed fitting the smoothed direct model with these HIV- and benchmark-adjusted direct estimates. \\

	\noindent \textbf{Fully Bayesian: Rejection sampler:}
	
	\noindent We obtain $k = 1, \dots, K$ posterior draws ${}_{60}\hat{q}_{0it}^{(k)}$ from the unbenchmarked model for each area $i$ and year $t$, and apply the algorithm described in the proposed approach section of the main body of the paper to obtain benchmarked posterior draws ${}_{60}\hat{q}^{FB2}_{0it}$. 
	
	\noindent \textbf{Fully Bayesian: Metropolis-Hastings algorithm:}
	
	\noindent We obtain $k = 1, \dots, K$ posterior draws $\hat{\theta}_{i}^{(k)}$ from an \textit{adjusted} unbenchmarked model for each area $i$, and apply the algorithm described in the proposed approach section of the main body of the paper using weights $w_i$, and the benchmark $y_2$ with associated standard error $\sigma^2_{y_2}$, to obtain benchmarked posterior draws $\hat{\theta}_i^{FB3(k)}$. The adjusted unbenchmarked model uses the prior $\pi^+(\beta_0) \sim \text{Normal}(\text{logit}(y_2), 0.001^{-1}).$
	
	\subsection{Software}
	
	We implement the unbenchmarked and raking benchmarked U5MR models in INLA via the R package \texttt{SUMMER} \autocite{rue2009approximate, li2020space}, and the proposed benchmarking approach and Bayes estimate approach using a combination of INLA and R. INLA is a popular tool for conducting Bayesian, space-time analyses, and provides great computational gains compared to traditional MCMC methods. While nonlinear predictors can be incorporated via the R package \texttt{inlabru}, INLA itself cannot make use of nonlinear predictors by design \autocite{bachl2019inlabru}. As such, the fully Bayesian benchmarking approach described in \cite{zhang2020fully} cannot be fit in INLA. Since the unbenchmarked U5MR model is straightforward to fit in INLA using the \texttt{SUMMER} package, this application serves as an example of how the proposed rejection sampling approach to fully Bayesian benchmarking may be particularly useful for modelers who have already produced unbenchmarked estimates in an a programme that doesn't allow them to incorporate additional likelihood(s) for the benchmarks. 
	
	\subsection{U5MR Results}

\begin{figure}[H]
		\centering
		\includegraphics[scale = 0.5]{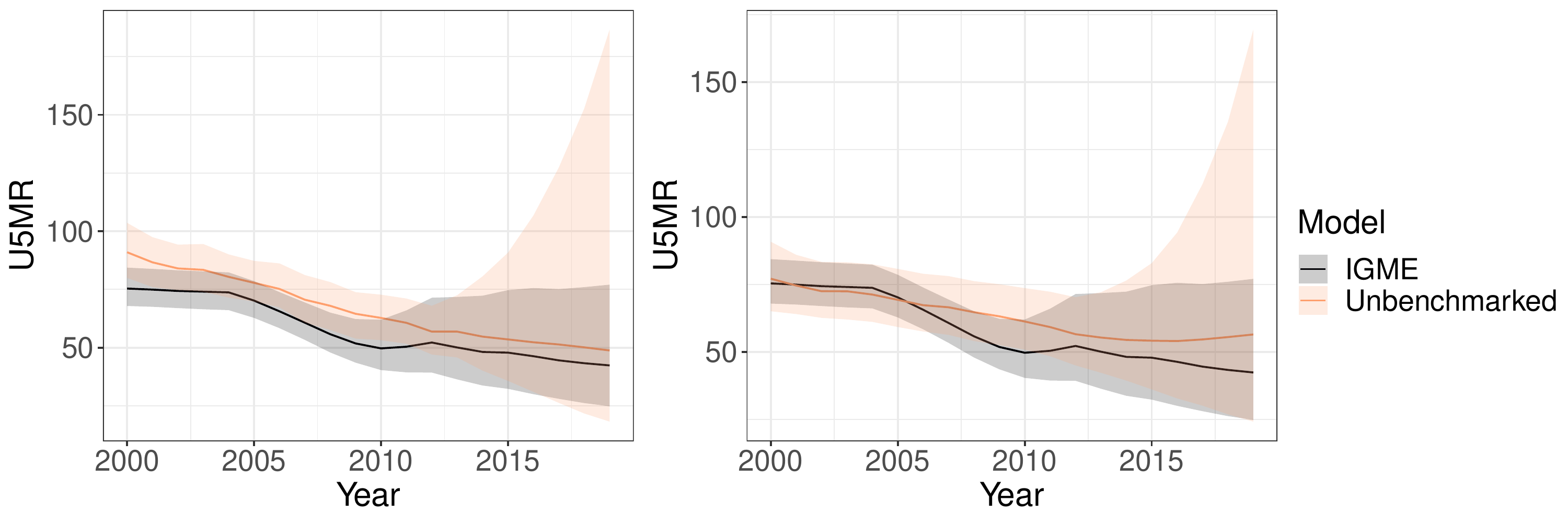}
		\caption{Aggregated national unbenchmarked medians from the area-level model (left) and unit-level model (right) compared to IGME national medians across time.}
		\label{fig:natlovertime}
	\end{figure}

\begin{figure}[H]
		\centering
		\includegraphics[scale = 0.3]{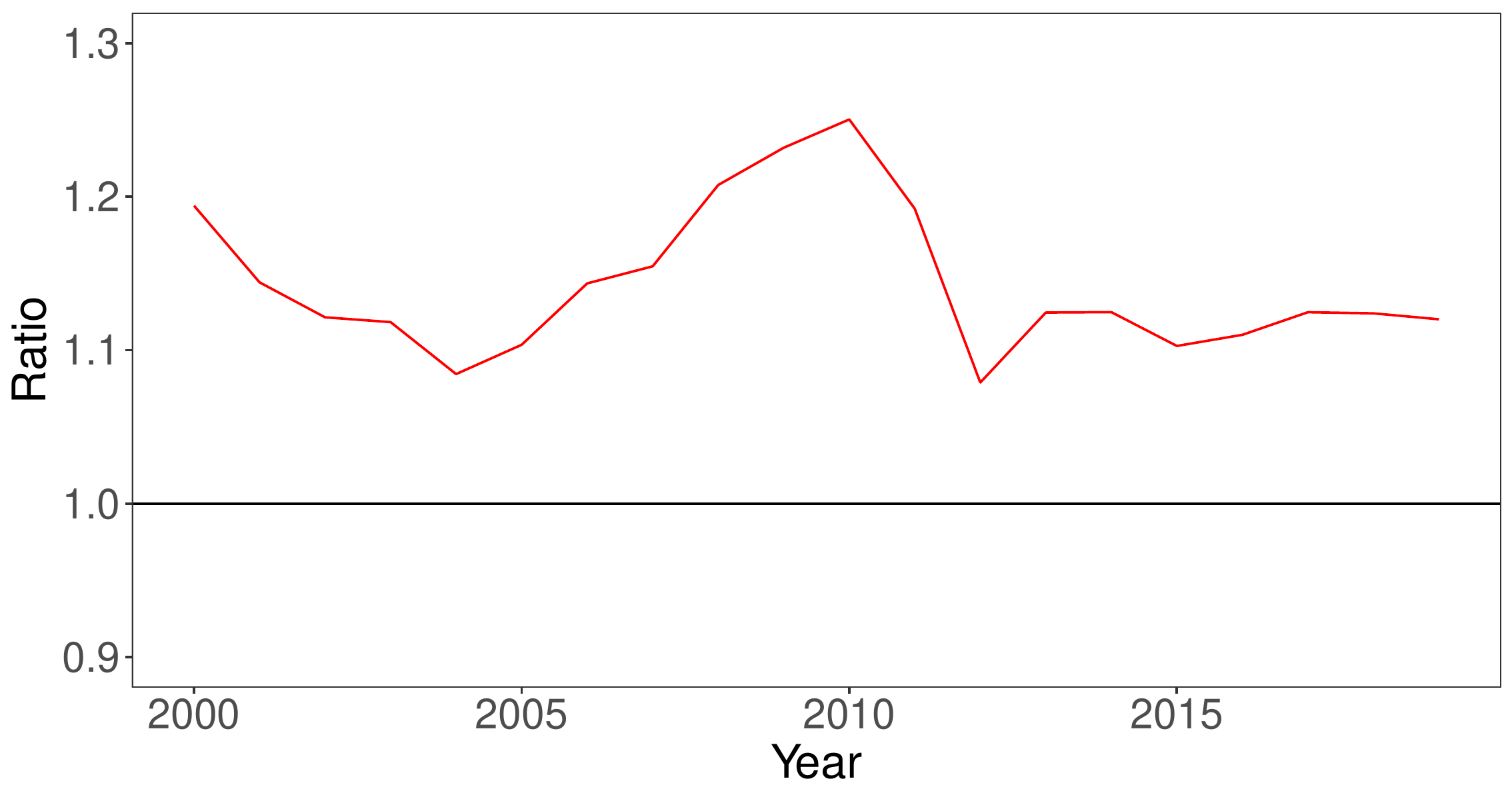} \hspace{1cm}
		\includegraphics[scale = 0.3]{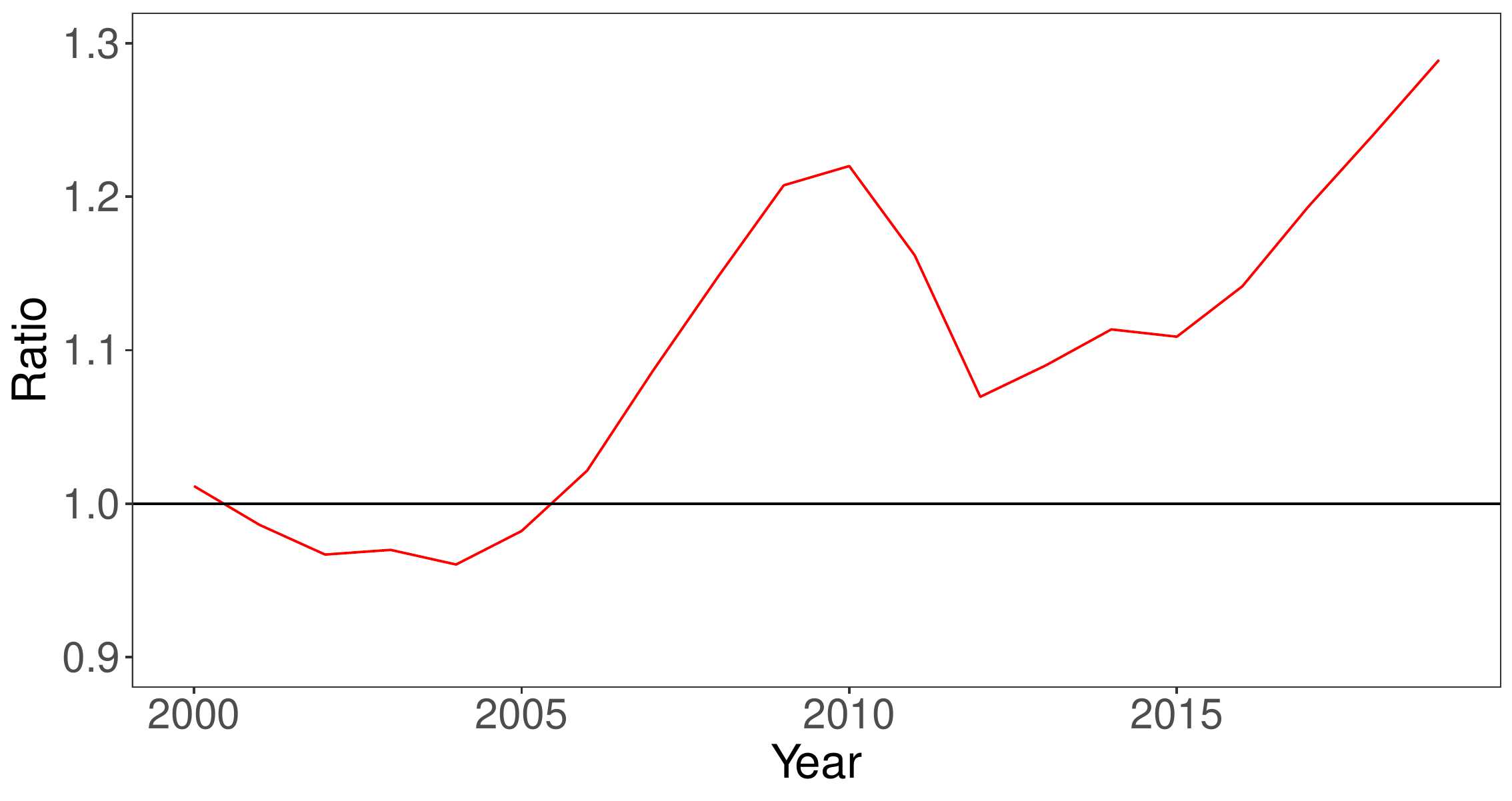}
		\caption{Ratio of aggregated national unbenchmarked medians from the area-level model (left) and unit-level model (right) to IGME national medians across time. A value greater than 1 corresponds to the unbenchmarked estimate being higher than the IGME estimate.}
		\label{fig:benchratio}
	\end{figure}

	The proposed rejection sampling approach to fully Bayesian benchmarking performs similarly in the U5MR example to the HIV prevalence example in that the posterior density of the aggregated national level estimate is a compromise between the national IGME density and the unbenchmarked density for both the unit-level and area-level models. Figure \ref{fig:benchratio} displays the ratio of unbenchmarked medians from the unit-level and area-level models to IGME national medians across time, with estimates and their confidence intervals shown in \ref{fig:natlovertime}. During ``prediction" years where there is no data informing estimates (2014 onward), the ratio for the unit-level model is particularly large, and we expect to see the greatest effect in benchmarking during these years. The ratio is particularly large likely due to the flexibility of the RW2 in time in this model, and the ratio is less large in prediction years for the area-level model. The ratio for the area-level model overall is greater for almost all years than the ratio for the unit-level model. 
	
\begin{figure}[H]
		\centering
		\includegraphics[scale = 0.3]{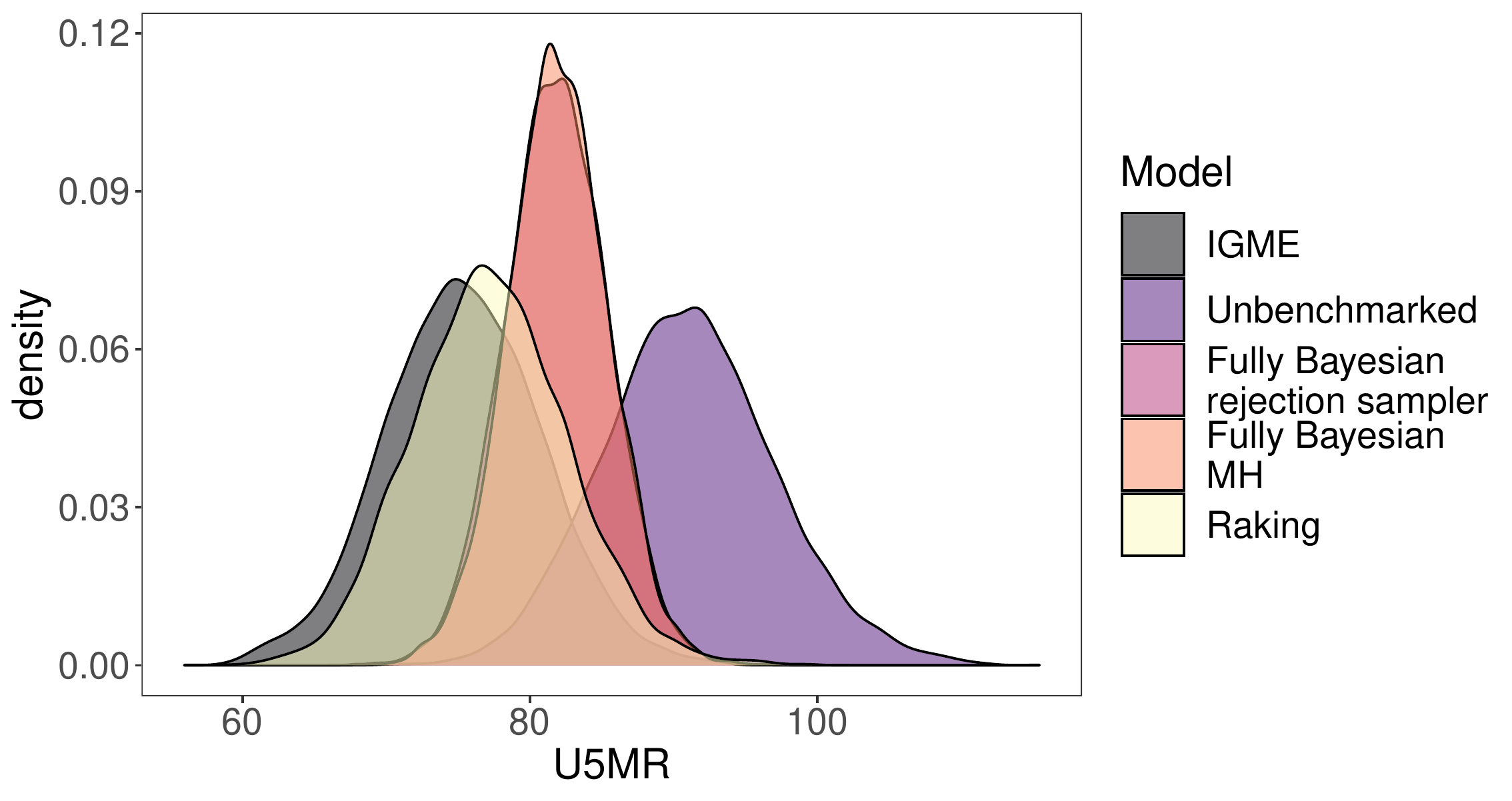}
		\includegraphics[scale = 0.3]{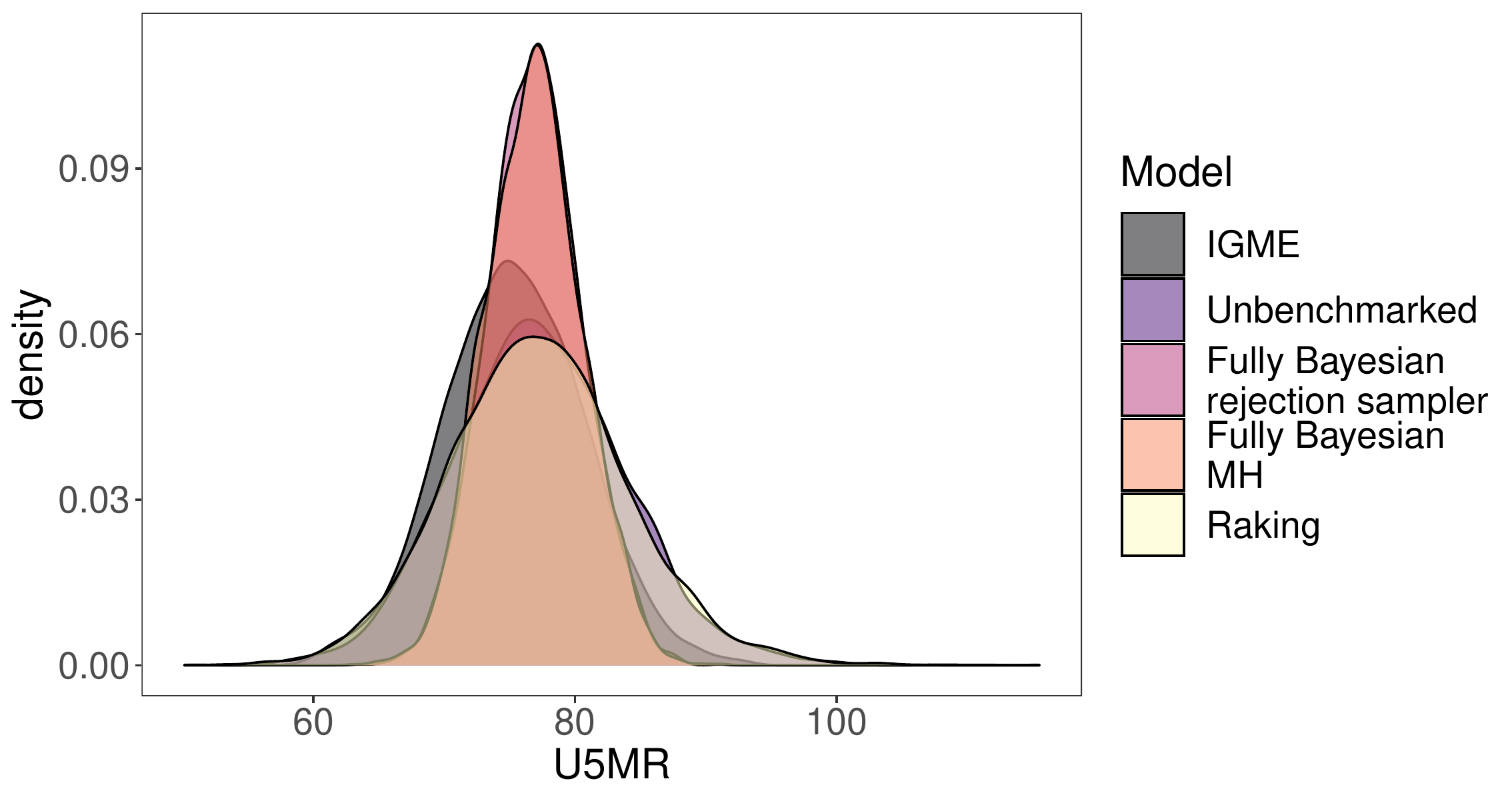}
		\caption{Aggregated national level U5MR estimates from IGME, unbenchmarked, and benchmarked models for area-level (left) and unit-level (right) models, for 2000. All densities are based on 5000 samples. U5MR is reported as deaths per 1000 live births.}
		\label{fig:u5mrnatlresults2000}
	\end{figure}

\begin{figure}[H]
		\centering
		\includegraphics[scale = 0.3]{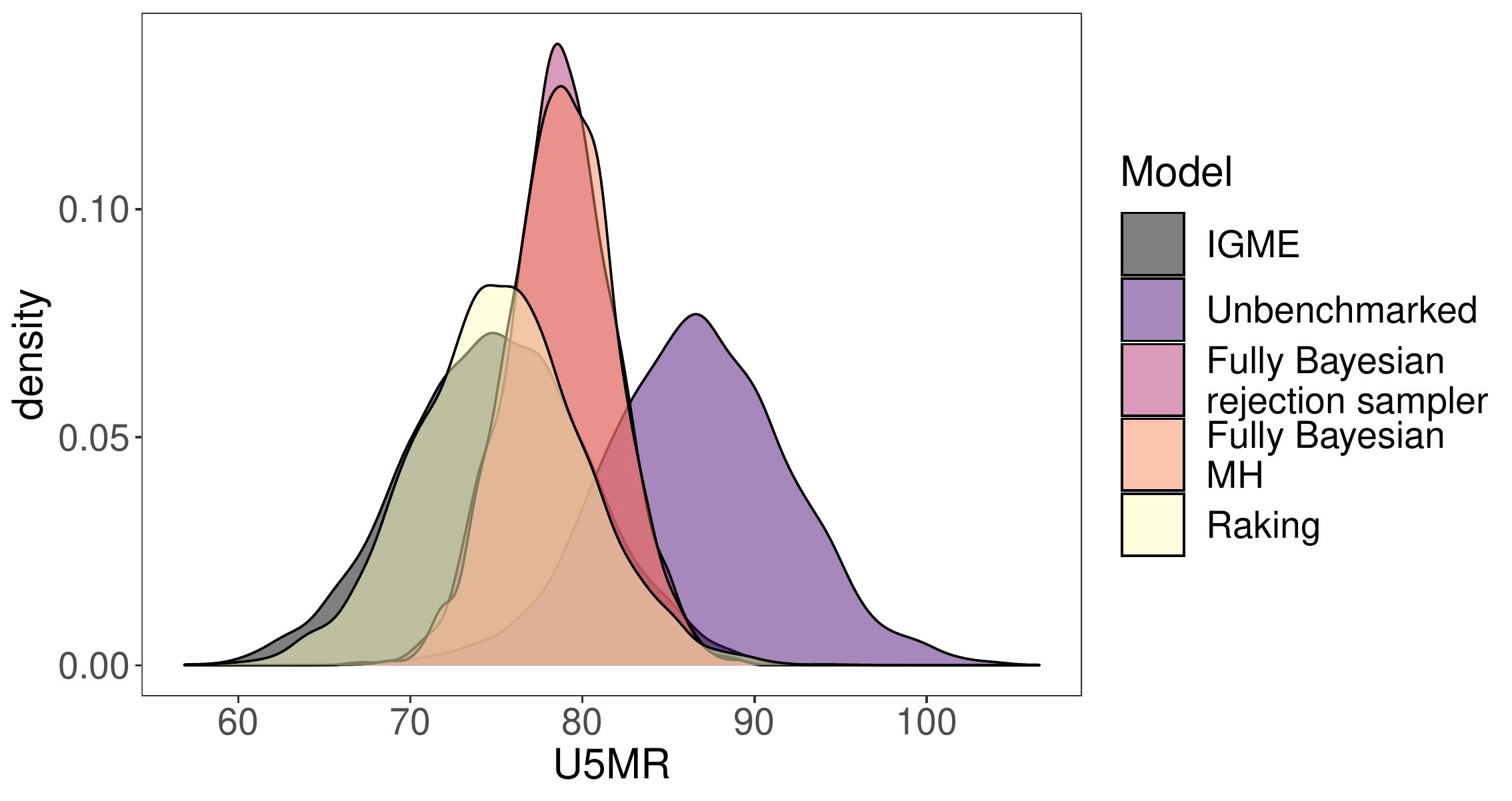}
		\includegraphics[scale = 0.3]{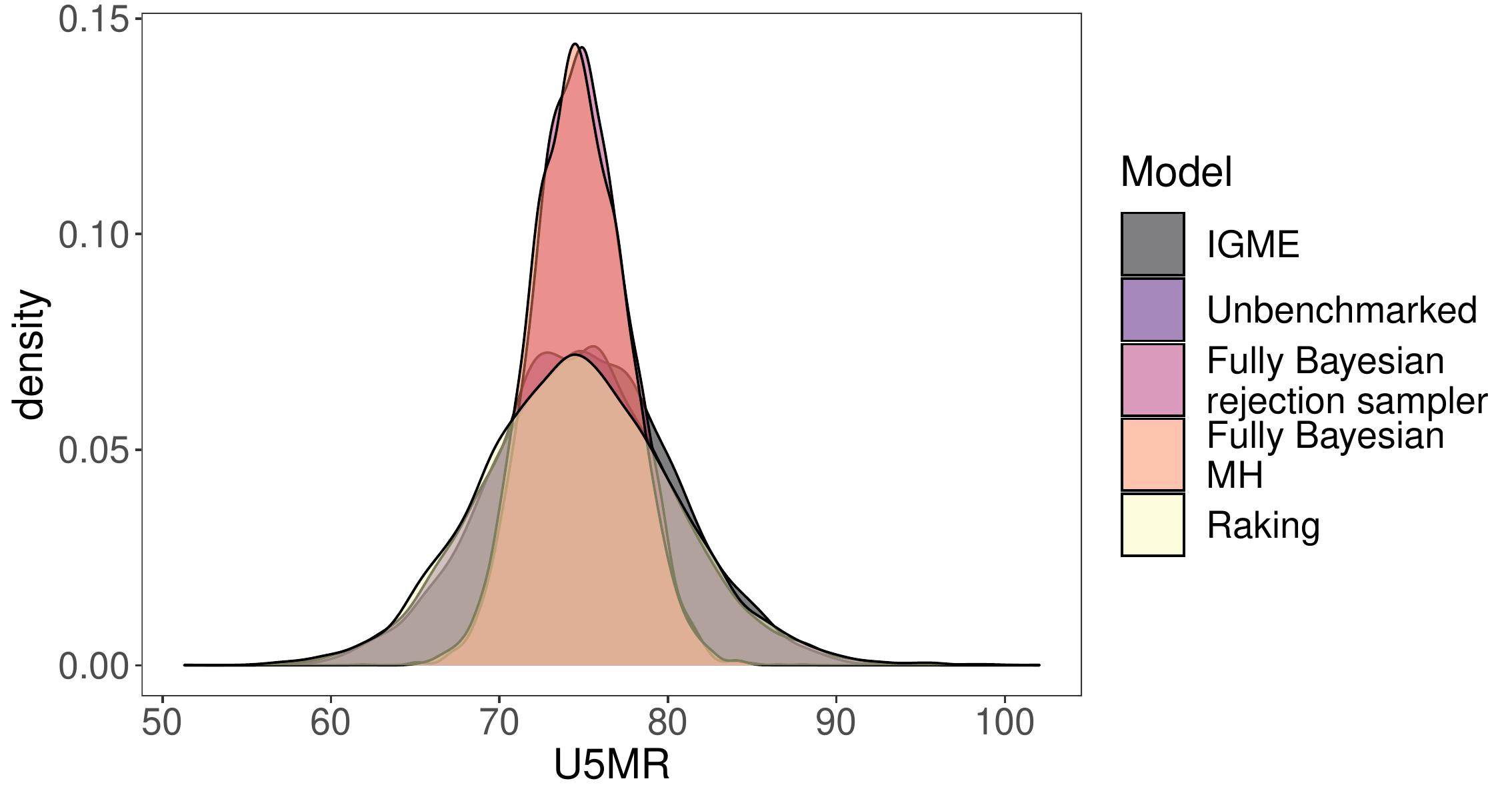}
		\caption{Aggregated national level U5MR estimates from IGME, unbenchmarked, and benchmarked models for area-level (left) and unit-level (right) models, for 2001. All densities are based on 5000 samples. U5MR is reported as deaths per 1000 live births.}
		\label{fig:u5mrnatlresults2001}
	\end{figure}

\begin{figure}[H]
		\centering
		\includegraphics[scale = 0.3]{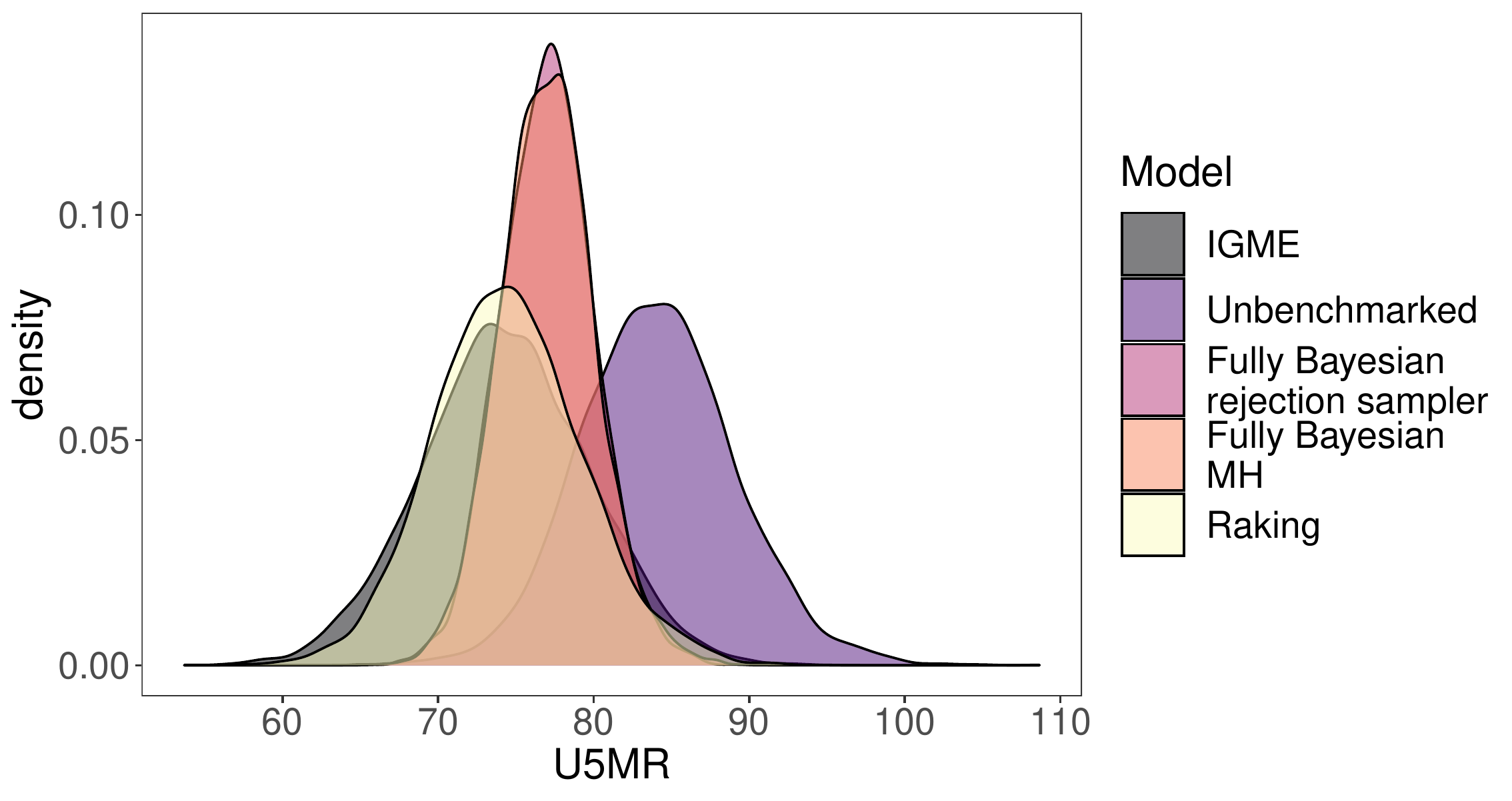}
		\includegraphics[scale = 0.3]{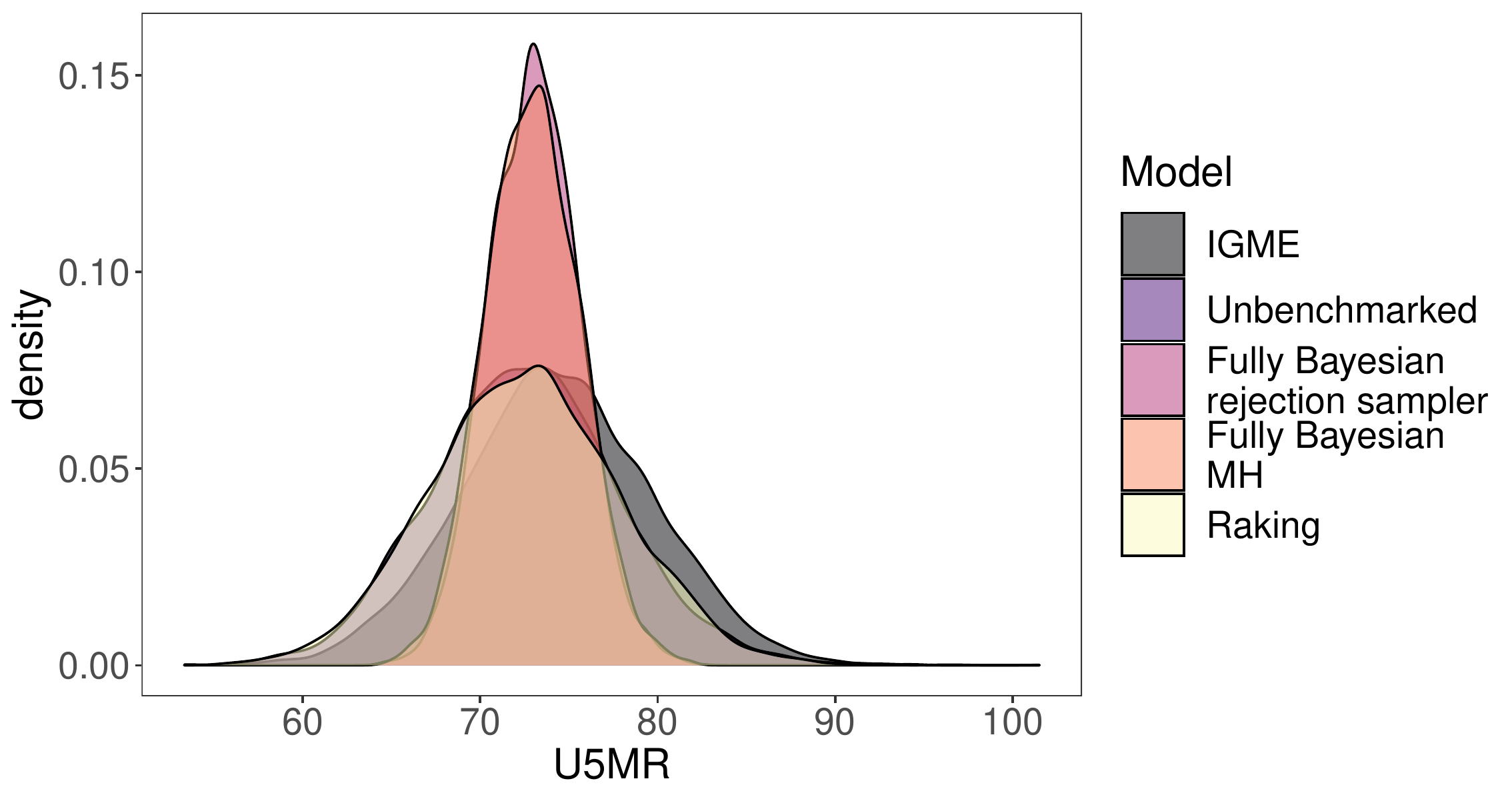}
		\caption{Aggregated national level U5MR estimates from IGME, unbenchmarked, and benchmarked models for area-level (left) and unit-level (right) models, for 2002. All densities are based on 5000 samples. U5MR is reported as deaths per 1000 live births.}
		\label{fig:u5mrnatlresults2002}
	\end{figure}

\begin{figure}[H]
		\centering
		\includegraphics[scale = 0.3]{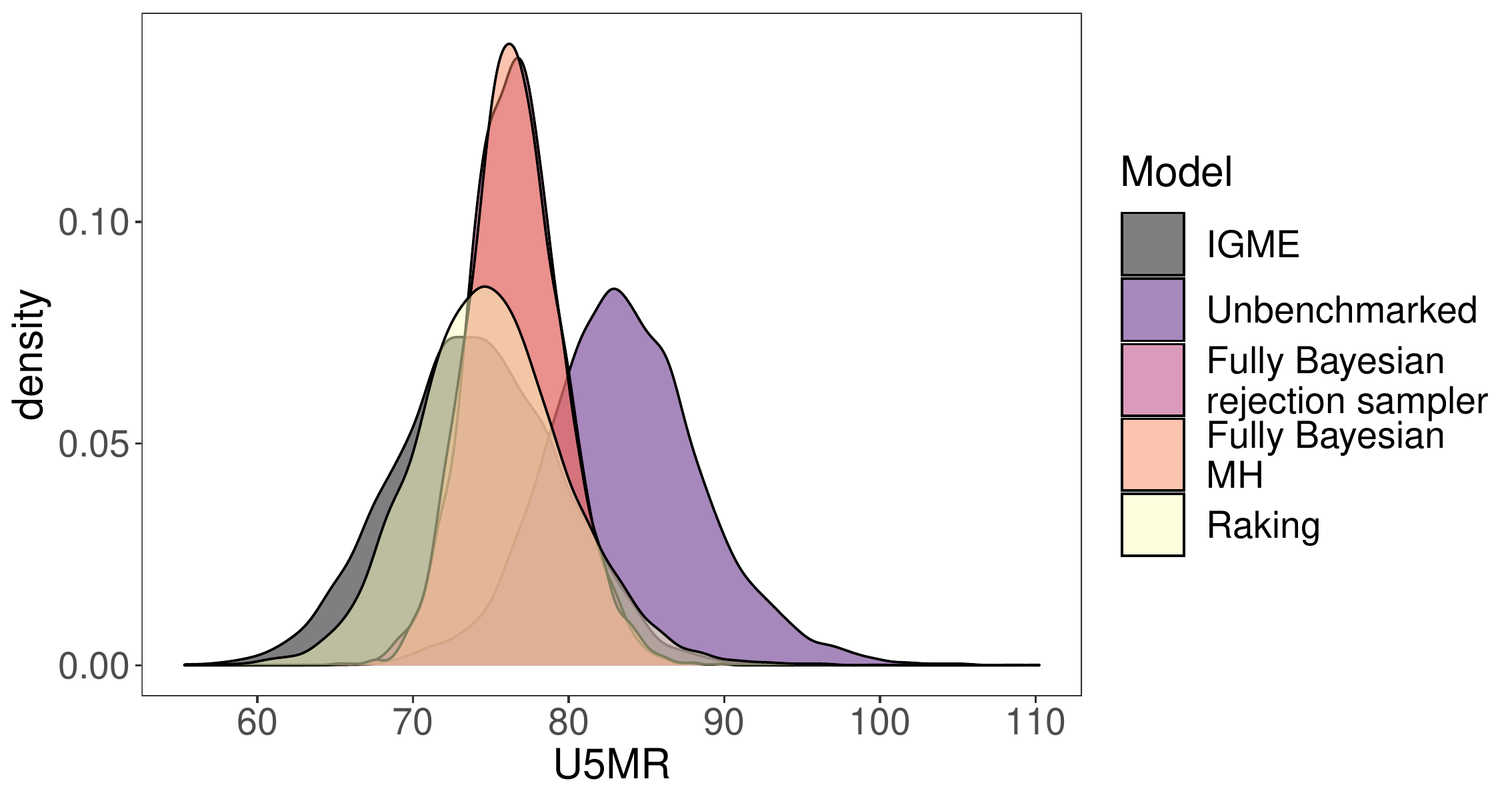}
		\includegraphics[scale = 0.3]{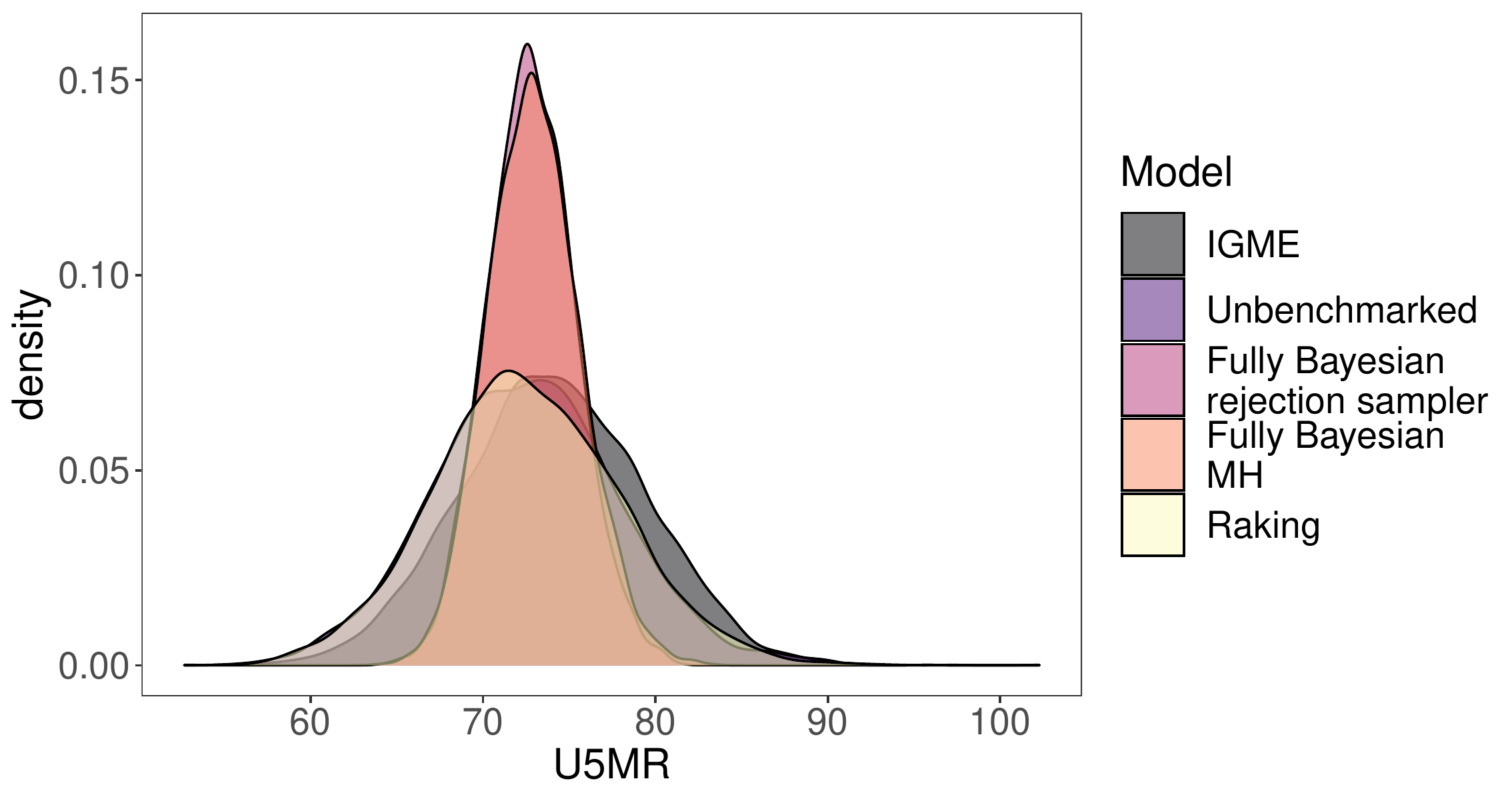}
		\caption{Aggregated national level U5MR estimates from IGME, unbenchmarked, and benchmarked models for area-level (left) and unit-level (right) models, for 2003. All densities are based on 5000 samples. U5MR is reported as deaths per 1000 live births.}
		\label{fig:u5mrnatlresults2003}
	\end{figure}

\begin{figure}[H]
		\centering
		\includegraphics[scale = 0.3]{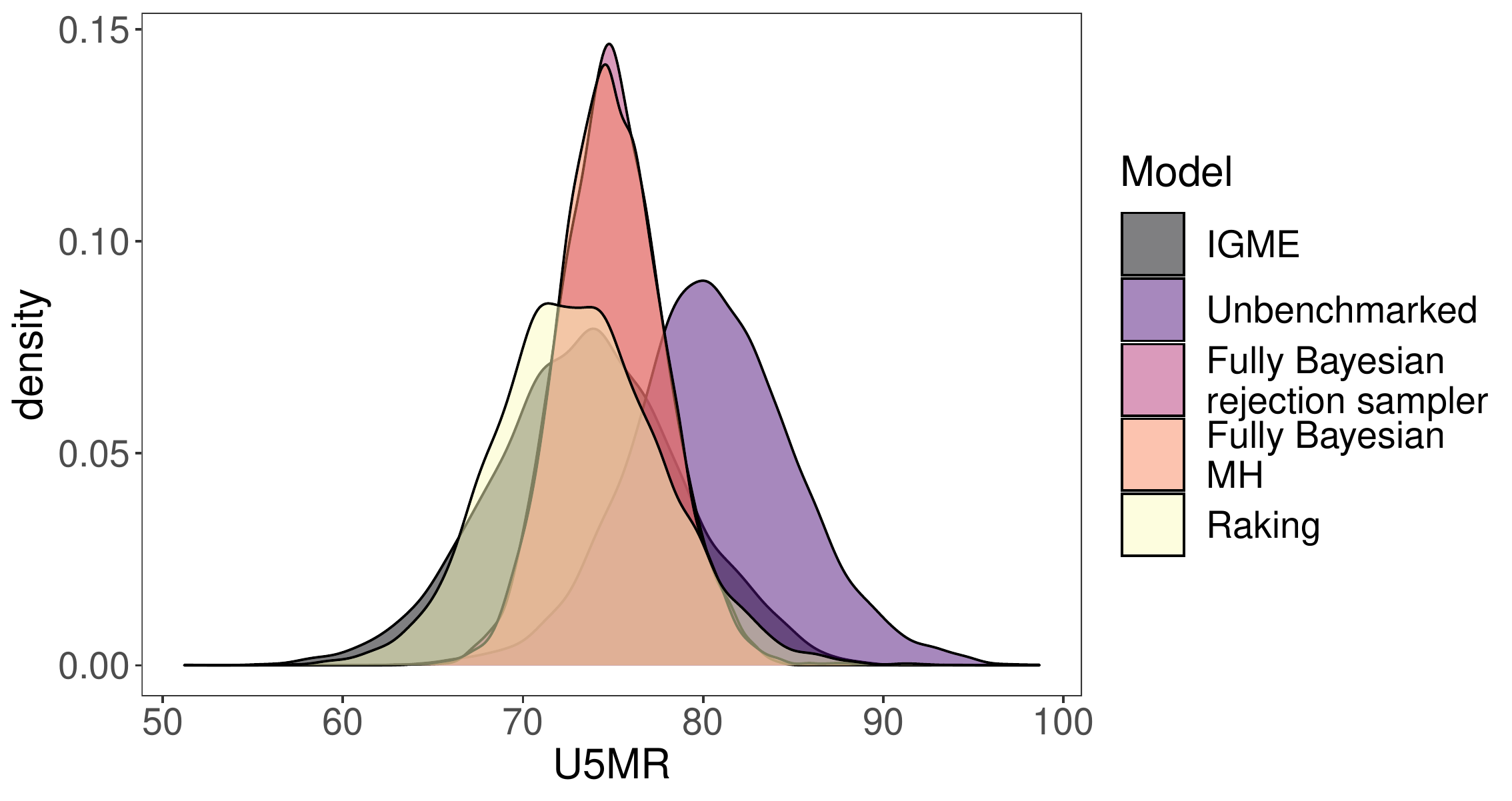}
		\includegraphics[scale = 0.3]{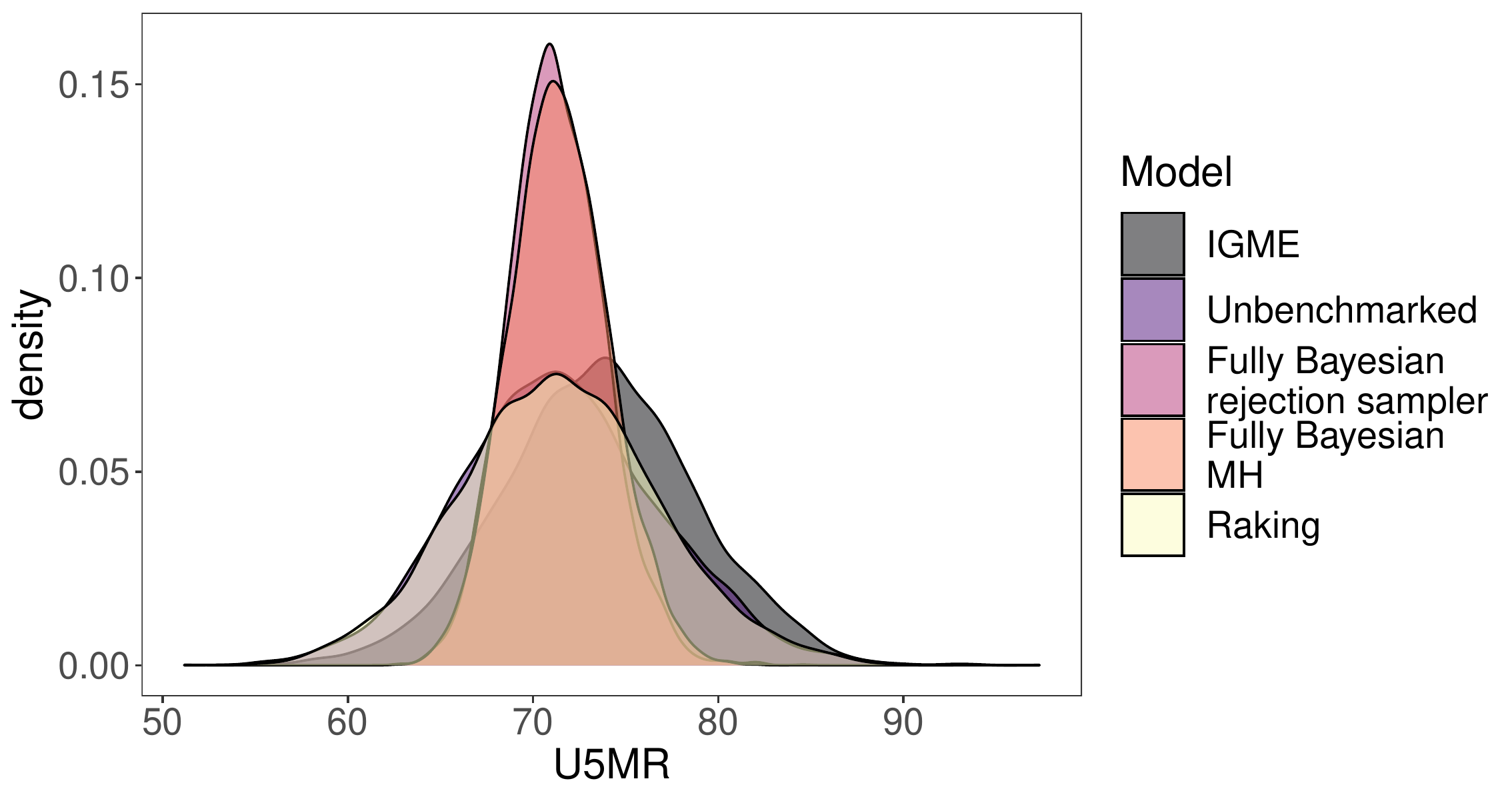}
		\caption{Aggregated national level U5MR estimates from IGME, unbenchmarked, and benchmarked models for area-level (left) and unit-level (right) models, for 2004. All densities are based on 5000 samples. U5MR is reported as deaths per 1000 live births.}
		\label{fig:u5mrnatlresults2004}
	\end{figure}

\begin{figure}[H]
		\centering
		\includegraphics[scale = 0.3]{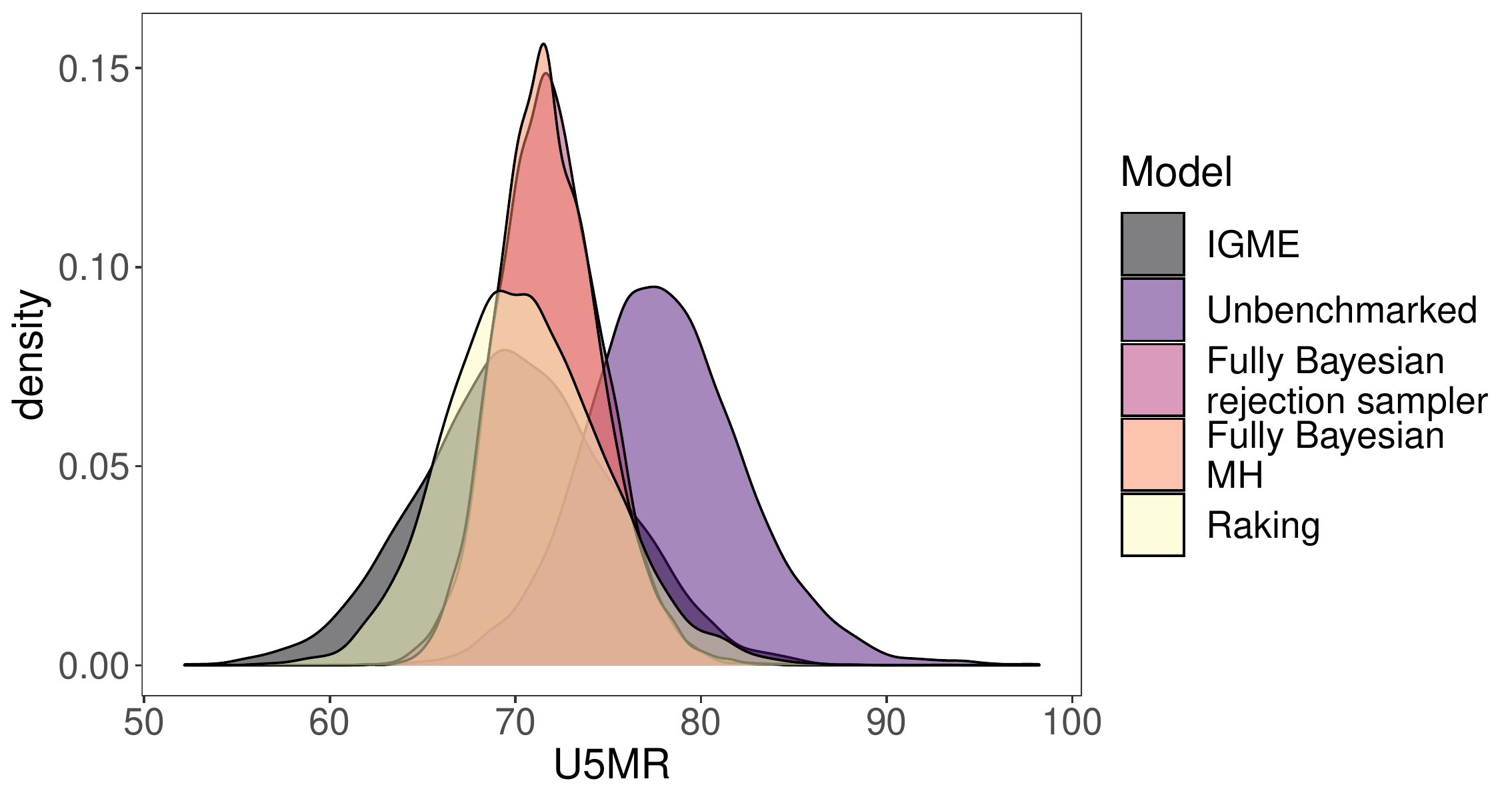}
		\includegraphics[scale = 0.3]{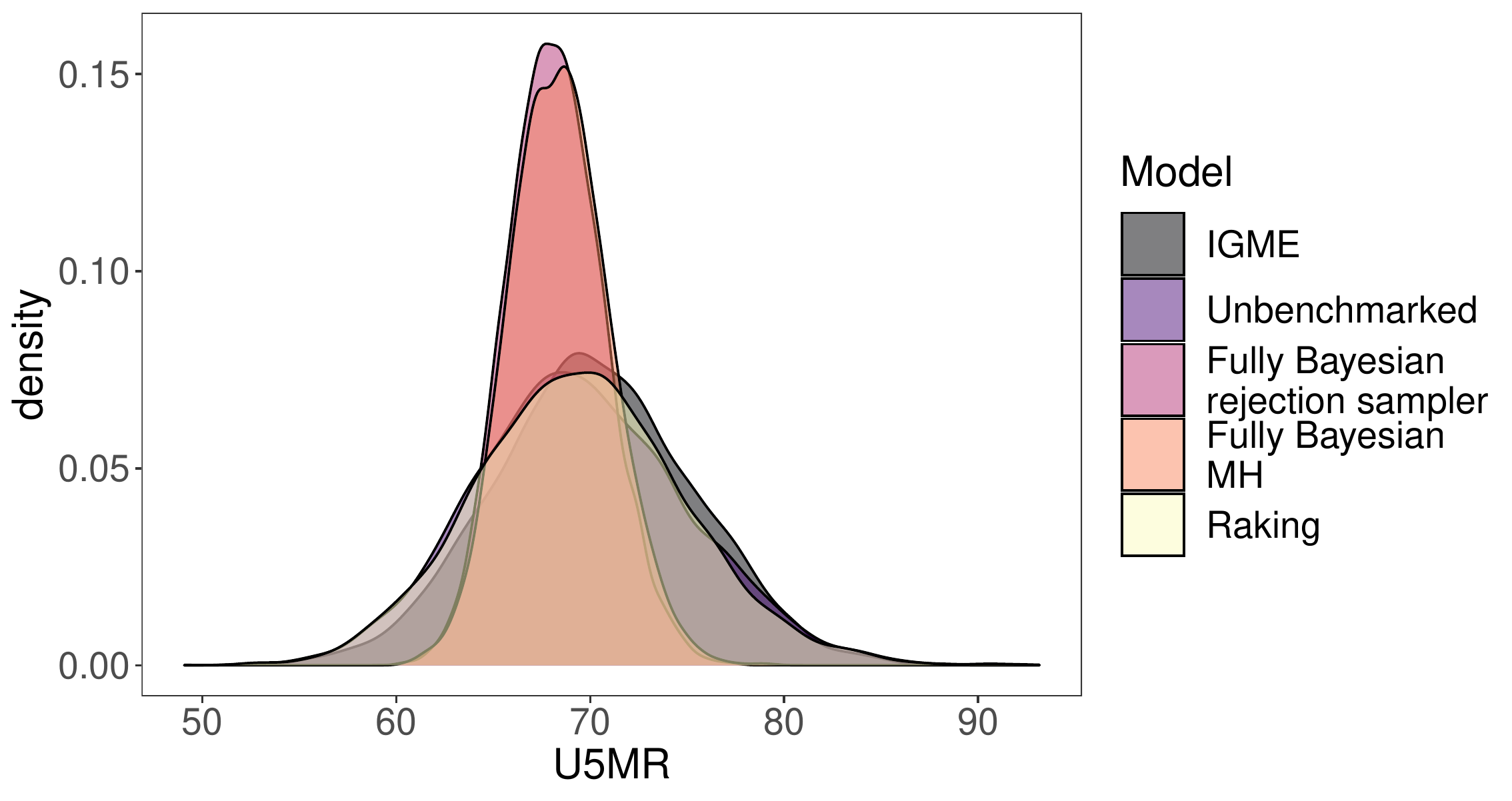}
		\caption{Aggregated national level U5MR estimates from IGME, unbenchmarked, and benchmarked models for area-level (left) and unit-level (right) models, for 2005. All densities are based on 5000 samples. U5MR is reported as deaths per 1000 live births.}
		\label{fig:u5mrnatlresults2005}
	\end{figure}

\begin{figure}[H]
		\centering
		\includegraphics[scale = 0.3]{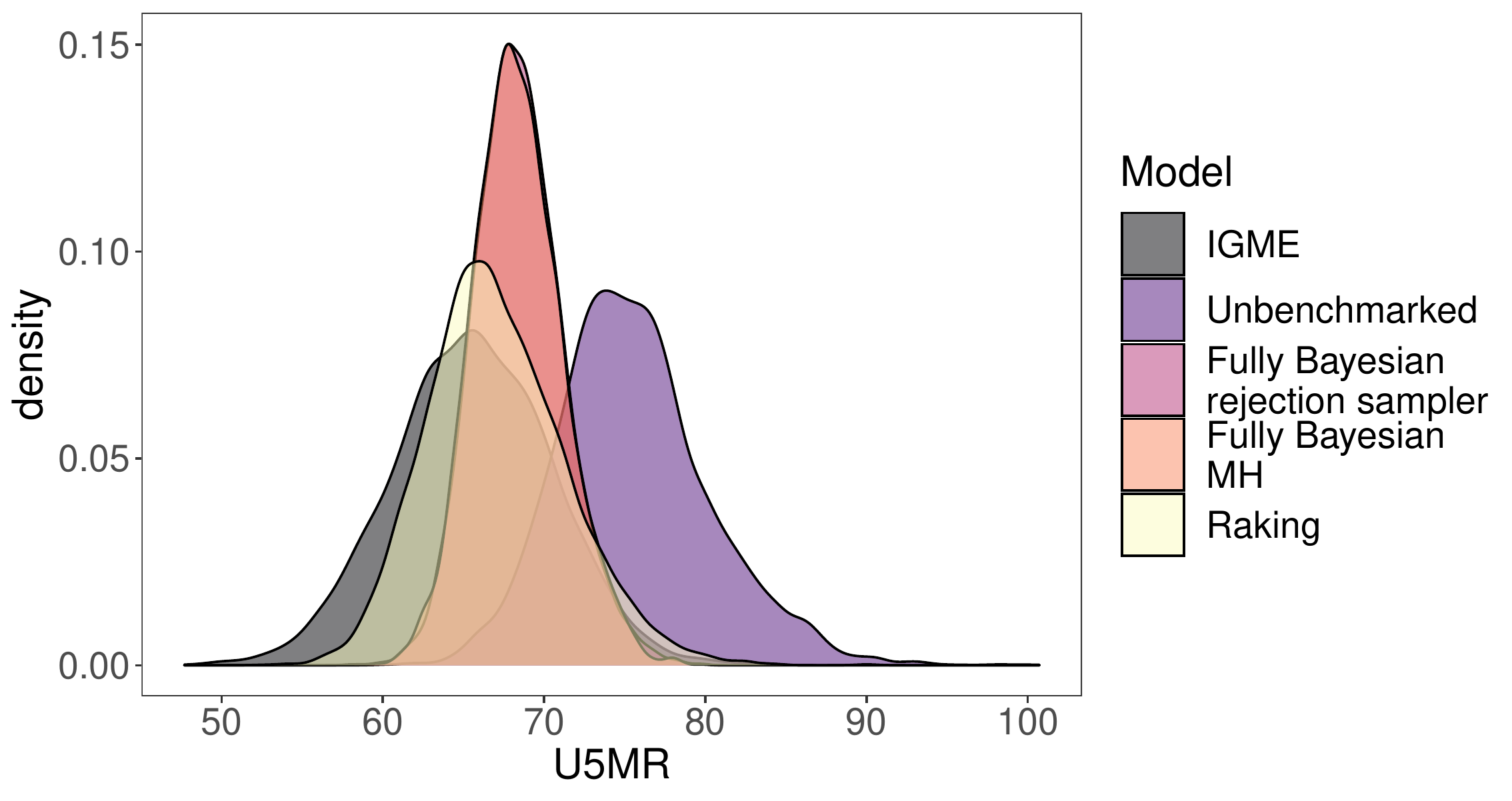}
		\includegraphics[scale = 0.3]{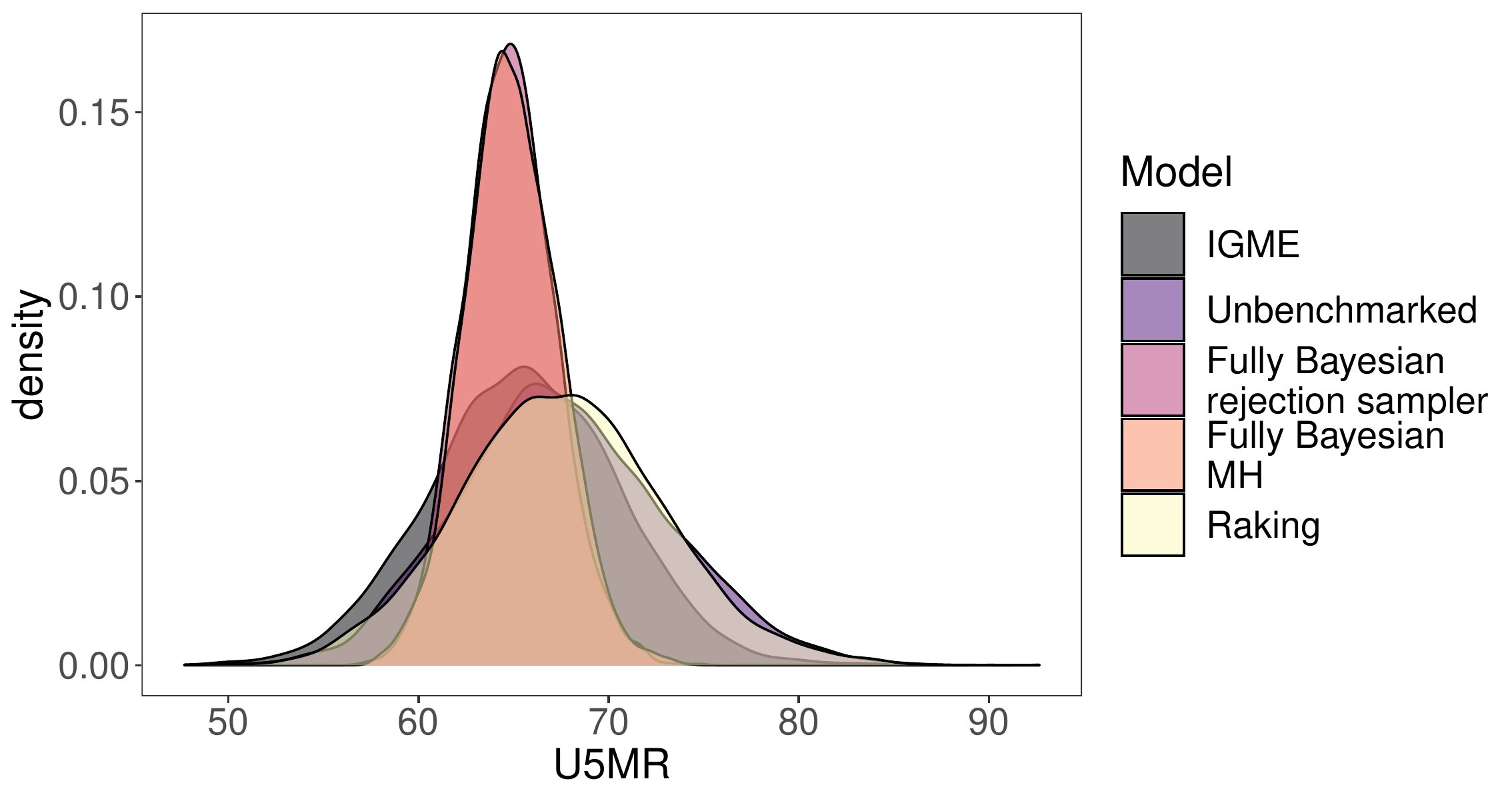}
		\caption{Aggregated national level U5MR estimates from IGME, unbenchmarked, and benchmarked models for area-level (left) and unit-level (right) models, for 2006. All densities are based on 5000 samples. U5MR is reported as deaths per 1000 live births.}
		\label{fig:u5mrnatlresults2006}
	\end{figure}

\begin{figure}[H]
		\centering
		\includegraphics[scale = 0.3]{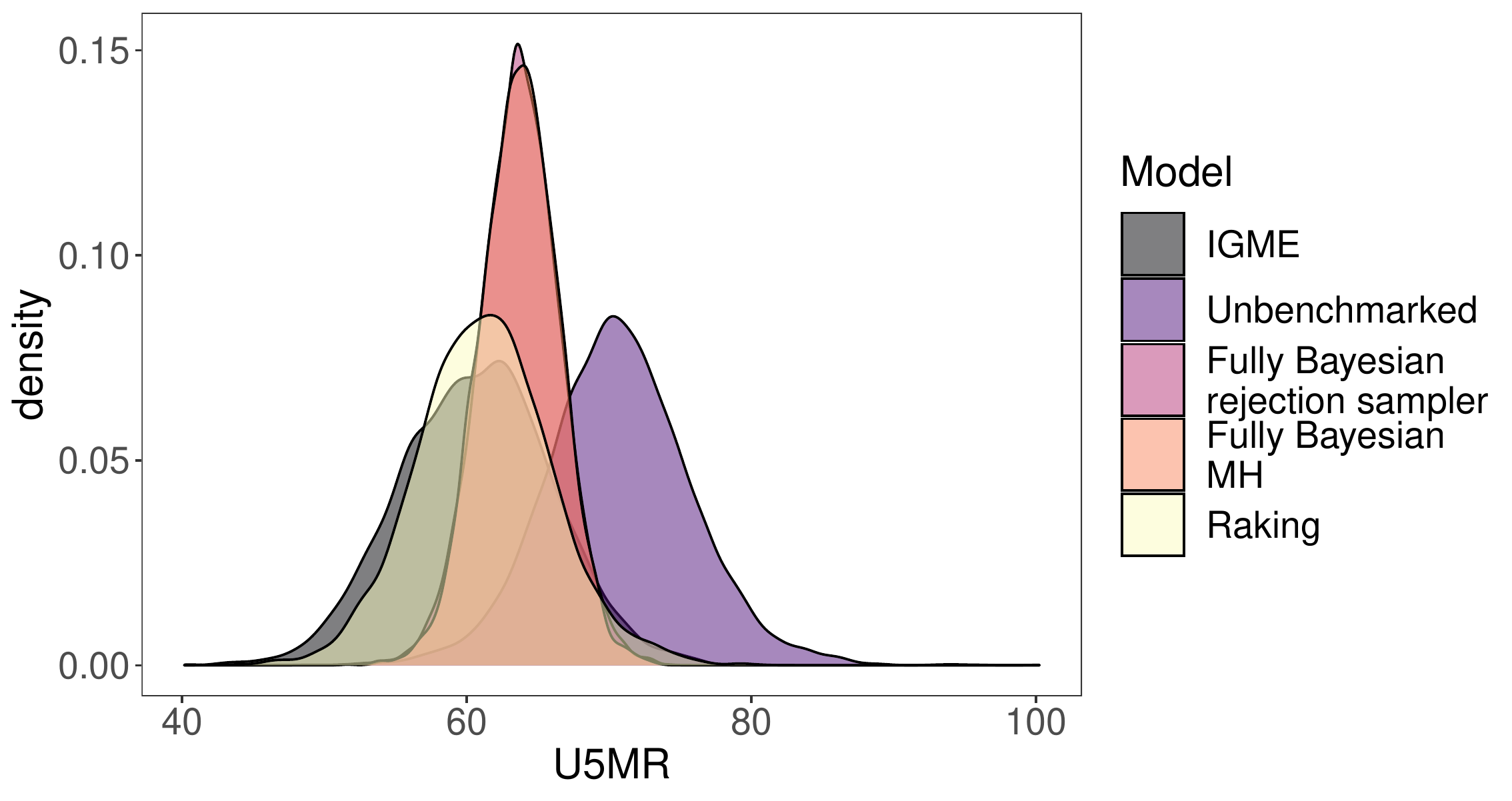}
		\includegraphics[scale = 0.3]{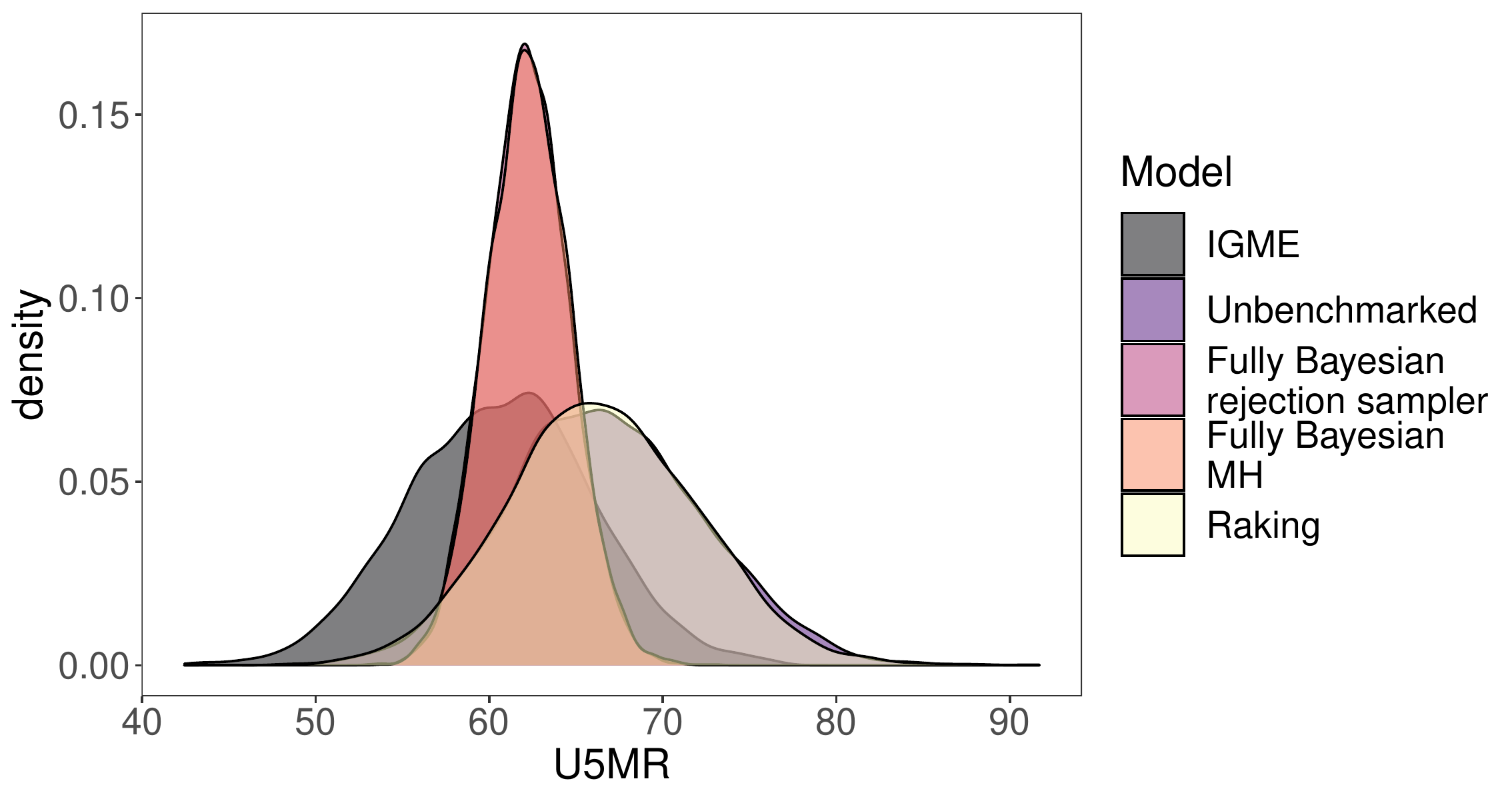}
		\caption{Aggregated national level U5MR estimates from IGME, unbenchmarked, and benchmarked models for area-level (left) and unit-level (right) models, for 2007. All densities are based on 5000 samples. U5MR is reported as deaths per 1000 live births.}
		\label{fig:u5mrnatlresults2007}
	\end{figure}

\begin{figure}[H]
		\centering
		\includegraphics[scale = 0.3]{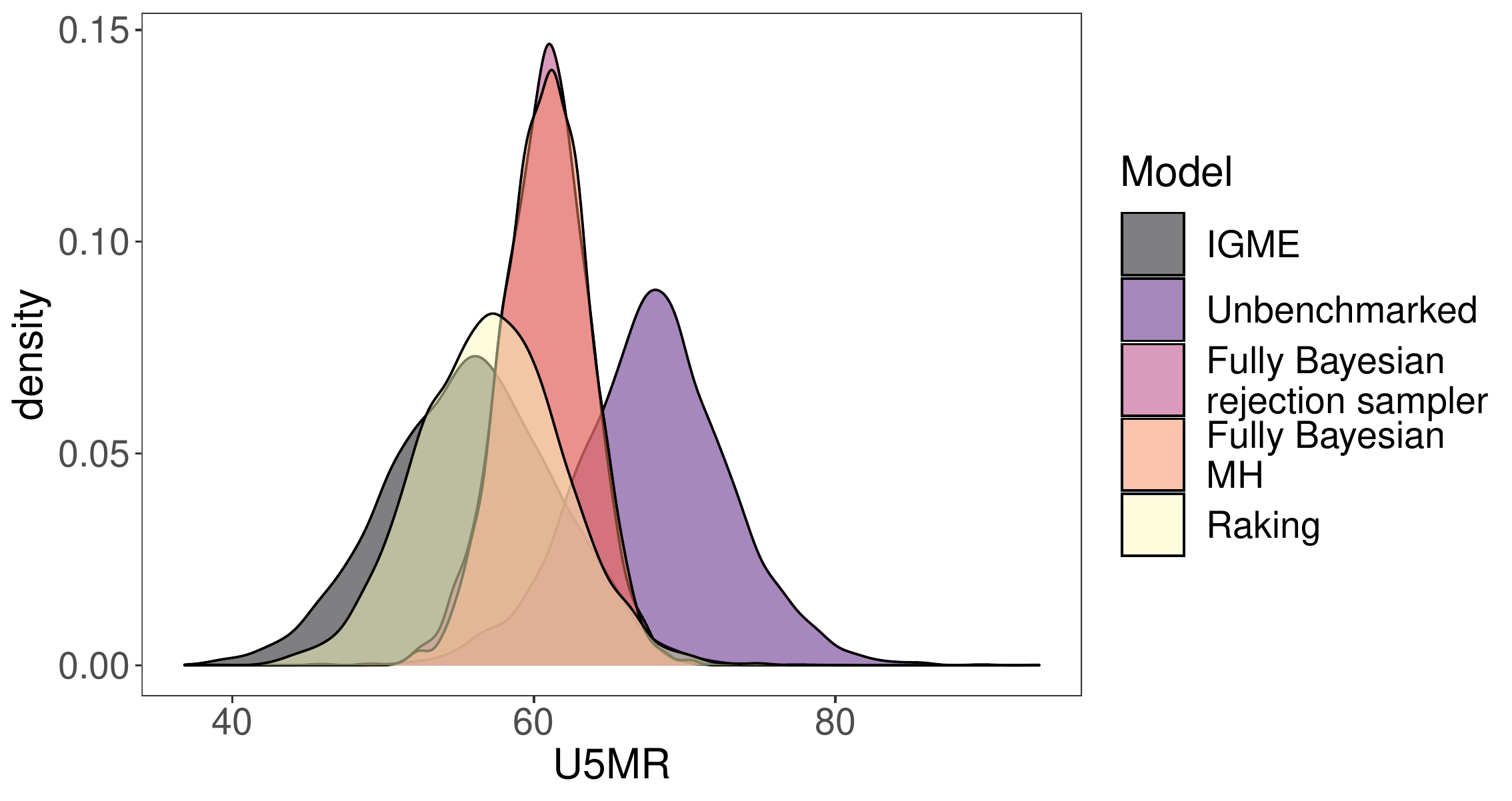}
		\includegraphics[scale = 0.3]{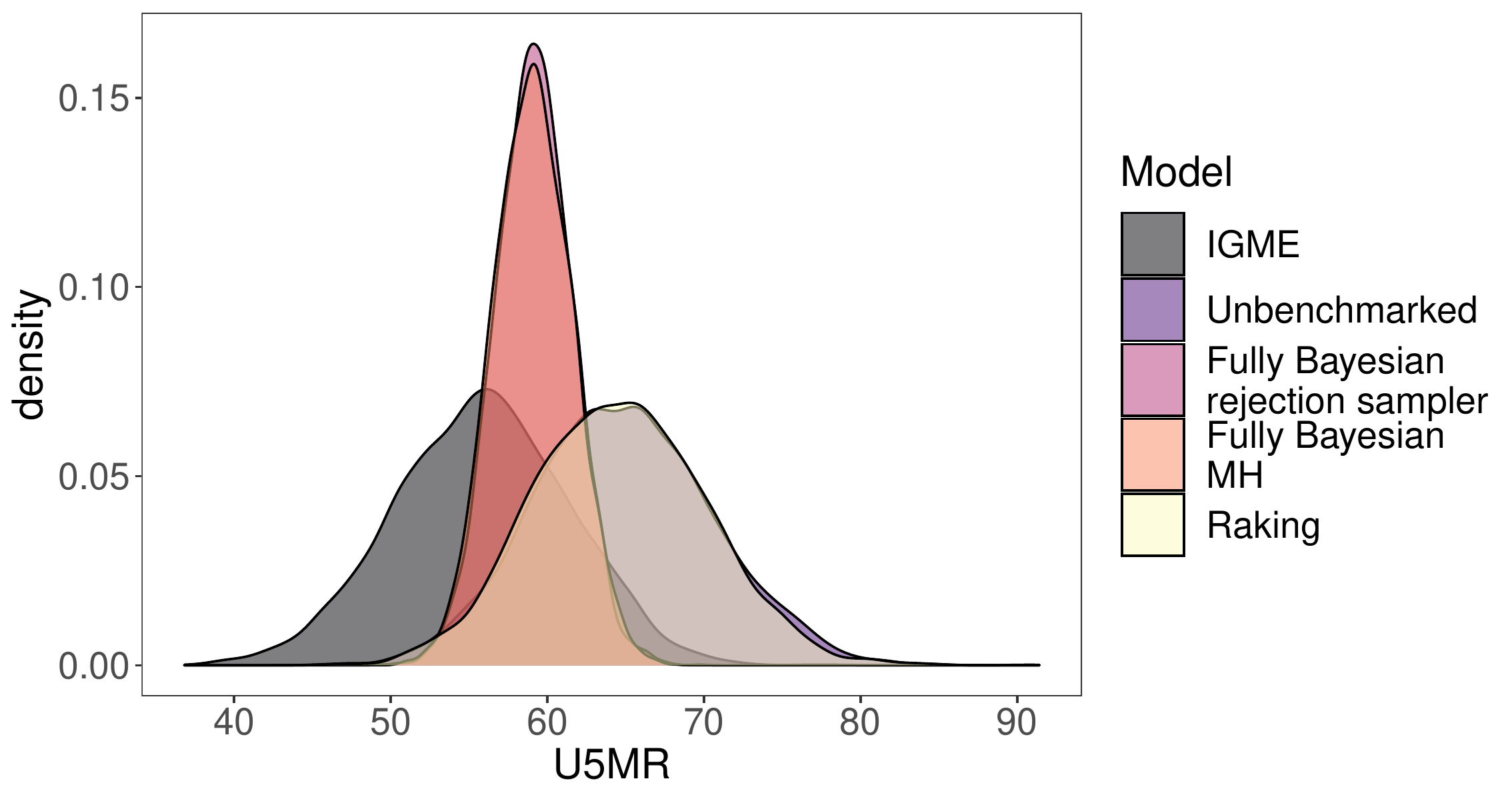}
		\caption{Aggregated national level U5MR estimates from IGME, unbenchmarked, and benchmarked models for area-level (left) and unit-level (right) models, for 2008. All densities are based on 5000 samples. U5MR is reported as deaths per 1000 live births.}
		\label{fig:u5mrnatlresults2008}
	\end{figure}

\begin{figure}[H]
		\centering
		\includegraphics[scale = 0.3]{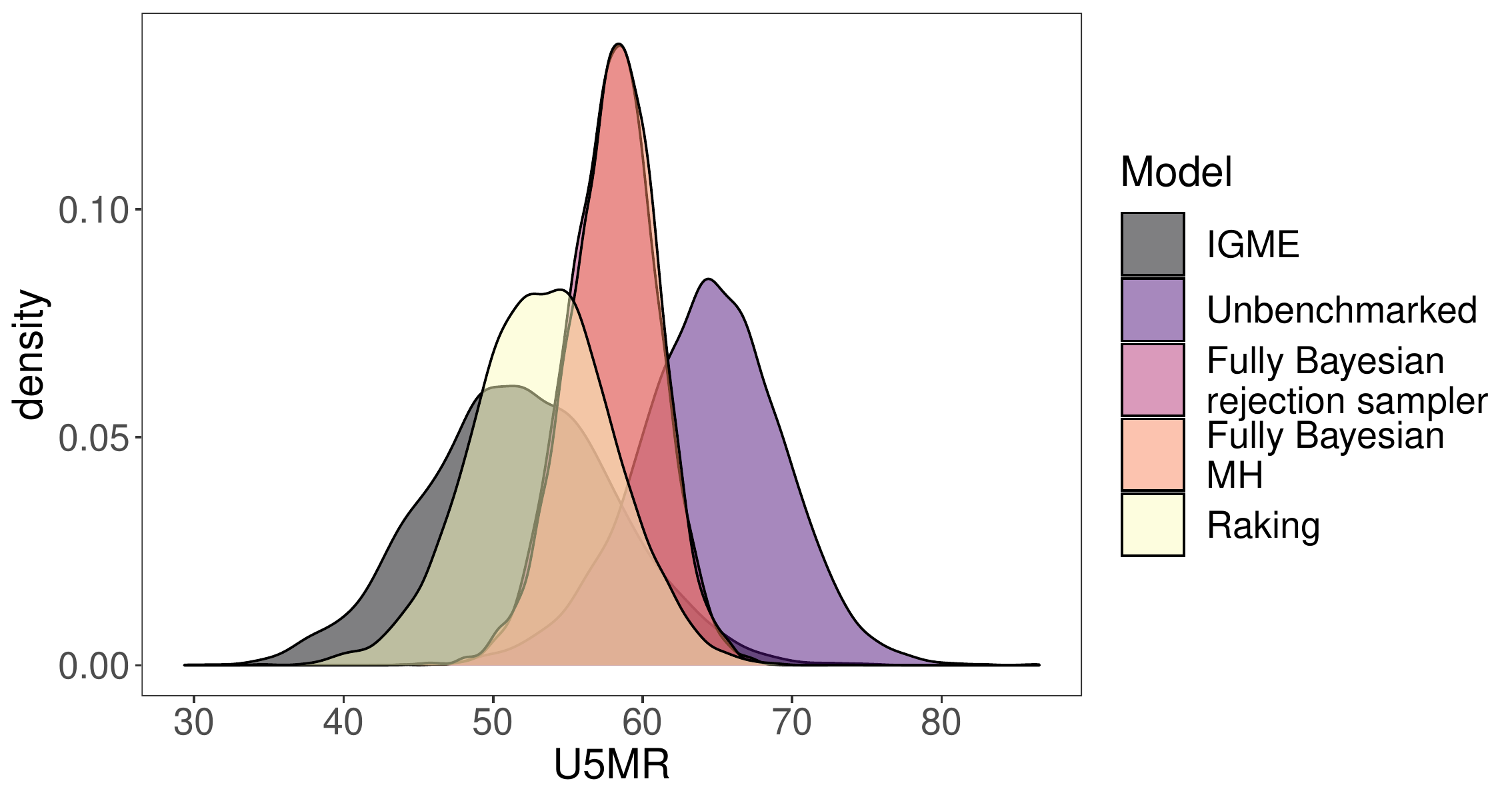}
		\includegraphics[scale = 0.3]{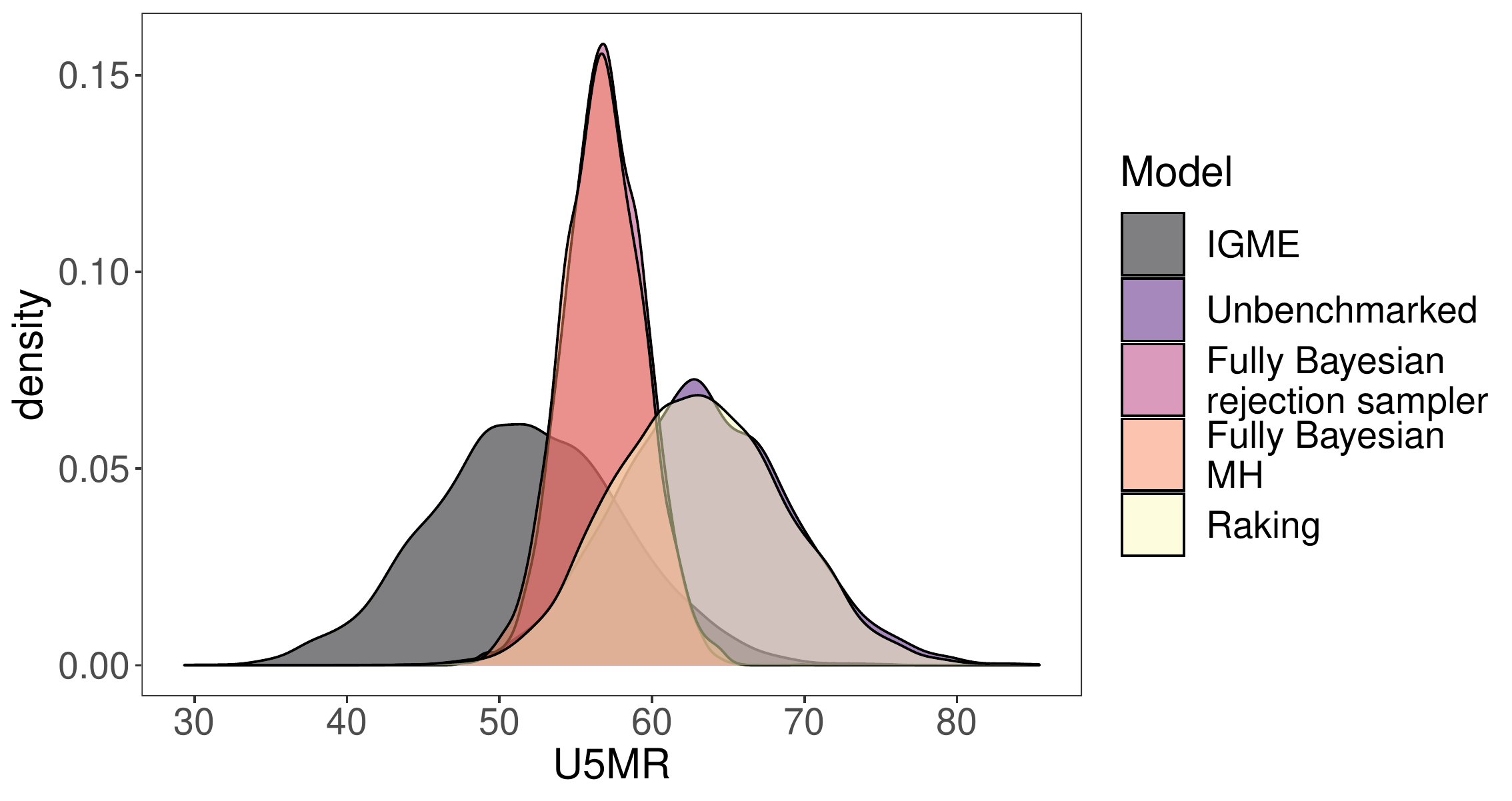}
		\caption{Aggregated national level U5MR estimates from IGME, unbenchmarked, and benchmarked models for area-level (left) and unit-level (right) models, for 2009. All densities are based on 5000 samples. U5MR is reported as deaths per 1000 live births.}
		\label{fig:u5mrnatlresults2009}
	\end{figure}

\begin{figure}[H]
		\centering
		\includegraphics[scale = 0.3]{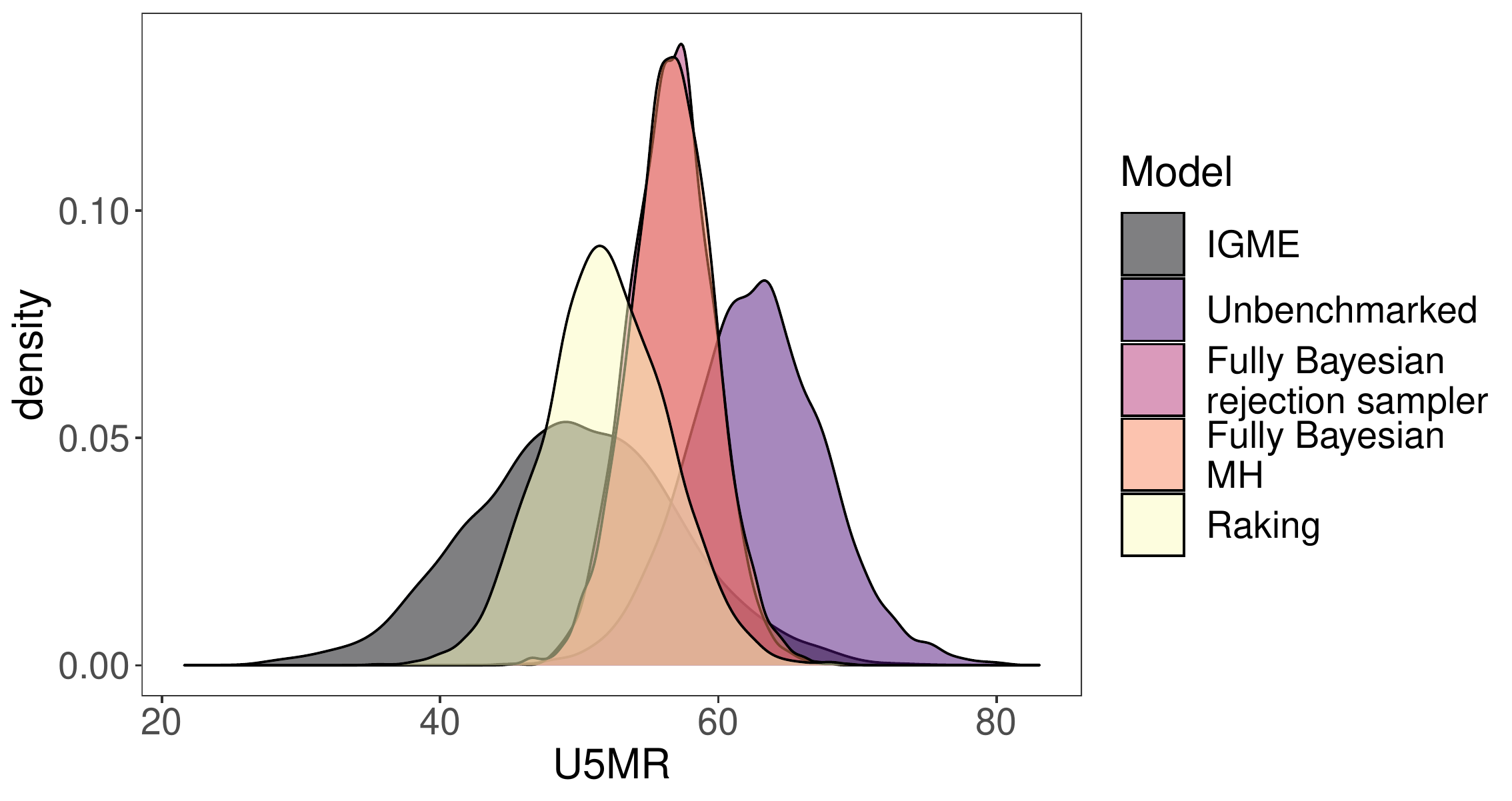}
		\includegraphics[scale = 0.3]{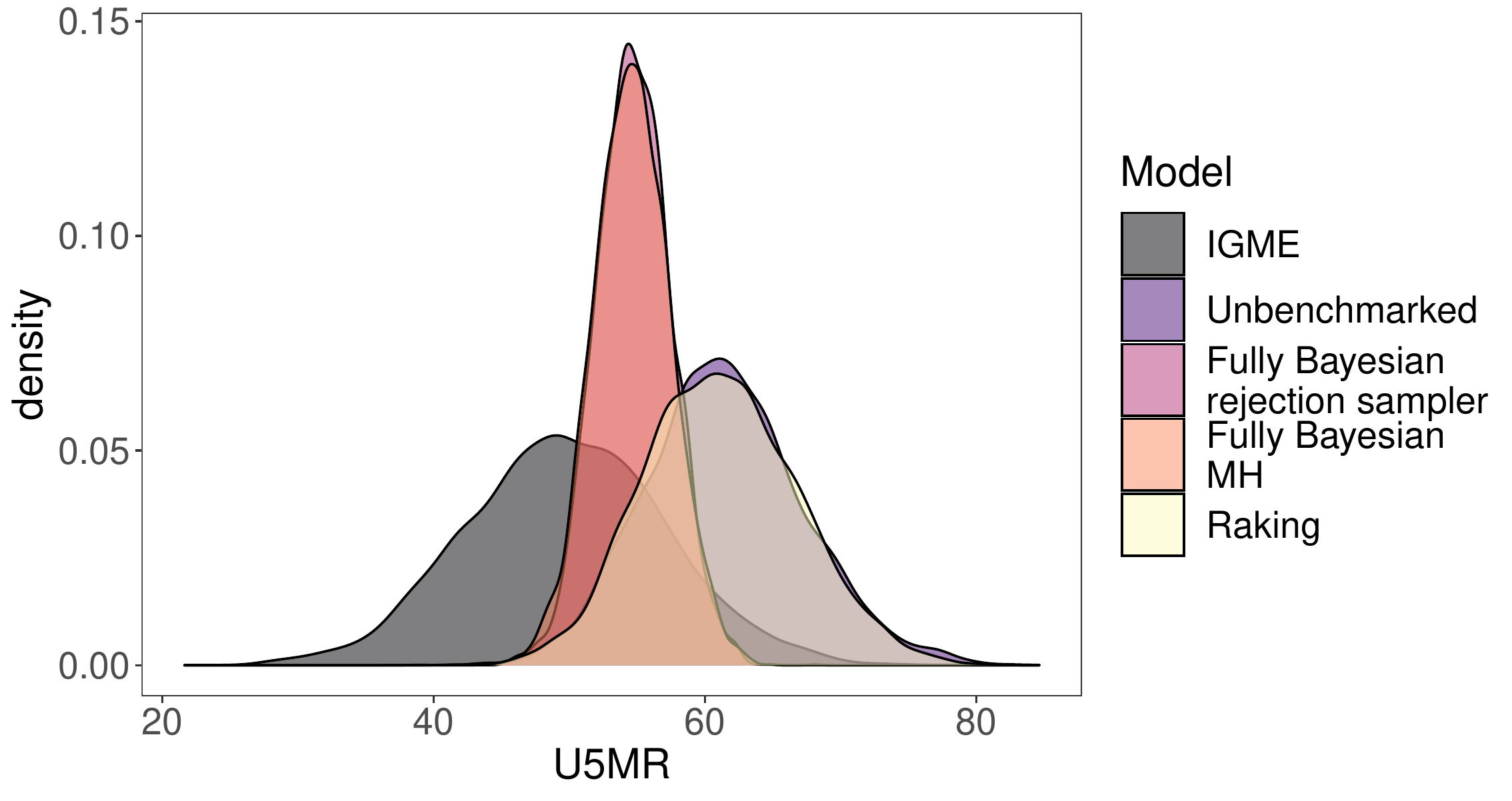}
		\caption{Aggregated national level U5MR estimates from IGME, unbenchmarked, and benchmarked models for area-level (left) and unit-level (right) models, for 2010. All densities are based on 5000 samples. U5MR is reported as deaths per 1000 live births.}
		\label{fig:u5mrnatlresults2010}
	\end{figure}

\begin{figure}[H]
		\centering
		\includegraphics[scale = 0.3]{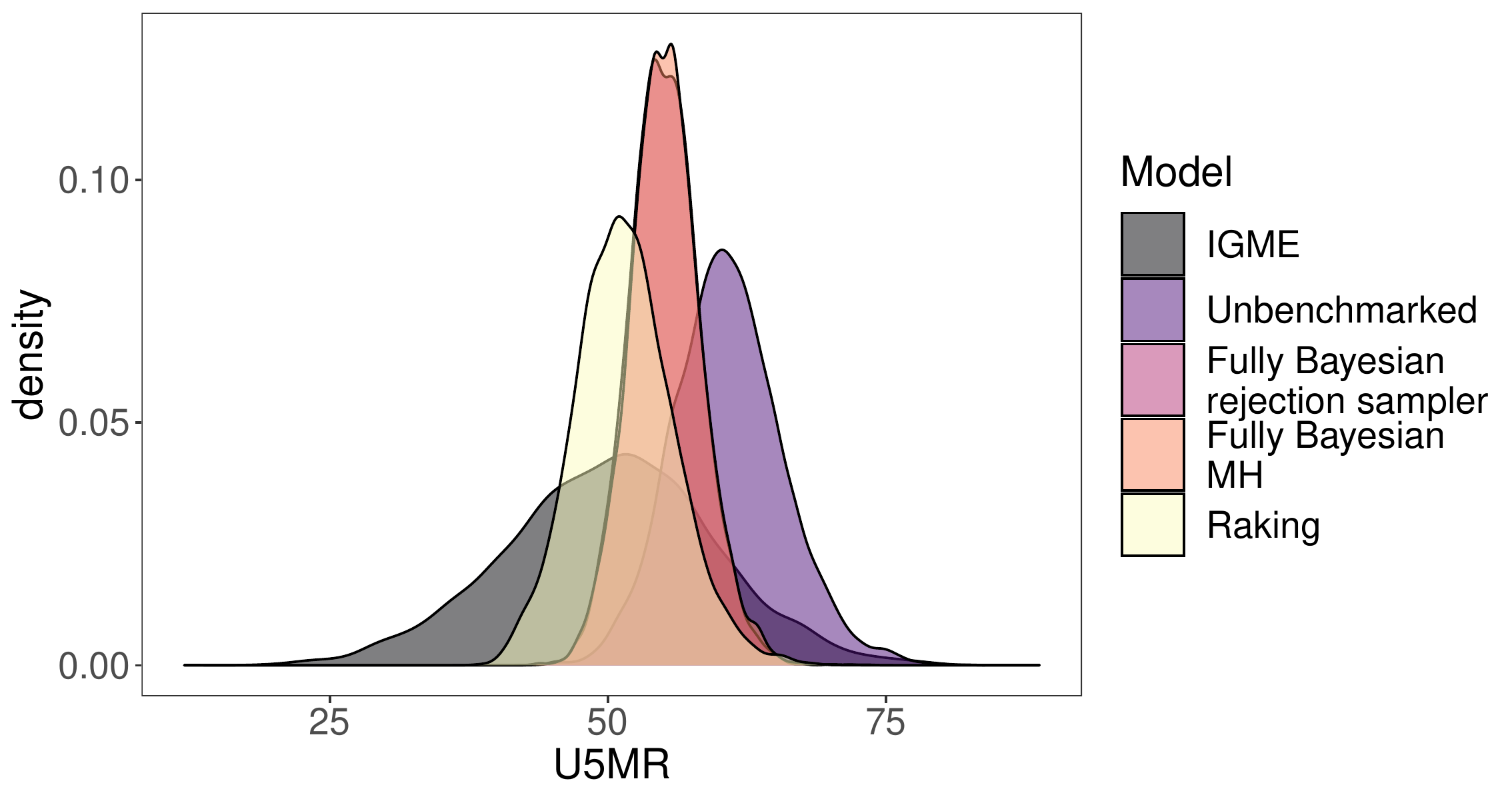}
		\includegraphics[scale = 0.3]{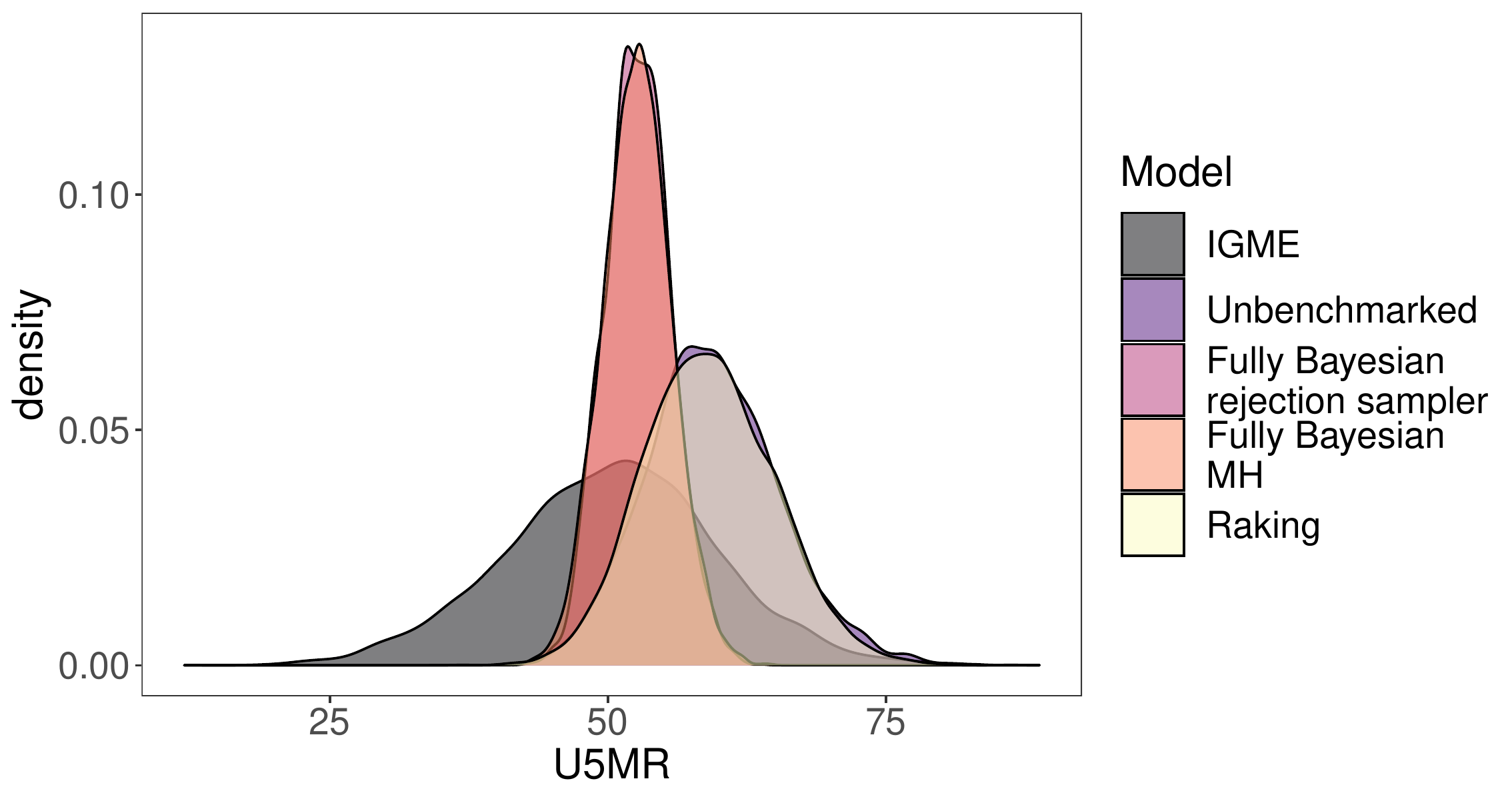}
		\caption{Aggregated national level U5MR estimates from IGME, unbenchmarked, and benchmarked models for area-level (left) and unit-level (right) models, for 2011. All densities are based on 5000 samples. U5MR is reported as deaths per 1000 live births.}
		\label{fig:u5mrnatlresults2011}
	\end{figure}

\begin{figure}[H]
		\centering
		\includegraphics[scale = 0.3]{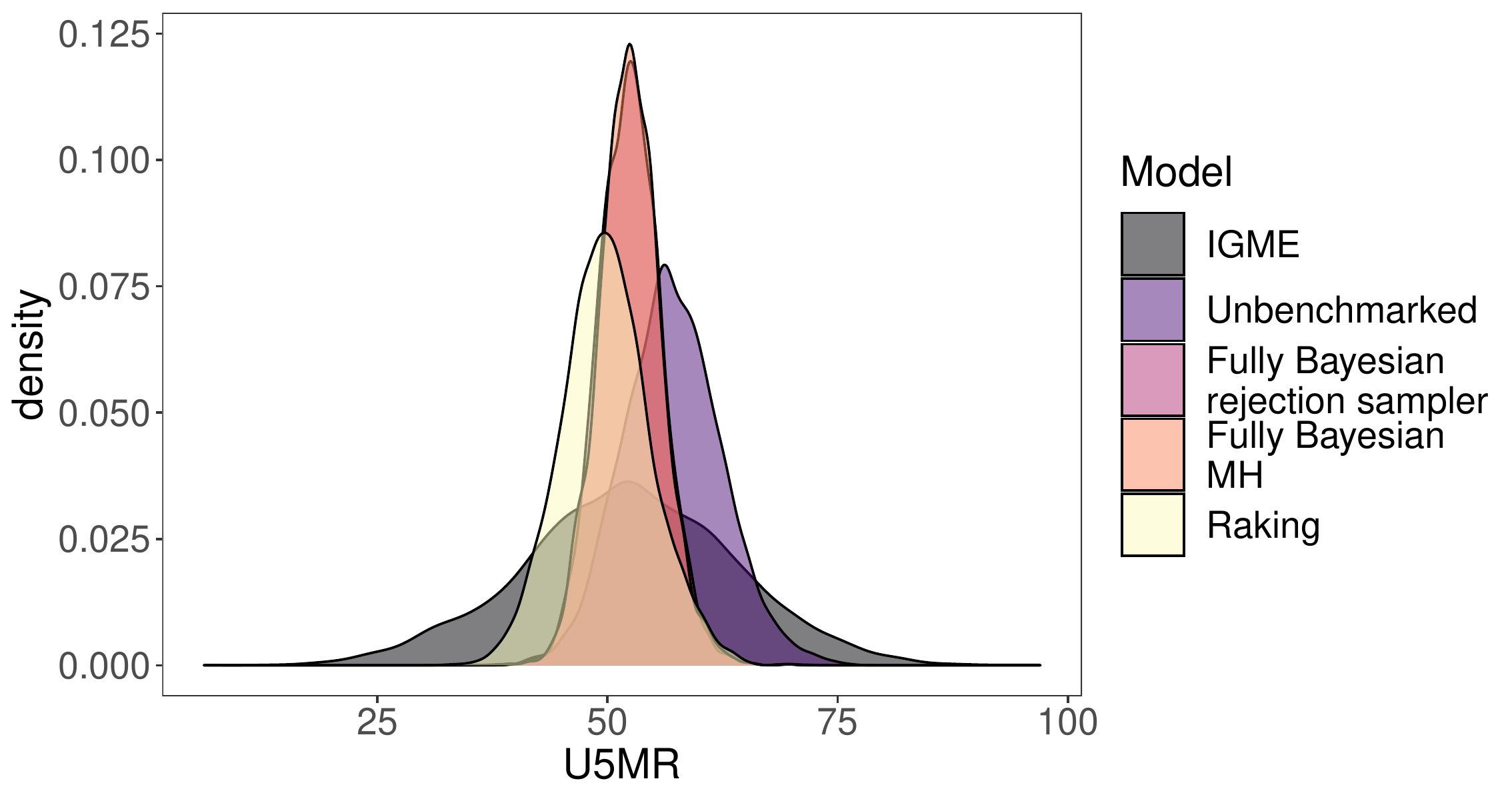}
		\includegraphics[scale = 0.3]{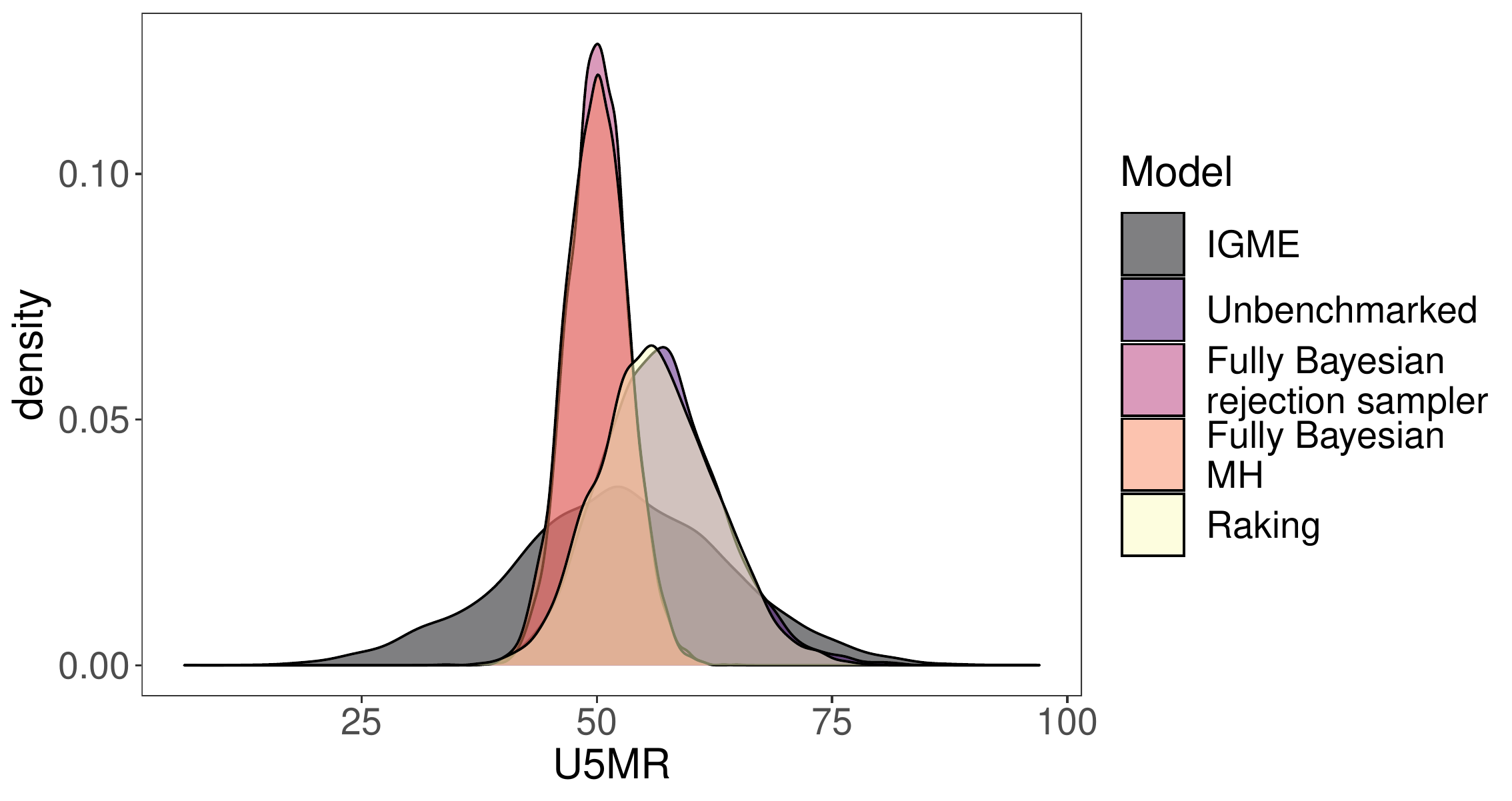}
		\caption{Aggregated national level U5MR estimates from IGME, unbenchmarked, and benchmarked models for area-level (left) and unit-level (right) models, for 2012. All densities are based on 5000 samples. U5MR is reported as deaths per 1000 live births.}
		\label{fig:u5mrnatlresults2012}
	\end{figure}

\begin{figure}[H]
		\centering
		\includegraphics[scale = 0.3]{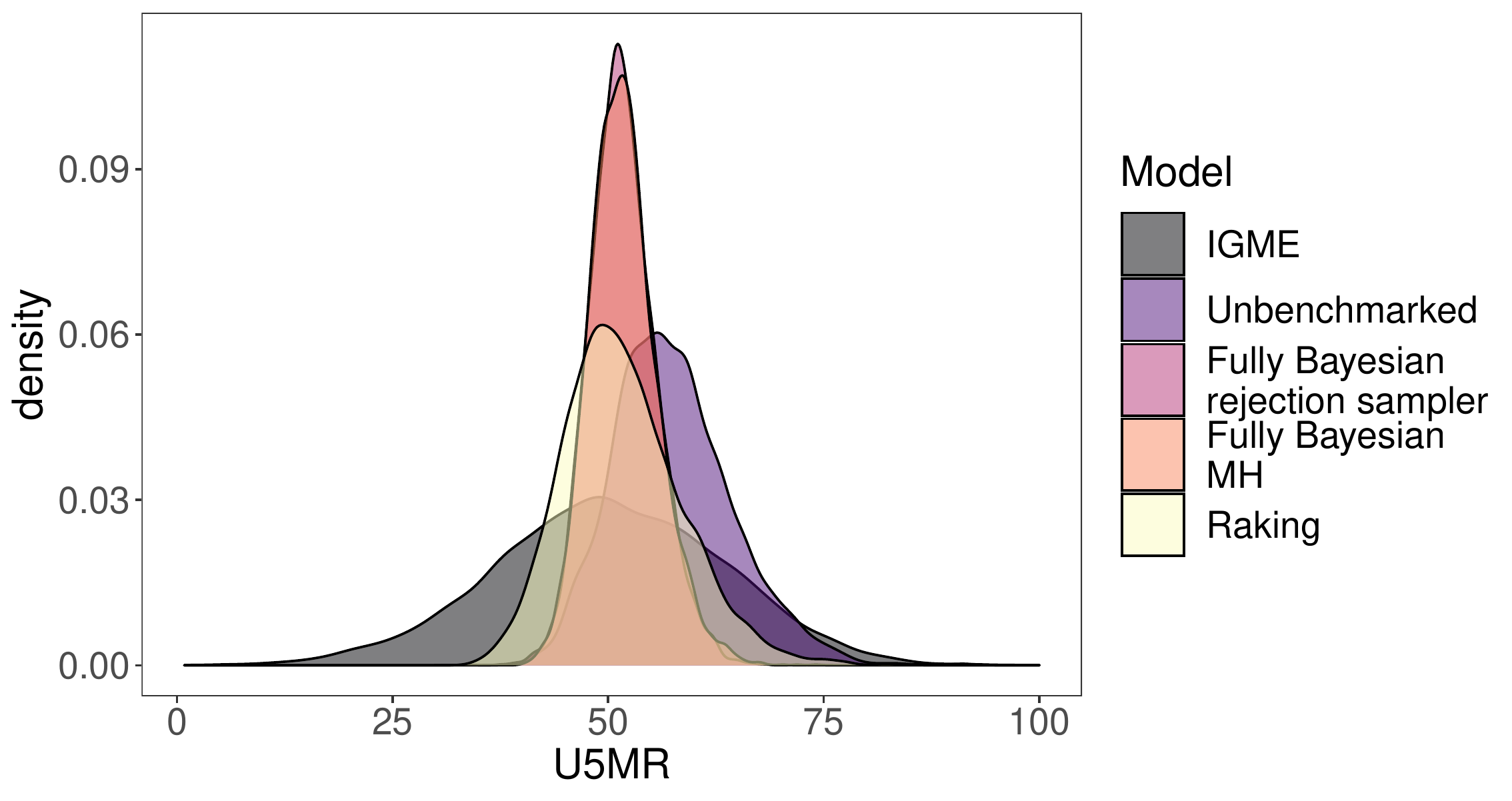}
		\includegraphics[scale = 0.3]{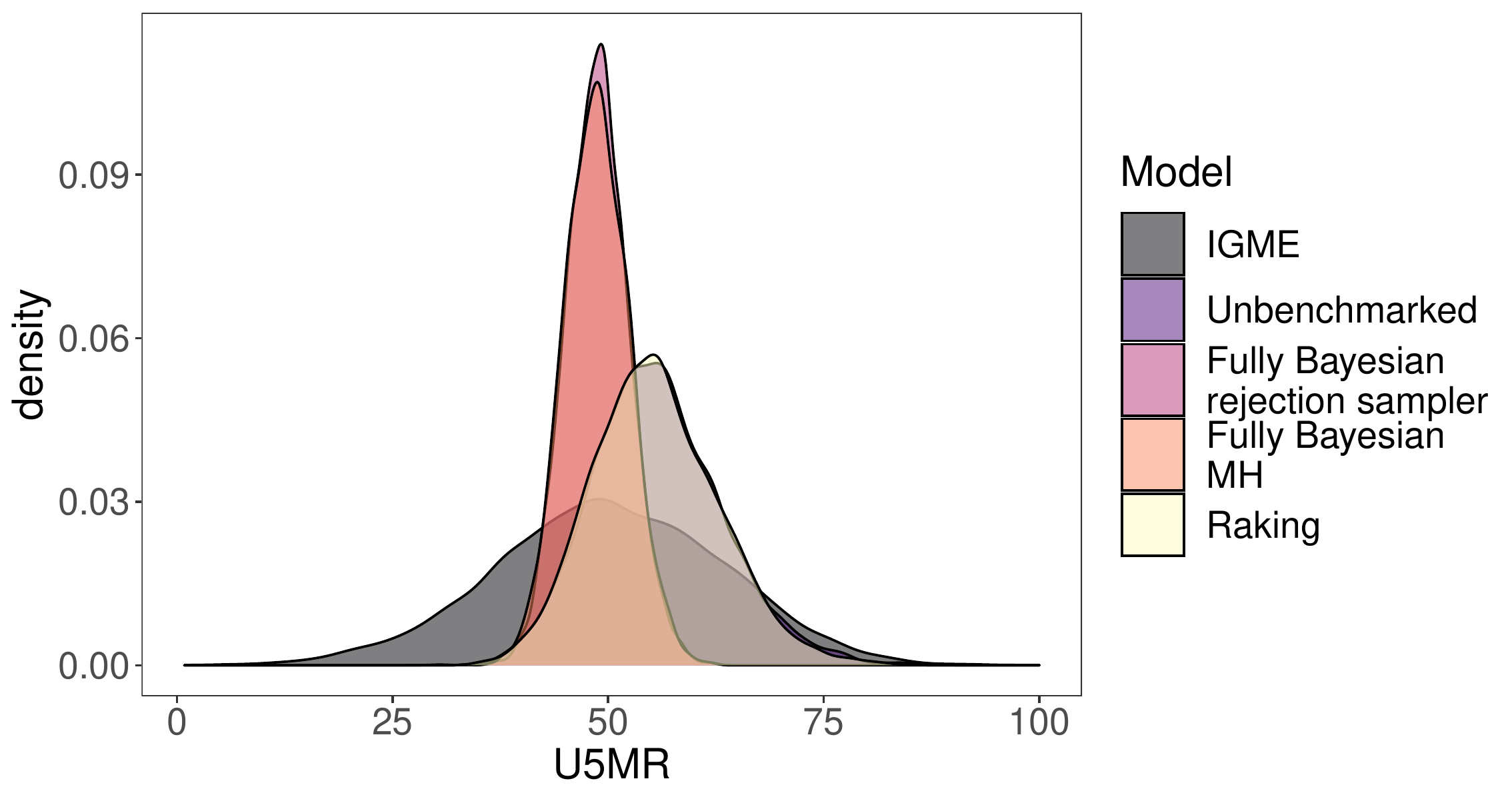}
		\caption{Aggregated national level U5MR estimates from IGME, unbenchmarked, and benchmarked models for area-level (left) and unit-level (right) models, for 2013. All densities are based on 5000 samples. U5MR is reported as deaths per 1000 live births.}
		\label{fig:u5mrnatlresults2013}
	\end{figure}

\begin{figure}[H]
		\centering
		\includegraphics[scale = 0.3]{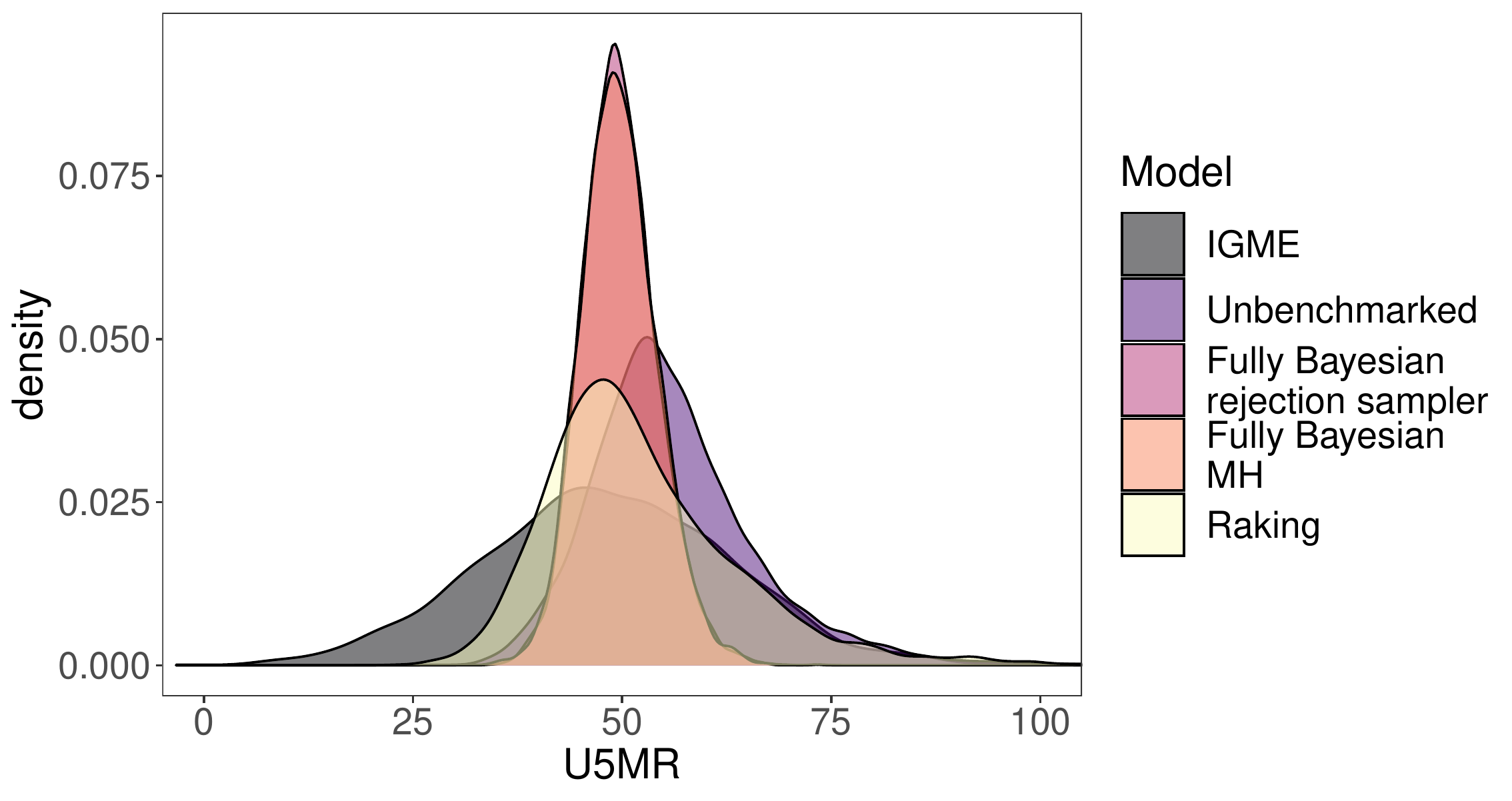}
		\includegraphics[scale = 0.3]{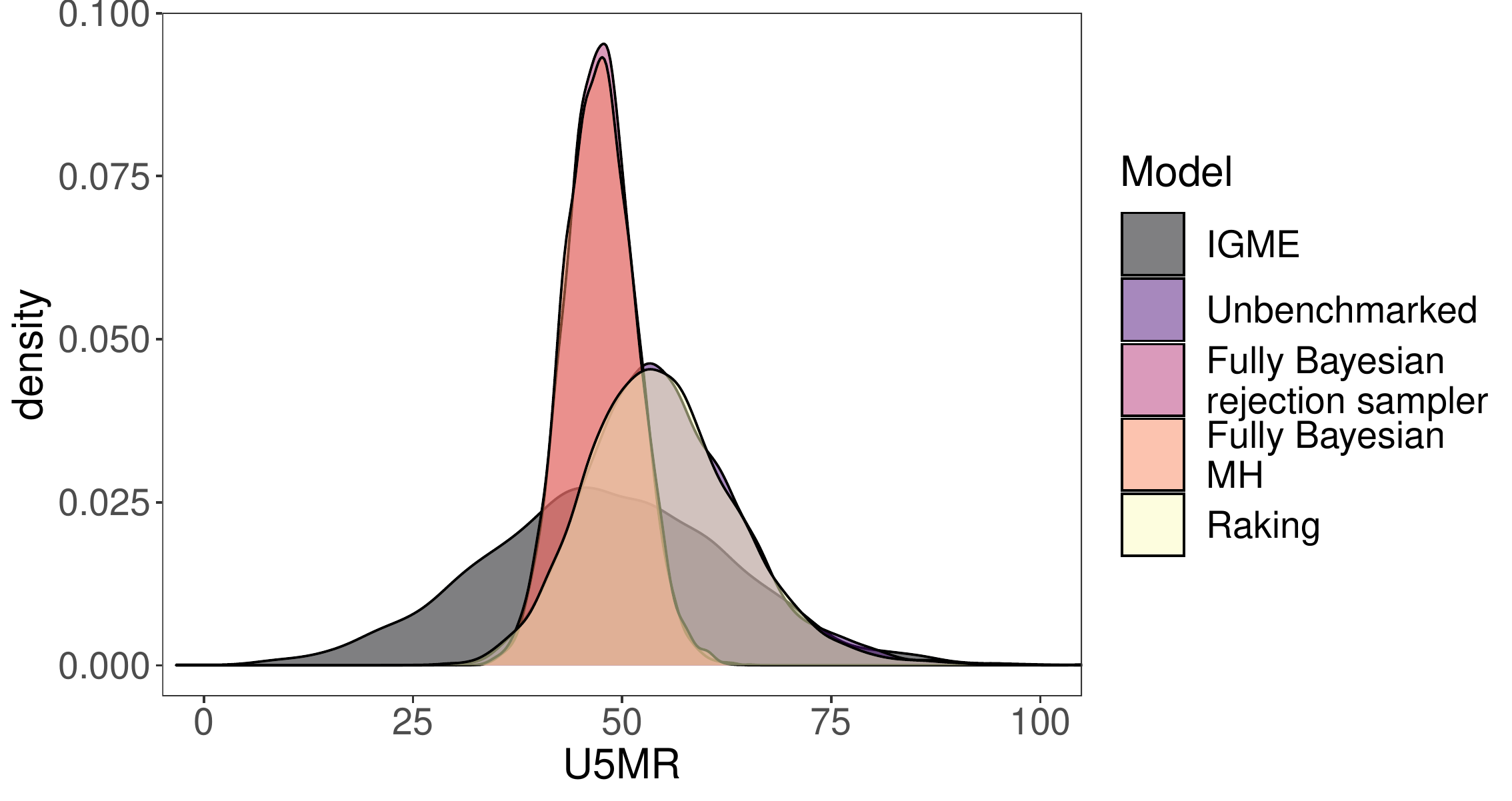}
		\caption{Aggregated national level U5MR estimates from IGME, unbenchmarked, and benchmarked models for area-level (left) and unit-level (right) models, for 2014. All densities are based on 5000 samples. U5MR is reported as deaths per 1000 live births.}
		\label{fig:u5mrnatlresults2014}
	\end{figure}

\begin{figure}[H]
		\centering
		\includegraphics[scale = 0.3]{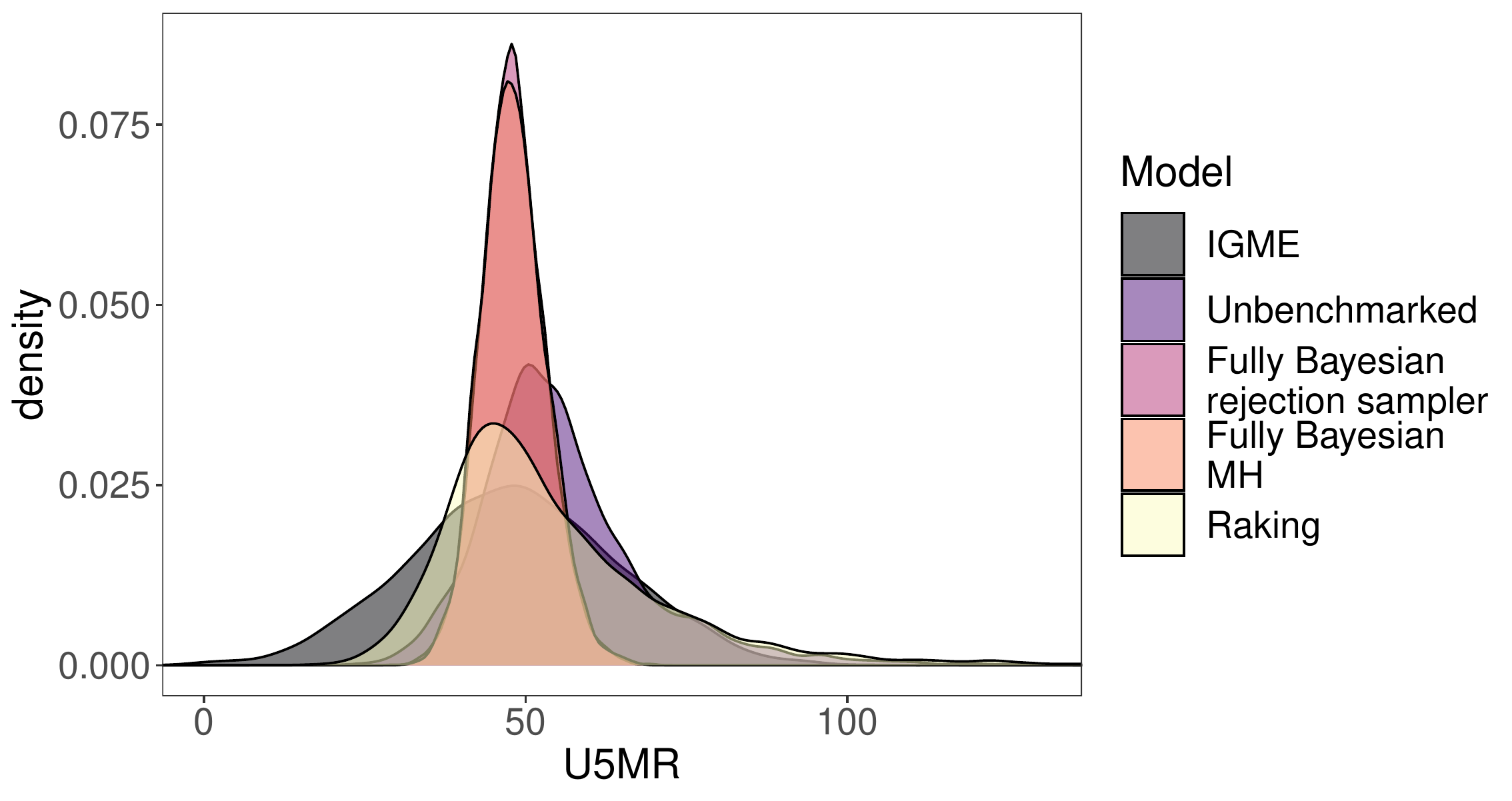}
		\includegraphics[scale = 0.3]{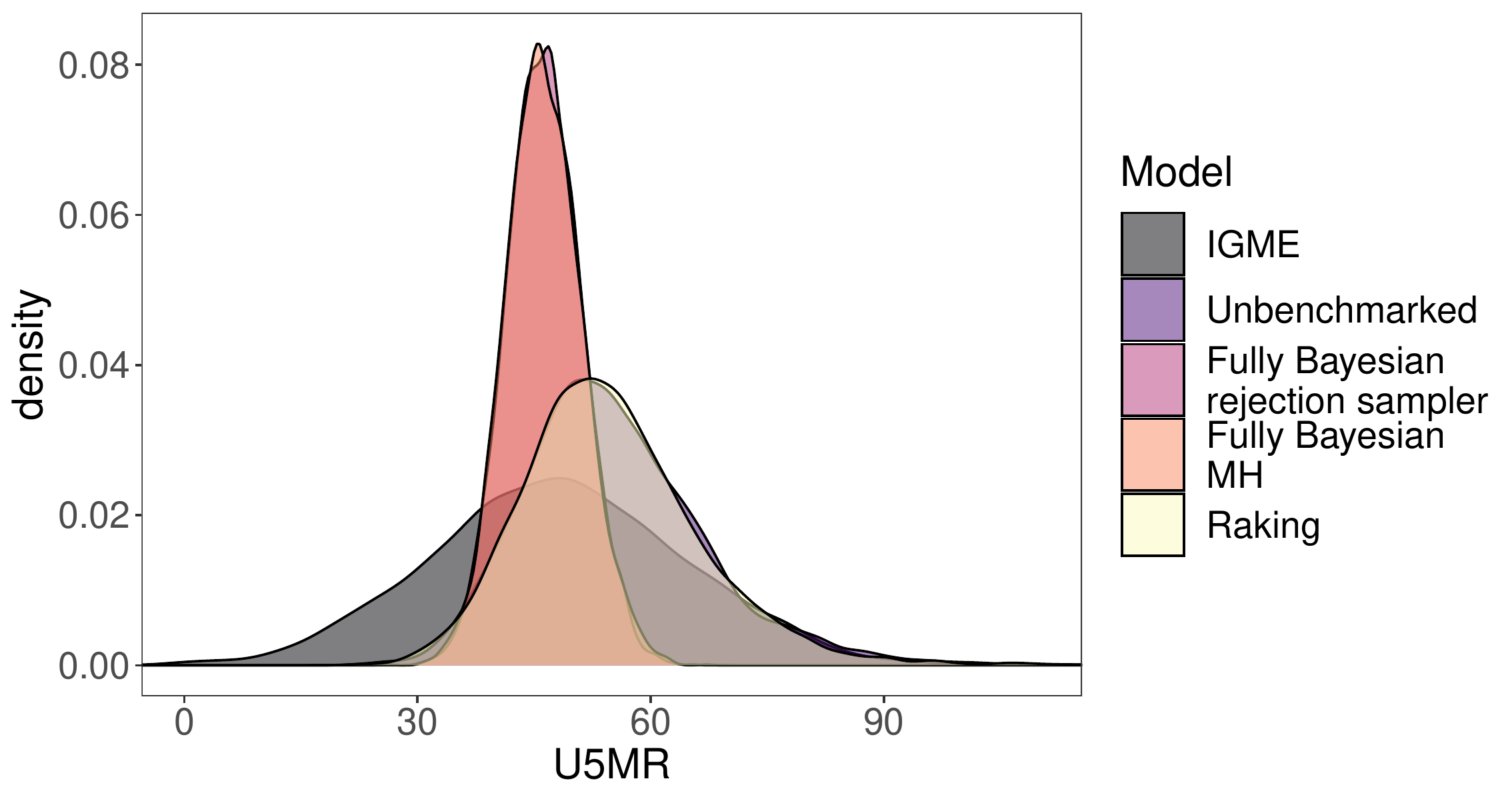}
		\caption{Aggregated national level U5MR estimates from IGME, unbenchmarked, and benchmarked models for area-level (left) and unit-level (right) models, for 2015. All densities are based on 5000 samples. U5MR is reported as deaths per 1000 live births.}
		\label{fig:u5mrnatlresults2015}
	\end{figure}

\begin{figure}[H]
		\centering
		\includegraphics[scale = 0.3]{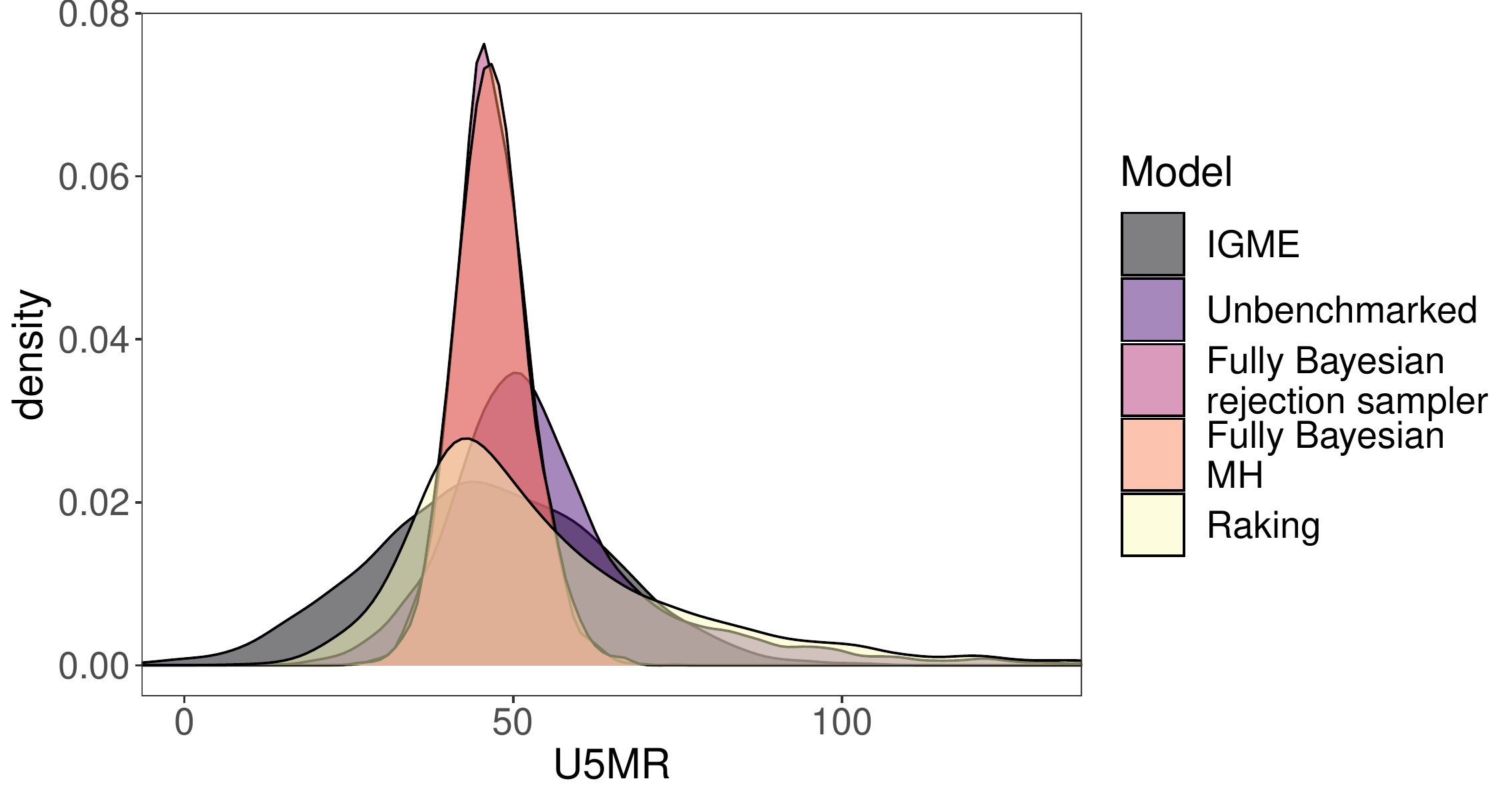}
		\includegraphics[scale = 0.3]{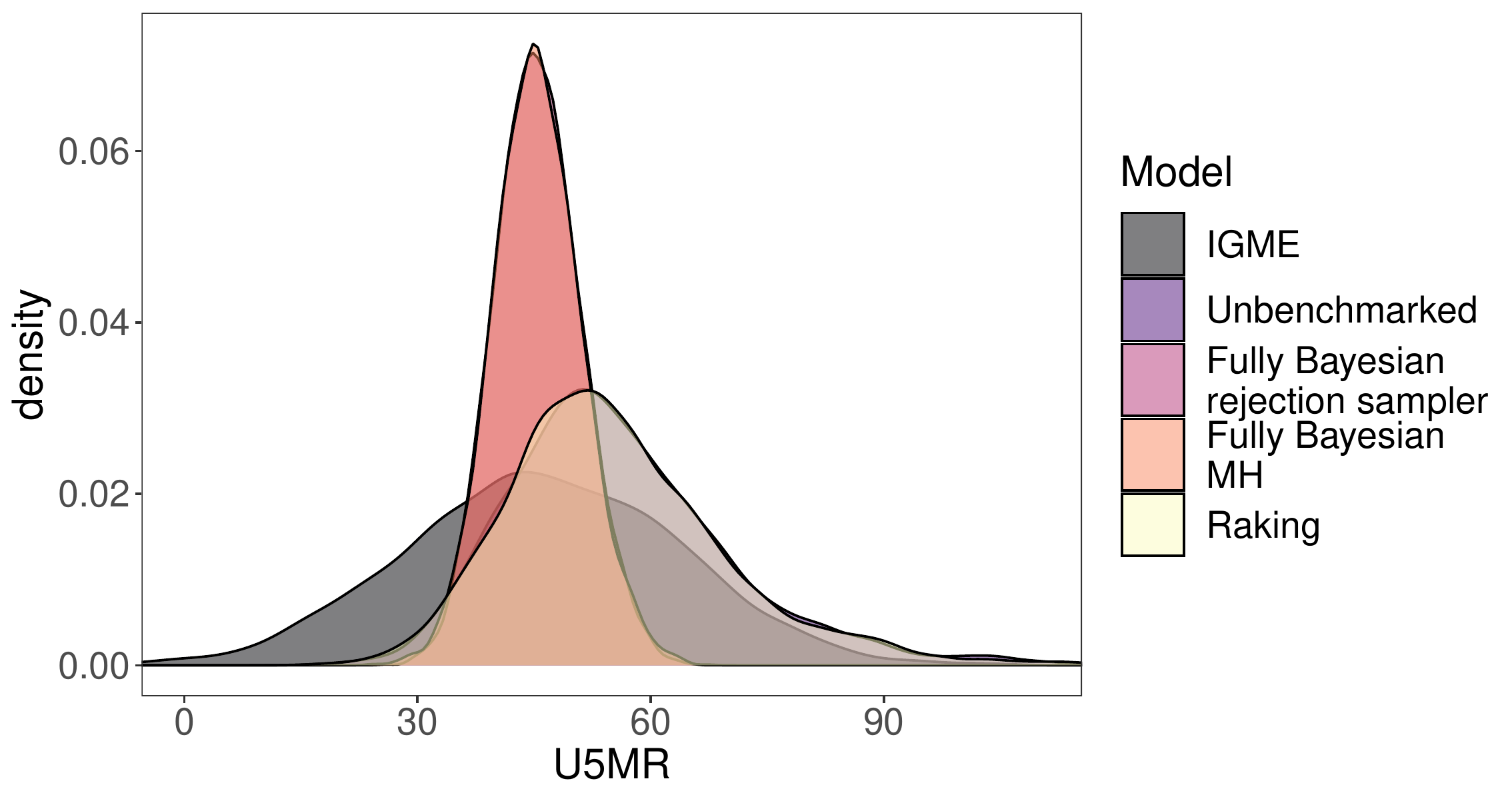}
		\caption{Aggregated national level U5MR estimates from IGME, unbenchmarked, and benchmarked models for area-level (left) and unit-level (right) models, for 2016. All densities are based on 5000 samples. U5MR is reported as deaths per 1000 live births.}
		\label{fig:u5mrnatlresults2016}
	\end{figure}

\begin{figure}[H]
		\centering
		\includegraphics[scale = 0.3]{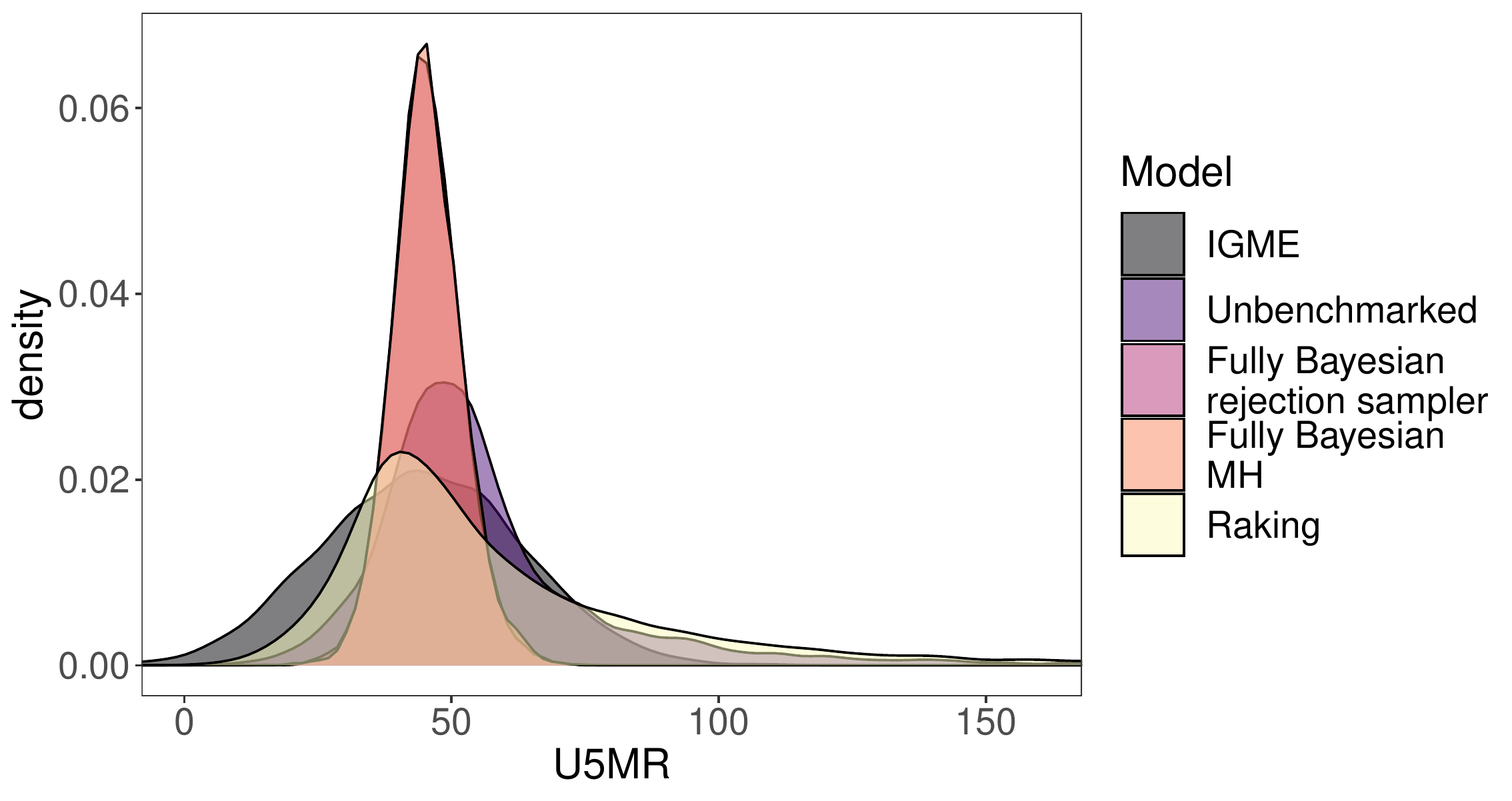}
		\includegraphics[scale = 0.3]{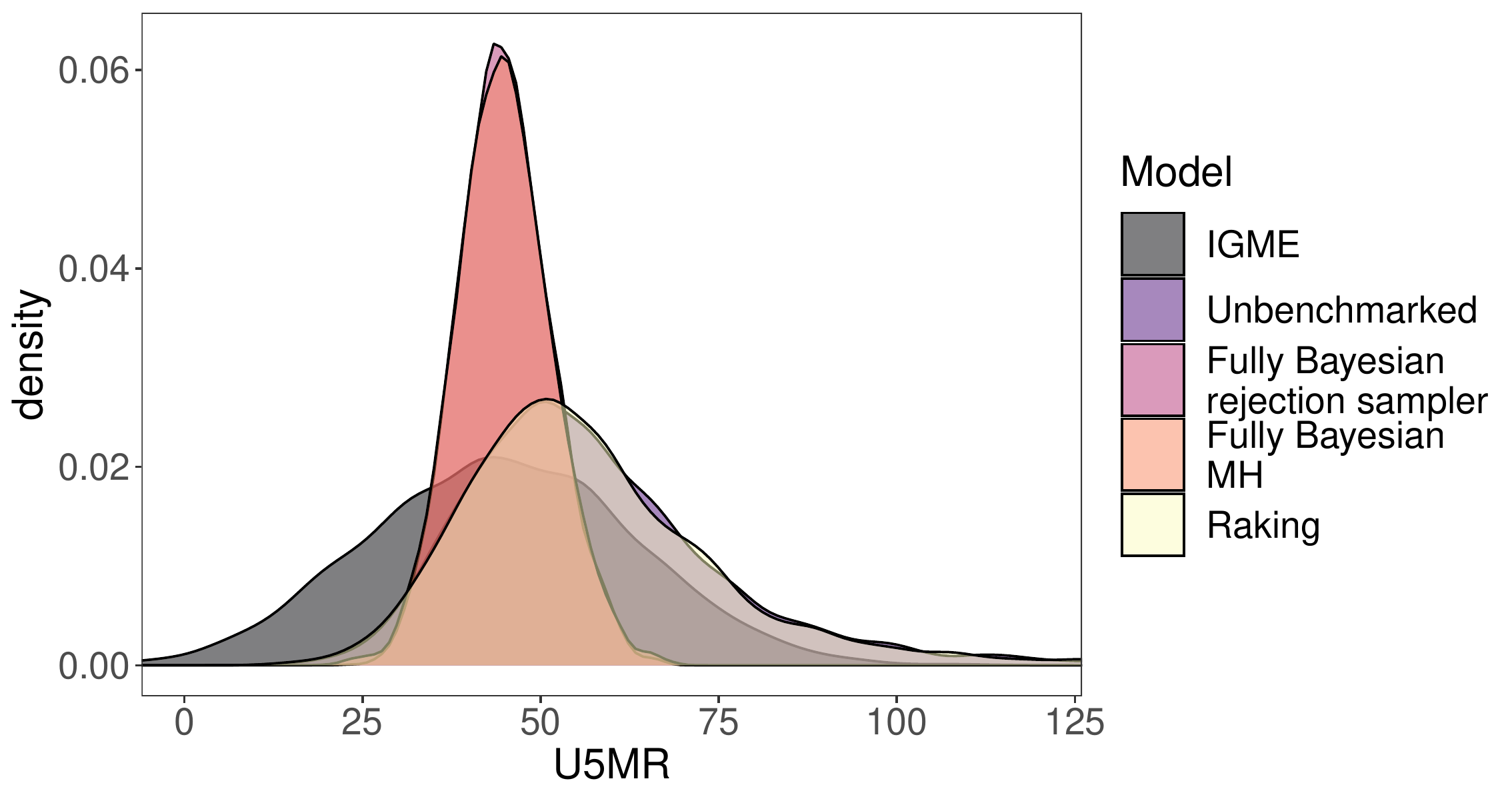}
		\caption{Aggregated national level U5MR estimates from IGME, unbenchmarked, and benchmarked models for area-level (left) and unit-level (right) models, for 2017. All densities are based on 5000 samples. U5MR is reported as deaths per 1000 live births.}
		\label{fig:u5mrnatlresults2017}
	\end{figure}

\begin{figure}[H]
		\centering
		\includegraphics[scale = 0.3]{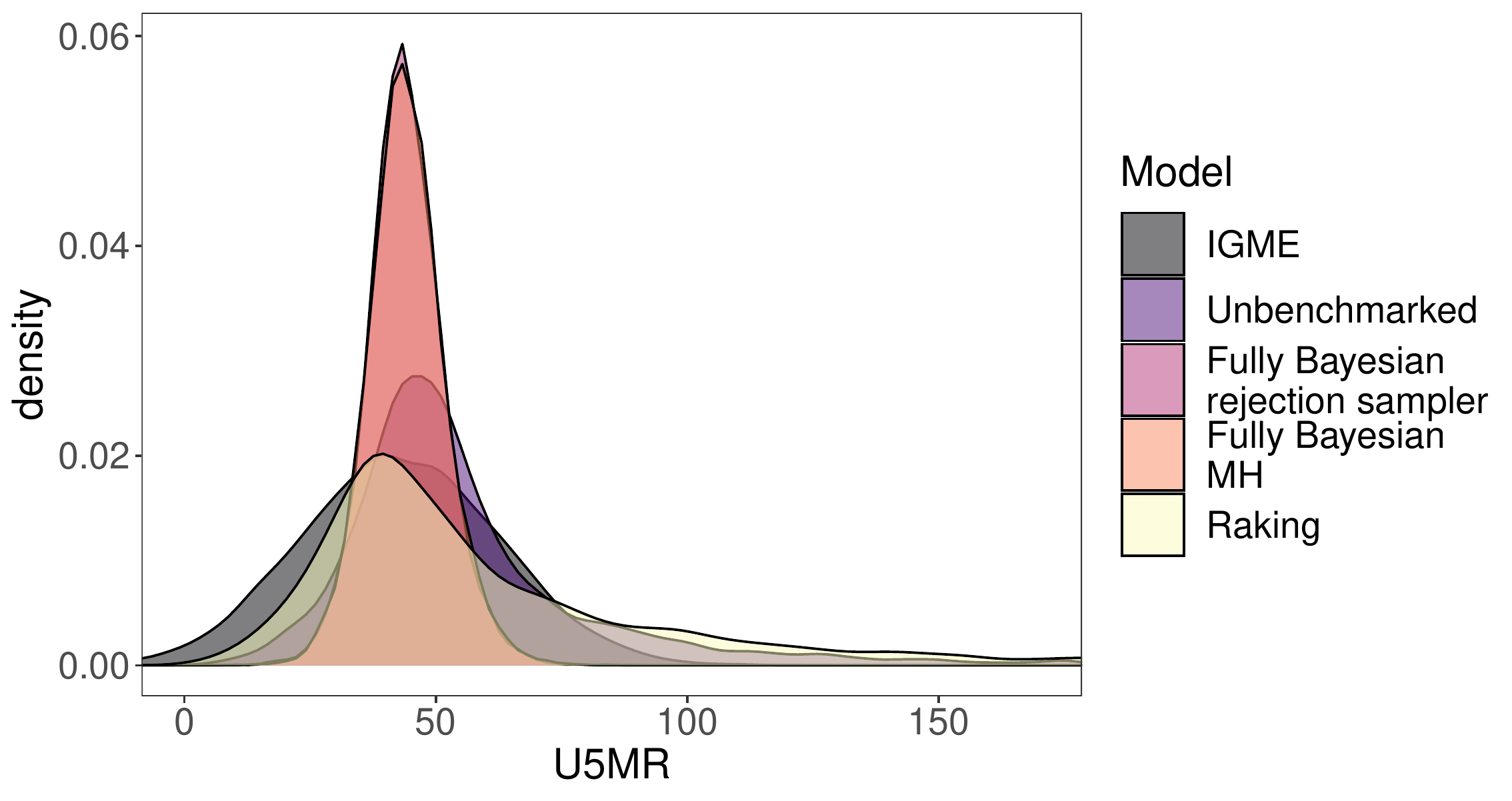}
		\includegraphics[scale = 0.3]{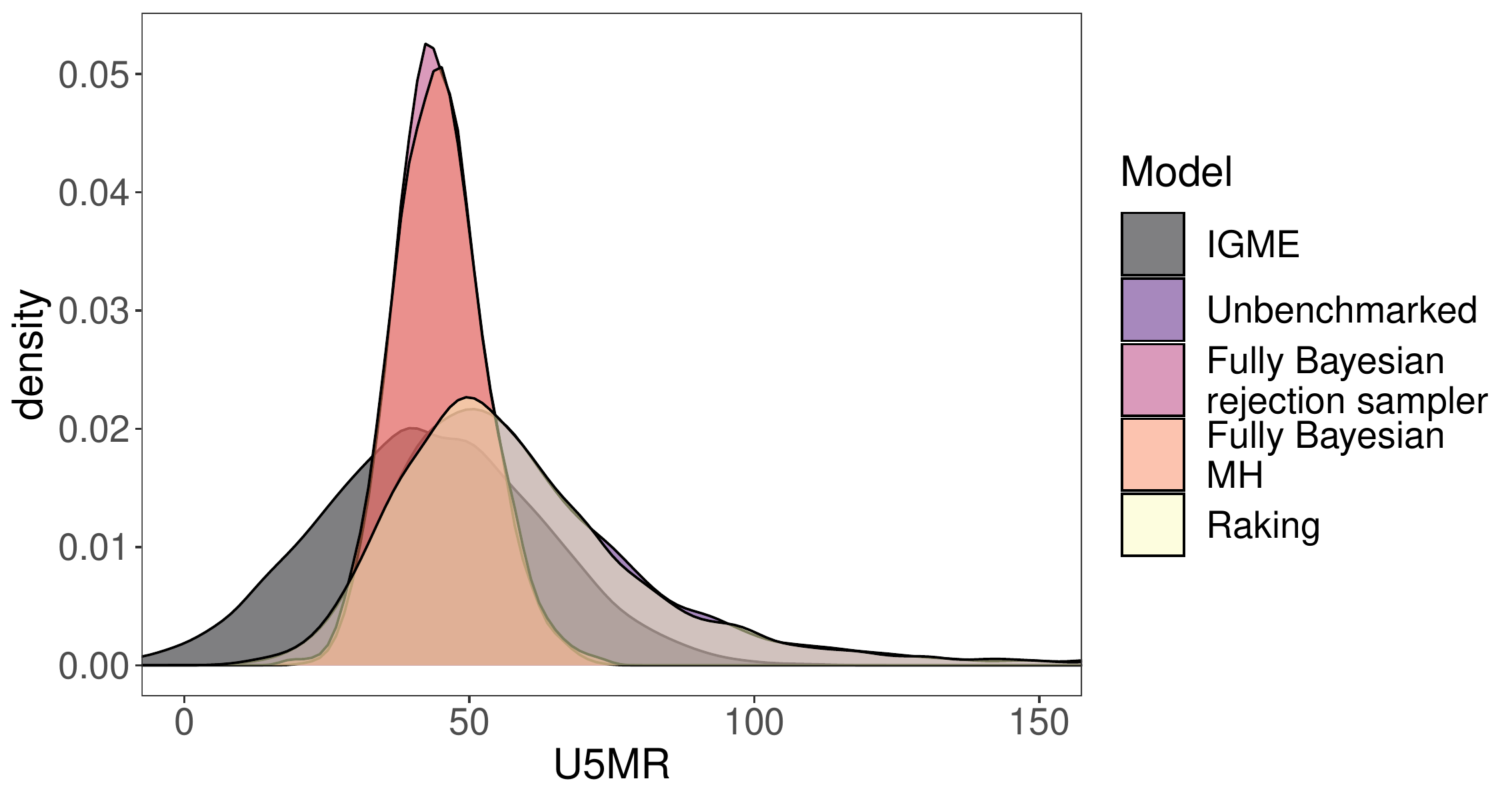}
		\caption{Aggregated national level U5MR estimates from IGME, unbenchmarked, and benchmarked models for area-level (left) and unit-level (right) models, for 2018. All densities are based on 5000 samples. U5MR is reported as deaths per 1000 live births.}
		\label{fig:u5mrnatlresults2018}
	\end{figure}

\begin{figure}[H]
		\centering
		\includegraphics[scale = 0.3]{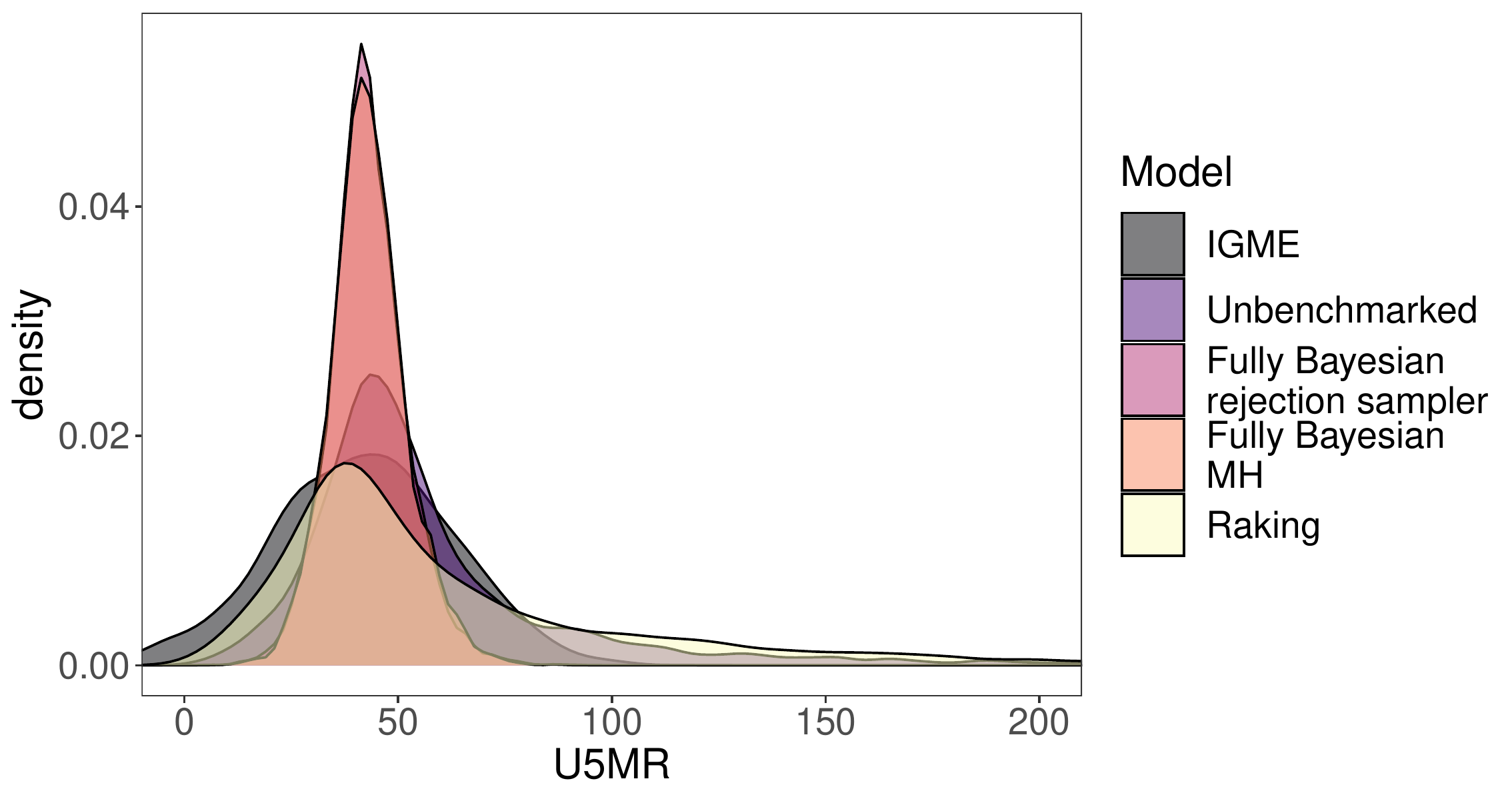}
		\includegraphics[scale = 0.3]{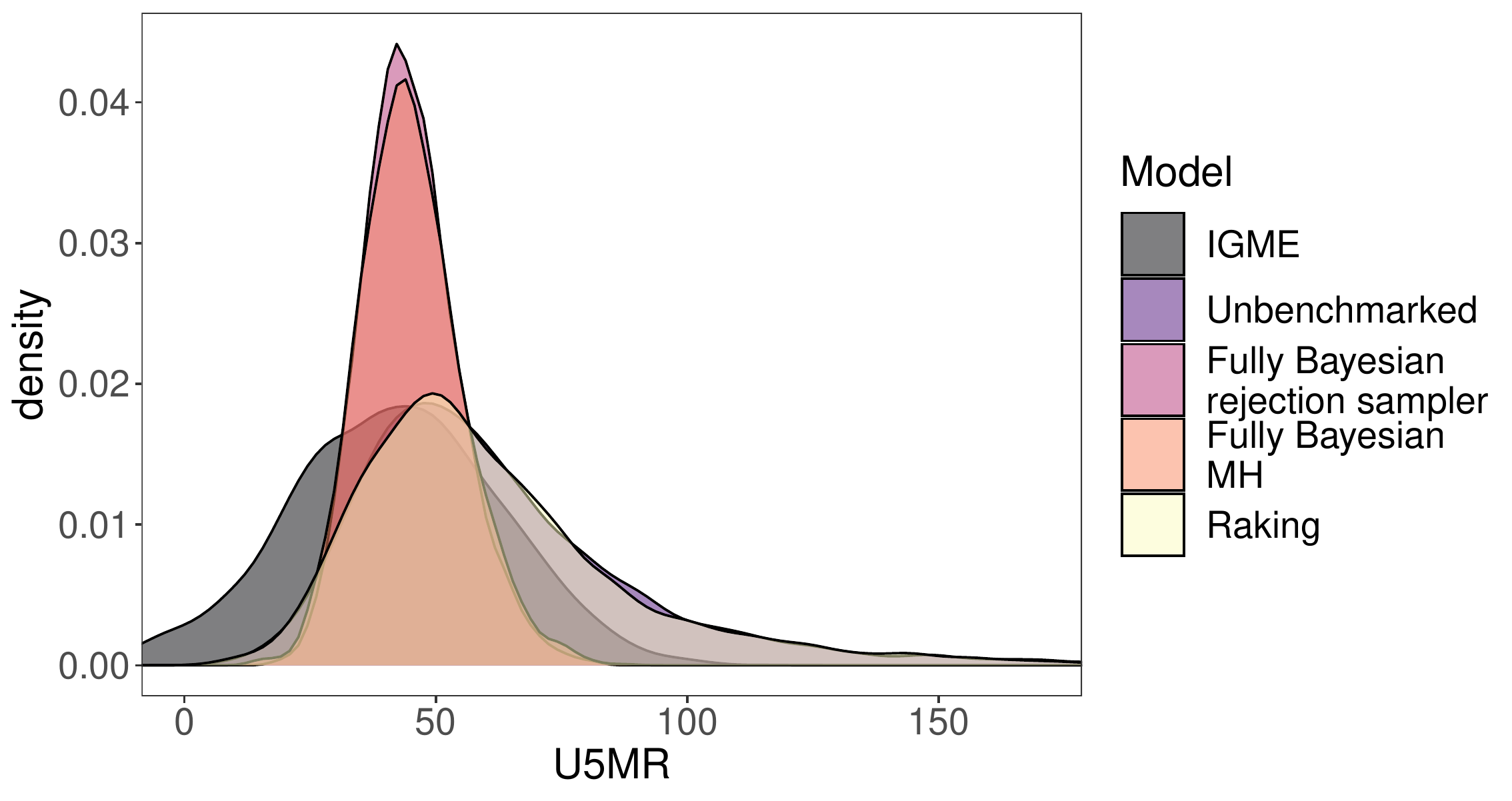}
		\caption{Aggregated national level U5MR estimates from IGME, unbenchmarked, and benchmarked models for area-level (left) and unit-level (right) models, for 2019. All densities are based on 5000 samples. U5MR is reported as deaths per 1000 live births.}
		\label{fig:u5mrnatlresults2019}
	\end{figure}

	Figures \ref{fig:u5mrnatlresults2000} through \ref{fig:u5mrnatlresults2019} display the aggregated national level estimates produced from each benchmarked and unbenchmarked method for both unit- and area-level models. The posterior densities from the fully Bayesian rejection sampling approach are less uncertain than the posterior densities from the unbenchmarked model due to the additional information contained in the benchmarking constraint being incorporated in the benchmarked approach. The overlap between the IGME density and the Bayes estimate approach is expected as the Bayes estimate approach performs exact benchmarking. A table of these national estimates for all years and models can be found in Table \ref{tab:u5mrnatlresults}. 
	
	The proportion of samples accepted in the fully Bayesian rejection sampler approach was 0.5\% for the unit-level model, and 0.001\% for the area-level model. The proportion of samples accepted in the fully  Bayesian MH approach was 7\% for the unit-level model, and 0.3\% for the area-level model.
	
	\begin{table}[H]
		\centering
		\caption{Aggregated national level U5MR estimates from IGME, unbenchmarked, and benchmarked models for 2010. 95\% credible intervals are given next to posterior medians. U5MR is reported as deaths per 1000 live births. AL = Area-level, UL = Unit-level.}
		\begin{tabular}{r|l|l|l|r|r}
			\textbf{Year}         & \textbf{Model}        & \textbf{AL: Median} & \textbf{UL: Median} & \textbf{AL: SD} & \textbf{UL: SD} \\ \hline
			\multicolumn{1}{c|}{\multirow{5}{*}{2000}} & IGME                  & 75.4 (67.9, 84.4)   & 75.4 (67.9, 84.4)   & 5.5             & 5.5             \\ \cline{2-6} 
			\multicolumn{1}{c|}{}                      & Unbenchmarked         & 91 (79.9, 103.6)    & 77.1 (65, 90.8)     & 6.0             & 6.6             \\ \cline{2-6} 
			\multicolumn{1}{c|}{}                      & FB: Rejection Sampler & 81.9 (75, 88.7)     & 76.9 (70, 84.1)     & 3.5             & 3.6             \\ \cline{2-6} 
			\multicolumn{1}{c|}{}                      & FB: MH                & 81.9 (75.1, 88.7)   & 76.9 (70, 84)       & 3.5             & 3.7             \\ \cline{2-6} 
			\multicolumn{1}{c|}{}                      & Raking                & 77.2 (67.2, 87.9)   & 77.1 (64.2, 91.1)   & 5.4             & 6.8             \\ \hline
			\multirow{5}{*}{2001}                      & IGME                  & 74.9 (67.5, 83.8)   & 74.9 (67.5, 83.8)   & 5.4             & 5.4             \\ \cline{2-6} 
			& Unbenchmarked         & 86.6 (76, 97.4)     & 74.6 (64, 86)       & 5.4             & 5.5             \\ \cline{2-6} 
			& FB: Rejection Sampler & 78.8 (72.7, 85.1)   & 74.8 (69.6, 80.2)   & 3.2             & 2.8             \\ \cline{2-6} 
			& FB: MH                & 78.9 (72.6, 85)     & 74.6 (69.3, 80.3)   & 3.1             & 2.8             \\ \cline{2-6} 
			& Raking                & 75.2 (65.6, 85.1)   & 74.6 (64, 86.2)     & 4.9             & 5.7             \\ \hline
			\multirow{5}{*}{2002}                      & IGME                  & 74.3 (67, 83.2)     & 74.3 (67, 83.2)     & 5.4             & 5.4             \\ \cline{2-6} 
			& Unbenchmarked         & 84 (74.7, 94.2)     & 72.5 (62.6, 83.3)   & 5.0             & 5.3             \\ \cline{2-6} 
			& FB: Rejection Sampler & 77.1 (71.4, 83.1)   & 73 (68.2, 78.1)     & 3.0             & 2.6             \\ \cline{2-6} 
			& FB: MH                & 77 (71.3, 83.4)     & 73 (67.9, 78.2)     & 3.0             & 2.7             \\ \cline{2-6} 
			& Raking                & 74.3 (65.5, 84.4)   & 72.6 (62.3, 83)     & 4.8             & 5.3             \\ \hline
			\multirow{5}{*}{2003}                      & IGME                  & 74 (66.5, 82.8)     & 74 (66.5, 82.8)     & 5.3             & 5.3             \\ \cline{2-6} 
			& Unbenchmarked         & 83.4 (74, 94.5)     & 72.5 (62, 83.2)     & 5.1             & 5.4             \\ \cline{2-6} 
			& FB: Rejection Sampler & 76.6 (71, 82.8)     & 72.8 (68.1, 78)     & 3.0             & 2.5             \\ \cline{2-6} 
			& FB: MH                & 76.5 (71.1, 82.8)   & 72.9 (68, 78.4)     & 3.0             & 2.7             \\ \cline{2-6} 
			& Raking                & 74.8 (66, 85.3)     & 72.4 (62.2, 83.4)   & 4.9             & 5.4             \\ \hline
			\multirow{5}{*}{2004}                      & IGME                  & 73.7 (66.1, 82.3)   & 73.7 (66.1, 82.3)   & 5.2             & 5.2             \\ \cline{2-6} 
			& Unbenchmarked         & 80.4 (71.5, 90.1)   & 71.2 (61.1, 82.4)   & 4.6             & 5.5             \\ \cline{2-6} 
			& FB: Rejection Sampler & 74.9 (69.5, 80.8)   & 71.1 (66.6, 76.5)   & 2.9             & 2.5             \\ \cline{2-6} 
			& FB: MH                & 74.8 (69.5, 80.9)   & 71.4 (66.5, 76.8)   & 2.9             & 2.7             \\ \cline{2-6} 
			& Raking                & 73 (64.5, 82.5)     & 71.4 (61.1, 82.5)   & 4.6             & 5.4             \\ \hline
			\multirow{5}{*}{2005}                      & IGME                  & 70.2 (62.8, 78.6)   & 70.2 (62.8, 78.6)   & 5.1             & 5.1             \\ \cline{2-6} 
			& Unbenchmarked         & 77.8 (69.8, 87.3)   & 69.3 (59.2, 80.7)   & 4.4             & 5.5             \\ \cline{2-6} 
			& FB: Rejection Sampler & 71.8 (66.7, 77.6)   & 68.1 (63.7, 73.3)   & 2.8             & 2.5             \\ \cline{2-6} 
			& FB: MH                & 71.7 (66.8, 77.7)   & 68.4 (63.7, 73.7)   & 2.8             & 2.6             \\ \cline{2-6} 
			& Raking                & 70.4 (62.7, 79.6)   & 69.4 (59.2, 80.4)   & 4.3             & 5.4             \\ \hline
			\multirow{5}{*}{2006}                      & IGME                  & 65.6 (58.4, 73.8)   & 65.6 (58.4, 73.8)   & 5.0             & 5.0             \\ \cline{2-6} 
			& Unbenchmarked         & 75.2 (67.3, 86.2)   & 67.3 (57.5, 78.9)   & 4.7             & 5.5             \\ \cline{2-6} 
			& FB: Rejection Sampler & 68.3 (63.5, 74.3)   & 64.8 (60.3, 69.8)   & 2.7             & 2.4             \\ \cline{2-6} 
			& FB: MH                & 68.2 (63.4, 74.2)   & 64.9 (60.2, 69.8)   & 2.7             & 2.4             \\ \cline{2-6} 
			& Raking                & 66.7 (59.4, 76.4)   & 67.5 (57.2, 78.6)   & 4.3             & 5.4             \\ \hline
			\multirow{5}{*}{2007}                      & IGME                  & 60.7 (53.3, 69.4)   & 60.7 (53.3, 69.4)   & 5.3             & 5.3             \\ \cline{2-6} 
			& Unbenchmarked         & 70.6 (60.8, 81.2)   & 66.5 (56.3, 78.1)   & 5.1             & 5.6             \\ \cline{2-6} 
			& FB: Rejection Sampler & 63.7 (58.1, 69.3)   & 62.2 (57.7, 67)     & 2.8             & 2.4             \\ \cline{2-6} 
			& FB: MH                & 63.8 (58.4, 69)     & 62.3 (57.6, 67.1)   & 2.7             & 2.4             \\ \cline{2-6} 
			& Raking                & 61.2 (52.1, 71)     & 66.4 (56.1, 77.5)   & 4.7             & 5.5             \\ \hline
			\multirow{5}{*}{2008}                      & IGME                  & 55.8 (47.9, 65)     & 55.8 (47.9, 65)     & 5.6             & 5.6             \\ \cline{2-6} 
			& Unbenchmarked         & 68 (57.6, 78.2)     & 64.7 (54, 76.2)     & 5.0             & 5.7             \\ \cline{2-6} 
			& FB: Rejection Sampler & 60.8 (54.8, 66.2)   & 59.1 (54.3, 63.8)   & 2.9             & 2.4             \\ \cline{2-6} 
			& FB: MH                & 60.9 (55.1, 66.2)   & 59.1 (54.2, 64)     & 2.8             & 2.5             \\ \cline{2-6} 
			& Raking                & 57.1 (48, 66.5)     & 64.6 (54, 75.5)     & 4.8             & 5.5             \\ \hline
			\multirow{5}{*}{2009}                      & IGME                  & 51.8 (43.5, 62.3)   & 51.8 (43.5, 62.3)   & 6.4             & 6.4             \\ \cline{2-6} 
			& Unbenchmarked         & 64.5 (53.9, 73.8)   & 63.2 (52.5, 75.1)   & 5.0             & 5.7             \\ \cline{2-6} 
			& FB: Rejection Sampler & 58.1 (52, 63.8)     & 56.9 (51.9, 61.9)   & 3.0             & 2.6             \\ \cline{2-6} 
			& FB: MH                & 58.2 (51.8, 63.7)   & 56.7 (51.6, 62.1)   & 3.0             & 2.7             \\ \cline{2-6} 
			& Raking                & 53.3 (44, 62.3)     & 63 (52.8, 74.2)     & 4.7             & 5.6             \\
			
		\end{tabular}
		\label{tab:u5mrnatlresults}
	\end{table}
	
	\begin{table}[H]
		\centering
		\begin{tabular}{r|l|l|l|r|r}
			\textbf{Year}         & \textbf{Model}        & \textbf{AL: Median} & \textbf{UL: Median} & \textbf{AL: SD} & \textbf{UL: SD} \\ \hline
			\multirow{5}{*}{2010}                      & IGME                  & 49.7 (40.4, 62)     & 49.7 (40.4, 62)     & 7.5             & 7.5             \\ \cline{2-6} 
			& Unbenchmarked         & 62.8 (53.2, 72.7)   & 61.2 (50.8, 73.6)   & 4.9             & 5.8             \\ \cline{2-6} 
			& FB: Rejection Sampler & 56.6 (50.8, 62.4)   & 54.7 (49.5, 60.1)   & 3.0             & 2.7             \\ \cline{2-6} 
			& FB: MH                & 56.7 (50.7, 62.7)   & 54.6 (49, 60.3)     & 3.0             & 2.8             \\ \cline{2-6} 
			& Raking                & 51.9 (43.5, 60.7)   & 61 (50.6, 72.7)     & 4.4             & 5.6             \\ \hline
			\multirow{5}{*}{2011}                      & IGME                  & 50.4 (39.4, 66)     & 50.4 (39.4, 66)     & 9.5             & 9.5             \\ \cline{2-6} 
			& Unbenchmarked         & 60.7 (51.7, 71.1)   & 59.2 (48.4, 72.3)   & 4.9             & 6.0             \\ \cline{2-6} 
			& FB: Rejection Sampler & 55.1 (49.2, 61.6)   & 52.7 (47.3, 58.6)   & 3.2             & 2.9             \\ \cline{2-6} 
			& FB: MH                & 55.2 (49.1, 61.9)   & 52.7 (46.8, 58.8)   & 3.2             & 3.1             \\ \cline{2-6} 
			& Raking                & 51.4 (43.3, 60.8)   & 58.9 (48.3, 70.9)   & 4.4             & 5.8             \\ \hline
			\multirow{5}{*}{2012}                      & IGME                  & 52.2 (39.3, 71.4)   & 52.2 (39.3, 71.4)   & 11.7            & 11.7            \\ \cline{2-6} 
			& Unbenchmarked         & 56.9 (47, 68)       & 56.5 (45, 70.3)     & 5.3             & 6.5             \\ \cline{2-6} 
			& FB: Rejection Sampler & 52.4 (45.9, 59.3)   & 50.1 (44, 56.6)     & 3.4             & 3.2             \\ \cline{2-6} 
			& FB: MH                & 52.4 (46, 59.5)     & 50 (43.6, 56.5)     & 3.4             & 3.3             \\ \cline{2-6} 
			& Raking                & 49.7 (40.7, 60)     & 56.2 (45, 69.2)     & 4.8             & 6.3             \\ \hline
			\multirow{5}{*}{2013}                      & IGME                  & 50.1 (36.3, 71.8)   & 50.1 (36.3, 71.8)   & 13.2            & 13.2            \\ \cline{2-6} 
			& Unbenchmarked         & 56.9 (45.8, 72.5)   & 55.3 (42.3, 72.1)   & 6.9             & 7.5             \\ \cline{2-6} 
			& FB: Rejection Sampler & 51.5 (44.9, 60)     & 48.7 (41.9, 56)     & 3.8             & 3.6             \\ \cline{2-6} 
			& FB: MH                & 51.6 (44.8, 60.3)   & 48.6 (41.4, 56.1)   & 3.9             & 3.7             \\ \cline{2-6} 
			& Raking                & 51 (40, 66.7)       & 55.2 (42.3, 70.9)   & 6.9             & 7.3             \\ \hline
			\multirow{5}{*}{2014}                      & IGME                  & 48.1 (33.7, 72.3)   & 48.1 (33.7, 72.3)   & 14.7            & 14.7            \\ \cline{2-6} 
			& Unbenchmarked         & 54.7 (40.1, 80.6)   & 54.4 (39.4, 76.4)   & 10.2            & 9.4             \\ \cline{2-6} 
			& FB: Rejection Sampler & 49.5 (41.7, 59.1)   & 47.4 (39.5, 55.7)   & 4.4             & 4.1             \\ \cline{2-6} 
			& FB: MH                & 49.5 (41.5, 59.2)   & 47.3 (39.3, 55.8)   & 4.5             & 4.3             \\ \cline{2-6} 
			& Raking                & 50.2 (35.3, 80.8)   & 54.3 (39, 74.7)     & 11.7            & 9.2             \\ \hline
			\multirow{5}{*}{2015}                      & IGME                  & 47.8 (32.3, 74.7)   & 47.8 (32.3, 74.7)   & 16.3            & 16.3            \\ \cline{2-6} 
			& Unbenchmarked         & 53.5 (35.7, 90.8)   & 54.1 (36.1, 82.9)   & 14.9            & 12.2            \\ \cline{2-6} 
			& FB: Rejection Sampler & 48 (38.9, 58.4)     & 46.2 (37.2, 56)     & 4.9             & 4.8             \\ \cline{2-6} 
			& FB: MH                & 47.9 (39, 58.7)     & 46.1 (36.9, 56.4)   & 5.0             & 4.9             \\ \cline{2-6} 
			& Raking                & 50 (31, 101.7)      & 53.9 (35.6, 81.1)   & 18.9            & 12.1            \\ \hline
			\multirow{5}{*}{2016}                      & IGME                  & 46.3 (30, 75.6)     & 46.3 (30, 75.6)     & 17.8            & 17.8            \\ \cline{2-6} 
			& Unbenchmarked         & 52.3 (31, 106.8)    & 54 (32.8, 94.2)     & 21.5            & 16.5            \\ \cline{2-6} 
			& FB: Rejection Sampler & 46.4 (36, 58)       & 45.3 (35.2, 56.7)   & 5.6             & 5.5             \\ \cline{2-6} 
			& FB: MH                & 46.6 (36.3, 58.9)   & 45.3 (34.6, 57.1)   & 5.7             & 5.6             \\ \cline{2-6} 
			& Raking                & 49.7 (25.9, 132.1)  & 53.8 (32.3, 92.1)   & 30.7            & 16.7            \\ \hline
			\multirow{5}{*}{2017}                      & IGME                  & 44.5 (28, 75.1)     & 44.5 (28, 75.1)     & 18.6            & 18.6            \\ \cline{2-6} 
			& Unbenchmarked         & 51.3 (26.3, 127.4)  & 54.6 (30.1, 112.2)  & 30.1            & 23.0            \\ \cline{2-6} 
			& FB: Rejection Sampler & 45.1 (33.1, 58.7)   & 44.7 (32.9, 58.4)   & 6.4             & 6.4             \\ \cline{2-6} 
			& FB: MH                & 45.1 (33.4, 59.3)   & 44.7 (32.2, 58.7)   & 6.5             & 6.7             \\ \cline{2-6} 
			& Raking                & 49.4 (21.6, 169.1)  & 54.3 (29.4, 109.8)  & 46.4            & 23.6            \\ \hline
			\multirow{5}{*}{2018}                      & IGME                  & 43.3 (26.2, 76)     & 43.3 (26.2, 76)     & 19.9            & 19.9            \\ \cline{2-6} 
			& Unbenchmarked         & 50.1 (21.7, 152.7)  & 55.5 (26.9, 135.3)  & 41.2            & 31.7            \\ \cline{2-6} 
			& FB: Rejection Sampler & 43.7 (29.7, 59.9)   & 44.2 (30.7, 60.9)   & 7.4             & 7.7             \\ \cline{2-6} 
			& FB: MH                & 43.9 (29.8, 60.1)   & 44.5 (29.6, 61.9)   & 7.6             & 8.1             \\ \cline{2-6} 
			& Raking                & 49.1 (17, 220.1)    & 55 (26.7, 133.2)    & 64.8            & 33.1            \\ \hline
			\multirow{5}{*}{2019}                      & IGME                  & 42.4 (24.8, 77)     & 42.4 (24.8, 77)     & 21.1            & 21.1            \\ \cline{2-6} 
			& Unbenchmarked         & 48.8 (18.2, 186.6)  & 56.4 (24.2, 169.5)  & 53.6            & 43.0            \\ \cline{2-6} 
			& FB: Rejection Sampler & 42.5 (26.3, 62.2)   & 44.2 (28.4, 65.6)   & 8.9             & 9.4             \\ \cline{2-6} 
			& FB: MH                & 42.6 (26.4, 62.9)   & 44.4 (27.1, 67.1)   & 8.9             & 10.2            \\ \cline{2-6} 
			& Raking                & 48.4 (13.5, 284.7)  & 55.9 (23.8, 164.4)  & 83.9            & 45.7           
		\end{tabular}
	\end{table}
	
\begin{figure}[H]
		\centering
		\includegraphics[scale = 0.55]{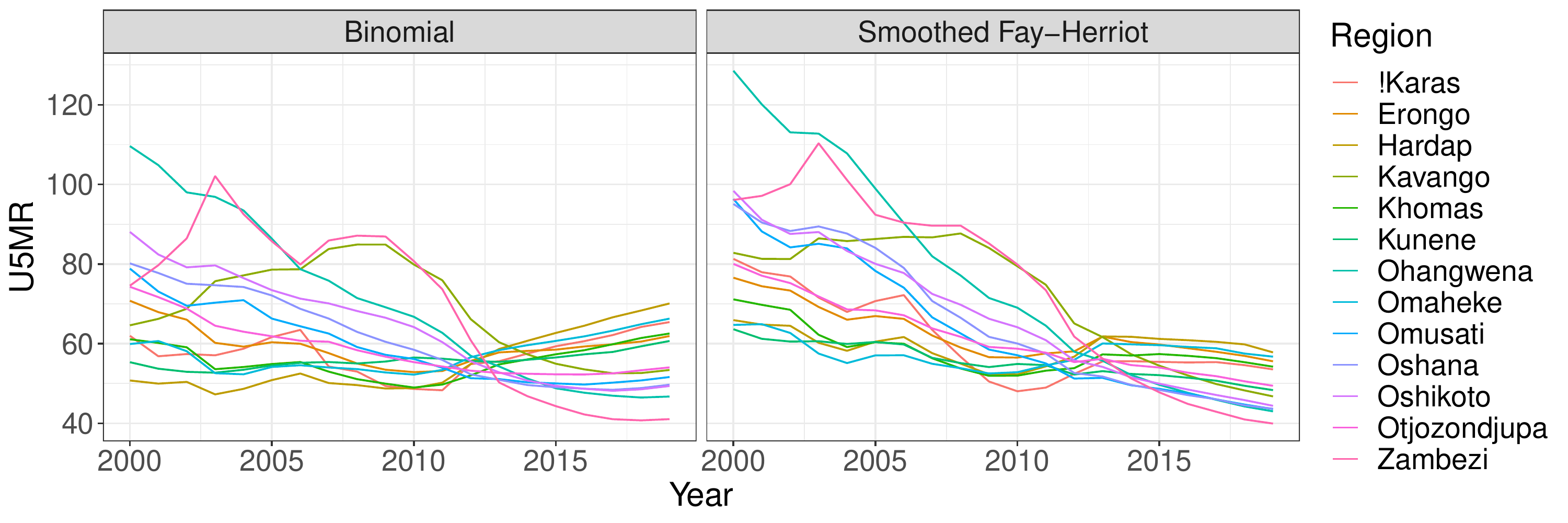}
		\caption{Comparison of median, unbenchmarked U5MR estimates from unit- and area-level models across time. U5MR is reported as deaths per 1000 live births.}
		\label{fig:u5mr_spagunbenched}
	\end{figure}

\begin{figure}[H]
		\centering
		\includegraphics[scale = 0.55]{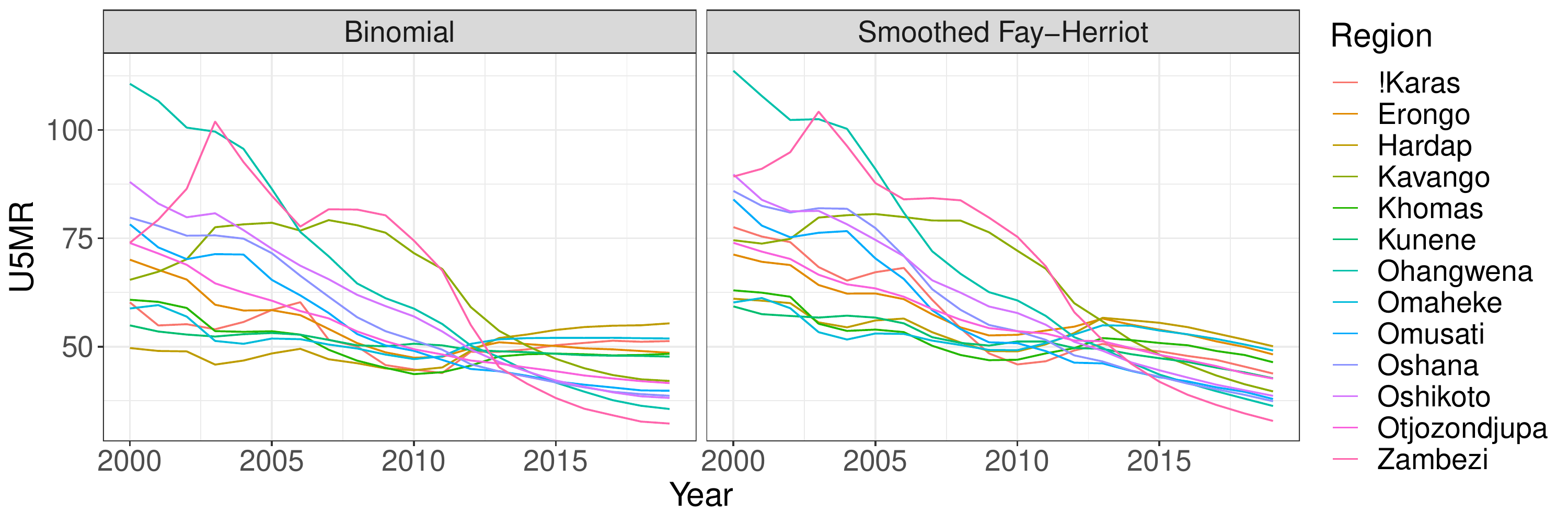}
		\caption{Comparison of median, benchmarked U5MR estimates from unit- and area-level models across time for the Fully Bayesian Metropolis-Hastings approach. U5MR is reported as deaths per 1000 live births.}
		\label{fig:u5mr_spagmh}
	\end{figure}

\begin{figure}[H]
		\centering
		\includegraphics[scale = 0.55]{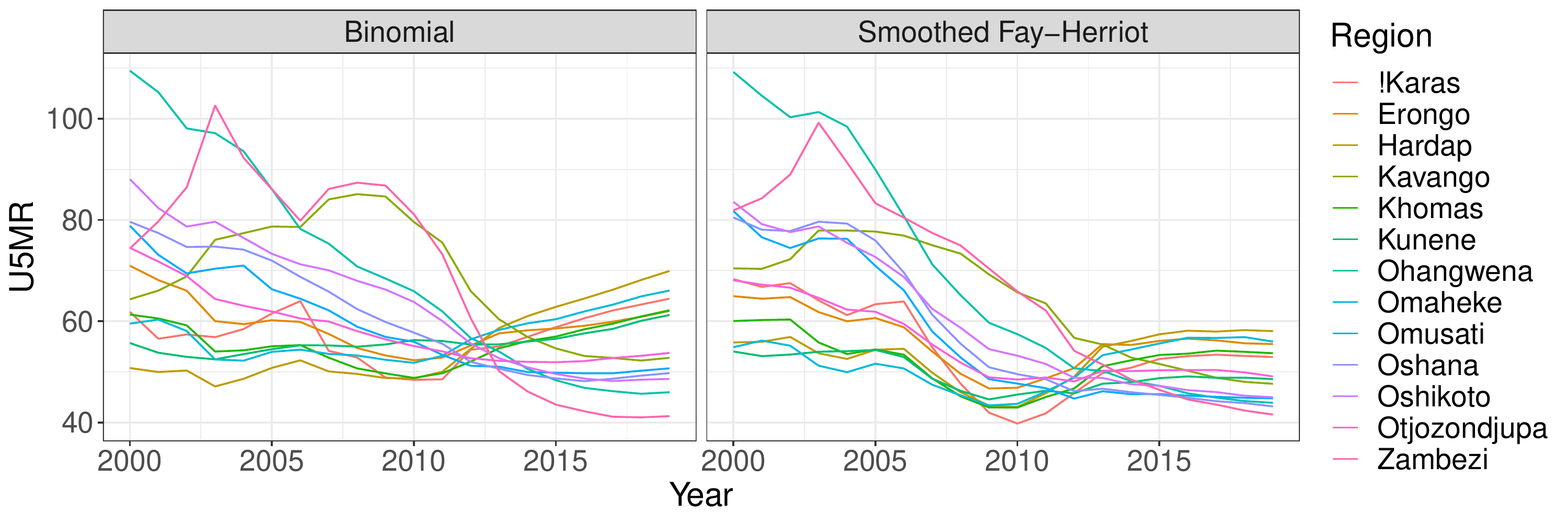}
		\caption{Comparison of median, benchmarked U5MR estimates from unit- and area-level models across time for the raking approach. U5MR is reported as deaths per 1000 live births.}
		\label{fig:u5mr_spagsimulrake}
	\end{figure}

\begin{figure}[H]
		\centering
		\includegraphics[scale = 0.55]{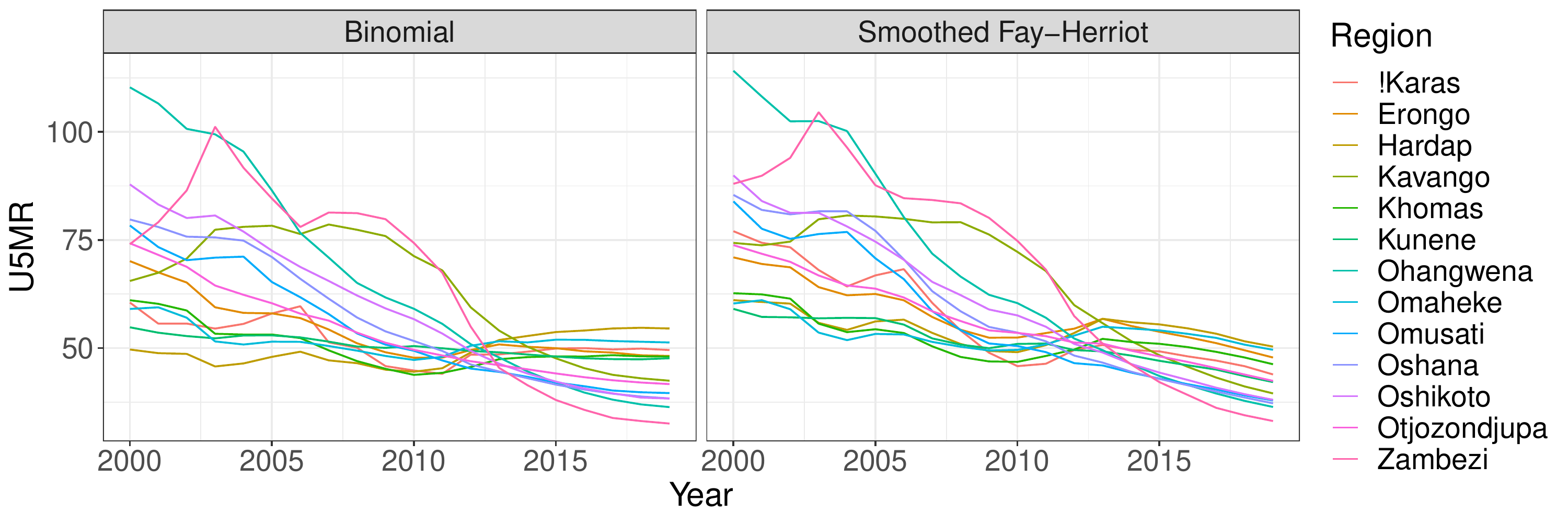}
		\caption{Comparison of median, benchmarked U5MR estimates from unit- and area-level models across time for the rejection sampler approach. U5MR is reported as deaths per 1000 live births.}
		\label{fig:u5mr_spagrejectsamp}
	\end{figure}

	Comparisons between unit-level and area-level models for each admin1 region across time can be found in Figures \ref{fig:u5mr_spagunbenched} through \ref{fig:u5mr_spagrejectsamp}.
	
\begin{figure}[H]
		\centering
		\includegraphics[scale = 0.75]{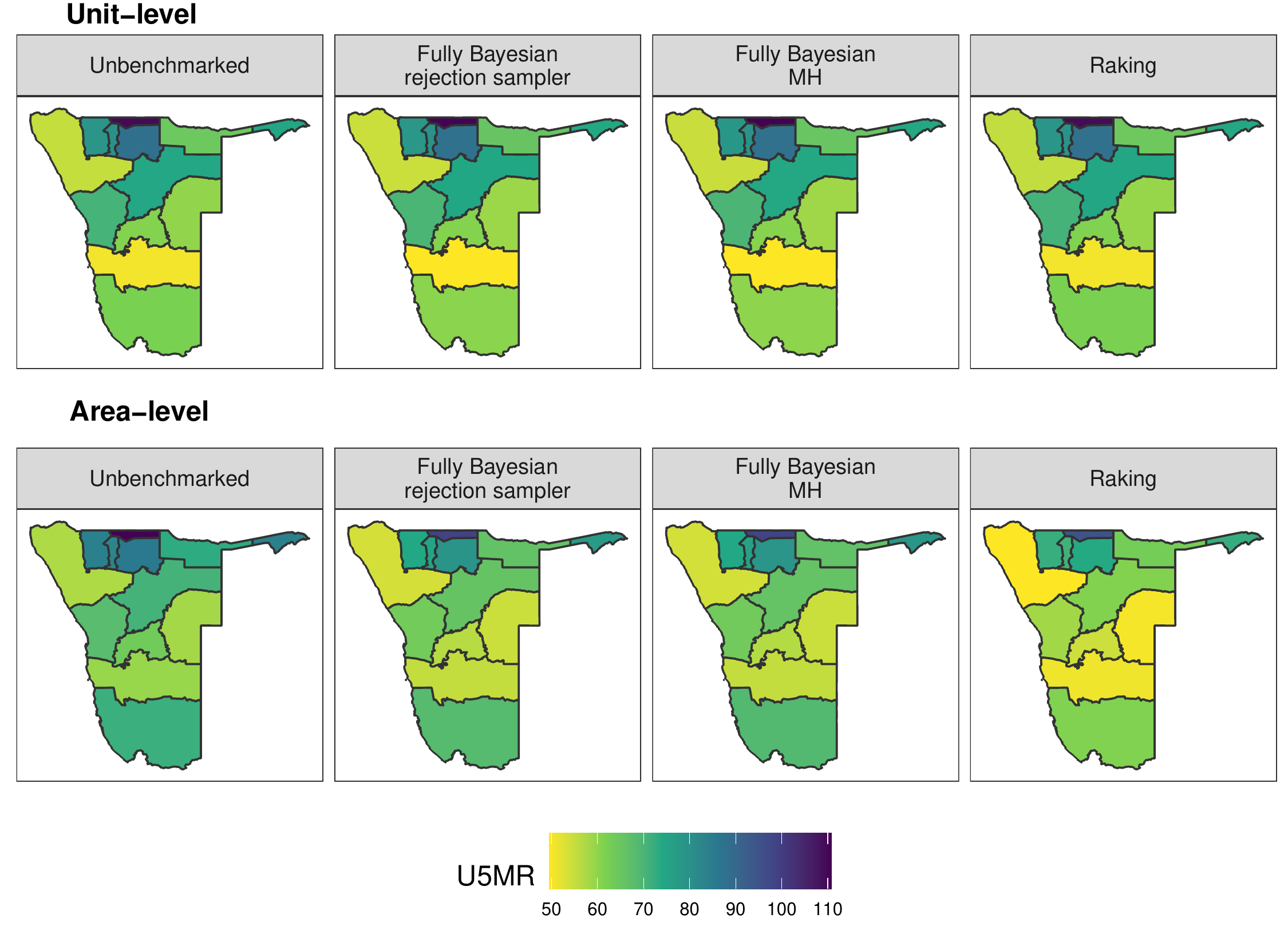}
		\caption{Comparison of median U5MR estimates from benchmarked and unbenchmarked unit- and area-level models for 2000. U5MR is reported as deaths per 1000 live births.}
		\label{fig:u5mrmedmaps2000}
	\end{figure}

\begin{figure}[H]
		\centering
		\includegraphics[scale = 0.75]{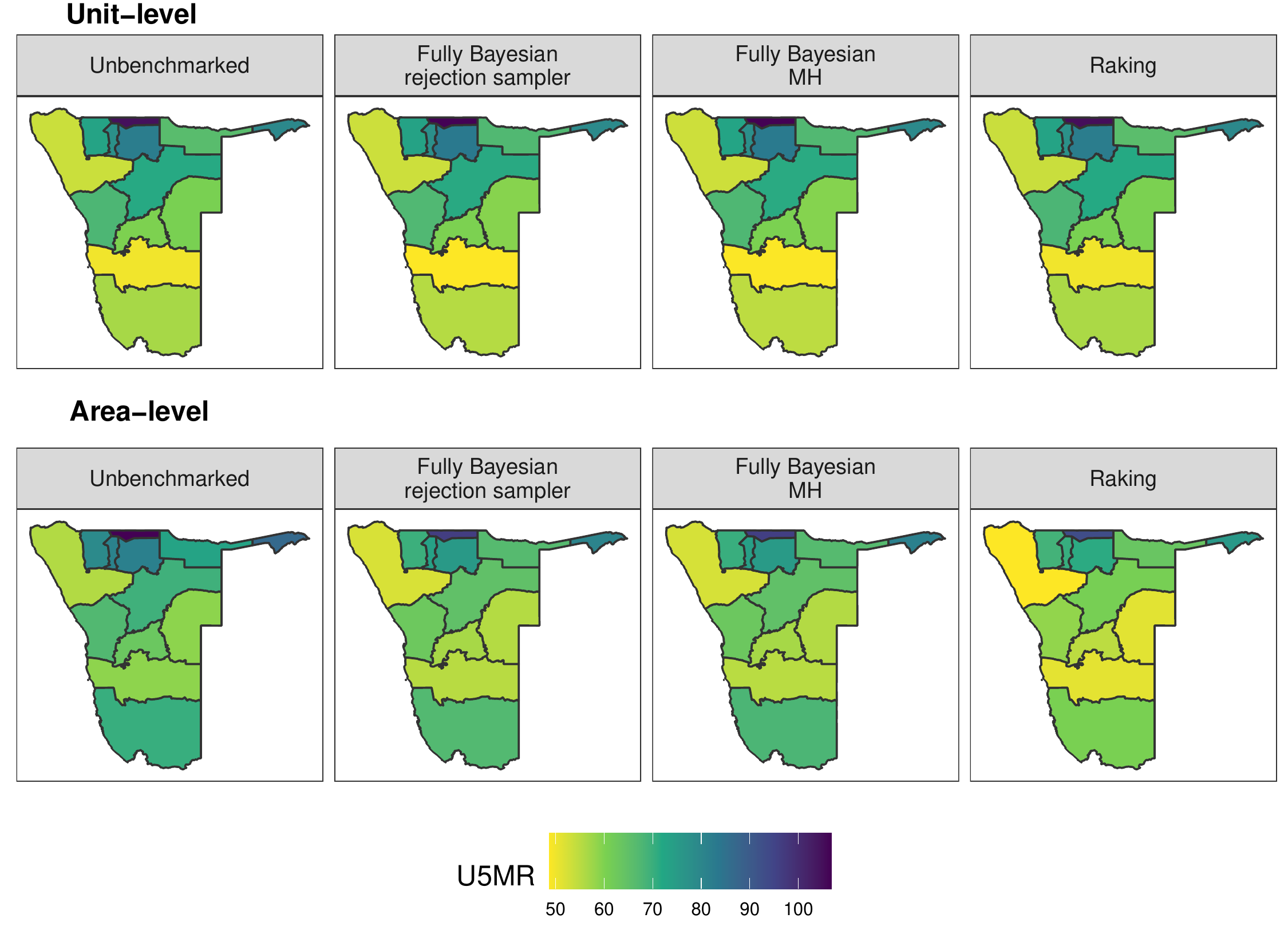}
		\caption{Comparison of median U5MR estimates from benchmarked and unbenchmarked unit- and area-level models for 2001. U5MR is reported as deaths per 1000 live births.}
		\label{fig:u5mrmedmaps2001}
	\end{figure}

\begin{figure}[H]
		\centering
		\includegraphics[scale = 0.75]{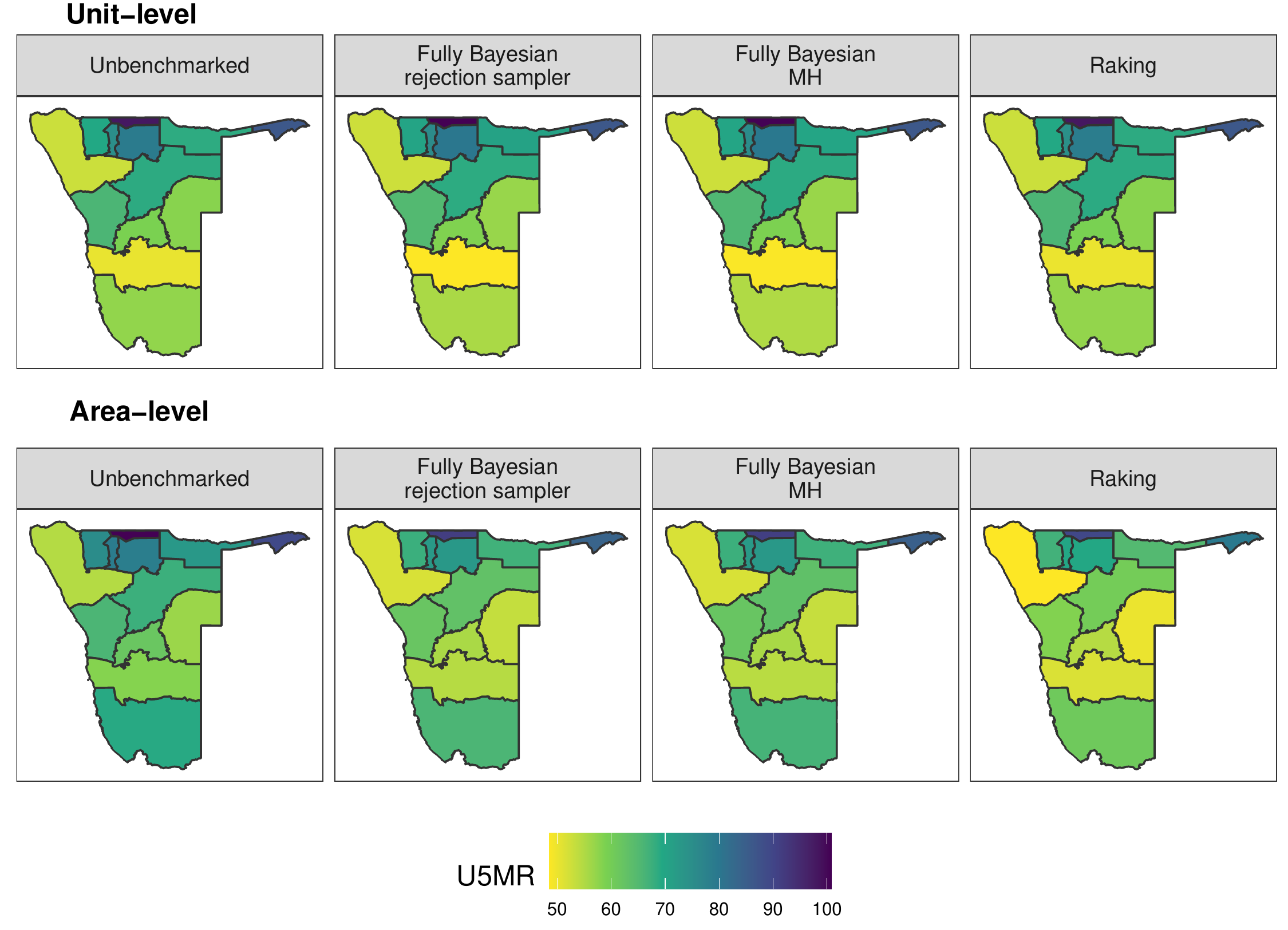}
		\caption{Comparison of median U5MR estimates from benchmarked and unbenchmarked unit- and area-level models for 2002. U5MR is reported as deaths per 1000 live births.}
		\label{fig:u5mrmedmaps2002}
	\end{figure}

\begin{figure}[H]
		\centering
		\includegraphics[scale = 0.75]{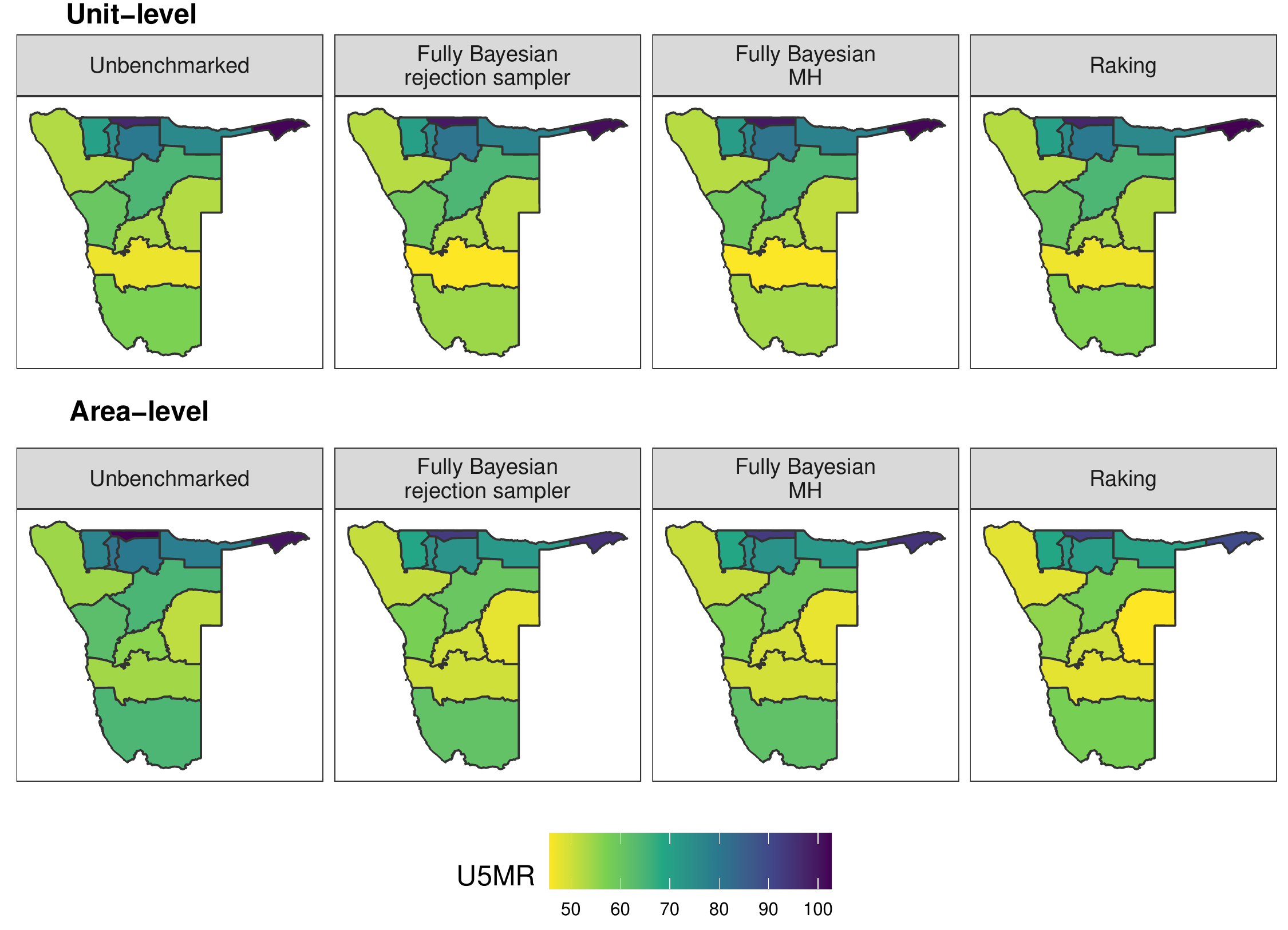}
		\caption{Comparison of median U5MR estimates from benchmarked and unbenchmarked unit- and area-level models for 2003. U5MR is reported as deaths per 1000 live births.}
		\label{fig:u5mrmedmaps2003}
	\end{figure}

\begin{figure}[H]
		\centering
		\includegraphics[scale = 0.75]{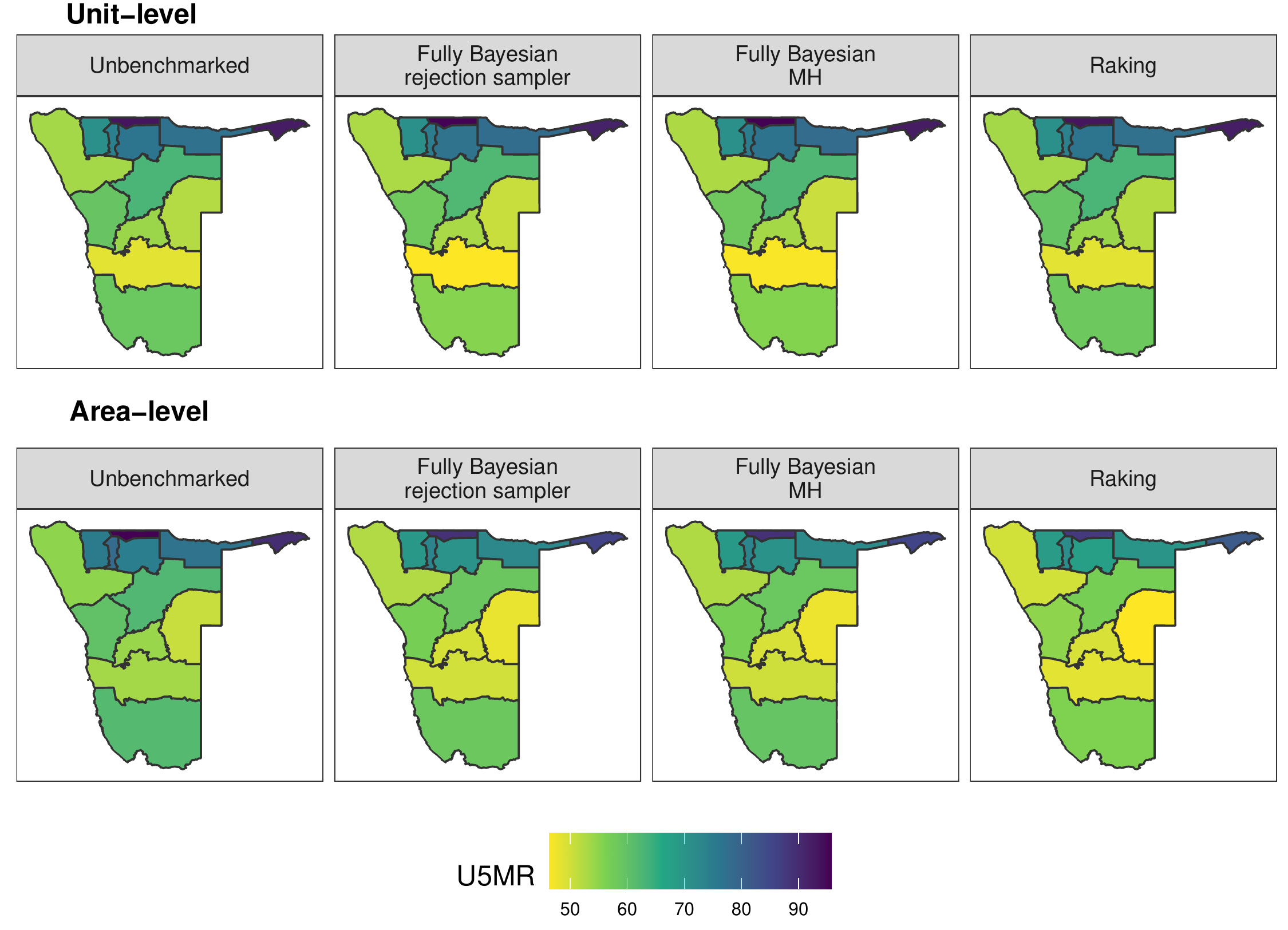}
		\caption{Comparison of median U5MR estimates from benchmarked and unbenchmarked unit- and area-level models for 2004. U5MR is reported as deaths per 1000 live births.}
		\label{fig:u5mrmedmaps2004}
	\end{figure}

\begin{figure}[H]
		\centering
		\includegraphics[scale = 0.75]{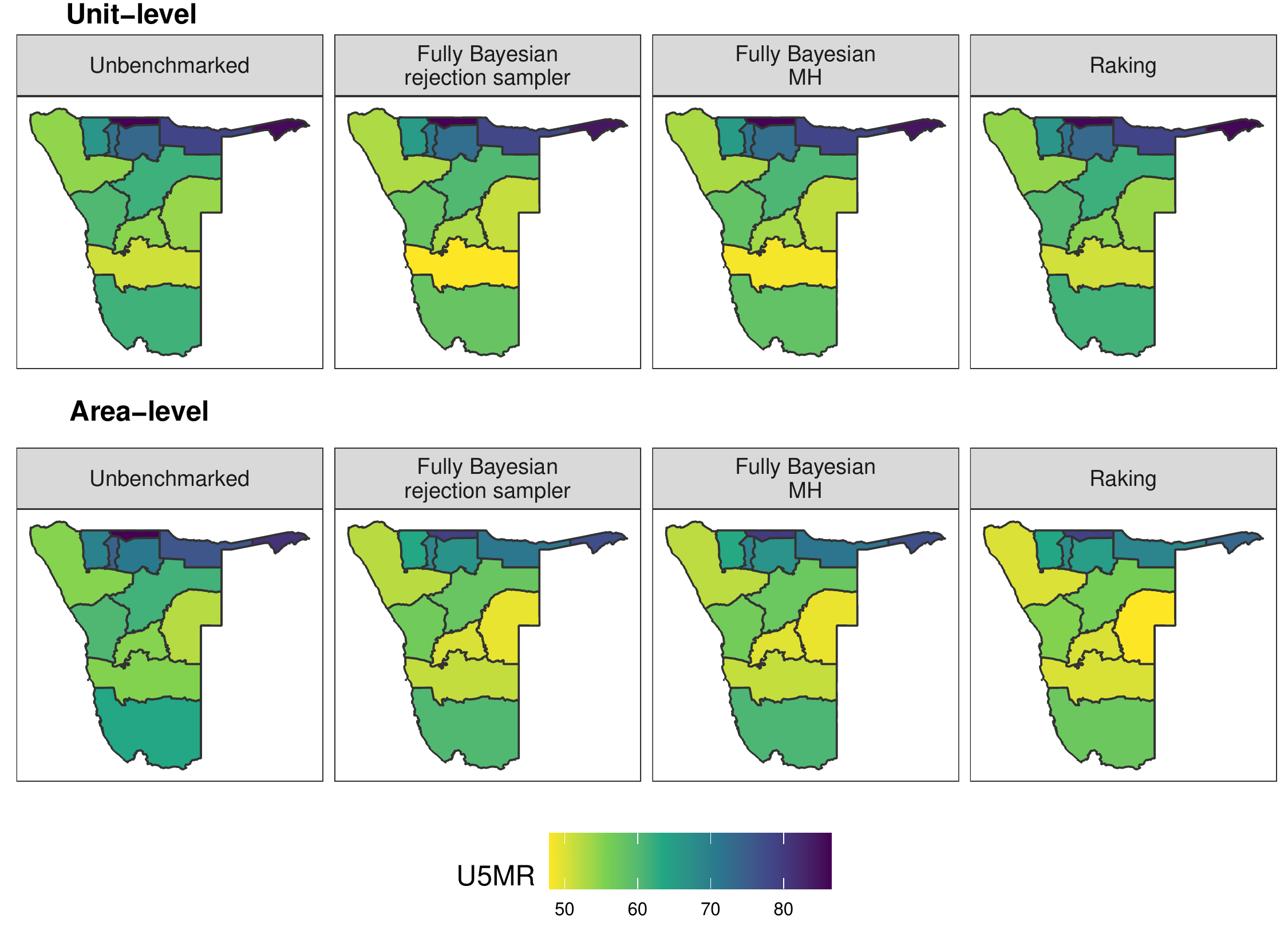}
		\caption{Comparison of median U5MR estimates from benchmarked and unbenchmarked unit- and area-level models for 2005. U5MR is reported as deaths per 1000 live births.}
		\label{fig:u5mrmedmaps2005}
	\end{figure}

\begin{figure}[H]
		\centering
		\includegraphics[scale = 0.75]{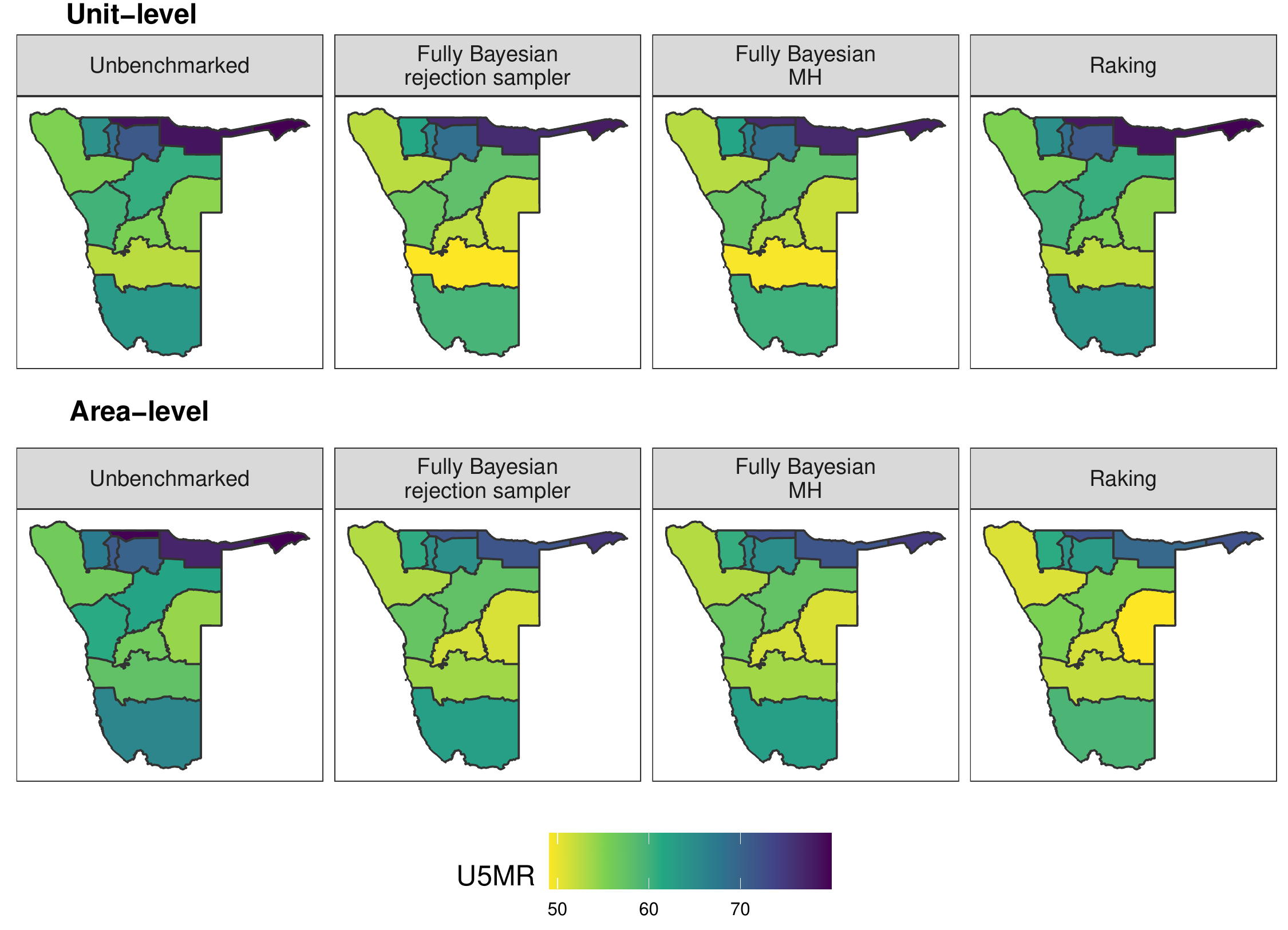}
		\caption{Comparison of median U5MR estimates from benchmarked and unbenchmarked unit- and area-level models for 2006. U5MR is reported as deaths per 1000 live births.}
		\label{fig:u5mrmedmaps2006}
	\end{figure}

\begin{figure}[H]
		\centering
		\includegraphics[scale = 0.75]{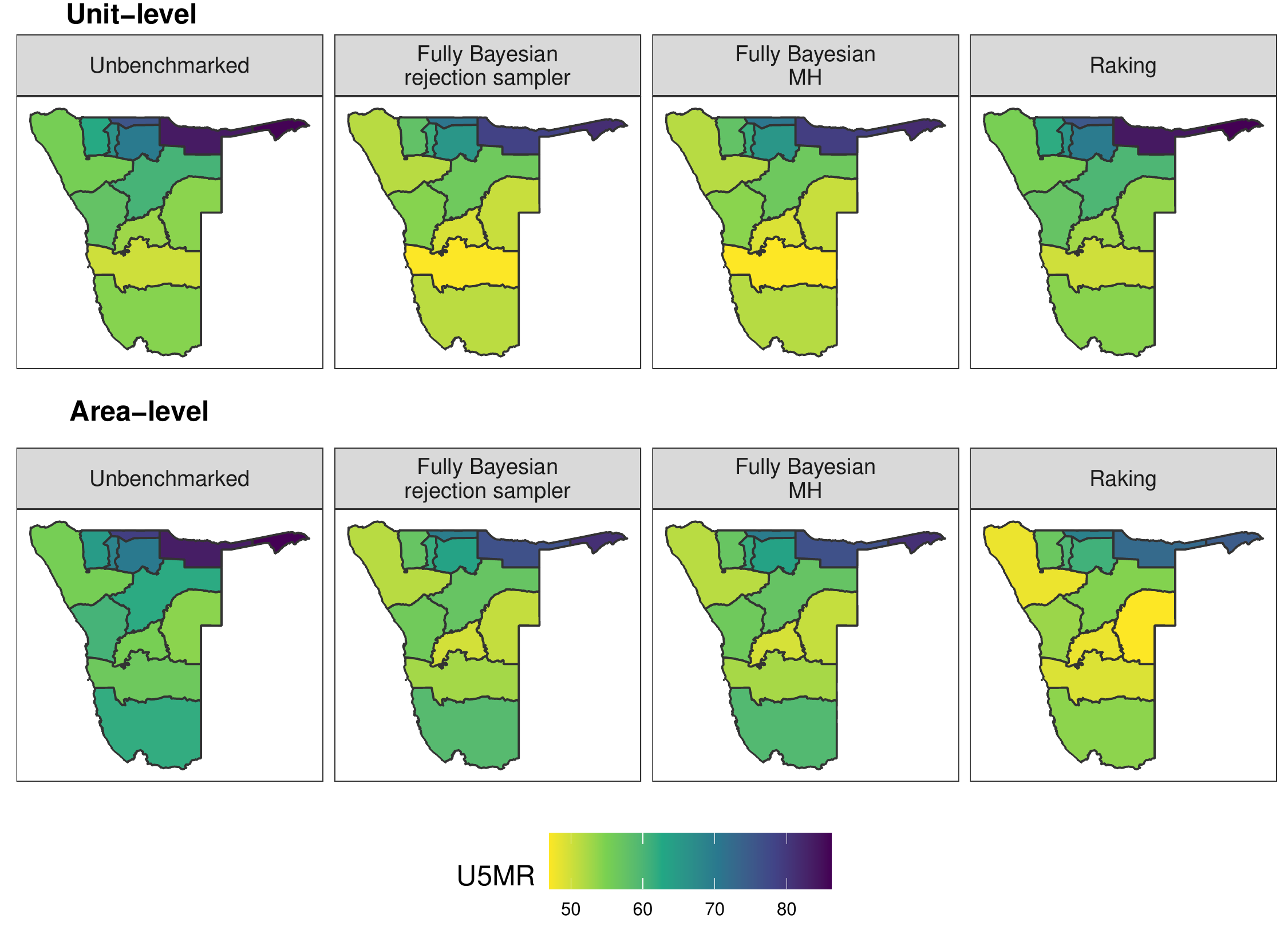}
		\caption{Comparison of median U5MR estimates from benchmarked and unbenchmarked unit- and area-level models for 2007. U5MR is reported as deaths per 1000 live births.}
		\label{fig:u5mrmedmaps2007}
	\end{figure}

\begin{figure}[H]
		\centering
		\includegraphics[scale = 0.75]{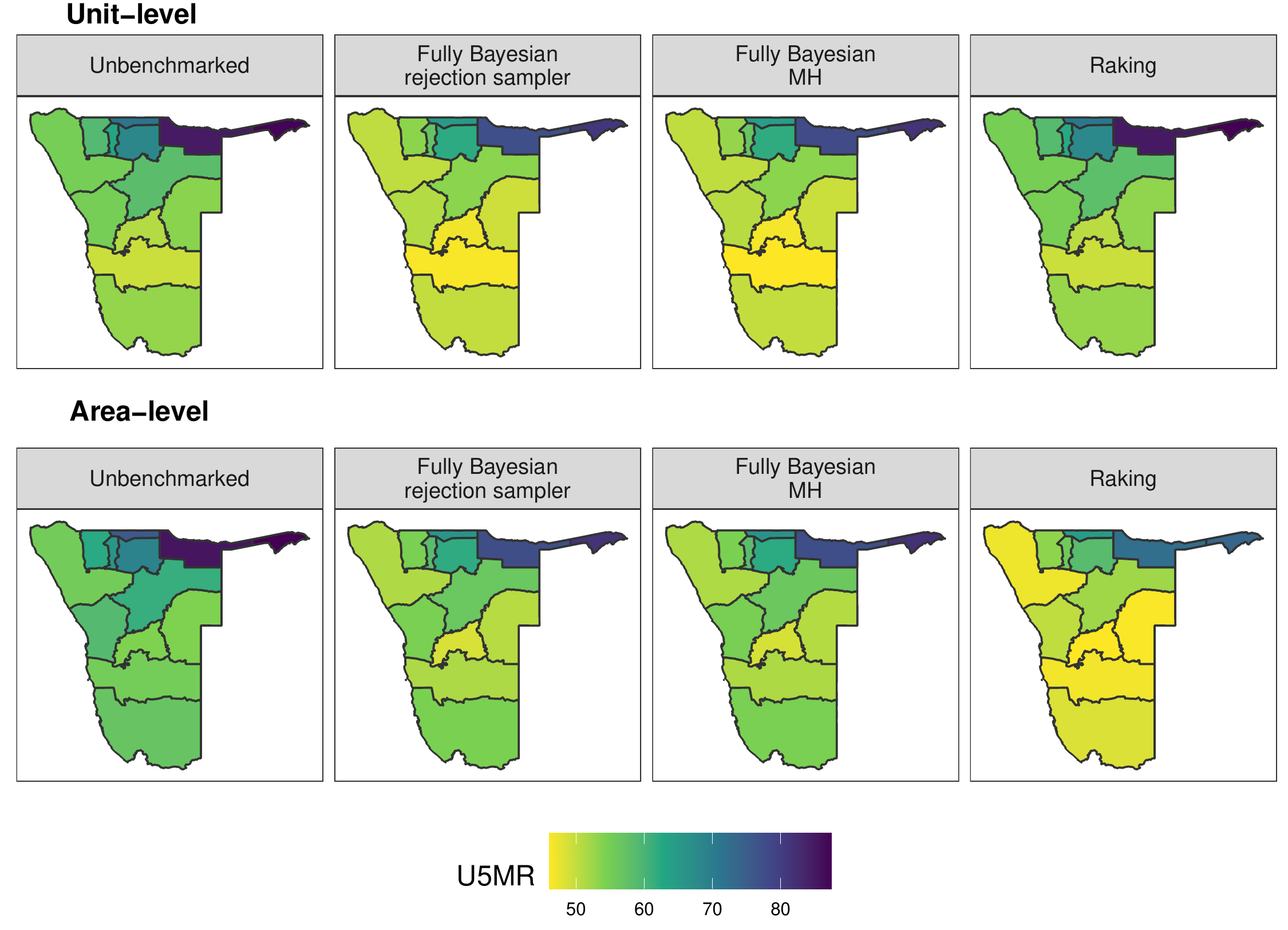}
		\caption{Comparison of median U5MR estimates from benchmarked and unbenchmarked unit- and area-level models for 2008. U5MR is reported as deaths per 1000 live births.}
		\label{fig:u5mrmedmaps2008}
	\end{figure}

\begin{figure}[H]
		\centering
		\includegraphics[scale = 0.75]{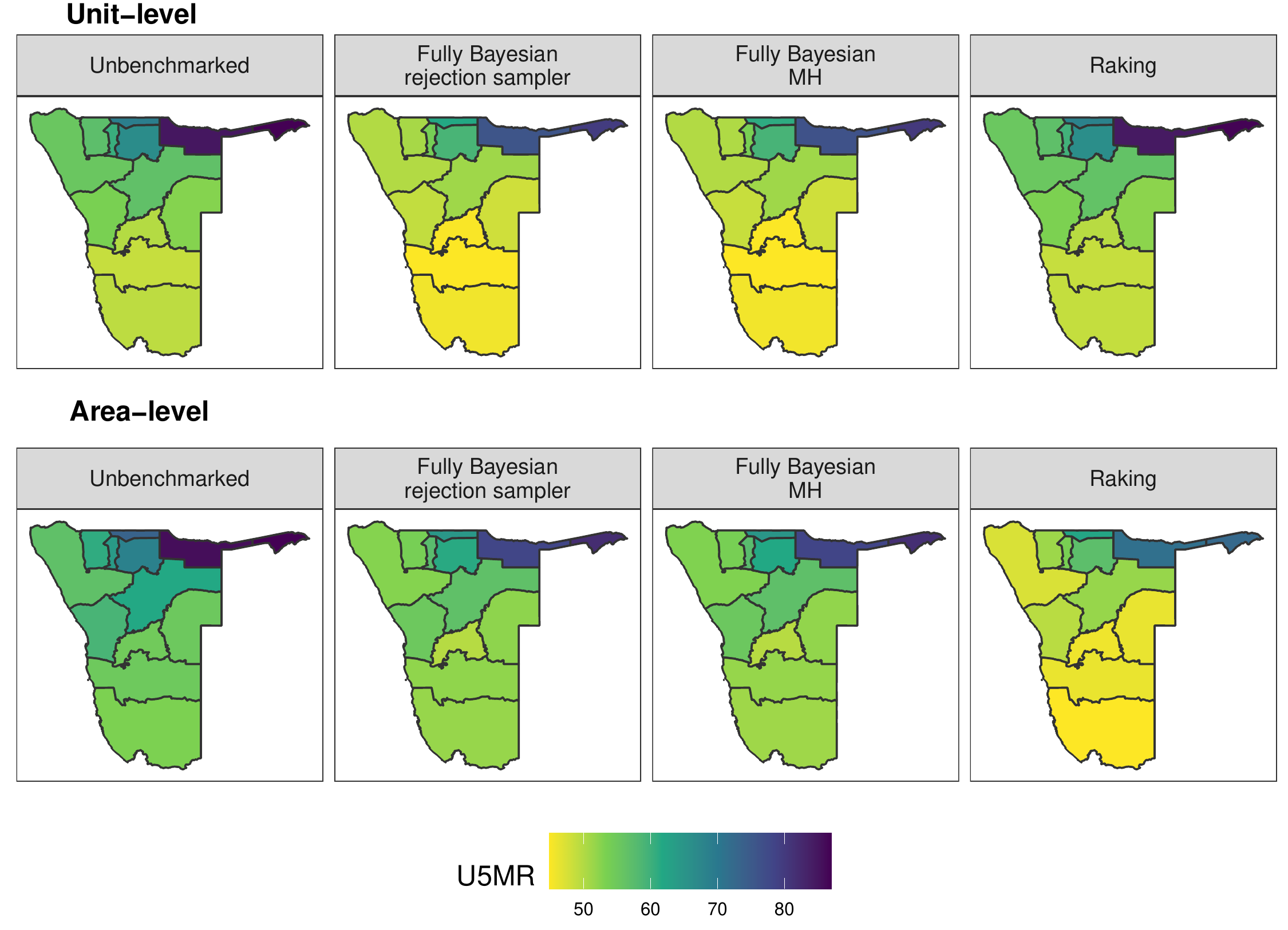}
		\caption{Comparison of median U5MR estimates from benchmarked and unbenchmarked unit- and area-level models for 2009. U5MR is reported as deaths per 1000 live births.}
		\label{fig:u5mrmedmaps2009}
	\end{figure}

\begin{figure}[H]
		\centering
		\includegraphics[scale = 0.75]{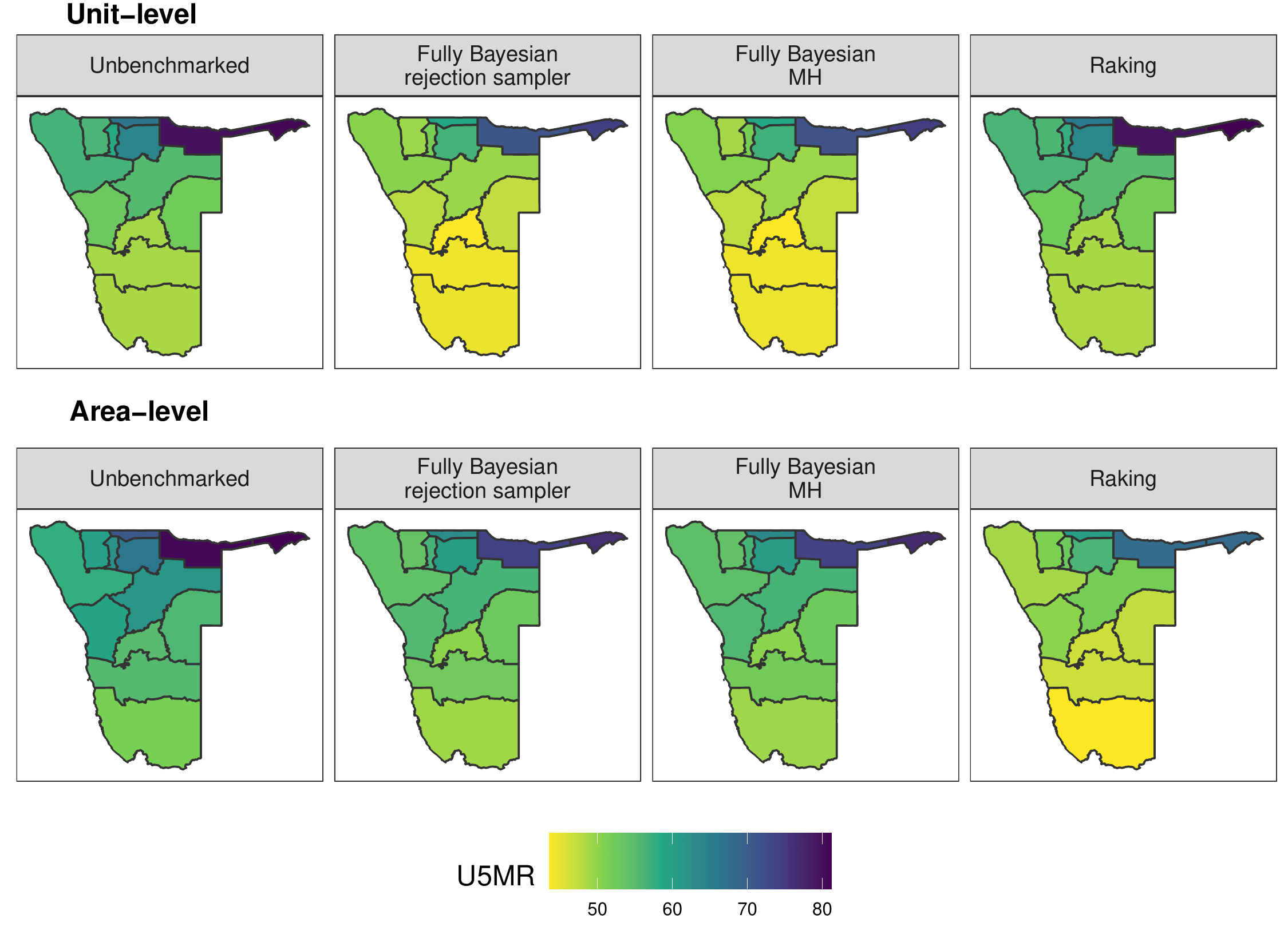}
		\caption{Comparison of median U5MR estimates from benchmarked and unbenchmarked unit- and area-level models for 2010. U5MR is reported as deaths per 1000 live births.}
		\label{fig:u5mrmedmaps2010}
	\end{figure}

\begin{figure}[H]
		\centering
		\includegraphics[scale = 0.75]{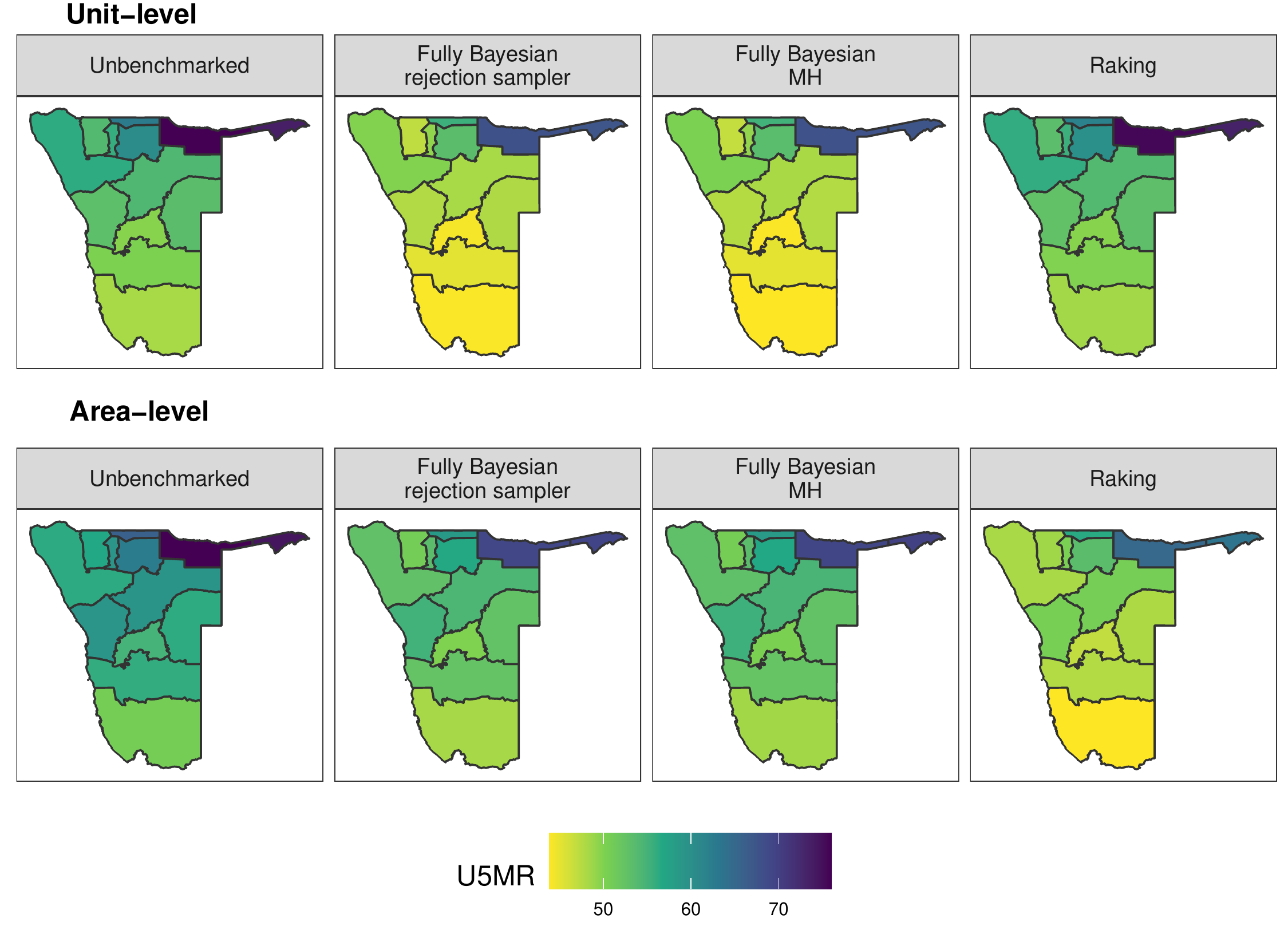}
		\caption{Comparison of median U5MR estimates from benchmarked and unbenchmarked unit- and area-level models for 2011. U5MR is reported as deaths per 1000 live births.}
		\label{fig:u5mrmedmaps2011}
	\end{figure}

\begin{figure}[H]
		\centering
		\includegraphics[scale = 0.75]{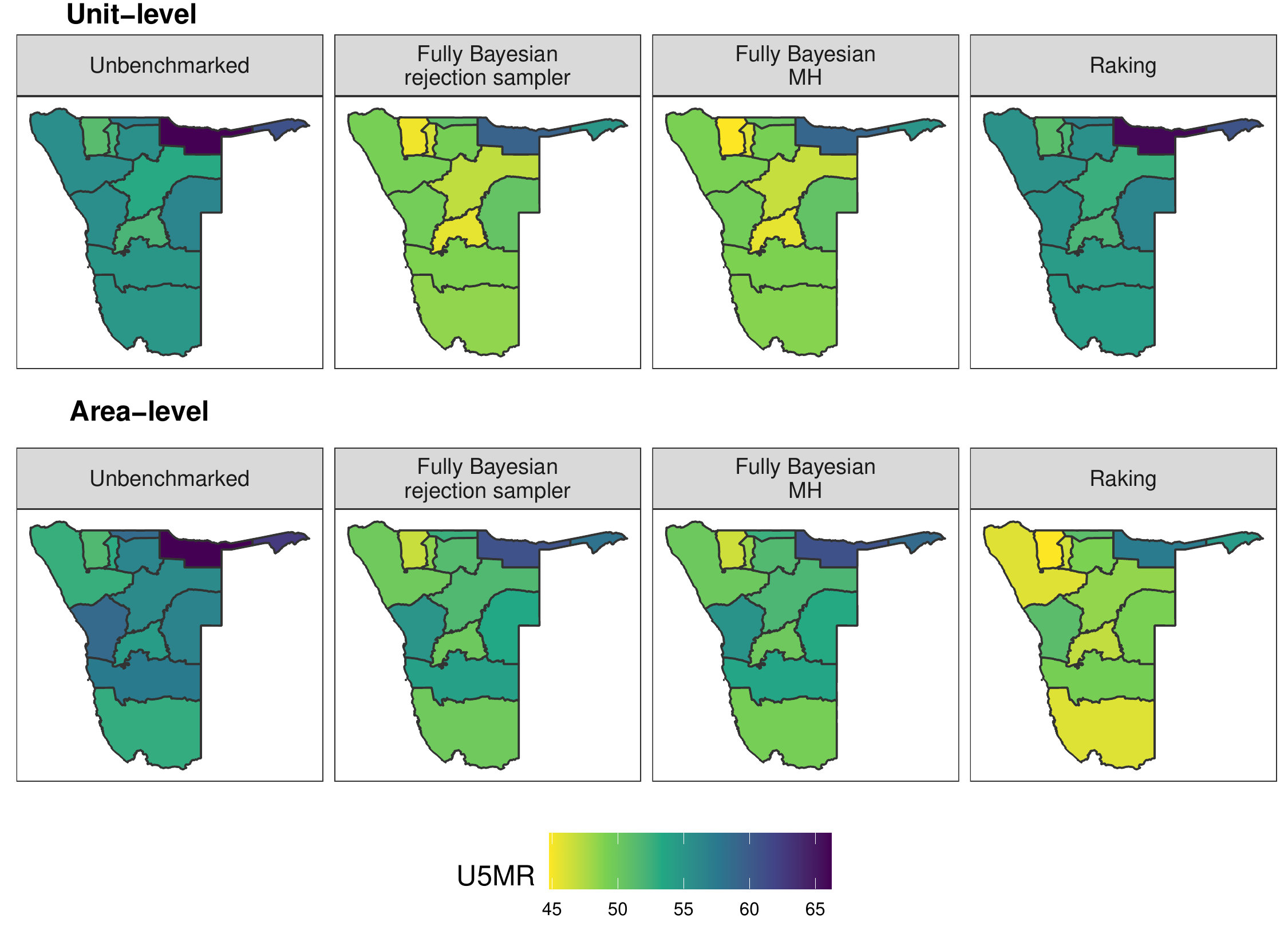}
		\caption{Comparison of median U5MR estimates from benchmarked and unbenchmarked unit- and area-level models for 2012. U5MR is reported as deaths per 1000 live births.}
		\label{fig:u5mrmedmaps2012}
	\end{figure}

\begin{figure}[H]
		\centering
		\includegraphics[scale = 0.75]{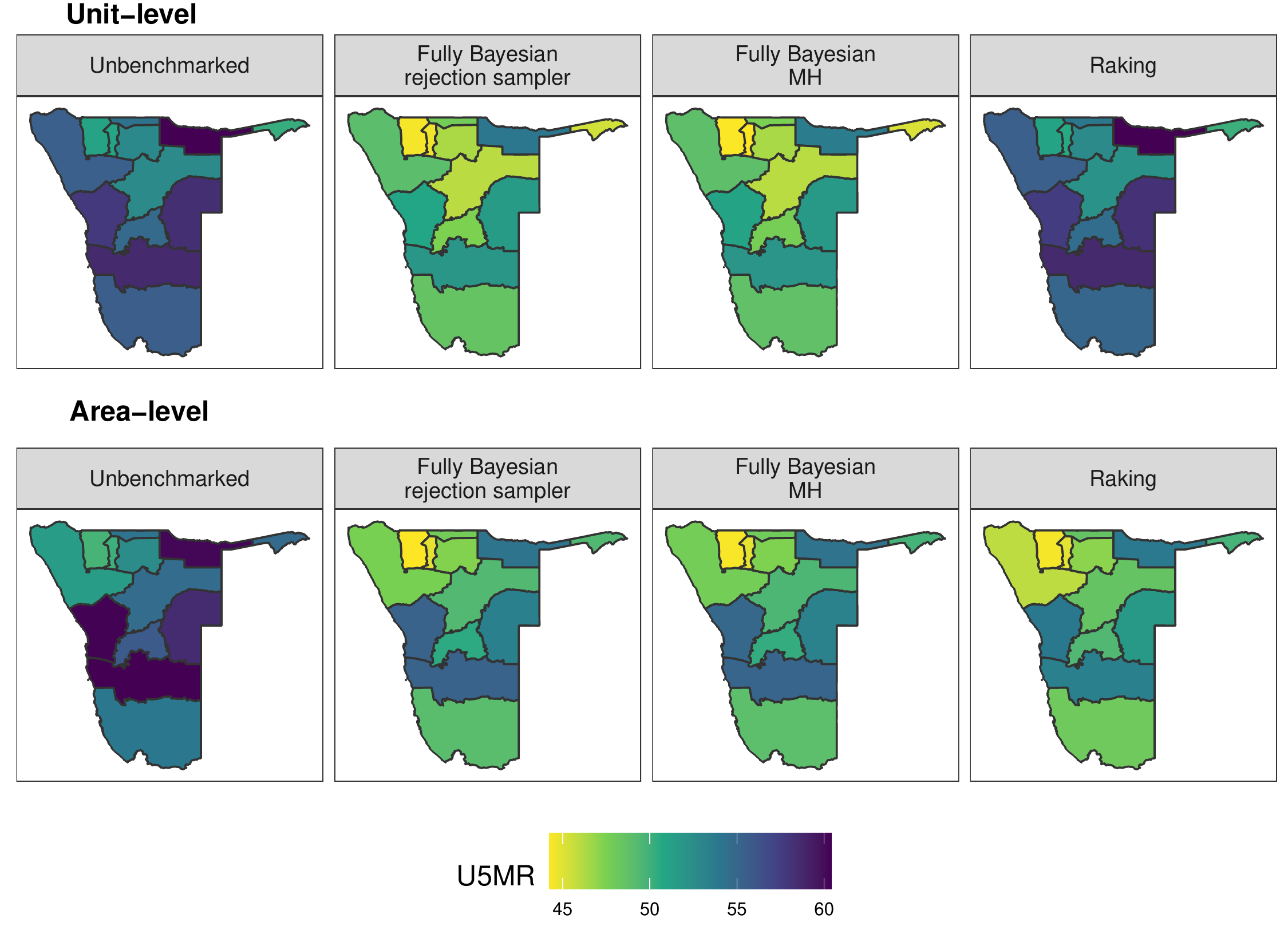}
		\caption{Comparison of median U5MR estimates from benchmarked and unbenchmarked unit- and area-level models for 2013. U5MR is reported as deaths per 1000 live births.}
		\label{fig:u5mrmedmaps2013}
	\end{figure}

\begin{figure}[H]
		\centering
		\includegraphics[scale = 0.75]{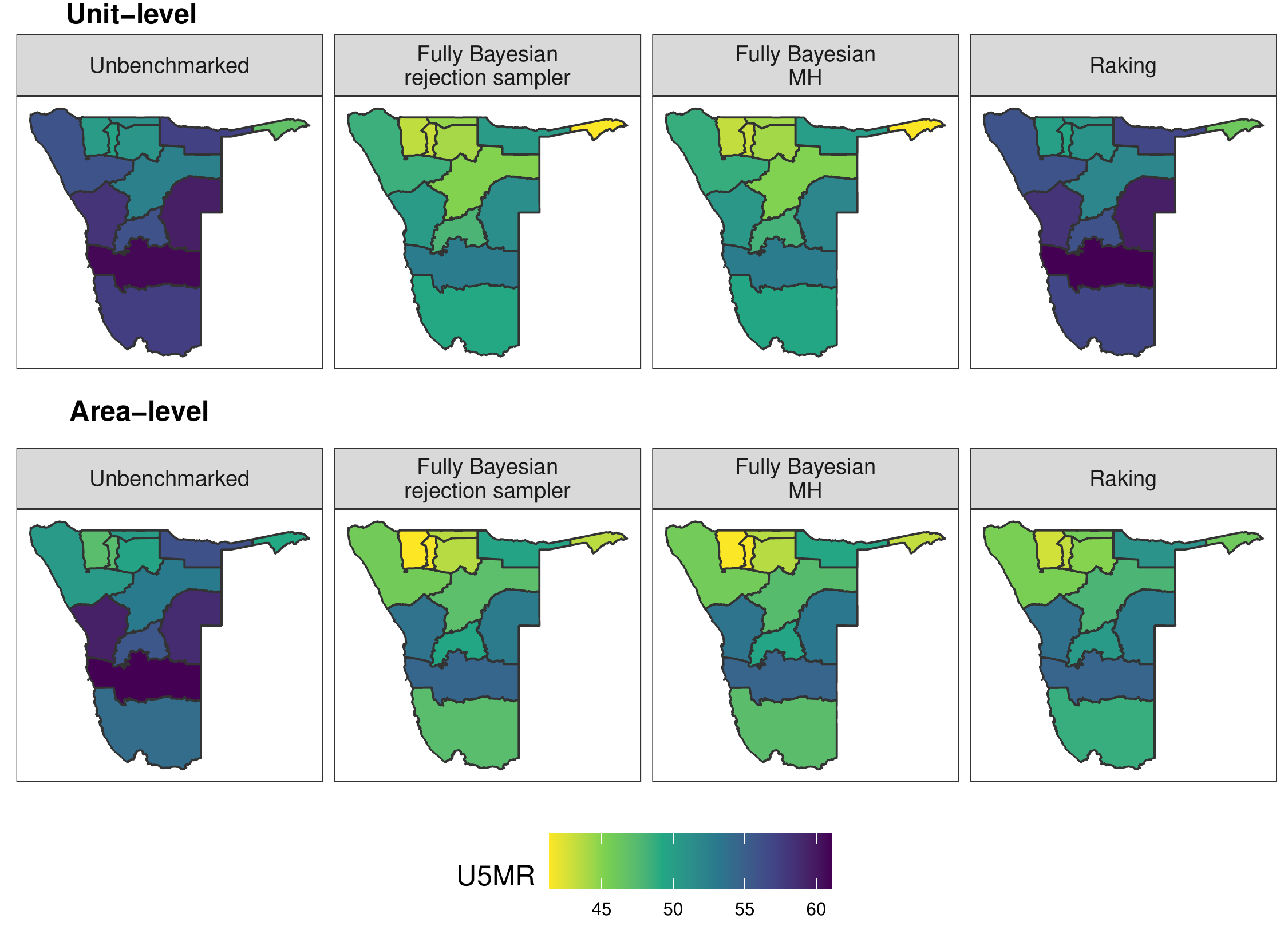}
		\caption{Comparison of median U5MR estimates from benchmarked and unbenchmarked unit- and area-level models for 2014. U5MR is reported as deaths per 1000 live births.}
		\label{fig:u5mrmedmaps2014}
	\end{figure}

\begin{figure}[H]
		\centering
		\includegraphics[scale = 0.75]{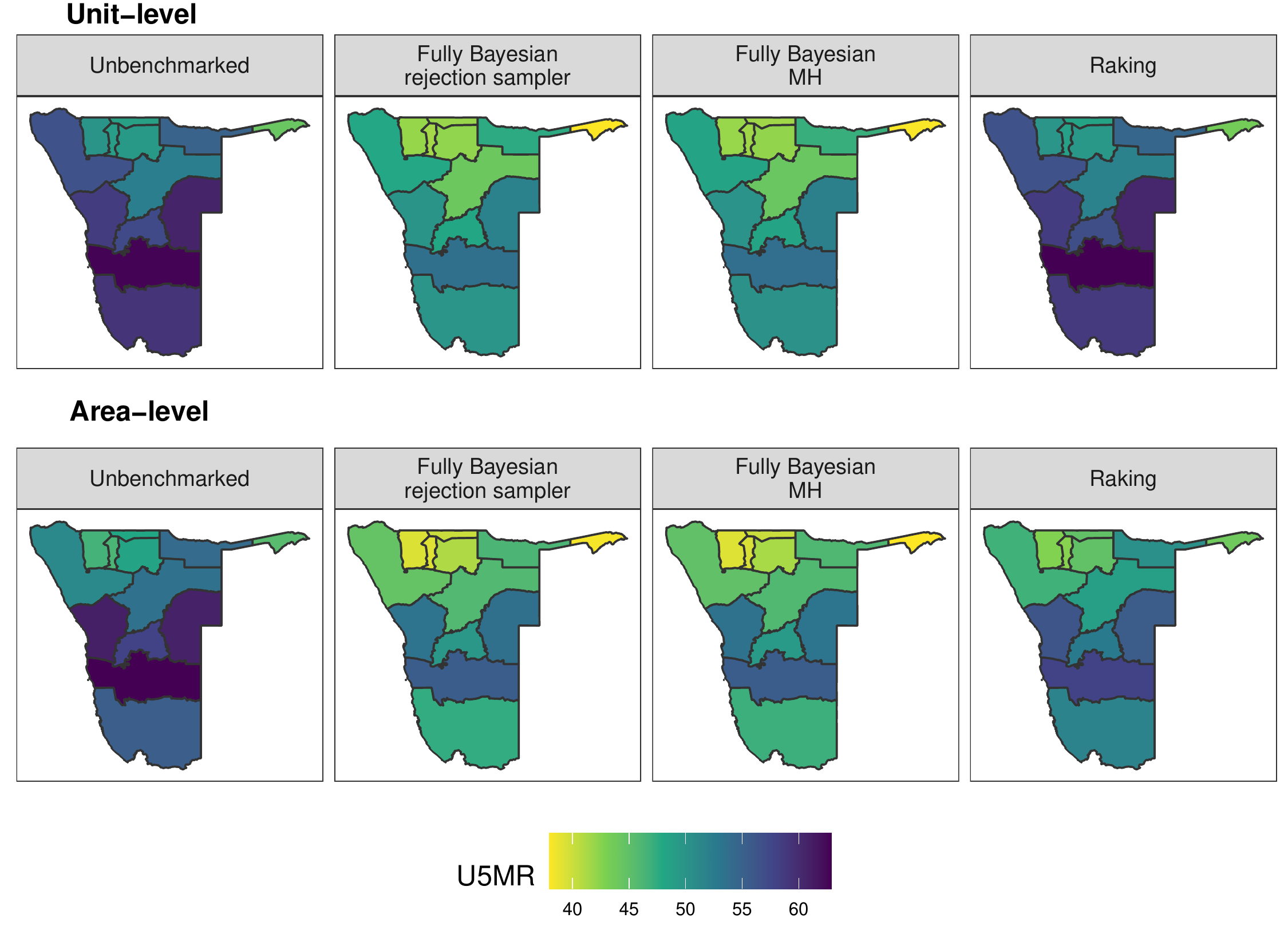}
		\caption{Comparison of median U5MR estimates from benchmarked and unbenchmarked unit- and area-level models for 2015. U5MR is reported as deaths per 1000 live births.}
		\label{fig:u5mrmedmaps2015}
	\end{figure}

\begin{figure}[H]
		\centering
		\includegraphics[scale = 0.75]{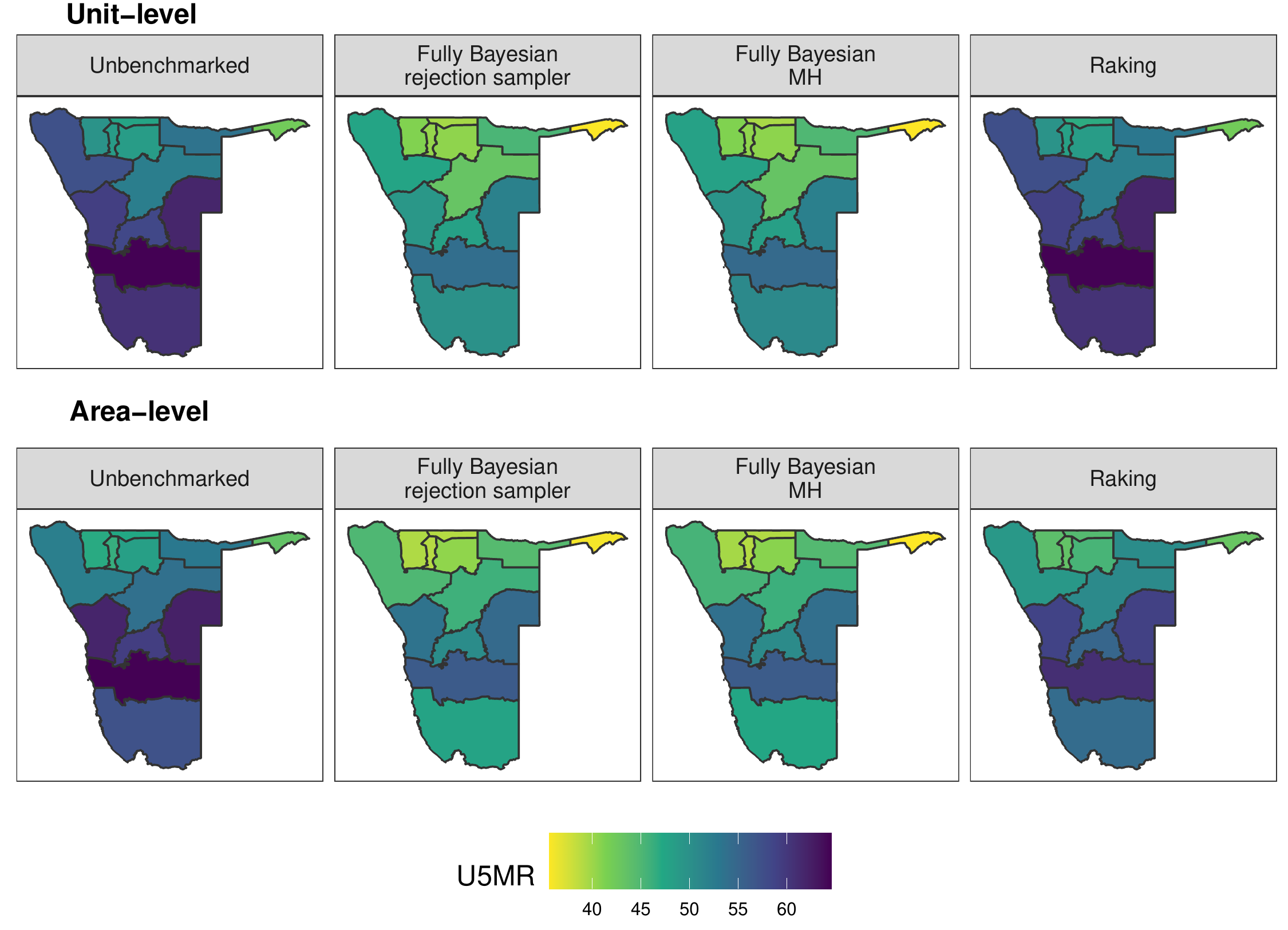}
		\caption{Comparison of median U5MR estimates from benchmarked and unbenchmarked unit- and area-level models for 2016. U5MR is reported as deaths per 1000 live births.}
		\label{fig:u5mrmedmaps2016}
	\end{figure}

\begin{figure}[H]
		\centering
		\includegraphics[scale = 0.75]{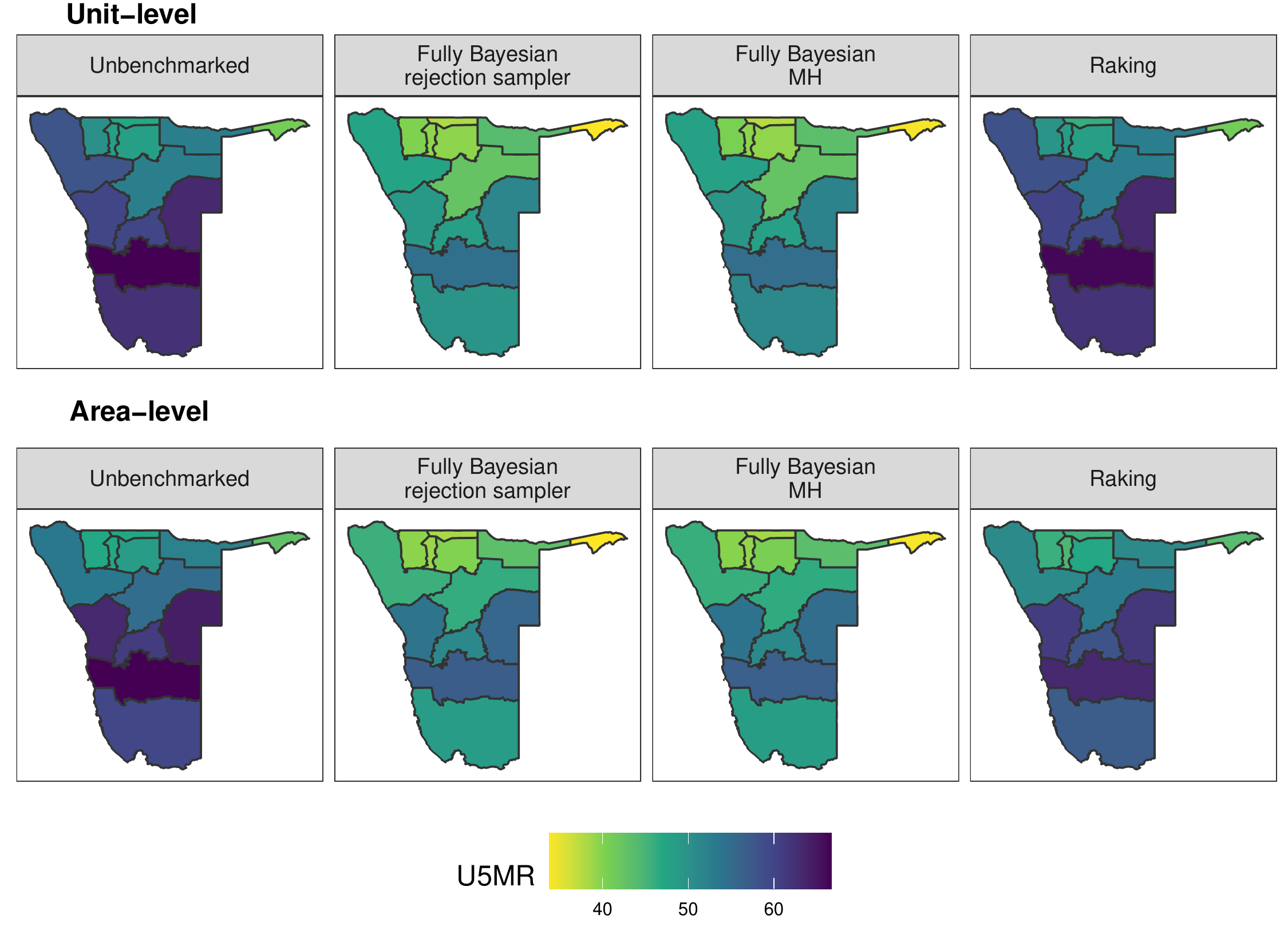}
		\caption{Comparison of median U5MR estimates from benchmarked and unbenchmarked unit- and area-level models for 2017. U5MR is reported as deaths per 1000 live births.}
		\label{fig:u5mrmedmaps2017}
	\end{figure}

\begin{figure}[H]
		\centering
		\includegraphics[scale = 0.75]{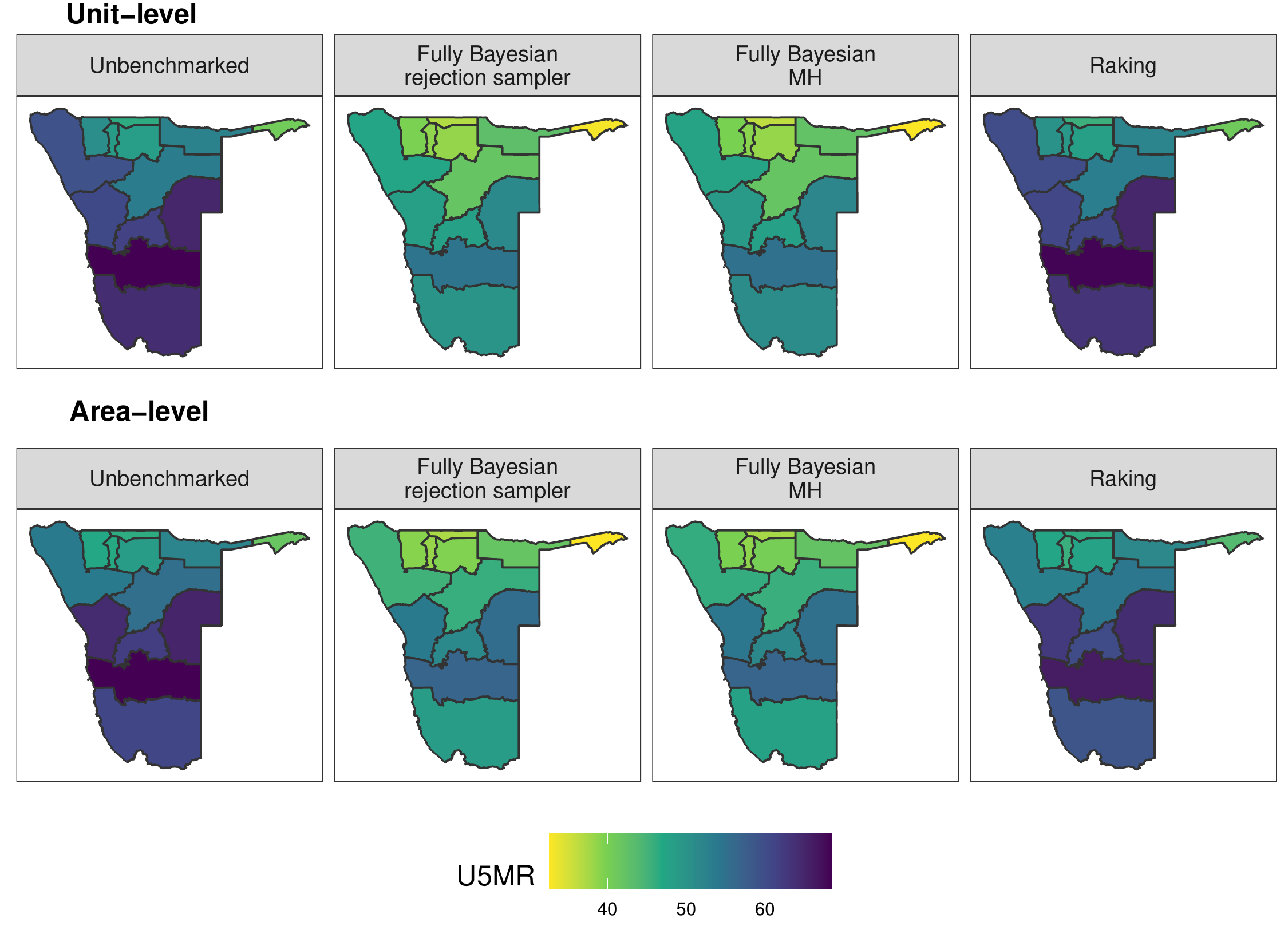}
		\caption{Comparison of median U5MR estimates from benchmarked and unbenchmarked unit- and area-level models for 2018. U5MR is reported as deaths per 1000 live births.}
		\label{fig:u5mrmedmaps2018}
	\end{figure}

\begin{figure}[H]
		\centering
		\includegraphics[scale = 0.75]{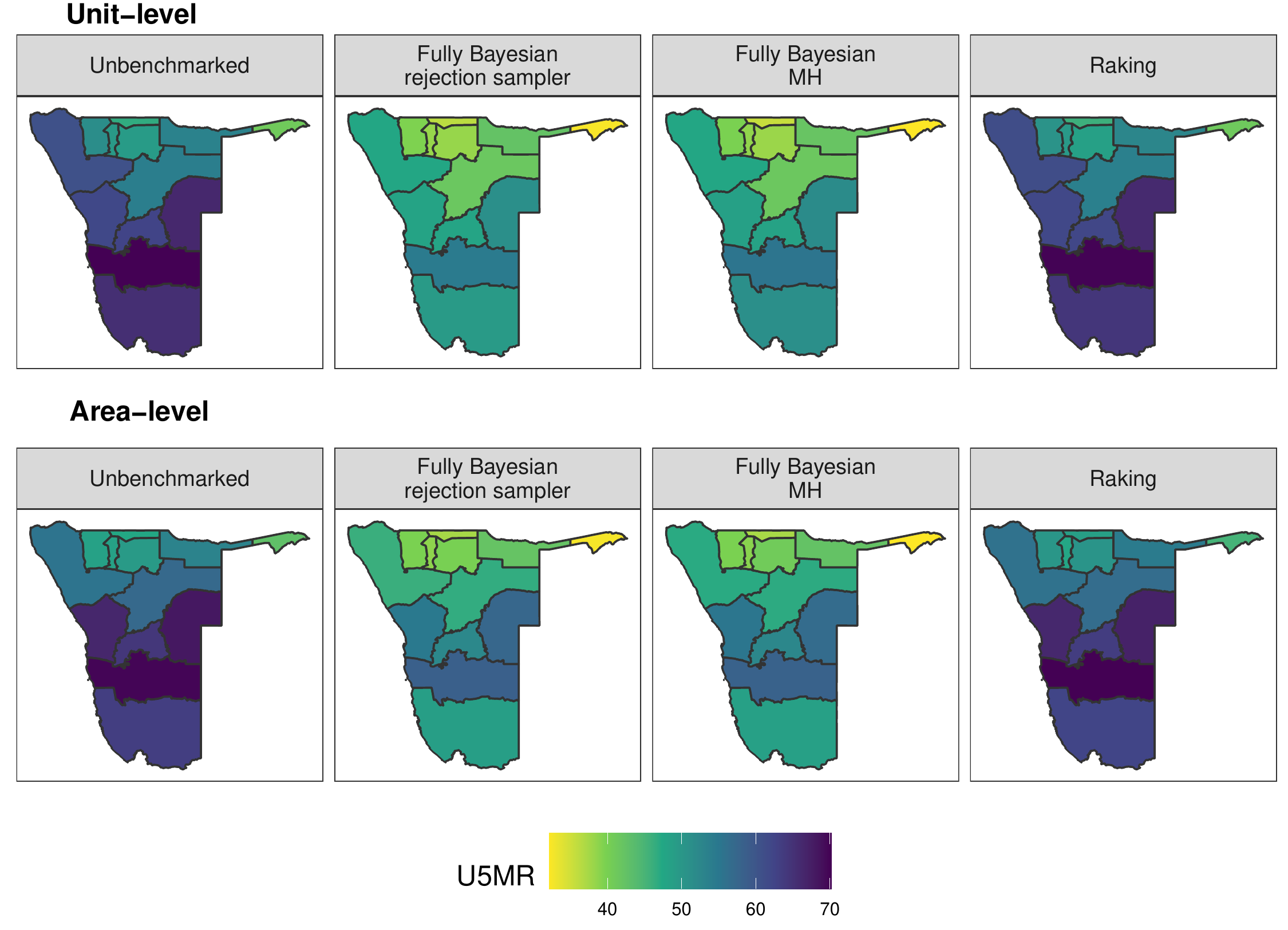}
		\caption{Comparison of median U5MR estimates from benchmarked and unbenchmarked unit- and area-level models for 2019. U5MR is reported as deaths per 1000 live births.}
		\label{fig:u5mrmedmaps2019}
	\end{figure}

	Subnational maps of unbenchmarked and benchmarked estimates from 2000 to 2019 can be found in Figures \ref{fig:u5mrmedmaps2000} through \ref{fig:u5mrmedmaps2019}.
	
	\subsection{Model Validation}
	
	Below we display the model validation results for the HIV application for both unit- and area-level models. We compare posterior medians of the predictive distribution in each area (having left that area out of model fitting) to the direct estimate in each area, respectively. 
	
	\begin{figure}[H]
		\centering
		\includegraphics[scale = 0.6]{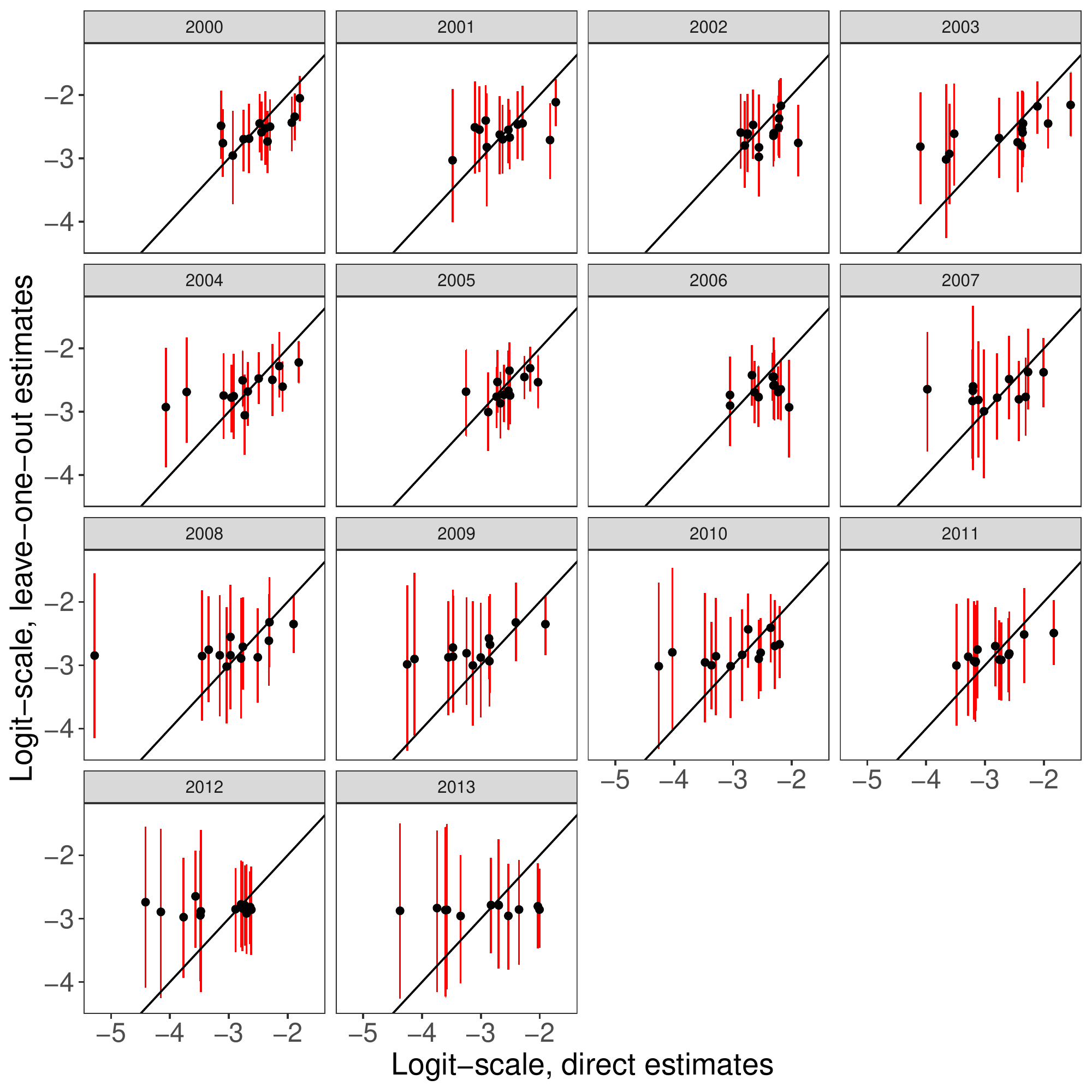}
		\caption{Scatterplot comparing leave-one-out posterior predictive estimates and direct estimates on the logit scale for the unit-level U5MR model, with 80\% confidence intervals.}
		\label{fig:loounit80_u5mr}
	\end{figure}
	
	\begin{figure}[H]
		\centering
		\includegraphics[scale = 0.6]{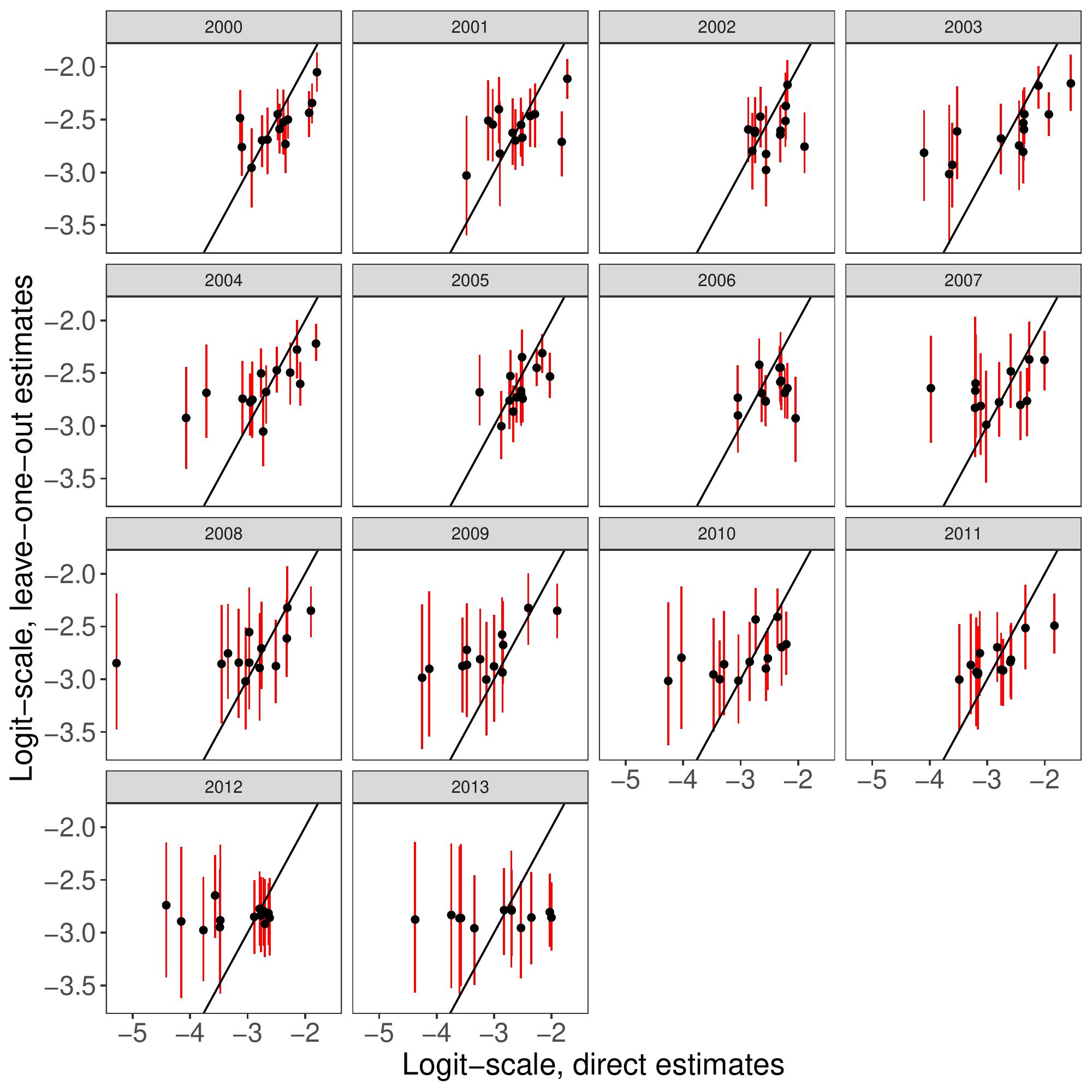}
		\caption{Scatterplot comparing leave-one-out posterior predictive estimates and direct estimates on the logit scale for the unit-level U5MR model, with 50\% confidence intervals.}
		\label{fig:loounit50_u5mr}
	\end{figure}
	
	\begin{figure}[H]
		\centering
		\includegraphics[scale = 0.6]{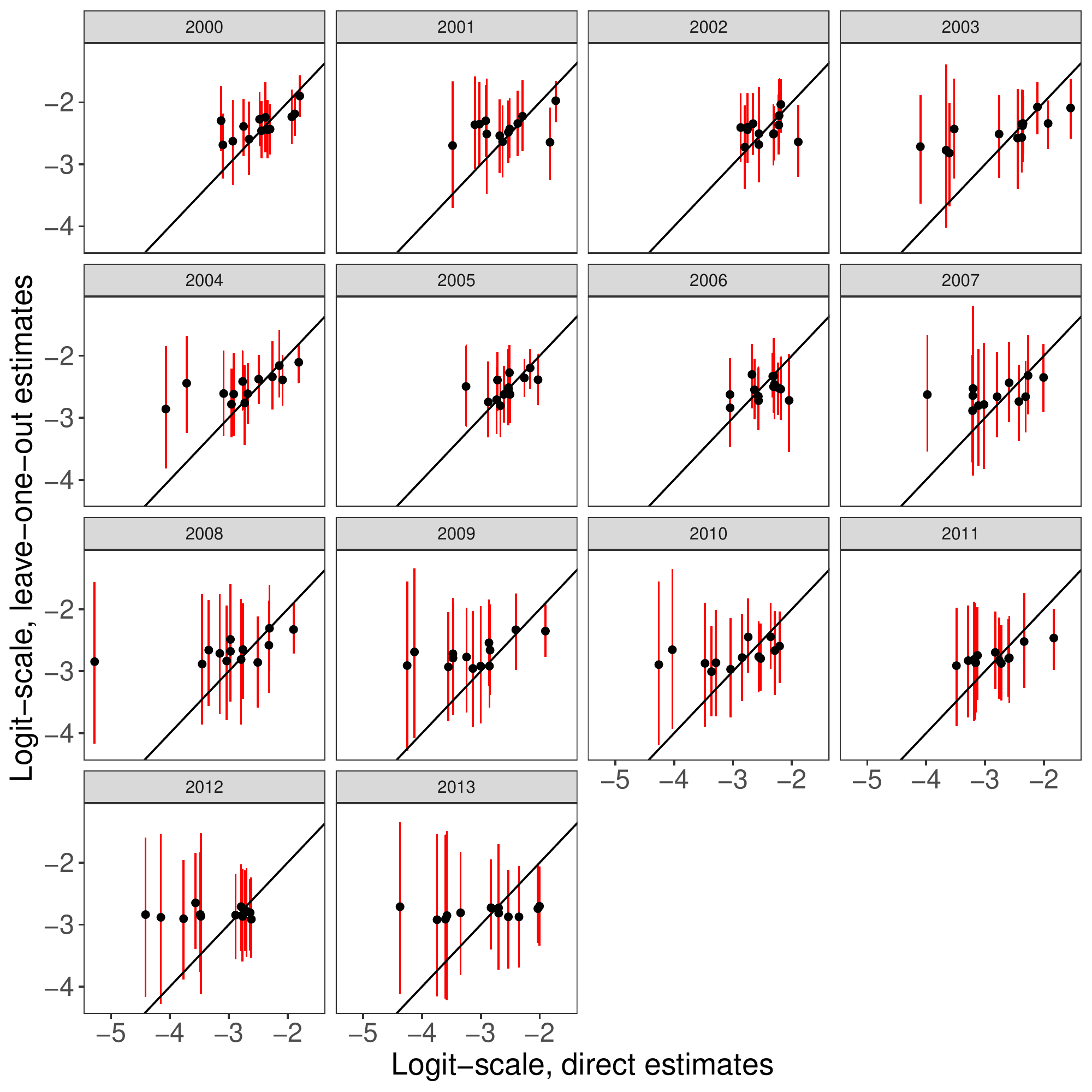}
		\caption{Scatterplot comparing leave-one-out posterior predictive estimates and direct estimates on the logit scale for the area-level U5MR model, with 80\% confidence intervals.}
		\label{fig:looarea80_u5mr}
	\end{figure}
	
	\begin{figure}[H]
		\centering
		\includegraphics[scale = 0.6]{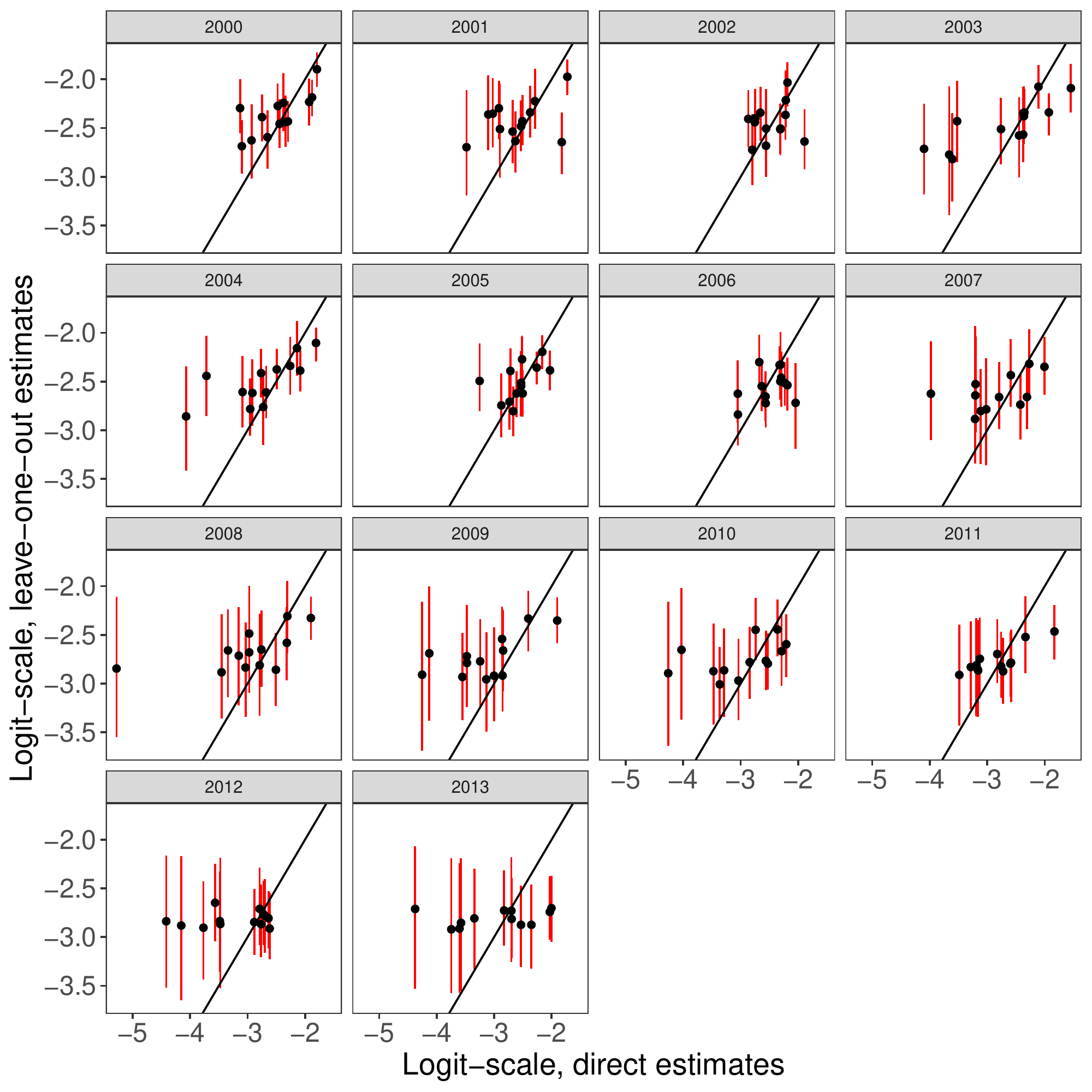}
		\caption{Scatterplot comparing leave-one-out posterior predictive estimates and direct estimates on the logit scale for the area-level U5MR model, with 50\% confidence intervals.}
		\label{fig:looarea50_u5mr}
	\end{figure}
	
	\section{Simulation}
	
	To demonstrate the improved computational speed and flexibility of our approach compared to \textcite{zhang2020fully}, we compare run-times for our approach used with INLA to that of \textcite{zhang2020fully} implemented in STAN; a probabilistic programming language that uses a variant of Hamiltonian Monte Carlo to do full Bayesian inference \autocite{carpenter2017stan}. We chose not to compare the computational speed of our approach to that of \textcite{datta2011bayesian} or the raking approach, as both of those benchmarking approaches target different benchmarked estimates than our proposed method. Note however that both the raking and benchmarked Bayes estimate approach will be the fastest approach to benchmarking in general, as they involve a very quick step to adjusting unbenchmarked draws/estimates, and do not rely on acceptance rates (as our methods do to achieve a reasonable number of effective samples) or MCMC methods.
	
	We simulate unit-level (cluster) binomial observations, using the nine provinces of South Africa as our spatial structure. For each simulation setting, binomial probabilities were given by $p_i = 0.28, 0.29, \dots, 0.35, 0.36$, with $100$ binomial trials in each cluster. Equal probability weights were given to each province. We varied: the number of samples taken in each area, $n = \{5, 10, 100, 1000\}$, the national-level benchmark, $y_2 = \{0.29, 0.3\}$, and the variance of the national-level benchmark, $\sigma^2_{y_2} = \{0.01, 0.0001\}$. For each simulation setting, we generated 10 unique datasets using the given parameters.
	
	The unbenchmarked model we fit to the generated data is the unit-level model used later in our application to South Africa, where we have binomial observations for clusters $c$ within area $i$. Let $y_{1i[c]}$ be the number of cases observed in $n_{i[c]}$ total observations in cluster $c$ within area $i$. We consider the unbenchmarked, unit-level model
	\begin{align*}
		y_{1i[c]}  \mid n_{i[c]}, \theta_{i[c]} & \sim \text{Binomial}(n_{i[c]}, \theta_{i[c]}), \\
		\eta_{i[c]} = \text{logit}(\theta_{i[c]}) & = \beta_0 + b_i + e_{i[c]}
	\end{align*} 
	where $\theta_{i[c]}$ is case prevalence in area $i$ and cluster $c$, $\beta_0$ is an intercept term, $b_i$ follows a BYM2 model \autocite{riebler2016intuitive}, denoted $b_i \sim \text{BYM2}(\tau_{b}, \phi)$, and $e_i[c] \overset{iid}{\sim} \text{N}(0, \sigma^2_e)$ is an iid cluster-level random effect. We used hyperpriors $\phi \sim \text{Beta}(0.5, 0.5)$, $\tau_b = \text{PC}(U = 1, \alpha = 0.01)$, $1/\sigma^2_e \sim \text{loggamma}(0.1, 0.1)$ in model fitting. Note that although the model we fit was not used to generate the data, as we are interested only in comparing run times from each method, and ensuring that the benchmarked distributions are the same from each method, this does not affect the validity of our simulation results. 
	
	The modified unbenchmarked model, used for the Metropolis-Hastings algorithm is the same as above with the exception of the prior for the intercept being $\pi^+(\beta_0) \overset{d}{=} \text{N}(\text{logit}(y_2), \sqrt{0.1})$, where $y_2$ is the national-level benchmark. The benchmarked model we fit using the \textcite{zhang2020fully} approach includes the additional likelihood
	$$
	y_2 \mid \boldsymbol{\theta}, \sigma^2_{y_2} \sim \text{N} \left( \bar{\boldsymbol{\theta}}, \sigma^2_{y_2} \right),
	$$
	and the proposed rejection sampler and Metropolis-Hastings algorithm approaches are carried out as described in subsections 3.4.1 and 3.4.2. Code for reproducing the simulation can be found at github.com/taylorokonek/benchmarking-paper-sim.
	
	To fairly compare run-times, we compare the time it takes to:
	\begin{enumerate}
		\item Fit the fully Benchmarked model per \textcite{zhang2020fully} in STAN, and obtain a bulk-ESS of $1000$.
		\item Fit the unbenchmarked model in INLA, draw posterior samples, and obtain $1000$ accepted samples using the rejection sampler.
		\item Fit the modified unbenchmarked model in INLA, draw posterior samples, and obtain a bulk-ESS of $1000$.
	\end{enumerate}
	For both STAN and the Metropolis-Hastings algorithm, we run four chains with $1000$ burn-in samples each, and the appropriate number of samples after to obtain the desired bulk-ESS. 
	
\begin{figure}[H]
		\centering
		\includegraphics[scale = 0.4]{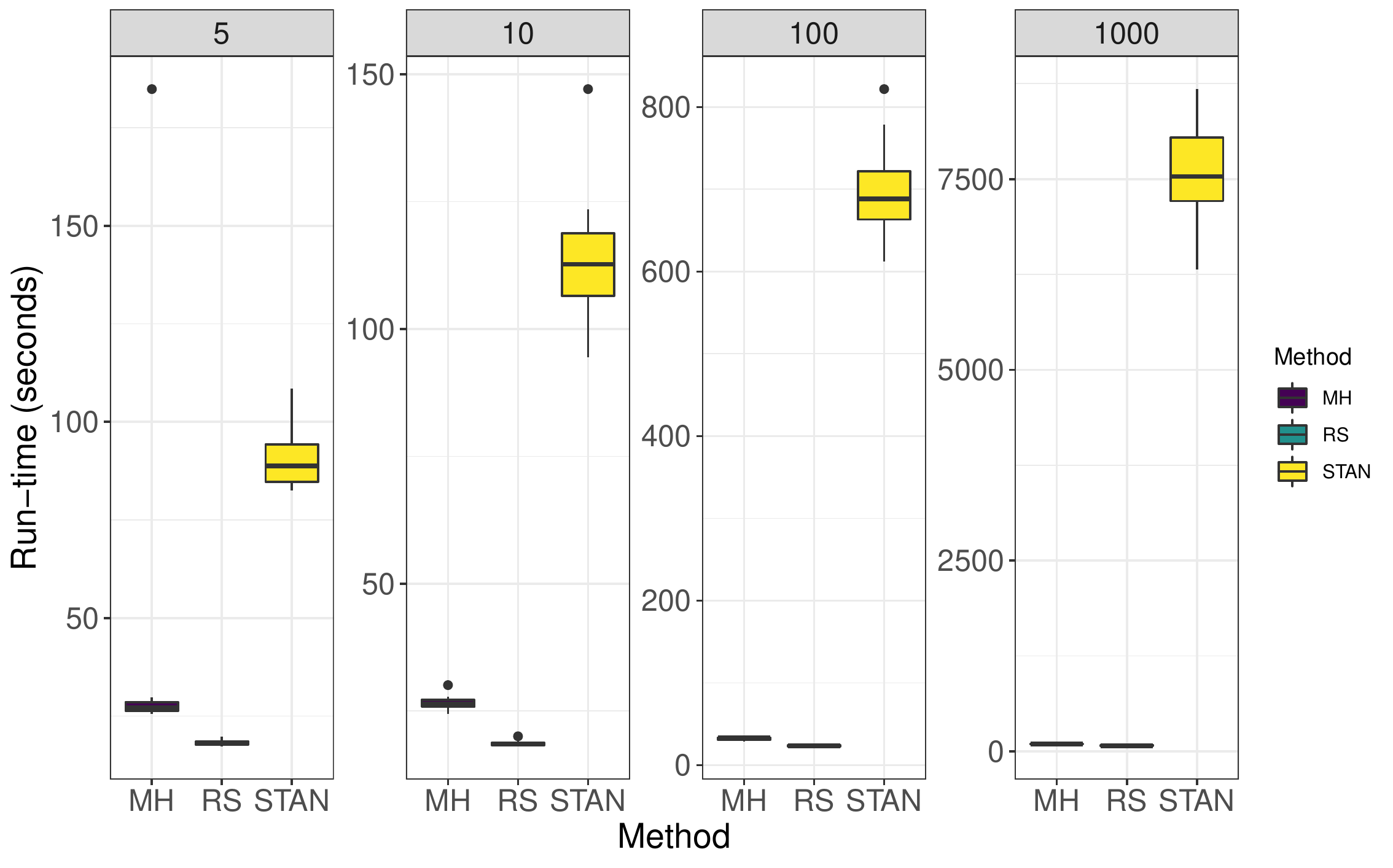}
		\caption{Comparative, total run-time (in seconds) needed to obtain $1000$ samples from the rejection sampling (RS) approach, or $1000$ bulk-ESS from the Metropolis-Hastings (MH) approach or the approach of \textcite{zhang2020fully} (STAN). Each boxplot contains $10$ observations, with data generated under the given simulation setting for $10$ different seeds. Setting: $y_2 = 0.29$, $\sigma^2_{y_2} = 0.01$}
		\label{fig:runtimes29_01}
	\end{figure}

\begin{figure}[H]
		\centering
		\includegraphics[scale = 0.4]{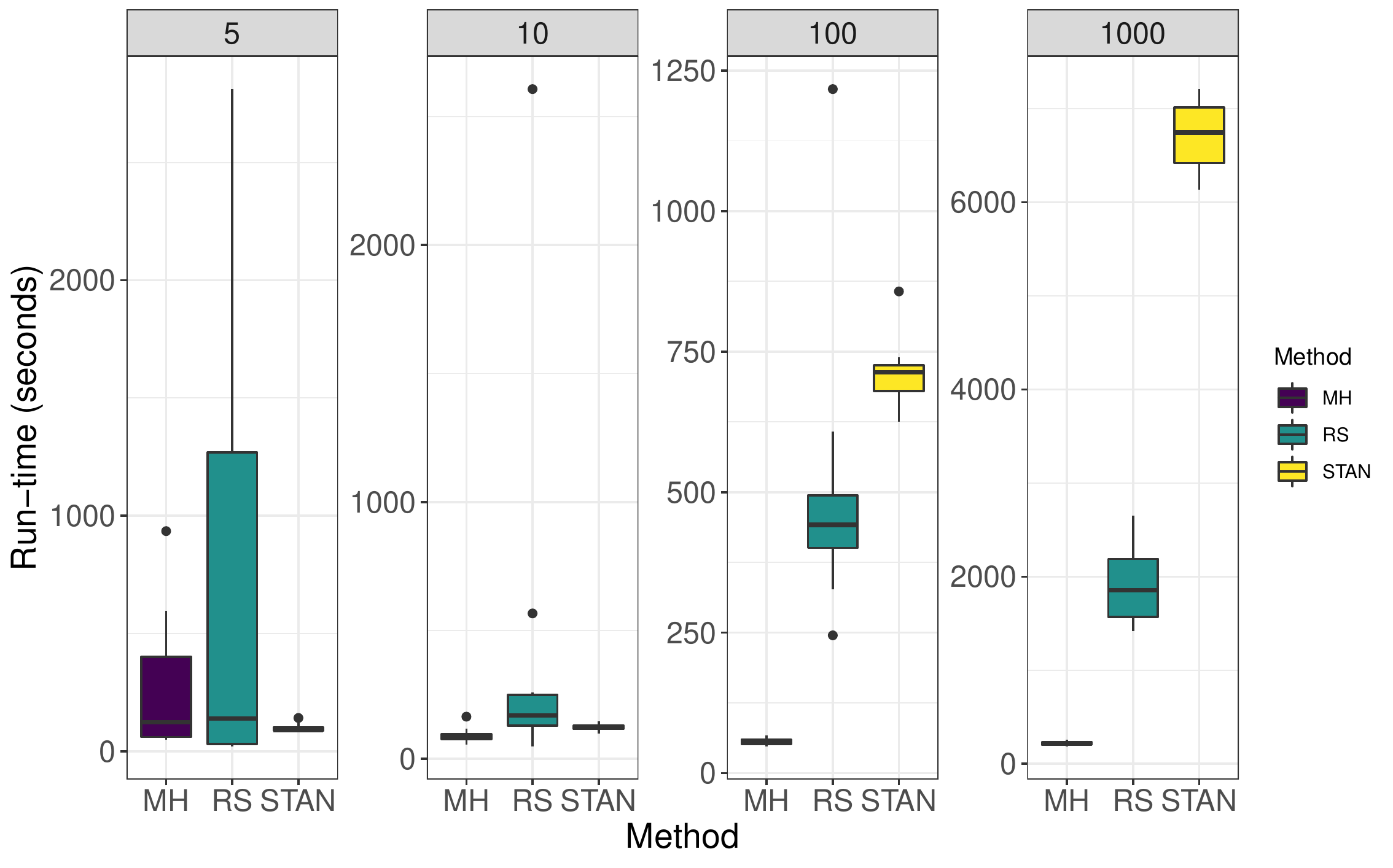}
		\caption{Comparative, total run-time (in seconds) needed to obtain $1000$ samples from the rejection sampling (RS) approach, or $1000$ bulk-ESS from the Metropolis-Hastings (MH) approach or the approach of \textcite{zhang2020fully} (STAN). Each boxplot contains $10$ observations, with data generated under the given simulation setting for $10$ different seeds. Setting: $y_2 = 0.29$, $\sigma^2_{y_2} = 0.0001$}
		\label{fig:runtimes29_0001}
	\end{figure}

	As expected, our proposed approaches outperform that of \textcite{zhang2020fully} in most simulation settings, as noted in Figures \ref{fig:runtimes29_01} and \ref{fig:runtimes29_0001}. In particular, the amount of time needed to obtain $1000$ samples from the benchmarked posterior distribution is much lower for both of our proposed approaches than that of \textcite{zhang2020fully} when the number of samples in each area is large. When the national variance is smaller, as in Figure \ref{fig:runtimes29_0001}, the \textcite{zhang2020fully} tends to outperform both the rejection sampler and MH algorithm at low sample sizes ($5$ and $10$ per area), but not at larger sample sizes. Finally, note that, in settings with smaller national variance, the Metropolis-Hastings algorithm outperforms the rejection sampler approach in terms of computational speed, while the reverse is true when national variance is large. The speed of the MH algorithm could potentially be optimized by modifying prior for the intercept, $\pi^+(\beta_0)$, and we suggest that when possible, multiple priors should be tested if the acceptance rate in the MH algorithm is lower than desired.

	\begin{figure}[H]
		\centering
		\includegraphics[scale = 0.4]{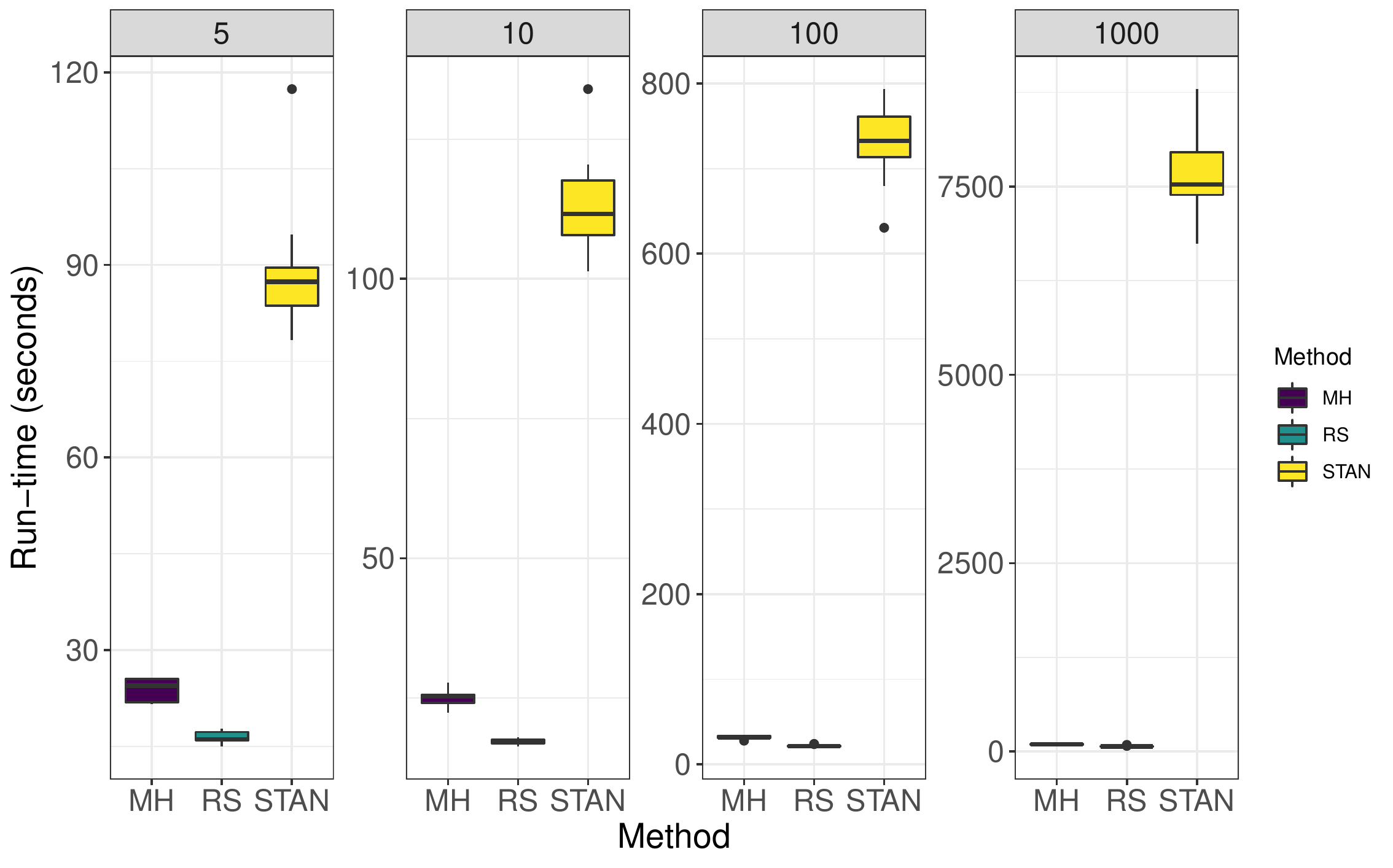}
		\caption{Comparative, total run-time (in seconds) needed to obtain $1000$ samples from the rejection sampling (RS) approach, or $1000$ bulk-ESS from the Metropolis-Hastings (MH) approach or the approach of \textcite{zhang2020fully} (STAN). Each boxplot contains $10$ observations, with data generated under the given simulation setting for $10$ different seeds. Setting: $y_2 = 0.30$, $\sigma^2_{y_2} = 0.01$}
		\label{fig:runtimes30_01}
	\end{figure}
	
	\begin{figure}[H]
		\centering
		\includegraphics[scale = 0.4]{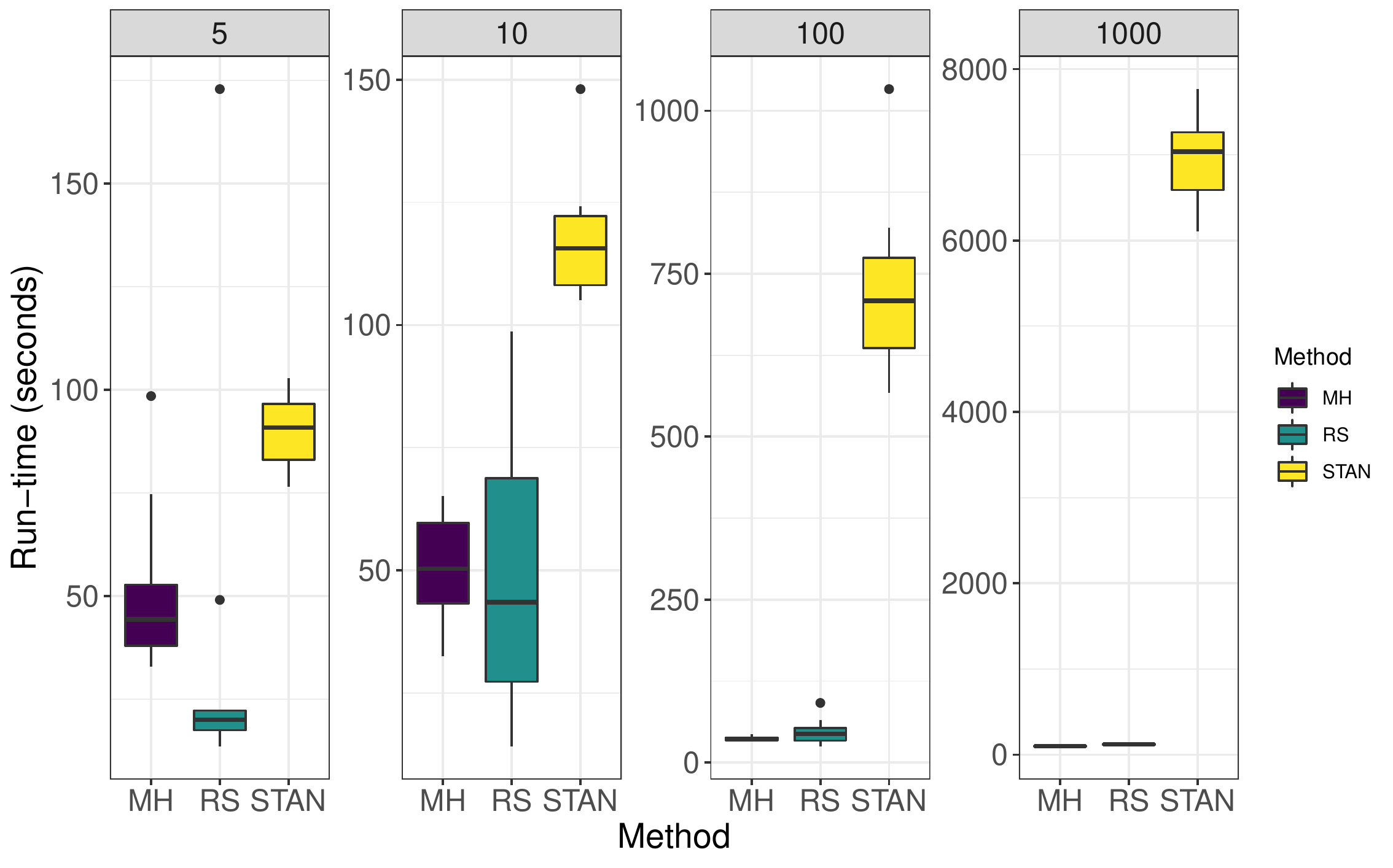}
		\caption{Comparative, total run-time (in seconds) needed to obtain $1000$ samples from the rejection sampling (RS) approach, or $1000$ bulk-ESS from the Metropolis-Hastings (MH) approach or the approach of \textcite{zhang2020fully} (STAN). Each boxplot contains $10$ observations, with data generated under the given simulation setting for $10$ different seeds. Setting: $y_2 = 0.30$, $\sigma^2_{y_2} = 0.0001$}
		\label{fig:runtimes30_0001}
	\end{figure}
	
	{\raggedright \printbibliography}